\documentclass{lmcs} %%% last changed 2014-08-20
\pdfoutput=1
\usepackage[utf8]{inputenc}

\usepackage{lastpage}
\lmcsdoi{21}{4}{6}
\lmcsheading{}{\pageref{LastPage}}{}{}%
{Jun.~06,~2024}{Oct.~08,~2025}{}


\usepackage{hyperref}
\theoremstyle{plain} %\crefname{satz}{Satz}{S\"atze}
\usepackage{tikz}
\usetikzlibrary{automata}
\usetikzlibrary{positioning,arrows,fit,calc}
\usetikzlibrary{decorations.pathreplacing}
\usetikzlibrary{matrix}

\tikzset{>=latex, 
	point/.style = {circle,draw,thick,minimum size=2mm,inner sep=0pt},
	point1/.style = {circle,draw,thick,minimum size=6mm,inner sep=0pt},
	hm/.style = {dotted,semithick},
	role/.style = {thick},
	tree/.style = {rounded corners=10pt, dashed, fill opacity=0.5, fill=nullscolour},
	wiggly/.style={thick,%decorate,decoration={snake,amplitude=0.3mm,segment length=2mm,post length=1mm}
	},
	query/.style={thick},
}

\pgfdeclarelayer{background}
\pgfdeclarelayer{foreground}
\pgfsetlayers{background,main,foreground}

\newcommand{\avec}[1]{\boldsymbol{#1}}

\newcommand{\md}{\textit{md}}

\newcommand{\diag}{\fform{e}}

\newcommand{\LTL}{\textsl{LTL}}

\newcommand{\sub}{\textit{sub}}

\newcommand{\sig}{\mathit{sig}}

\newcommand{\tiling}{\mathfrak T}
\newcommand{\bbvarup}{\Zvar^\uparrow}
\newcommand{\bbvarright}{\Zvar^\to}
\newcommand{\Dht}[1]{\Diamond_h{#1}}
\newcommand{\Dvt}[1]{\Diamond_v{#1}}
\newcommand{\Bht}[1]{\Box_h{#1}}
\newcommand{\Bvt}[1]{\Box_v{#1}}
\newcommand{\ssvar}{\fform{s}}

\newcommand{\FOM}{\ensuremath{\smash{\mathsf{FOM^1}}}}
\newcommand{\FO}{\ensuremath{\mathsf{FO}}}
\newcommand{\FOO}{\ensuremath{\mathsf{FO^1}}}
\newcommand{\FOT}{\ensuremath{\mathsf{FO^2}}}
\newcommand{\SF}{\ensuremath{\mathsf{S5}}}
\newcommand{\K}{\ensuremath{\mathsf{K}}}
\newcommand{\QSF}{\ensuremath{\smash{\mathsf{Q^1S5}}}}
\newcommand{\QK}{\ensuremath{\smash{\mathsf{Q^1K}}}}
\newcommand{\FOMK}{\ensuremath{\mathsf{Q^1K}}}
\newcommand{\FOMS}{\ensuremath{\mathsf{Q^1S5}}}

\newcommand{\ALCS}{\ensuremath{\mathsf{S5}_{\mathcal{ALC}^u}}}
\newcommand{\ALC}{\ensuremath{\mathcal{ALC}}}

\newcommand{\BS}{$\mathsf{S5}$-\text{bisimulation}}

\newcommand{\wt}{\mathsf{wt}}
\newcommand{\dt}{\mathsf{dt}}
\newcommand{\ft}{\mathsf{ft}}

\newcommand{\wm}{\mathsf{wm}}
\newcommand{\dm}{\mathsf{dm}}

\newcommand{\iwp}{\mathsf{wp}_i}
\newcommand{\idp}{\mathsf{dp}_i}

\newcommand{\owp}{\mathsf{wp}_1}
\newcommand{\twp}{\mathsf{wp}_2}
\newcommand{\odp}{\mathsf{dp}_1}
\newcommand{\tdp}{\mathsf{dp}_2}

\newcommand{\pet}{\ft}
\newcommand{\pam}{\mathsf{fm}}
\newcommand{\ipp}{\mathsf{fp}_i}
\newcommand{\et}{\ft}

\newcommand{\bs}{\boldsymbol{\beta}}

\newcommand{\fform}[1]{\boldsymbol{#1}}
\newcommand{\pvar}{\fform{p}}
\newcommand{\qvar}{\fform{q}}
\newcommand{\svar}{\fform{s}}
\newcommand{\avar}{\fform{a}}
\newcommand{\varset}{\mathcal{P}}

\newcommand{\gvar}{\fform{g}}
\newcommand{\bvar}{\fform{b}}
\newcommand{\tvar}{\fform{t}}
\newcommand{\xvar}{\fform{x}}
\newcommand{\yvar}{\fform{y}}
\newcommand{\ovar}{\fform{o}}
\newcommand{\hvar}{\fform{h}}
\newcommand{\vvar}{\fform{v}}
\newcommand{\Zvar}{\fform{z}}

\newcommand{\blank}{\flat}
\newcommand{\cell}{\mathsf{cells}}
\newcommand{\cA}{A}
\newcommand{\cR}{B}
\newcommand{\cU}{U}
\newcommand{\cV}{V}
\newcommand{\cvf}[3]{[{#1}\mathop{#2}{#3}]}
\newcommand{\inpw}{\overline{a}}
\newcommand{\inpww}{a}
\newcommand{\nextR}{\fform{n}^\cR}

\newcommand{\nextU}{\fform{n}^\cU}
\newcommand{\nextV}{\scc^\cV}
\newcommand{\cinit}{\config_{\textit{init}}}
\newcommand{\Bv}{\forall}
\newcommand{\Bh}{\Box}
\newcommand{\Dv}{\exists}
\newcommand{\Dh}{\Diamond}
\newcommand{\config}{c}
\newcommand{\start}{\fform{r}}
\newcommand{\nextt}[1]{\mathsf{next}({#1})}
\newcommand{\grid}{\mathsf{equ}}
\newcommand{\scc}{\mathsf{succ}}
\newcommand{\tree}{\fform{t}}
\newcommand{\qA}{\qvar_\forall}
\newcommand{\qE}{\qvar_\exists}
\newcommand{\treemod}{\mathfrak T}
\newcommand{\treedom}{T}
\newcommand{\treerel}{S}
\newcommand{\treet}{t}

\newcommand{\ALCIOS}{\ensuremath{\mathsf{S5}_{\mathcal{ALC}^u}}}
\newcommand{\ALCIOSnou}{\ensuremath{\mathsf{S5}_{\mathcal{ALC}}}}
\newcommand{\ALCIO}{\ensuremath{\mathcal{ALC}}}

\begin{document}
		
\title[Interpolants and Definitions in First-Order Modal Logic]{Deciding the Existence of Interpolants and Definitions in First-Order Modal Logic}
%\texorpdfstring{\MakeLowercase{\texttt{lmcs.cls}}}{lmcs.cls}\rsuper*\\Version of 		%  2022-04-01}

\titlecomment{{\lsuper*}The extended abstract~\cite{DBLP:conf/kr/KuruczWZ23} of this article was presented at the 20th International Conference on Principles of Knowledge Representation and Reasoning (KR 2023).}

%	\thanks{thanks, optional.}	%optional
	
	% affiliations are numbered automatically with a, b, c (see below)
	% use the optional argument to indicate the affiliation(s) of each author
	% omit the argument if there is only one author, or only one affiliation
	\author[A.~Kurucz]{Agi Kurucz\lmcsorcid{0000-0002-6233-6277}}[a]
	\author[F.~Wolter]{Frank Wolter\lmcsorcid{0000-0002-4470-606X}}[b]
	\author[M.~Zakharyaschev]{Michael Zakharyaschev\lmcsorcid{0000-0002-2210-5183}}[c]
	
	% affiliation 1 (automatically numbered a)
	\address{Department of Informatics, King's College London, U.K.}
	\email{agi.kurucz@kcl.ac.uk}	
	
	%optional
	% write emails for all authors having that affiliation
	%\email{name1@email1, name2@email1, name3@email1}  %optional
	
	% affiliation 2 (automatically numbered b)
	\address{Department of Computer Science, University of Liverpool, U.K.}  %optional
	\email{wolter@liverpool.ac.uk}

	\address{School of Computing and Mathematical Sciences, Birkbeck, University of London, U.K.}
	\email{m.zakharyaschev@bbk.ac.uk}
		
	%% etc.
	
	%% required for running head on odd and even pages, use suitable
	%% abbreviations in case of long titles and many authors:
	
	%%%%%%%%%%%%%%%%%%%%%%%%%%%%%%%%%%%%%%%%%%%%%%%%%%%%%%%%%%%%%%%%%%%%%%%%%%%
	
	%% the abstract has to PRECEDE the command \maketitle:
	%% be sure not to issue the \maketitle command twice!
	
\begin{abstract}
None of the first-order modal logics between $\mathsf{K}$ and $\SF{}$ under the constant domain semantics enjoys Craig interpolation or projective Beth definability, even in the language restricted to a single individual variable. It follows that deciding the existence of a Craig interpolant for a given implication or of an explicit definition for a given predicate cannot be directly reduced to deciding the validity of an implication, as in classical first-order and many other logics. Our concern here is the decidability and computational complexity of the interpolant and definition  existence problems. We first consider two decidable fragments of first-order modal logic \SF{}: the one-variable fragment \FOMS{} and its extension \ALCIOS{} that combines \SF{} and the description logic \ALCIO{} with the universal role. We prove that interpolant and definition existence in $\QSF$ and \ALCIOS{} is decidable in \textsc{coN2ExpTime}, being 2\textsc{ExpTime}-hard, while uniform interpolant existence is undecidable. These results transfer to the two-variable fragment $\FOT$ of classical first-order logic without equality. We also show that interpolant and definition existence in the one-variable fragment \FOMK{} of first-order modal logic $\mathsf{K}$ is non-elementary decidable, while uniform interpolant existence is again undecidable.  
%{\color{red} No first-order modal logic with constant domain between $\mathsf{K}$ and $\SF{}$ enjoys Craig interpolation or projective Beth definability. This also applies to any fragment containing the one-variable fragment. It follows that neither the existence of a Craig interpolant for an input implication nor of an explicit definition for a predicate can be directly reduced to validity. Our concern here is the decidability and computational complexity of the problem whether (uniform) interpolants and definitions exist. We first consider two decidable fragments of first-order modal logic \SF{}: the one-variable fragment \FOMS{} and its extension \ALCIOS{} that combines \SF{} and the description logic \ALCIO{} with the universal role and prove that interpolant and definition existence in $\QSF$ and \ALCIOS{} are decidable in \textsc{coN2ExpTime}, being 2\textsc{ExpTime}-hard, while uniform interpolant existence is undecidable. We also show that interpolant and definition existence in the one-variable fragment \FOMK{} of first-order modal logic $\mathsf{K}$ is nonelementary decidable, while uniform interpolant existence is undecidable.} 
\end{abstract}
	
\maketitle

%*******************

\section{Introduction}\label{intro}

First-order modal logics and their fragments are well-established formalisms in computational logic. For many decades, they have been used, e.g., as first-order temporal logics in program verification, policy monitoring, and databases~\cite{DBLP:series/eatcs/Kroger87,DBLP:conf/cav/BasinKM10,DBLP:conf/rv/HavelundP18,DBLP:reference/db/ChomickiT18a}, as epistemic, temporal, and standpoint description logics \cite{DBLP:journals/ai/DoniniLNNS98,DBLP:conf/time/LutzWZ08,DBLP:conf/time/ArtaleKKRWZ17,DBLP:conf/semweb/AlvarezRS22}, as spatio-temporal logics \cite{DBLP:reference/spatial/KontchakovKWZ07}, and as logics of knowledge and belief \cite{DBLP:journals/ai/BelardinelliL09,DBLP:journals/corr/Wang17d,liu_et_al:LIPIcs.MFCS.2022.70,DBLP:journals/apal/WangWS22}. By now, significant progress has been made in understanding entailment in these `two-dimensional' logics, in particular its computational complexity; see, e.g.,~\cite{GabEtAl03,DBLP:journals/tocl/HampsonK15,DBLP:journals/iandc/LiuPRW23,DBLP:journals/tocl/ArtaleMO24} and references therein. However, very little is known about the decidability and complexity of fundamental algorithmic problems that can go beyond entailment. For example, the following reasoning tasks have been used in different areas of computer science:
\begin{description} 
\item[\emph{definition existence}] Given a formula $\varphi$, a predicate $P$, and a signature $\sigma$, does there exist an \emph{explicit definition} $\chi$ of $P$ modulo $\varphi$ in $\sigma$ in the sense that \mbox{$\varphi\models \forall \avec{x}\, (P(\avec{x}) \leftrightarrow \chi(\avec{x}))$} and $\sig(\chi)\subseteq \sigma$, where $\sig(\chi)$ comprises the non-logical symbols in $\chi$. Such definitions can, for instance, support query evaluation in databases~\cite{DBLP:series/synthesis/2011Toman,DBLP:series/synthesis/2016Benedikt}.

\item[\emph{interpolant existence}] Given formulas $\varphi$ and $\psi$, does there exist a \emph{Craig interpolant} $\chi$ for $\varphi$ and $\psi$ in the sense 
that $\models \varphi \rightarrow \chi$, $\models \chi \rightarrow \psi$, and
\mbox{$\sig(\chi)\subseteq \sig(\varphi) \cap \sig(\psi)$}. Craig interpolants are applied, for instance, in model checking and concept learning~\cite{DBLP:reference/mc/McMillan18,DBLP:journals/ai/JungLPW22}.

\item[\emph{conservative extensions/uniform interpolant verification}] Given formulas $\varphi$ and $\chi$ such that $\models \varphi \rightarrow \chi$ and $\sigma=\sig(\chi)\subseteq \sig(\varphi)$, is $\chi$ a \emph{$\sigma$-uniform interpolant} of $\varphi$ in the sense that $\chi$ is an interpolant for $\varphi$ and $\psi$ whenever $\models \varphi\rightarrow \psi$ and $\sig(\varphi)\cap \sig(\psi) \subseteq \sigma$? In this case, $\varphi$ is known as a \emph{conservative extension} of $\chi$. These notions are used, for instance, for modular knowledge base design and modularisation~\cite{DBLP:conf/rweb/BotoevaKLRWZ16}.  

\item[\emph{uniform interpolant existence}] Given a formula $\varphi$ and a signature $\sigma$, does there exist a $\sigma$-uniform interpolant 
of $\varphi$? Uniform interpolants are, for instance, a mechanism for forgetting symbols from a knowledge base~\cite{DBLP:journals/ki/EiterK19,DBLP:journals/tplp/GoncalvesKL23,DBLP:series/lncs/KonevLWW09,DBLP:journals/ki/Koopmann20}.
\end{description} 
%These and related problems have been studied extensively for many formalisms, including (quantified) propositional logic~\cite{DBLP:journals/ai/LangM08,DBLP:reference/mc/McMillan18,DBLP:conf/cav/Slivovsky20}, modal logic~\cite{MGabbay2005-MGAIAD,DBLP:journals/apal/Kuznets18}, fragments of first-order logic~\cite{DBLP:journals/tocl/BenediktCB16,DBLP:conf/icalp/JungLM0W17}, answer set programming~\cite{DBLP:journals/ki/EiterK19,DBLP:journals/tplp/GoncalvesKL23}, and description logics~\cite{DBLP:series/lncs/KonevLWW09,TenEtAl13,DBLP:conf/rweb/BotoevaKLRWZ16}, 
%
Investigating these problems for first-order modal logics (FOMLs) poses particular challenges. In contrast to many other logical formalisms, FOMLs typically do not enjoy the Craig interpolation property (CIP) as $\models \varphi \rightarrow \psi$ does not necessarily entail the existence of an interpolant $\chi$ for $\varphi$ and $\psi$. Nor do they enjoy the projective Beth definability property (BDP) according to which implicit definability of a predicate in a given signature  implies its explicit definability as required in definition existence. For logics with the CIP, 
the interpolant existence problem trivially reduces to checking validity of implications. Similarly, for logics with the BDP, checking explicit definability reduces to checking the validity of implications representing implicit definability. These trivial reductions of existence problems to validity problems do not work for logics without the CIP or BDP. 
%Forgetting and conservative extensions in FOMLs become dependent on predicates %that do not occur in the original theory. 
Fine~\cite{DBLP:journals/jsyml/Fine79} showed that no FOML with constant domain models (a standard assumption) between the first-order modal logic $\mathsf{K}$ of all Kripke frames and the first-order modal logic $\mathsf{S5}$ of all universal Kripke frames enjoys the CIP or BDP.

\begin{exa}[based on~\cite{DBLP:journals/jsyml/Fine79}]\label{Ex:Fine}
Interpreting $\Box$ as the $\mathsf{S5}$-modality `always'\!, let $T$ consist of the following axioms, where $\mathsf{rep}$ stands for the proposition `replaceable':
\begin{align*}
\mathsf{rep} &\rightarrow \Diamond \forall x\, \big(\mathsf{inPower}(x) \rightarrow \Box (\mathsf{rep} \rightarrow \neg \mathsf{inPower}(x)) \big),\\
\neg \mathsf{rep} &\rightarrow \Box \exists x\, \big( \mathsf{inPower}(x) \wedge \Box (\neg \mathsf{rep} \rightarrow \mathsf{inPower}(x)) \big).
\end{align*}
Then $\mathsf{rep}$ is true at a world $w$ satisfying $T$ iff there is a world $w'$ where all those who were in power at $w$ lose it. It follows that $\mathsf{rep}$ is implicitly defined via $\mathsf{inPower}$. However, we shall see in Example~\ref{ex-fine1} that there is no explicit definition of $\mathsf{rep}$ 
via $\mathsf{inPower}$ in FOML.  \hfill $\dashv$
\end{exa}

Fine's example shows that the CIP and BDP fail already in typical decidable fragments of FOML lying between the one-variable fragment and full FOML.  Because of their wide use, `repairing' the CIP and BDP has become a major research challenge. For instance, it is shown in~\cite{DBLP:journals/jsyml/Fitting02,DBLP:journals/apal/ArecesBM03} that by adding second-order quantifiers or the machinery of hybrid logic constructors to FOML, one obtains logics with the CIP and BDP. The price, however, is that these extensions are undecidable even if applied to decidable fragments of FOML.  
%and so are not directly relevant for KR applications.        

In this article, we take a fundamentally different, non-uniform  approach. Instead of repairing the CIP and BDP by enriching the language, we stay within its original boundaries and explore the possibility of checking the existence of interpolants and definitions even though the reduction to validity via the CIP and BDP is blocked. 

%We\nz{omit sentence?} conjecture that, in real-world applications, %interpolants and definitions often do exist, so the failure of the %CIP and BDP will have a limited effect on the users.

We first focus on two decidable fragments of first-order \SF: its one-variable fragment \FOMS{} illustrated in Example~\ref{Ex:Fine} and 
\ALCIOS{}, the FOML obtained by combining \SF{} and the description logic (DL) $\mathcal{ALC}^{u}$, which extends  the basic DL $\mathcal{ALC}$---a notational variant of multimodal logic $\K$---with the universal role. In $\ALCIOS{}$, we admit the application of modal operators to concepts but not to roles, and so consider a typical monodic fragment of FOML, in which modal operators are only applied to formulas with at most one free variable~\cite{DBLP:journals/apal/HodkinsonWZ00,DBLP:journals/jsyml/WolterZ01}. \FOMS{} is a fragment of \ALCIOS{}, and satisfiability is \textsc{NExpTime}-complete for both languages~\cite{GabEtAl03}.

Our first main result is:

\begin{thm}\label{thm1intro}
The interpolant and definition existence problems in \FOMS{} and \ALCS{} are decidable in \textsc{coN2ExpTime}, being $2$\textsc{ExpTime}-hard.
\end{thm}

Thus, interpolant and definition existence is still decidable but harder than satisfiability by about one exponential. The proof is based on `component-wise' bisimulations that replace standard FOML bisimulations in our characterisation of interpolant and definition existence. For the upper bound, we show that whenever there are component-wise bisimilar models 
witnessing non-existence of interpolants/definitions, then there are component-wise bisimilar models of at most double-exponential size. The proof is inspired by the recent upper bound proofs of interpolant existence in the two-variable first-order logic \FOT{}~\cite{DBLP:conf/lics/JungW21} but requires a very different notion of type reflecting the two-dimensional flavour of first-order modal logic. 
%For didactic\nb{??} purposes we first give the proof for \FOMS{} and then extend it to \ALCS{}. 
The lower bound proof combines the interpolation counterexample  of~\cite{DBLP:journals/ndjfl/MarxA98}, the exponential grid generation from~\cite{DBLP:conf/time/HodkinsonKKWZ03,DBLP:journals/tocl/GollerJL15}, and the representation of exponential-space bounded alternating Turing machines from~\cite{DBLP:conf/lics/JungW21}. 

Our result can be used to solve an open problem for $\FOT{}$ without equality. The known 2\textsc{ExpTime}-lower bound proof for interpolant existence in $\FOT{}$ uses equality in a critical way. Utilising a close link
between \FOMS{} and equality-free \FOT{}, we obtain the lower bound of the following theorem as a corollary to Theorem~\ref{thm1intro}, answering an open question of~\cite{DBLP:conf/lics/JungW21}
(the upper bound is a straightforward consequence of the proof in~\cite{DBLP:conf/lics/JungW21}):

\begin{thm}\label{thm2intro}
The interpolant and definition existence problems in equality-free \FOT{} 
are decidable in \textsc{coN2\-Exp\-Time}, being $2$\textsc{ExpTime}-hard.
\end{thm}

We then consider uniform interpolant existence and conservative extension and show that they behave rather differently to interpolation existence:

\begin{thm}\label{thm3intro}
The	uniform interpolant existence and conservative extension problems in \FOMS{} and \ALCS{} are both  undecidable.
\end{thm}

The proof extends a reduction proving undecidability of conservative extensions for \FOT{} (with and without equality) from~\cite{DBLP:conf/icalp/JungLM0W17}. As a corollary of our proof, we obtain that uniform interpolant existence in \FOT{} (with and without equality) is 
undecidable, settling an open problem from~\cite{DBLP:conf/icalp/JungLM0W17}.

Finally, we consider the one-variable fragment \FOMK{} of first-order modal logic $\mathsf{K}$ determined by the class of all Kripke frames and prove the following:

\begin{thm}\label{thm4intro}
$(i)$ The interpolant and definition existence problems in \FOMK{} are decidable in non-elementary time. 
	
$(ii)$ The uniform interpolant existence and conservative extensions problems in \QK{} are both undecidable.
\end{thm}

	The non-elementary upper bound is established using the fact that \FOMK{} has finitely many non-equivalent formulas of bounded modal depth. The undecidability result is proved by adapting the undecidability proof for \FOMS{}.
%\nb{not to forget:
%Undecidability: relationship to French: bisimulation quantifiers.} 

%********

\paragraph{Related work on interpolant existence.}
%Interpolation has been applied in many areas, including verification~\cite{DBLP:conf/cav/McMillan03}, databases~\cite{DBLP:series/synthesis/2011Toman,DBLP:series/synthesis/2016Benedikt}, and philosophy~\cite{?}.\nz{??}
%
The non-uniform approach to Craig interpolants and explicit definitions has only very recently been studied for a small number of modal and description logics, and also related decidable fragments of first-order logic such as the guarded and two-variable fragment~\cite{DBLP:conf/lics/JungW21}, classical description and modal logics with nominals~\cite{DBLP:journals/tocl/ArtaleJMOW23},
various Horn logics~\cite{DBLP:series/synthesis/2016Benedikt,DBLP:journals/corr/abs-2202-07186}, modal logics of linear frames~\cite{DBLP:journals/corr/abs-2312-05929}, and modal logics with the derivative operator~\cite{DBLP:journals/corr/abs-2403-11255}. For all these logics either the complexity of interpolant and explicit definition existence goes up by one exponential compared to validity or stays the same as validity even without the Craig interpolation property. An example of a decidable fragment of FO with undecidable interpolant and explicit definition existence is given in~\cite{DL24:undecidability}. 
Some decidability results on separating disjoint regular languages using FO-definable languages~\cite{henkell1,henkell2,DBLP:journals/corr/PlaceZ14} can be interpreted as results about interpolant existence in linear temporal logic \LTL{}. An overview of the non-uniform approach to Craig interpolants and the relationship to separation of formal languages is given in~\cite{chapter:separation}.

The non-uniform investigation of \emph{uniform} interpolants started much earlier in description logic with complexity results for uniform interpolant existence~\cite{DBLP:conf/kr/LutzSW12,DBLP:conf/ijcai/LutzW11} and size upper bounds for uniform interpolants if they exist~\cite{DBLP:journals/ai/NikitinaR14}. The practical computation of uniform interpolants is an active research area for many years~\cite{DBLP:conf/ijcai/KonevWW09,DBLP:conf/aaai/KoopmannS15,DBLP:conf/ijcai/ZhaoS16}; see~\cite{DBLP:conf/dlog/ZhaoFADS18,DBLP:journals/ki/Koopmann20} for recent system descriptions.
%The existence problem for uniform interpolants in logics without the uniform interpolation property has been an active research area for many years. 
%Recall that uniform interpolants generalize Craig interpolants in the sense that
%a uniform interpolant is an interpolant for a fixed $\varphi$ and all $\psi$ which are entailed by $\varphi$ and share with $\varphi$ a fixed set of symbols.  First-order logic enjoys the CIP but not the UIP, but propositional intuitionistic logic, local modal logic, and the modal mu-calculus all enjoy the UIP~\cite{DBLP:journals/jsyml/Pitts92,visser1996uniform,DBLP:conf/aiml/DAgostinoH96}, see \cite{kowalski2019uniform,iemhoff2019uniform} for more recent investigations. In description logic, uniform interpolants of ontologies (extending what we call CI-interpolants in this article) are of particular importance but do not always exist for any standard basic description logic. 

%\cite{DBLP:conf/dlog/CalvaneseGLLR06}, \cite{DBLP:books/daglib/0005127}
%among other areas, verification~\cite{DBLP:conf/cav/McMillan03}, databases~\cite{DBLP:series/synthesis/2011Toman,DBLP:series/synthesis/2016Benedikt}, and philosophy~\cite{?}, there have been various attempts to `repair' CIP and BDP in FOML.

%***********

\paragraph{Structure.}
The article is organised as follows. Section~\ref{Sec:prelims} reminds the reader of the syntax and semantics of the one-variable fragments $\FOMK$ and $\FOMS$ of two basic first-order modal logics $\K$ and $\SF$, as well as defines bisimulations between their models.  Section~\ref{main-notions} introduces and illustrates the main notions we are concerned with here---Craig and uniform interpolants, explicit definitions, conservative extensions---and provides their model-theoretic characterisations. Section~\ref{sec:s5xs5} establishes the upper and lower bounds for the interpolant and explicit definition existence problems in $\QSF$ and Section~\ref{sec: CEP-S5} proves that the conservative extension and uniform interpolant existence
problems in $\QSF$ are undecidable. Section~\ref{sec:ALCIO} extends the results of the previous two sections to the modal description logic $\ALCIOS$. Section~\ref{sec:K} establishes decidability of the interpolant and definition existence in $\FOMK$ and undecidability of conservative extension and uniform interpolant existence.  Finally, Section~\ref{outlook} discusses further research and some open problems  that arise from this work. 

Some technical details are omitted from the main part of the article and can be found in the appendix. Appendix~\ref{s:FO} discusses the connections between \FOMS{} and \FOT{}, and uses them to prove the lower bound of Theorem~\ref{thm2intro}. 
Appendix~\ref{s:sec6proofs} gives polynomial-time reductions of various   interpolant existence problems modulo an ontology to the IEP.

%***********

\section{Preliminaries}\label{Sec:prelims}

\paragraph{Logics.} 
The formulas of the \emph{one-variable fragment \FOM{} of first-order modal logic} are built from unary predicate symbols $\pvar\in\varset$ in a countably-infinite set $\varset$  
%$i < \omega$, 
and a single individual variable $x$ using $\top$, $\neg$, $\land$, $\exists x$, and the possibility operator $\Diamond$. The  other Booleans, $\forall x$, and the necessity operator $\Box$ are defined as standard abbreviations.
A \emph{signature} is any finite set $\sigma\subseteq\varset$; 
the signature $\sig(\varphi)$ of a formula $\varphi$ comprises the predicate symbols in $\varphi$.  If $\sig(\varphi) \subseteq \sigma$, we call $\varphi$ a \emph{$\sigma$-formula}. By $\sub(\varphi)$ we denote the closure under single negation of the set of subformulas of $\varphi$, and by $|\varphi|$ the cardinality of $\sub(\varphi)$.

We interpret \FOM-formulas in (\emph{Kripke}) \emph{models} with \emph{constant domains} of the form $\mathfrak M = (W,R,D,I)$, where $W \ne \emptyset$ is a set of \emph{worlds}, $R\subseteq W \times W$ an \emph{accessibility relation} on $W$, $D \ne \emptyset$ an  (FO-)\emph{domain} of $\mathfrak{M}$, and $I(w)$ is an \emph{interpretation} of the $\pvar\in\varset$ over $D$, for each $w\in W$, that is, $\smash{\pvar^{I(w)}\subseteq D}$. The \emph{truth-relation} $\mathfrak M, w, d \models \varphi$, for any $w \in W$, $d \in D$ and \FOM-formula $\varphi$, is defined inductively by taking
\begin{itemize}
\item[--] $\mathfrak M, w, d \models \top$, 

\item[--] $\mathfrak M, w, d \models \pvar(x)$ iff $\smash{d \in \pvar^{I(w)}}$, for all $\pvar\in\varset$, 

\item[--] $\mathfrak M, w, d \models \exists x\, \varphi$ iff there is $d' \in D$ with $\mathfrak M, w, d' \models \varphi$,

\item[--] $\mathfrak M, w, d \models \Diamond \varphi$ iff there is $w' \in W$ with 
%$R(w,w')$ 
$(w,w')\in R$ 
and $\mathfrak M, w', d \models \varphi$,
\end{itemize}
and the standard clauses for $\neg$ and $\land$. 
%If $\mathfrak{M},w,d\models \varphi$ we say that $\varphi$ is \emph{true} at $w,d$ in $\mathfrak M$. 
%and also write\nz{do we need this?} $\mathfrak{M},w\models \varphi(d)$. 
%If $\mathfrak{M},w,d\models \varphi$ for all $w$ and $d$, we call $\mathfrak{M}$ is a \emph{model}\nz{?} of $\varphi$ and write $\mathfrak{M}\models \varphi$. 
%
If $\varphi$ is a sentence (i.e., every occurrence of $x$ in $\varphi$ is in the scope of $\exists$), then $\mathfrak{M},w,d\models \varphi$ iff $\mathfrak{M},w,d'\models \varphi$, for any $d,d'\in D$, and so we can omit $d$ and write $\mathfrak{M},w \models \varphi$. Similarly, we can write $\mathfrak{M},d\models \psi$ if every $\pvar$ in $\psi$ is in the scope of $\Diamond$.

The set of formulas $\varphi$ with $\mathfrak M,w,d \models \varphi$, for all $\mathfrak M$, $w$, $d$, is denoted by $\FOMK$; it is the \FOM{}-extension of the modal logic $\mathsf{K}$. Those $\varphi$ that are true everywhere in all models $\mathfrak M$ with $R = W \times W$ comprise $\FOMS$, the \FOM{}-extension of the modal logic $\mathsf{S5}$. 
Let $L$ be one of these two logics. A \emph{knowledge base} (KB), $K$, is any finite set of sentences. 
We say that $K$ (\emph{locally}) \emph{entails $\varphi$ in $L$} and write $K\models_{L}\varphi$ if $\mathfrak M,w \models K$ implies $\mathfrak M,w,d \models \varphi$, for any $L$-model $\mathfrak M$ and any $w$ and $d$ in it. Shortening $\emptyset \models_{L} \varphi$ to $\models_L \varphi$ (i.e.,  $\varphi \in L$), we note that \mbox{$K\models_L \varphi$} iff $\models_L  (\bigwedge_{\psi\in K} \psi  \rightarrow \varphi)$, reducing KB-entailment in $L$ to $L$-validity, which is known to be \textsc{coNExpTime}-complete~\cite{DBLP:journals/logcom/Marx99}. 
%Global KB-entailment in $\FOMS$ can  also be expressed via validity: %$\models_{\FOMS}(\bigwedge_{\psi\in K}\Box\forall x \psi \rightarrow %\varphi)$.  
%{\color{red} Let see what we say about K where this does not hold.}\nb{??}
%
%
%We say that \emph{$K$ entails $\varphi$ in $L$} and write $K\models_{L}\varphi$ if any model of $K$ is also a model of $\varphi$. We use $\models_L \varphi$ synonymously with $\emptyset \models_{L} \varphi$. Note that $K\models_{\FOMS} \varphi$ iff $\models_{\FOMS}(\bigwedge_{\psi\in K}\Box\forall x \psi) \rightarrow \varphi$ and so checking entailment from KBs reduces to checking validy. Deciding $K\models_{\FOMS} \varphi$ is known to be \textsc{coNExpTime}-complete~\cite{DBLP:journals/logcom/Marx99}. 

%Further, we write $\models_\mathsf{K} \varphi$ if $\varphi$ is true in all models; the set of such formulas $\varphi$ is the \FOM{}-extension of the basic modal logic $\mathsf{K}$. By $\models_\mathsf{S5} \varphi$ we mean that $\varphi$ is true in all models $\mathfrak M$ with $R = W \times W$; the set of such $\varphi$ is the \FOM{}-extension of the modal logic $\mathsf{S5}$.
%Deciding $\models_L \varphi$, for $L \in \{\mathsf{K}, \mathsf{S5}\}$, is known to be \textsc{coNExpTime}-complete~\cite{DBLP:journals/logcom/Marx99}.

We refer the reader to~\cite{DBLP:books/el/07/BraunerG07,FitMen12}
for detailed introductions to first-order modal logic in general and to~\cite{GabEtAl03,DBLP:books/el/07/Kurucz07} for decidable fragments of first-order modal logics.

\paragraph{Bisimulations.}
Given two models $\mathfrak{M}= (W,R,D,I)$ with $w,d$ and $\mathfrak{M}'= (W',R',D',I')$ with $w',d'$,  we write \mbox{$\mathfrak{M},w,d  \equiv_{\sigma} \mathfrak{M}',w',d'$}, for a signature $\sigma$, if the same $\sigma$-formulas are true at $w,d$ in $\mathfrak M$ and at $w',d'$ in $\mathfrak M'$. 
%
%We\nz{remove} write $\mathfrak{M}_{1},w_{1}\equiv_{\sigma} \mathfrak{M}_{2},w_{2}$ when $\mathfrak{M}_{1}, w_1 \models \exists x \varphi$ iff $\mathfrak{M}_{2}, w_2 \models \exists x \varphi$, for all $\sigma$-formulas $\varphi$; and we write $\mathfrak{M}_{1},d_{1}\equiv_{\sigma} \mathfrak{M}_{2},d_{2}$ when $\mathfrak{M}_{1},d_1 \models \Diamond \varphi$ iff $\mathfrak{M}_{2},d_2 \models \Diamond \varphi$, for all $\sigma$-formulas $\varphi$.
%
We characterise $\equiv_{\sigma}$ using bisimulations. 
Namely, a relation 
$$
\bs \subseteq (W\times D) \times (W'\times D')
$$
is called a $\sigma$-\emph{bisimulation} between models $\mathfrak M$ and $\mathfrak M'$  if the following conditions hold for all $((w,d),(w',d'))\in \bs$ and $\pvar\in \sigma$:  
\begin{enumerate}[style=multiline,leftmargin=0.7cm]
\item[\textbf{(a)}] $\mathfrak M,w,d \models \pvar$ iff $\mathfrak M',w',d' \models \pvar$;
	
\item[\textbf{(w)}] if $(w,v)\in R$, then there is $v'$ such that $(w',v')\in R'$ and $((v,d),(v',d'))\in \bs$, and if $(w',v')\in R'$, then there is $v$ such that $(w,v)\in R$ and $((v,d),(v',d'))\in \bs$;
	
\item[\textbf{(d)}] for every $e\in D$, there is $e'\in D'$ such that $((w,e),(w',e'))\in \bs$ and, for every $e'\in D'$, there is $e\in D$ such that $((w,e),(w',e'))\in \bs$.
\end{enumerate}
We say that $\mathfrak{M},w,d$ and $\mathfrak{M}',w',d'$ are $\sigma$-\emph{bisimilar} and write 
$\mathfrak{M},w,d  \sim_{\sigma} \mathfrak{M}',w',d'$
if there is a $\sigma$-bisimulation $\bs$ between $\mathfrak{M}$ and $\mathfrak{M}'$ with $((w,d),(w',d')) \in \bs$. 
The next characterisation is proved in a standard way using $\omega$-saturated models~\cite{modeltheory,goranko20075}:

\begin{lem}\label{bisim-lemma}
For any signature $\sigma$ and $\omega$-saturated models $\mathfrak{M}$ with $w,d$ and $\mathfrak{M}'$ with $w',d'$, 
	%
	%$$
$$
\mathfrak{M},w,d  \equiv_{\sigma} \mathfrak{M}',w',d' \quad \text{ iff } \quad \mathfrak{M},w,d  \sim_{\sigma} \mathfrak{M}',w',d'.
$$
%		
%\item[--] $\mathfrak{M},w \equiv_{\sigma} \mathfrak{M}',w'$ \ iff \ $\mathfrak{M},w \sim_{\sigma} \mathfrak{M}',w'$\textup{;}\nz{remove}
%\item[--] $\mathfrak{M},d  \equiv_{\sigma} \mathfrak{M}',d'$ \ iff \ $\mathfrak{M},d  \sim_{\sigma} \mathfrak{M}',d'$.\nz{remove}
%\end{itemize}
%
The direction from right to left holds for arbitrary models.
\end{lem}

%Observe that $\sigma$-bisimulations between the same models are preserved under set-theoretic union: if $\Gamma$ is a set of $\sigma$-bisimulations, then $\bigcup_{\bs\in \Gamma} \bs$ is a $\sigma$-bisimulation too. It follows that if there is a $\sigma$-bisimulation between $\omega$-saturated models $\mathfrak M$ and $\mathfrak M'$, then $\equiv_\sigma$ is always a $\sigma$-bisimulation between them (the largest one).

%****

\paragraph{Modal products and succinct notation.}
As observed by \cite{Wajsberg33}, $\mathsf{S5}$ is a notational variant of the one-variable fragment $\FOO$ of first-order logic, $\FO$: just drop $x$ from $\exists x$ and $\pvar(x)$ in \FOO-formulas, treating $\exists$ as a possibility operator and $\pvar$ as a \emph{propositional variable}. The same operation transforms \FOM-formulas into more succinct \emph{bimodal formulas} with $\Diamond$ interpreted over the $(W,R)$ `dimension' and $\exists$ over the $(D, D\times D)$ `dimension'\!. 
This way we view the \FOM-extensions of $\mathsf{S5}$ and $\mathsf{K}$ as two-dimensional \emph{products} of modal logics: $\mathsf{S5}\times \mathsf{S5}$ and $\mathsf{K}\times \mathsf{S5}$.   
The former is known to be the `equality and substitution-free' fragment of two-variable fragment $\FOT$ of $\FO$~\cite{GabEtAl03} (see Appendix~\ref{s:FO} for details); the latter is embedded into \FO{} by the \emph{standard translation} $^*$ defined inductively by taking  
$\pvar^* = \qvar(z,x)$, $(\neg \varphi)^* = \neg \varphi^*$,  $(\varphi \land \psi)^* = \varphi^* \land \psi^*$,  
$(\exists \varphi)^* = \exists x\, \varphi^*$, $(\Diamond \varphi)^* = \exists y\, ( R(z,y) \land \varphi^*\{y/z\})$,
where $y$ is a fresh variable not occurring in $\varphi^*$ and $\{y/z\}$ means a substitution of $y$ in place of $z$.

From now on, we write \FOM-formulas as \emph{bimodal} ones: for example, $\exists\Box\pvar $ stands for $\exists x\, \Box \pvar(x)$. By a formula we mean an \FOM-formula unless indicated otherwise; a logic, $L$, is one of $\QSF$ and $\QK$, again unless stated otherwise.

%**********

\section{Main Notions and Characterisations}\label{main-notions}
%
%\tagi{Given a pair $(\varphi,\psi)$ of formulas and a signature $\sigma$, 
%a \emph{$\sigma$-interpolant for $(\varphi,\psi)$ in a logic $L$} is a formula $\chi$ such that
%$\sig(\chi)\subseteq\sigma$,
%$\models_{L}\varphi\to\chi$ and $\models_{L}\chi\to\psi$.
%The  \emph{$\sigma$-interpolant existence} problem for $L$ is to decide, for any pair $(\varphi,\psi)$ of formulas and any signature $\sigma$, whether there exists a $\sigma$-interpolant for $(\varphi,\psi)$ in $L$.}\nb{do we need $\sigma=\cap$?}
%

We now introduce the main notions studied in this article and provide their model-theoretic characterisations. 
We start with interpolants and explicit definitions.

\paragraph{Craig interpolants.}
A formula $\chi$ is  an \emph{interpolant} of formulas $\varphi$ and $\psi$ \emph{in a logic} $L$ if $\sig(\chi) \subseteq \sig(\varphi) \cap \sig(\psi)$, $\models_L \varphi \to \chi$ and \mbox{$\models_L \chi \to \psi$}. 
$L$ enjoys the \emph{Craig interpolation property} (\emph{CIP}) if an interpolant for $\varphi$ and $\psi$ exists whenever $\models_{L}\varphi\rightarrow \psi$.
One of our main concerns here is the \emph{interpolant existence problem} (\emph{IEP}) for $L$: decide if given $\varphi$ and $\psi$ have an interpolant in $L$. For logics with the CIP, the IEP reduces to validity, and so is not interesting. This is the case for many logics including propositional $\SF$ and $\mathsf{K}$, but  not for FOMLs with constant domain between $\FOMK$ and $\FOMS$~\cite{DBLP:journals/jsyml/Fine79,DBLP:journals/ndjfl/MarxA98}.
%Interpolants\nz{TBE} are closely related to definitions which we introduce next. 

We note that besides the interpolants introduced above, one can also consider interpolants for the \emph{global} consequence relation $\models_{L}^{g}$ that is defined by setting $\varphi\models_{L}^{g}\psi$ if, for all $L$-models $\mathfrak{M}$, whenever $\mathfrak{M},w,d\models \varphi$ for all $w$, $d$ in $\mathfrak{M}$, then $\mathfrak{M},w,d\models \psi$ for all $w$, $d$ in $\mathfrak{M}$. A formula $\chi$ is a \emph{global deductive interpolant} of $\varphi$ and $\psi$ in $L$ if $\sig(\chi) \subseteq \sig(\varphi) \cap \sig(\psi)$, $\varphi\models_L^{g} \chi$ and $\chi \models_L^{g} \psi$. While the relationship between the respective CIPs is well understood~\cite{chapter:nonclassical}, the relationship between the respective IEPs remains to be investigated. In this paper, we are only concerned with `local' interpolants, but note that, for $\FOMS$, the global IEP can be polynomially reduced to the local IEP because $\varphi\models_{\FOMS}^{g}\psi$ iff $\Box\forall \varphi \models_{\FOMS}\psi$; see also the discussion of interpolants modulo ontologies in Section~\ref{sec:ALCIO}.

\paragraph{Explicit definitions.} Given formulas $\varphi$, $\psi$ and a signature $\sigma$, 
an \emph{explicit $\sigma$-definition of $\psi$ modulo $\varphi$ in $L$} is a $\sigma$-formula $\chi$ such that 
%$\sig(\chi)\subseteq\sigma$ and 
$\models_L\varphi\to(\psi\leftrightarrow\chi)$. 
The \emph{explicit $\sigma$-definition existence problem} (\emph{EDEP}) for $L$ is to decide, given $\varphi$, $\psi$ and $\sigma$, whether there exists an explicit $\sigma$-definition of $\psi$ modulo $\varphi$ in $L$. The 
EDEP reduces trivially to entailment for logics enjoying the \emph{projective Beth definability property} (\emph{BDP})  according to which $\psi$ is explicitly $\sigma$-definable modulo $\varphi$ in $L$ iff it is implicitly $\sigma$-definable modulo $\varphi$ in the sense that $\{\varphi,\varphi'\}\models_{L} \psi \leftrightarrow \psi'$, where $\varphi'$ and $\psi'$ result from $\varphi$ and $\psi$ by uniformly replacing all non-$\sigma$-symbols with fresh ones~\cite{modeltheory}. Again, many logics including propositional $\SF$ and $\mathsf{K}$ enjoy the BDP while FOMLs with constant domains between $\FOMK$ and $\FOMS$ do not. 
%~\cite{DBLP:journals/jsyml/Fine79,DBLP:journals/ndjfl/MarxA98}. 

Note that, in many applications, $\varphi$ in our formulation of the EDEP corresponds to a KB $K$ and $\psi$ is a predicate $\pvar$. Then the problem whether there exists an explicit $\sigma$-definition of $\pvar$ modulo $K$ is the problem of deciding whether there is $\chi$ with $\sig(\chi) \subseteq \sigma$ and  $K\models_{L}\forall x(\pvar(x) \leftrightarrow \chi(x))$. This problem trivially translates to the EDEP using our discussion of KBs above. In more detail, this view of the EDEP is discussed in Section~\ref{sec:ALCIO} in the context of $\ALCIOS$.

Lemma~\ref{bisim-lemma} together with the fact that \FOM{} is a fragment of \FO{} are used to obtain, again in a standard way, the following criterion of interpolant existence. We call formulas $\varphi$ and $\psi$ \emph{$\sigma$-bisimulation consistent} in $L$ if there exist $L$-models $\mathfrak{M}$ with $w,d$ and $\mathfrak{M}'$ with $w',d'$ such that $\mathfrak{M},w,d \models \varphi$, $\mathfrak{M}',w',d'\models \psi$ and $\mathfrak{M},w,d  \sim_{\sigma} \mathfrak{M}',w',d'$. 

\begin{thm}\label{int-crit}
For any $\FOM$-formulas $\varphi$ and $\psi$, the following conditions are equivalent\textup{:}
\begin{itemize}
\item[--] there does not exist an interpolant of $\varphi$ and $\psi$ in $L$\textup{;}

\item[--] $\varphi$ and $\neg \psi$ are $\sig(\varphi) \cap \sig(\psi)$-bisimulation consistent in $L$.
\end{itemize}
\end{thm}
\begin{proof} 
Suppose $\varphi$ and $\psi$ do not have an interpolant in $L$ and $\sigma = \sig(\varphi) \cap \sig(\psi)$. Consider the set $\Xi$ of $\sigma$-formulas $\chi$ with $\models_L \varphi \to \chi$. By compactness, we have an $\omega$-saturated model $\mathfrak M$ of $L$ with $w$ and $d$ such that $\mathfrak M,w,d \models \chi$, for all $\chi \in \Xi$, and $\mathfrak M,w,d \models  \neg \psi$. Take the set $\Xi'$ of $\sigma$-formulas $\chi$ with $\mathfrak M,w,d \models \chi$ and an $\omega$-saturated model $\mathfrak M'$ with $\mathfrak M',w',d' \models \Xi'$ and $\mathfrak M',w',d' \models \varphi$, for some $w'$ and $d'$. Then $\mathfrak{M},w,d  \equiv_{\sigma} \mathfrak{M}',w',d'$, and so $\mathfrak{M},w,d  \sim_{\sigma} \mathfrak{M}',w',d'$ by Lemma~\ref{bisim-lemma}. The converse implication is straightforward.
\end{proof}

\begin{exa}\label{ex1}
For any $n<\omega$, \cite{DBLP:journals/ndjfl/MarxA98} constructed two \FOM-formulas $\varphi$ and $\psi$ with $\models_{\QSF} \varphi \to \psi$ and $\sig(\varphi) \cap \sig(\psi) = \{\diag\}$ that have no interpolant in the $n$-variable $\mathsf{Q^{n}S5}$. For $n=1$, the formulas $\varphi$ and  $\psi$ look as follows: 
%The formulas $\varphi$, $\psi$ below with $\models_{\QSF} \varphi \to \psi$ and $\sig(\varphi) \cap \sig(\psi) = \{\diag\}$ have no  interpolant in $\QSF$
%---in fact, any $\mathsf{Q^nS5}$, for $n < \omega$---}
%any finite-variable fragment of quantified $\SF$}
%
\begin{align*}
& \varphi ~=~  \pvar_0 \land \Diamond \exists (\pvar_1 \land \Diamond \exists \pvar_2) \land{}   
 \Box \forall \big[ ( \diag \leftrightarrow  \pvar_0 \lor \pvar_1 \lor \pvar_2) \land{} \textstyle\bigwedge_{i \ne j} (\pvar_i \to \neg \pvar_j ) \land{}\\ 
& \mbox{}\hspace*{7cm} \bigwedge_{0 \le i \le 2} \bigl(\pvar_i \to  \Box (\diag \to \pvar_i) \land \forall (\diag\to \pvar_i) \bigr) \big],\\
& \psi ~=~  \Bh \Bv (\diag \leftrightarrow  \bvar_0 \lor \bvar_1) \to {} \Dh \Dv  \bigl(\bvar_0 \land \Dh (\neg \diag \land \Dv  \bvar_0)\bigr) \lor{} 
 \Dh\Dv \bigl(\bvar_1 \land \Dh (\neg \diag \land \Dv \bvar_1) \bigr).
\end{align*}
To see that $\varphi$ and $\neg\psi$ are $\{\diag\}$-bisimulation consistent in $\QSF$, take the models 
$\mathfrak{M}_1$ and $\mathfrak{M}_2$ depicted below with 
$\mathfrak M_1,u_0,d_0 \models \varphi$ and $\mathfrak M_2,v_0,c_0 \models\neg\psi$. (In our pictures, the worlds are always shown along the horizontal axis and the domain elements  along the vertical one, giving points of the form $(w,d)$.) The relation $\bs$ connecting
each $\diag$-point in $\mathfrak M_1$ with each $\diag$-point in $\mathfrak M_2$, and similarly for $\neg\diag$-points, 
is an $\{\diag\}$-bisimulation between $\mathfrak{M}_1$ and $\mathfrak{M}_2$ such that $\bigl((u_0,d_0),(v_0,c_0)\bigr)\in\bs$. \hfill $\dashv$

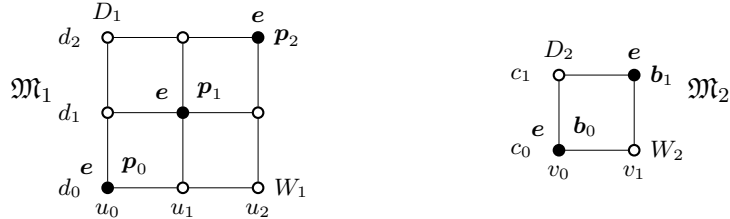
\centerline{
\begin{tikzpicture}[>=latex,line width=0.2pt]
%\node[point,fill,scale = 0.7,label=below:{\footnotesize $u_0$},label=above:{\footnotesize $\ \diag\quad\pvar_0$}] (u0d0) at (0,0) {};
\node[point,fill,scale = 0.7,label=below:{\footnotesize $u_0$},label=above left:{\footnotesize $\diag$},label=above right:{\footnotesize $\pvar_0$}] (u0d0) at (0,0) {};
\node[point,fill=white,scale = 0.7,label=below:{\footnotesize $u_1$}] (u1d0) at (1,0) {};
\node[point,fill=white,scale = 0.7,label=below:{\footnotesize $u_2$},label=right:{\footnotesize $W_1$}] (u2d0) at (2,0) {};
\node[point,fill=white,scale = 0.7] (u0d1) at (0,1) {};
\node[point,fill=black,scale = 0.7,label=above left:{\footnotesize $\diag$},label=above right:{\footnotesize $\pvar_1$}] (u1d1) at (1,1) {};
\node[point,fill=white,scale = 0.7] (u2d1) at (2,1) {};
\node[point,fill=white,scale = 0.7,label=above:{\footnotesize $D_1$}] (u0d2) at (0,2) {};
\node[point,fill=white,scale = 0.7] (u1d2) at (1,2) {};
\node[point,fill=black,scale = 0.7,label=above:{\footnotesize $\diag$},label=right:{\footnotesize $\pvar_2$}] (u2d2) at (2,2) {};
\node[]  at (-.5,0) {{\footnotesize $d_0$}};
\node[]  at (-.5,1) {{\footnotesize $d_1$}};
\node[]  at (-.5,2) {{\footnotesize $d_2$}};
\node[]  at (-1,1.3) {$\mathfrak M_1$};
\draw[-] (u0d0) to (u1d0);
\draw[-] (u1d0) to (u2d0);
\draw[-] (u0d1) to (u1d1);
\draw[-] (u1d1) to (u2d1);
\draw[-] (u0d2) to (u1d2);
\draw[-] (u1d2) to (u2d2);
\draw[-] (u0d0) to (u0d1);
\draw[-] (u0d1) to (u0d2);
\draw[-] (u1d0) to (u1d1);
\draw[-] (u1d1) to (u1d2);
\draw[-] (u2d0) to (u2d1);
\draw[-] (u2d1) to (u2d2);
\node[point,fill=black,scale = 0.7,label=below:{\footnotesize $v_0$},label=above left:{\footnotesize $\diag$},label=above right:{\footnotesize $\bvar_0$}] (v0c0) at (6,.5) {};
\node[point,fill=white,scale = 0.7,label=right:{\footnotesize $W_2$},label=below:{\footnotesize $v_1$}] (v1c0) at (7,.5) {};
\node[point,fill=white,scale = 0.7,label=above:{\footnotesize $D_2$}] (v0c1) at (6,1.5) {};
\node[point,fill=black,scale = 0.7,label=above:{\footnotesize $\diag$},label=right:{\footnotesize $\bvar_1$}] (v1c1) at (7,1.5) {};
\node[]  at (5.5,.5) {{\footnotesize $c_0$}};
\node[]  at (5.5,1.5) {{\footnotesize $c_1$}};
\node[]  at (8,1.3) {$\mathfrak M_2$};
\draw[-] (v0c0) to (v1c0);
\draw[-] (v0c1) to (v1c1);
\draw[-] (v0c0) to (v0c1);
\draw[-] (v1c0) to (v1c1);
\end{tikzpicture}
}
%As shown by~\cite{DBLP:journals/ndjfl/MarxA98}, the formulas $\varphi$ and $\psi$ below with $\sig(\varphi) \cap \sig(\psi) = \{A\}$ and $\models_{\SF} \varphi \to \psi$ do not have an interpolant in $\QSF$:
%%
%\begin{align*}
%& \varphi ~=~  \pvar_1 \land \Diamond \exists (\pvar_2 \land \Diamond \exists \pvar_3) \land{}  \\ 
%& \mbox{}\hspace*{1cm} \Box \forall \big[ ( \avar \leftrightarrow  \pvar_1 \lor \pvar_2 \lor \pvar_3 ) \land{} \textstyle\bigwedge_{i \ne j} (\pvar_i \to \neg \pvar_j ) \land{}\\ 
%& \mbox{}\hspace*{3cm} \textstyle\bigwedge_i [ \pvar_i \to  \Box (\avar \to \pvar_i) \land \forall (\avar \to \pvar_i)  ] \big],\\
%%
%& \psi ~=~  \Box \forall (\avar \leftrightarrow  \qvar_1 \lor \qvar_2) \to {}\\
%& \mbox{}\hspace*{1.2cm} \Diamond \exists  (\qvar_1 \land \exists (\neg \avar \land \Diamond  \qvar_1)) \lor{} 
% \Diamond \exists (\qvar_2 \land \exists (\neg \avar \land \Diamond \qvar_2) ).
%\end{align*}
%% 
%To see that $\varphi$ and $\neg\psi$ are $\{\avar\}$-bisimulation consistent, take the models $\mathfrak{M}_{1}$ and $\mathfrak{M}_{2}$ below with $\mathfrak M_1,u_1,d_1 \models \varphi$ and $\mathfrak M_2,v_1,e_1 \not\models \psi$. The depicted relation $\bs$ is an $\{\avar\}$-bisimulation containing $((u_1,d_1), (v_1,e_1))$.\nz{worlds horizontally?} 
%%
%\begin{center}
%\includegraphics[scale=0.7]{PICS/example1-1}
%\end{center}
\end{exa}

Similarly to Theorem~\ref{int-crit} we obtain the following criterion of explicit definition existence:

\begin{thm}\label{p:pbdpsem}
%Let $L \in \{\QSF, \QK\}$. 
For any $\varphi$, $\psi$, $\sigma$, the following are equivalent\textup{:}
\begin{itemize}
\item[--]
there is no explicit $\sigma$-definition of $\psi$ modulo $\varphi$ in $L$\textup{;}

\item[--]
$\varphi\land \psi$ and $\varphi\land\neg \psi$ are $\sigma$-bisimulation consistent in $L$. 
%there exist $L$-models $\mathfrak M$ with $w$, $d$ and $\mathfrak M'$ with  $w'$, $d'$ such that
%%
%$\mathfrak M,w,d\models\varphi\land\psi$, 
%$\mathfrak M',w',d'\models\varphi\land\neg\psi$, and 
%$\mathfrak M,w,d\sim_\sigma\mathfrak M',w',d'$. 
\end{itemize}
\end{thm}

\begin{exa}\label{ex-fine1}
Suppose $\varphi$ is the conjunction of the two KB axioms from Example~\ref{Ex:Fine}, $\sigma = \{\mathsf{inPower}\}$, and $\psi = \mathsf{rep}$.\footnote{As our \FOM{} has no 0-ary predicates, the \emph{proposition} $\mathsf{rep}$ is given as $\forall x\,\mathsf{rep}(x)$ assuming that we have $\models_{L}\forall x\, \mathsf{rep}(x) \vee \forall x \, \neg \mathsf{rep}(x)$.} Then the second condition of Theorem~\ref{p:pbdpsem} holds for the $\FOMS$-models shown below, in which $(w,d)$ in $\mathfrak M$ is $\sigma$-bisimilar to $(w',d')$ in $\mathfrak M'$ iff $(w,d)$ and $(w',d')$ agree on $\sigma$.\\
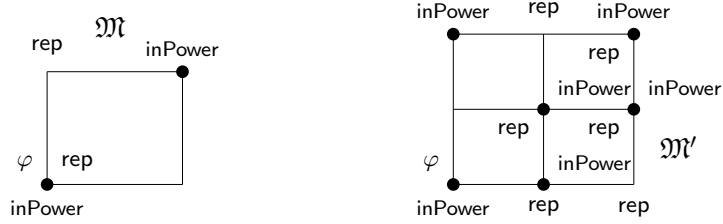
\centerline{
\begin{tikzpicture}[>=latex,line width=0.2pt,xscale = 1.2]
\node[]  at (6,1) {$\mathfrak M'$};
\node[point,fill=black,scale = 0.7,label=below:{\scriptsize $\mathsf{inPower}$},label=above left:{\footnotesize $\varphi$}] (aa1) at (3.5,.5) {};
\node[point,fill=black,scale = 0.7,label=below:{\footnotesize $\mathsf{rep}$},label=above right:{\scriptsize $\mathsf{inPower}$}] (ba1) at (4.5,.5) {};
\node[point,fill=black,scale = 0.7,label=above right:{\scriptsize $\mathsf{inPower}$},label=below left:{\footnotesize $\mathsf{rep}$}] (bb1) at (4.5,1.5) {};
\node[point,fill=black,scale = 0.7,label=above right:{\scriptsize $\mathsf{inPower}$},label=below left:{\footnotesize $\mathsf{rep}$}] (cb1) at (5.5,1.5) {};
\node[point,fill=black,scale = 0.7,label=above:{\scriptsize $\mathsf{inPower}$}] (ac1) at (3.5,2.5) {};
\node[point,fill=black,scale = 0.7,label=above:{\scriptsize $\mathsf{inPower}$},label=below left:{\footnotesize $\mathsf{rep}$}] (cc1) at (5.5,2.5) {};
\node[scale = 0.7,label=below:{\footnotesize $\mathsf{rep}$}] (ca1) at (5.5,.5) {};
\node[scale = 0.7,label=above:{\footnotesize $\mathsf{rep}$}] (bc1) at (4.5,2.5) {};
\draw[-] (3.5,.5) -- (5.5,.5);
\draw[-] (3.5,1.5) -- (5.5,1.5);
\draw[-] (3.5,2.5) -- (5.5,2.5);
\draw[-] (3.5,.5) -- (3.5,2.5);
\draw[-] (4.5,.5) -- (4.5,2.5);
\draw[-] (5.5,.5) -- (5.5,2.5);
\node[]  at (-.3,2.6) {$\mathfrak M$};
\node[point,fill=black,scale = 0.7,label=below:{\scriptsize $\mathsf{inPower}$},label=above left:{\footnotesize $\varphi$},label=above right:{\footnotesize $\mathsf{rep}$}] (aa) at (-1,.5) {};
\node[point,fill=black,scale = 0.7,label=above:{\scriptsize $\mathsf{inPower}$}] (bb) at (.5,2) {};
\node[scale = 0.7,label=above:{\footnotesize $\mathsf{rep}$}] (ab) at (-1,2) {};
\draw[-] (-1,.5) -- (.5,.5);
\draw[-] (-1,2) -- (.5,2);
\draw[-] (-1,.5) -- (-1,2);
\draw[-] (.5,.5) -- (.5,2);
\end{tikzpicture}
} 
\\
It follows that $\mathsf{rep}$ has no definition via $\mathsf{inPower}$ modulo $\varphi$ in $\FOMS$. \hfill $\dashv$
\end{exa}

The IEP and EDEP are closely related~\cite{MGabbay2005-MGAIAD}. Here, we only require the following:

\begin{thm}\label{p:cipvspbdp}
For any $L \in \{\QSF, \QK\}$, the EDEP for $L$ and the IEP for $L$ are polynomially reducible to each other.
\end{thm}
\begin{proof}
The	EDEP is polynomially reducible to the IEP by a standard trick~\cite{MGabbay2005-MGAIAD}: a formula $\psi$ has an explicit $\sigma$-definition modulo $\varphi$ in $L$ iff the formulas $\varphi \land \psi$ and $\varphi^\sigma \to \psi^\sigma$ have an interpolant in $L$, where $\varphi^\sigma$, $\psi^\sigma$ are obtained by replacing each variable $\pvar \notin \sigma$ with a fresh variable $\pvar^\sigma$. Indeed, any $\sigma$-formula $\chi$ with $\models_L\varphi\to(\psi\leftrightarrow\chi)$ is an interpolant of $\varphi \land \psi$ and $\varphi^\sigma \to \psi^\sigma$ in $L$. 
	Conversely, any interpolant of $\varphi \land \psi$ and $\varphi^\sigma \to \psi^\sigma$ is an explicit $\sigma$-definition of $\psi$ modulo $\varphi$ in $L$.
	
	For the other reduction, we observe first that the decision problem for $L$ is polynomially reducible to the EDEP because, for $\psi = \pvar \notin \sig(\varphi)$ and $\sigma = \emptyset$, we have $\models_L \neg \varphi$ iff there is an explicit $\sigma$-definition of $\psi$ modulo $\varphi$ in $L$.
	Then we use Theorems~\ref{int-crit} and~\ref{p:pbdpsem} to show that formulas $\varphi$, $\psi$ have an interpolant in $L$ iff $\models_{L}\varphi\to\psi$ and there is a $\sig(\varphi) \cap \sig(\psi)$-definition of $\psi$ modulo $\psi \to \varphi$ in $L$. 
	Indeed, it suffices to observe that, for any $L$-models $\mathfrak M$ with $w,d$ and $\mathfrak M'$ with $w',d'$, we have 
	$\mathfrak M,w,d\models\varphi$ and $\mathfrak M',w',d'\models\neg\psi$ iff  
	$\mathfrak M,w,d\models (\psi\to\varphi)\land\psi$ and $\mathfrak M',w',d' \models (\psi\to\varphi)\land\neg\psi$.
\end{proof}

We next define conservative extensions, an important notion in the context of ontology modules and modularisation~\cite{GrauHKS08,DBLP:conf/rweb/BotoevaKLRWZ16}.

\paragraph{Conservative extensions.} Given formulas $\varphi$ and $\psi$, we call $\varphi$ an \emph{$L$-conservative extension of $\psi$} if (a) $\models_{L}\varphi \rightarrow \psi$ and (b) $\models_L \varphi \rightarrow \chi$ implies $\models_{L}\psi \rightarrow \chi$, for any $\chi$ with $\sig(\chi)\subseteq \sig(\psi)$. In many applications, $\psi$ is given by a KB $K$ and $\varphi$ is obtained by adding fresh axioms to $K$ \cite{DBLP:conf/kr/GhilardiLW06,AIJ10}. (The translation of our results to the language of KBs is obvious.) The next example shows that this notion of conservative extension is 
%the standard one but 
syntax-dependent in the sense that it is not robust under the addition of fresh predicates. 
\begin{exa}\label{example9} 
Consider the formulas
\begin{align*}
&\varphi = \mathsf{rep} \wedge \Diamond \forall  \big(\mathsf{inPower} \rightarrow \Box (\mathsf{rep} \rightarrow \neg \mathsf{inPower}) \big),\\
& \psi = \Box \forall (\Diamond \mathsf{inPower} \wedge \Diamond \neg \mathsf{inPower} \wedge \exists \mathsf{inPower} \wedge \exists \neg \mathsf{inPower}). 
\end{align*}
We claim that $\varphi \wedge \psi$ is a conservative extension of $\psi$ in $\FOMS$.  %(as all models of $\psi$ are $\{\mathsf{inPower}\}$-bisimilar to $\mathfrak M$ in Example~\ref{ex-fine1}); see Appendix~\ref{s:ex9proof} for details.\nb{!} Now, let  
%% 
%$\psi' = \psi \land (\mathsf{p} \lor \neg \mathsf{p})$,
%%
%for a fresh proposition $\mathsf{p}$. Then the formula $\varphi \land \psi'$ is not a conservative extension of $\psi'$, which can be witnessed by the formula
%%
%$\chi = \neg \big(\mathsf{p} \wedge \Box \exists ( \mathsf{inPower} \wedge \Box (\mathsf{p} \rightarrow \mathsf{inPower}) )\big)$. Again, details can be found in Appendix~\ref{s:ex9proof}.\nb{!}  
%
%We claim first that, for the formulas
%%
%\begin{align*}
%&\varphi = \mathsf{rep} \wedge \Diamond \forall  \big(\mathsf{inPower} \rightarrow \Box (\mathsf{rep} \rightarrow \neg \mathsf{inPower}) \big),\\
%%
%& \psi = \Box \forall (\Diamond \mathsf{inPower} \wedge \Diamond \neg \mathsf{inPower} \wedge \exists \mathsf{inPower} \wedge \exists \neg \mathsf{inPower}),
%\end{align*}
%%
%$\varphi \wedge \psi$ is a conservative extension of $\psi$ in $\FOMS$. 
%
Indeed, condition (a) of the definition of conservative extension is trivial. To show (b), suppose $\models_{\FOMS} \varphi \land \psi \to \chi$, for some $\chi$ such that $\sig(\chi) \subseteq \{\mathsf{inPower}\} = \sigma$. We need to prove $\models_{\FOMS} \psi \to \chi$. Suppose $\psi$ is true somewhere in a \FOMS-model $\mathfrak N$. By the definition of $\psi$, it is true everywhere in $\mathfrak N$. Consider the relation $\bs$ that connects each $\mathsf{inPower}$-point in $\mathfrak N$ with each $\mathsf{inPower}$-point in $\mathfrak M$ from Example~\ref{ex-fine1}, and each $\neg\mathsf{inPower}$-point in $\mathfrak N$ with each $\neg\mathsf{inPower}$-point in $\mathfrak M$. It follows from the definition of $\psi$ and the structure of $\mathfrak M$ that $\bs$ is a $\sigma$-bisimulation between $\mathfrak N$ and $\mathfrak M$. The reader can check that $\mathfrak M,w \models \varphi \land \psi$, and so $\mathfrak M,w,d \models \chi$ and $\mathfrak M,w,e \models \chi$. As $\bs$ is a $\sigma$-bisimulation, we obtain that $\chi$ is true everywhere in $\mathfrak N$, establishing (b).

Now, let  
$\psi' = \psi \land (\mathsf{p} \lor \neg \mathsf{p})$,
for a fresh proposition $\mathsf{p}$. Then $\varphi \land \psi'$ is not a conservative extension of $\psi'$ as witnessed by the formula $ \chi$ below 
$$
\chi = \neg \big(\mathsf{p} \wedge \Box \exists ( \mathsf{inPower} \wedge \Box (\mathsf{p} \rightarrow \mathsf{inPower}) )\big).
$$  
Indeed, we have $\models_{\FOMS} \varphi \land \psi' \to \chi$. For suppose $\mathfrak M, w \models \varphi \land \psi'$, and so $w \models \mathsf{rep}$. Then, by $\varphi$, there is a world $u$ with $u \models \forall (\mathsf{inPower} \rightarrow \Box (\mathsf{rep} \rightarrow \neg \mathsf{inPower}))$. By $\psi'$, there is a domain element $d$ with $u,d \models \mathsf{inPower}$, from which $w,d \models \neg \mathsf{inPower}$. Moreover, this is the case for all $d$ with $u,d \models \mathsf{inPower}$. 
Now, if $w \models \neg \mathsf{p}$, we have $w \models \chi$. So let $w \models \mathsf{p}$. Then $u \not\models \exists ( \mathsf{inPower} \wedge \Box (\mathsf{p} \rightarrow \mathsf{inPower}) )$ because if we had $u,d' \models \mathsf{inPower}$ for some $d'$, then $w,d' \models \neg\mathsf{inPower}$, which is a contradiction. Thus, we obtain $w \models \chi$, which proves $\models_{\FOMS} \varphi \land \psi' \to \chi$. 
On the other hand, in the \FOMS-model shown in the picture below,\\
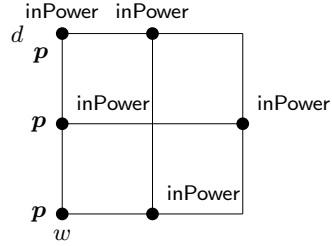
\centerline{
\begin{tikzpicture}[>=latex,line width=0.2pt,scale = 1.2]
\node[]  at (3,2.5) {{\footnotesize $d$}};
\node[point,fill=black,scale = 0.7,label=below:{\footnotesize $w$},label=left:{\footnotesize $\pvar$}] (aa1) at (3.5,.5) {};
\node[point,fill=black,scale = 0.7,label=above right:{\scriptsize $\mathsf{inPower}$}] (ba1) at (4.5,.5) {};
\node[point,fill=black,scale = 0.7,label=above right:{\scriptsize $\mathsf{inPower}$},label=left:{\footnotesize $\pvar$}] (ab1) at (3.5,1.5) {};
\node[point,fill=black,scale = 0.7,label=above right:{\scriptsize $\mathsf{inPower}$}] (cb1) at (5.5,1.5) {};
\node[point,fill=black,scale = 0.7,label=above:{\scriptsize $\mathsf{inPower}$},label=below left:{\footnotesize $\pvar$}] (ac1) at (3.5,2.5) {};
\node[point,fill=black,scale = 0.7,label=above:{\scriptsize $\mathsf{inPower}$}] (bc1) at (4.5,2.5) {};
\draw[-] (3.5,.5) -- (5.5,.5);
\draw[-] (3.5,1.5) -- (5.5,1.5);
\draw[-] (3.5,2.5) -- (5.5,2.5);
\draw[-] (3.5,.5) -- (3.5,2.5);
\draw[-] (4.5,.5) -- (4.5,2.5);
\draw[-] (5.5,.5) -- (5.5,2.5);
\end{tikzpicture}}\\
$\psi'$ is true at $w$ while $\chi$ is false, and so $\not\models_{\FOMS} \psi' \to \chi$. \hfill $\dashv$
%
%Details are provided in the full paper. 
%since is satisfiable but $\chi \wedge \psi\wedge \varphi$ is not, so the addition of $\mathsf{rep}'\rightarrow \mathsf{rep}'$ as a conjunct to $\psi$ results in $\varphi \wedge \psi$ that is not a conservative extension $\psi$
\end{exa}

If in the definition of conservative extension we require property (b) to hold for all $\chi$ with $\sig(\chi) \cap \sig(\varphi) \subseteq \sig(\psi)$, then $\varphi$ is called a \emph{strong $L$-conservative extension of $\psi$}. As observed by~\cite{DBLP:conf/icalp/JungLM0W17}, the difference between conservative and strong conservative extensions is closely related to the failure of the CIP: if $L$ enjoys the CIP, then $L$-conservative extensions coincide with strong  $L$-conservative extensions. The problem of deciding whether a given $\varphi$ is a (strong) conservative extension of a given $\psi$ will be referred to as (S)CEP. The study of the complexity of the (S)CEP for DLs and modal logics started with~\cite{DBLP:conf/kr/GhilardiLW06} and \cite{DBLP:conf/aiml/GhilardiLWZ06}; see~\cite{DBLP:journals/ai/BotoevaLRWZ19,DBLP:conf/kr/JungLM22} for more recent work.

\paragraph{Uniform interpolants.} Given a signature $\sigma$, we call a formula $\chi$ a \emph{$\sigma$-uniform interpolant of a formula $\varphi$ in} $L$ if $\sig(\chi)=\sigma$ and $\varphi$ is a strong $L$-conservative extension of $\chi$. Observe that $\chi$ is then an interpolant of $\varphi$ and $\psi$ in $L$ for any $\psi$ with $\models_{L}\varphi \rightarrow \psi$ and $\sig(\varphi) \cap \sig(\psi) = \sigma$. 
	
A logic $L$ has the \emph{uniform interpolation property} (\emph{UIP}) if, for any $\varphi$ and $\sigma$, there is a $\sigma$-uniform interpolant of $\varphi$ in $L$. The UIP entails the CIP  but not the other way round. For example, modal logic $\mathsf{S4}$ and description logic $\mathcal{ALC}^{u}$ enjoy the CIP but not the UIP~\cite{DBLP:journals/sLogica/GhilardiZ95,DBLP:conf/ijcai/LutzW11}.
This leads to the \emph{uniform interpolant existence problem} (\emph{UIEP}):  given $\varphi$ and $\sigma$, decide whether $\varphi$ has a $\sigma$-uniform interpolant in $L$. We refer the reader to the discussion of related work on interpolant existence in the introduction for a brief survey of the work done on the UIEP in description logic. This work is mostly motivated by the observation that a $\sigma$-uniform interpolant of a formula $\varphi$ can be seen as the result of forgetting all non-$\sigma$-symbols from $\varphi$. Forgetting was first introduced in~\cite{Lin94a}. 
%In\nb{omit} this case, one often drops the requirement that the conservative extension is strong.
%
Note that the SCEP is equivalent to verifying whether a given formula is a uniform interpolant.

%********

\section{Deciding the IEP and EDEP in $\QSF$}\label{sec:s5xs5}

In this section, we prove Theorem~\ref{thm1intro} for \FOMS{}, stating that IEP and EDEP in \FOMS{} are decidable in \textsc{coN2\-Exp\-Time} and 2\textsc{ExpTime}-hard. As a corollary (established in Appendix~\ref{s:FO}), we obtain the lower bound of  Theorem~\ref{thm2intro} stating that the IEP and EDEP in equality-free $\FOT{}$ are 2\textsc{ExpTime}-hard. 

%****

\subsection{Upper bound}\label{ssec:s5xs5up}

Suppose we want to check whether $\varphi$ and $\psi$ have an interpolant in $\FOMS$. By Theorem~\ref{int-crit}, this is not the case iff there are $\FOMS$-models $\mathfrak{M}_1$ with $w_1,d_1$ and $\mathfrak{M}_2$ with $w_2,d_2$ such that $\mathfrak{M}_1,w_1,d_1 \models \varphi$, $\mathfrak{M}_2,w_2, d_2 \models\neg \psi$, and $\mathfrak{M}_1,w_1,d_1  \sim_{\sigma} \mathfrak{M}_2,w_2,d_2$. We are going to show that if such $\mathfrak{M}_i$ do exist, they can be chosen to be of double-exponential size in $|\varphi|+|\psi|$. 

To begin with,
as $R = W \times W$ in any $\FOMS$-model $\mathfrak M = (W,R,D,I)$, in this section we drop $R$ and write simply $\mathfrak M = (W,D,I)$.
Fix $\varphi$, $\psi$ and $\sigma = \sig(\varphi) \cap \sig(\psi)$. Denote by $\sub_\exists(\varphi,\psi)$ the closure under single negation of the set of formulas of the form $\exists \xi$ in $\sub(\varphi,\psi) = \sub(\varphi)\cup \sub(\psi)$. 
The \emph{world-type} of $w\in W$ in $\mathfrak{M} = (W,D,I)$ is defined as
\begin{equation*}
\wt_{\mathfrak{M}}(w) = \{ \rho \in \sub_\exists(\varphi,\psi) \mid \mathfrak M,w\models \rho\}.
\end{equation*}
A set $\wt\subseteq \sub_\exists(\varphi,\psi)$ is called a  \emph{world-type} in $\mathfrak M$ if it is the world-type of some $w\in W$. 

Similarly, let $\sub_\Diamond(\varphi,\psi)$ be the closure under single negation of the set of formulas of the form $\Diamond \xi$ in $\sub(\varphi, \psi)$. The \emph{domain-type} of $d\in D$ in $\mathfrak{M}$ is the set
\begin{equation*}
\dt_{\mathfrak{M}}(d)= \{ \rho \in \sub_\Diamond(\varphi,\psi) \mid \mathfrak M,d \models \rho\}. 
\end{equation*}
A set $\dt\subseteq \sub_\Diamond(\varphi,\psi)$ is called a \emph{domain-type} in $\mathfrak M$ if it is the domain-type 
of some $d \in D$. 

The \emph{full type} of $(w,d) \in W \times D$ in $\mathfrak{M}$ is the set
\begin{equation*}
\ft_{\mathfrak{M}}(w,d)= \{ \rho \in \sub(\varphi, \psi) \mid \mathfrak M, w,d \models \rho\}. 
\end{equation*}
A set $\ft \subseteq \sub(\varphi, \psi)$ is called a \emph{full type} in $\mathfrak M$ if it is the full type of some $(w,d)$ in $\mathfrak M$. 

The main technique of this section generalises the following construction that shows how, given any $\FOMS$-model $\mathfrak M$ satisfying a formula $\varphi$, we can construct from the world and domain types in $\mathfrak M$ a model $\mathfrak M'$ satisfying $\varphi$ and having exponential size in $|\varphi|$. 
Intuitively, as a first approximation, we could start by taking the worlds $W'$ (domain $D'$) in $\mathfrak M'$ to comprise all the world- (domain-) types in $\mathfrak M$. But then we might have $w,w'$ and $d,d'$ with $\wt_{\mathfrak M}(w)=\wt_{\mathfrak M}(w')$, $\dt_{\mathfrak M}(d)=\dt_{\mathfrak M}(d')$ and different truth-values of some variables $\pvar$ at $(w,d)$ and $(w',d')$ in $\mathfrak M$. To deal with this issue, we introduce, as shown in the example below, sufficiently many copies of each world- and domain-type so that we can accommodate \emph{all} possible truth-values in $\mathfrak M$ of the $\pvar$ in $\varphi$. 
(It is to be noted that there are many alternative ways of introducing copies of the $\wt$ and $\dt$ to define $W'$ and $D'$ and the truth-value of $\pvar$ at pairs of such copies. For instance, one could  swap the role of $W$ and $D$ or give a symmetric construction introducing $2n$ copies of each $\wt$ and $\dt$. We opted for the representation below as it generalises well to $\sigma$-bisimulation-consistency for $\QSF$ and \ALCIOS{}, and it admits transparent inductive proofs.) 

\begin{exa}\label{s5sat}
Let $\mathfrak M, w,d \models \varphi$, for $\mathfrak M = (W,D,I)$, and let $n$ be the number of full types in $\mathfrak M$  (over $\sub(\varphi)$) and $[n] = \{1,\dots,n\}$. Define $D'$ to be a set that contains $n$ distinct copies of each $\dt$ in $\mathfrak M$ over $\sub_\Diamond(\varphi)$, denoting the $k$th copy by $\dt^k$. For any pair $\wt$, $\dt$ in $\mathfrak M$, let $\Pi_{\wt,\dt}$ denote the set of functions from $[n]$ \emph{onto} the set of full types $\ft$ in $\mathfrak{M}$ with $\wt = \ft \cap \sub_\exists(\varphi)$ and $\dt = \ft \cap \sub_\Diamond(\varphi)$. 
Let $\Pi$ denote the set of functions $\pi$ mapping each pair $\wt,\dt$ in $\mathfrak{M}$ to an element of $\Pi_{\wt,\dt}$. For $\pi\in \Pi$ we set 
$\pi_{\wt,\dt}=\pi(\wt,\dt)$. Then let $\Pi^{\dag}\subseteq \Pi$
be a smallest set for which the following condition holds: 
%$|\Pi|\leq n^{2}$ and, 
for any $\ft=\ft_{\mathfrak M}(u,e)$ and $k\in [n]$, there exists $\pi\in \Pi^{\dag}$ with $\pi_{\wt_{\mathfrak M}(u),\dt_{\mathfrak M}(e)}(k) = \ft$. 
We claim that $|\Pi^{\dag}|\leq n^{2}$. Indeed, for any $k\in [n]$ and any full type $\ft$ in $\mathfrak{M}$,
we will include just a single function $f^{k,\ft}\in \Pi$ in $\Pi^\dag$.
Assume $k$ and $\ft=\ft_{\mathfrak{M}}(u,e)$ are given. Then we can choose $f^{k,\ft}$ to be any function mapping pairs $\dt,\wt$ into $\Pi_{\wt,\dt}$ such that $f^{k,\ft}(\wt_{\mathfrak{M}}(u),\dt_{\mathfrak{M}}(e))(k)=\ft$. The resulting $\Pi^{\dag}$ is as required.

We set $W' = \{\wt^\pi_{\mathfrak M}(u) \mid u \in W,\ \pi \in \Pi^{\dag}\}$, treating each $\wt^\pi_{\mathfrak M}(u)$ as
a fresh $\pi$-copy of $\wt^\pi_{\mathfrak M}(u)$. Then both $|W'|$ and $|D'|$ are exponential in $|\varphi|$. Define a model $\mathfrak M' = (W',D',I')$ by taking $\mathfrak M', \wt^\pi, \dt^k \models \pvar$ iff $\pvar \in \pi_{\wt,\dt}(k)$, for all $\pi\in \Pi^{\dag}$ and $\wt,\dt$ in $\mathfrak{M}$. We show by induction that $\mathfrak M', \wt^\pi, \dt^k \models \rho$ iff $\rho \in \pi_{\wt,\dt}(k)$, for any $\rho \in \sub(\varphi)$. The basis of induction and the Boolean cases are straightforward.

\emph{Case} $\rho = \exists \xi$. If $\wt^\pi,\dt^k \models \rho$, there is $\dt'^{k'}$ with $\wt^\pi,\dt'^{k'} \models \xi$. By IH, $\xi \in \pi_{\wt,\dt'}(k')$, so $\rho \in \pi_{\wt,\dt'}(k')$ and $\rho \in \wt$, whence  $\rho \in \pi_{\wt,\dt}(k)$. Conversely, let $\rho \in \pi_{\wt,\dt}(k) = \ft_{\mathfrak{M}}(u,e)$. Then there is $e'$ with $\mathfrak{M}, u, e' \models \xi$, and so $\xi\in\ft(u,e')$. Let $\dt'=\dt_{\mathfrak M}(e')$. 
As $\pi_{\wt,\dt'}$ is surjective, there is $k'$ with $\pi_{\wt,\dt'}(k')=\ft(u,e')$, and so  $\xi\in \pi_{\wt,\dt'}(k')$. By IH, $\wt^\pi, \dt'^{k'} \models \xi$, and so $\wt^\pi, \dt^k \models \rho$. 
%$\wt^\pi \models \rho(\dt^k)$.

\emph{Case} $\rho = \Diamond \xi$. If $\wt^\pi, \dt^k \models \rho$, there exists $\wt'^{\pi'}$ with $\wt'^{\pi'}, \dt^k \models \xi$. By IH, $\xi \in \pi'_{\wt',\dt}(k)$, so $\Diamond \xi \in \pi'_{\wt',\dt}(k)$ and $\rho \in \dt$, whence $\rho \in \pi_{\wt,\dt}(k)$.
Conversely, if $\rho \in \pi_{\wt,\dt}(k) = \ft_{\mathfrak{M}}(u,e)$, there is $u'$ with $\mathfrak{M}, u', e \models \xi$. Let $\wt'=\wt_{\mathfrak M}(u')$.  
By the choice of $\Pi^{\dag}$, it has $\pi'$ with $\pi'_{\wt',\dt}(k) = \ft_{\mathfrak{M}}(u',e)$. Then $\xi \in \pi'_{\wt',\dt}(k)$, so $\wt'^{\pi'}, \dt \models \xi$  by IH and $\wt^{\pi}, \dt \models \Diamond \xi$. \hfill $\dashv$
\end{exa}

Note that Example~\ref{s5sat} is of interest beyond illustrating our method as it provides a new and short proof of the exponential finite model property of $\QSF$ \cite{DBLP:journals/bsl/GradelKV97} (equivalently, the exponential finite product model property of $\mathsf{S5}\times \mathsf{S5}$). 

In order to be able to introduce more complex `data structures' that allow us to extend the construction above from satisfiability to $\sigma$-bisimulation  consistency, we start 
by giving a simpler, yet equivalent, definition of bisimulation between $\QSF$-models.
% and then use it to show that, when checking bisimulation consistency in $\QSF$, it is enough to look for bisimilar models of double-exponential size in the size of the given formulas. 

%As $R = W \times W$ in any $\FOMS$-model $\mathfrak M = (W,R,D,I)$, we drop $R$ and write simply $\mathfrak M = (W,D,I)$. 
Given a signature $\sigma$ and $(w,d) \in W \times D$, the \emph{literal $\sigma$-type} $\ell^\sigma_{\mathfrak M}(w,d)$ of $(w,d)$ in $\mathfrak M$ is the set 
\begin{equation*}
\{ \pvar\in \sigma \mid \mathfrak M,w,d \models \pvar\} \cup \{ \neg \pvar \mid \pvar\in \sigma, \ \ \mathfrak M,w,d \not \models \pvar\}. 
\end{equation*}
A pair $(\bs_W,\bs_D)$  of relations $\bs_{W}\subseteq W_1\times W_2$ and $\bs_{D} \subseteq D_1 \times D_2$ is called a $\sigma$-\emph{\BS} between $\mathfrak M_1 = (W_1,D_1,I_1)$ and $\mathfrak M_2 = (W_2,D_2,I_2)$ when the following conditions hold: 
\begin{enumerate}[style=multiline,leftmargin=1.2cm]
\item[\textbf{($\mathsf{s5}_W$)}] if $(w_1,w_2)\in \bs_{W}$ then, for any $d_1 \in D_1$, there is $d_2\in D_2$ such that $(d_1,d_2)\in \bs_{D}$ and $\ell^\sigma_{\mathfrak M_1}(w_1,d_1) =\ell^\sigma_{\mathfrak M_2}(w_2,d_2)$, and the other way round;

\item[\textbf{($\mathsf{s5}_D$)}] if $(d_1,d_2)\in \bs_{D}$ then, for any $w_1 \in W_1$, there is $w_2\in W_2$ such that $(w_1,w_2) \in \bs_{W}$ and $\ell^\sigma_{\mathfrak M_1}(w_1,d_1) =\ell^\sigma_{\mathfrak M_2}(w_2,d_2)$, and the other way round.
\end{enumerate}
We say that $\mathfrak{M}_1,w_1,d_1$ and $\mathfrak{M}_2,w_2,d_2$ are \emph{$\sigma$-$\mathsf{S5}$-bisimilar}  and write $\mathfrak{M}_1,w_1,d_1  \sim_{\sigma}^{\mathsf{S5}} \mathfrak{M}_2,w_2,d_2$ if there exists a $\sigma$-\BS \ $(\bs_{W},\bs_{D})$ with $(w_1,w_2)\in \bs_{W}$, 
$(d_1,d_2)\in \bs_{D}$ and $\ell^\sigma_{\mathfrak M_1}(w_1,d_1) =\ell^\sigma_{\mathfrak M_2}(w_2,d_2)$. Note that in this case $\text{dom}(\bs_{W}) = W_1$, $\text{ran}(\bs_{W})= W_2$, $\text{dom}(\bs_{D})=D_1$, and $\text{ran}(\bs_{D})= D_2$. 

\begin{thm}\label{S5bisim}
$\mathfrak{M}_1,w_1,d_1  \sim_{\sigma}^{\mathsf{S5}} \mathfrak{M}_2,w_2,d_2$ \ iff \ 
$\mathfrak{M}_1,w_1,d_1  \sim_\sigma \mathfrak{M}_2,w_2,d_2$.
\end{thm}
\begin{proof}
	$(\Rightarrow)$ Suppose $\mathfrak{M}_1,w_1,d_1  \sim_\sigma^{\textbf{S5}} \mathfrak{M}_2,w_2,d_2$ is witnessed by $(\bs_{W},\bs_{D})$. Define $\bs$ by setting $((v_1,e_1),(v_2,e_2))\in \bs$ iff $(v_1,v_2)\in \bs_{W}$, $(e_1,e_2)\in \bs_{D}$ and $\ell^\sigma_{\mathfrak M_1}(v_1,e_1)=\ell^\sigma_{\mathfrak M_2}(v_2,e_2)$. It follows that $((w_1,d_1),(w_2,d_2))\in \bs$. We show that $\bs$ satisfies {\bf (a)}, {\bf (w)} and {\bf (d)}. Let $((v_1,e_1),(v_2,e_2))\in \bs$. Then {\bf (a)} follows from $\ell^\sigma_{\mathfrak M_1}(v_1,e_1)=\ell^\sigma_{\mathfrak M_2}(v_2,e_2)$. To show {\bf (w)}, take any $u_1 \in W_1$. As $(e_1,e_2) \in \bs_D$, there is $u_2$ with $(u_1,u_2) \in \bs_W$ and $\ell^\sigma_{\mathfrak M_1}(u_1,e_1) =\ell^\sigma_{\mathfrak M_2}(u_2,e_2)$ by \textbf{($\mathsf{s5}_D$)}, 
from which $((u_1,e_1),(u_2,e_2))\in \bs$. The other implication in {\bf (w)} is symmetric. Finally, consider any $c_1 \in D_1$. By \textbf{($\mathsf{s5}_W$)}, there exists $c_2$ such that $(c_1,c_2)\in \bs_{D}$ and $\ell^\sigma_{\mathfrak M_1}(v_1,c_1) =\ell^\sigma_{\mathfrak M_2}(v_2,c_2)$, from which $((v_1,c_1),(v_2,c_2))\in \bs$. This and a symmetric argument establish  {\bf (d)}.
	
	$(\Leftarrow)$ Let $\mathfrak{M}_1,w_1,d_1  \sim_\sigma \mathfrak{M}_2,w_2,d_2$ be witnessed by $\bs$. Set 
$$
\bs_{W}\!=\! \{(v_1,v_2) \mid \exists e_1,e_2\, ((v_1,e_1),(v_2,e_2))\!\in\!\bs\},\;
\bs_{D}\!=\! \{(e_1,e_2) \mid \exists v_1,v_2\, ((v_1,e_1),(v_2,e_2)) \!\in\! \bs\}.
$$
	Then $(w_1,w_2)\in \bs_{W}$, $(d_1,d_2)\in \bs_{D}$, 
	$\ell^\sigma_{\mathfrak M_1}(w_1,d_1)=\ell^\sigma_{\mathfrak M_2}(w_2,d_2)$. To show \textbf{($\mathsf{s5}_W$)}, suppose that $(v_1,v_2)\in \bs_{W}$ and $c_1\in D_1$. Then there are $e_1,e_2$ with $((v_1,e_1),(v_2,e_2))\in \bs$, and so, by {\bf (d)}, there is $c_2$ with $((v_1,c_1),(v_2,c_2)) \in \bs$, from which $(c_1,c_2) \in \bs_D$ and $\ell^\sigma_{\mathfrak M_1}(v_1,c_1) =\ell^\sigma_{\mathfrak M_2}(v_2,c_2)$. Condition ($\mathsf{s5}_D$) is proved similarly using {\bf (w)}. 
\end{proof}

%In this section, we only deal with $\sigma$-\BS{}s, and so omit explicit $\mathsf{S5}$ from the relevant notations. 
%,
For any $w_1\in W_1$, $w_2\in W_2$,
we write $\mathfrak M_1,w_1  \sim_{\sigma}\mathfrak M_2, w_2$ if there is a $\sigma$-\BS{} $(\bs_{W},\bs_{D})$ between $\mathfrak M_1$ and $\mathfrak M_2$ with $(w_1,w_2)\in \bs_{W}$. 
%By \textbf{($\mathsf{s5}_1$)}, $\mathfrak{M}_1,w_1  \sim_{\sigma} \mathfrak{M}_2,w_2$ entails that the interpretations $I_1(w_1)$ in $\mathfrak M_1$ and $I_2(w_2)$ in $\mathfrak M_2$ are \emph{globally $\sigma$-bisimilar} in the sense that, for any $d_1\in D_1$, there exists $d_2 \in D_2$ satisfying the same $\pvar\in \sigma$ in $I_1(w_1)$ and $I_2(w_2)$, and the other way round.
%
Similarly, for any $d_1\in D_1$, $d_2\in d_2$,
we write $\mathfrak M_1,d _1 \sim_{\sigma}\mathfrak M_2,d_2$ if there is a $\sigma$-\BS{} $(\bs_{W},\bs_{D})$ 
between $\mathfrak M_1$ and $\mathfrak M_2$  with $(d_1,d_2)\in \bs_{D}$. 
We omit $\mathfrak M_1$ and $\mathfrak M_2$ and write simply $(w_1,d_1) \sim_{\sigma} (w_2,d_2)$, $w_1 \sim_{\sigma} w_2$, $d_1  \sim_{\sigma} d_2$ if understood. 
%
%Observe that $\sigma$-bisimulations between the same models are preserved under set-theoretic union: if $\Gamma$ is a set of $\sigma$-bisimulations, then $(\bigcup_{(\bs_{1},\bs_{2})\in \Gamma} \bs_{1},\bigcup_{(\bs_{1},\bs_{2})\in \Gamma}\bs_{2})$ is a $\sigma$-bisimulation too. It follows that $(\bs_{1},\bs_{2})$ defined by taking $(w_{1},w_{2})\in \bs_{1}$ if $w_1 \sim_{\sigma} w_2$ and $(d_{1},d_{2}) \in \bs_{2}$ if $d_{1}  \sim_{\sigma} d_2$ is the maximal $\sigma$-bisimulation between the given models. 

\begin{exa}\label{ex2}
Consider $\mathfrak{M}_{1}$, $\mathfrak{M}_{2}$, and $\sigma = \{\diag\}$ from Example~\ref{ex1}. Then $(W_1 \times W_1, D_1 \times D_1)$ is a $\sigma$-\BS{} between $\mathfrak M_1$ and $\mathfrak M_1$  witnessing  
%
%\begin{equation*}
$(u_{i},d_{i})\sim_{\sigma}(u_{j},d_{j})$ and $ 
(u_{k},d_{l})\sim_{\sigma}(u_{m},d_{n})$, 
%\end{equation*}
%
for $i,j,k,l,m,n \in \{0,1,2\}$, $k \ne l$, $m \ne n$. 
The pair $(W_1 \times W_2, D_1 \times D_2)$ is a $\sigma$-\BS{} between $\mathfrak M_1$ and $\mathfrak M_2$ that witnesses 
%
%\begin{equation*}
$(u_{i},d_{i})\sim_{\sigma}(v_{j},c_{j})$ and  
$(u_{k},d_{l})\sim_{\sigma}(v_{m},c_{n})$, 
%\end{equation*}
%
for all $i,k,l \in \{0,1,2\}$ with $k \ne l$, and $j,m,n \in \{0,1\}$ with $m \ne n$ (cf.\ $\bs$ in Example~\ref{ex1}). \hfill $\dashv$
% We also have 
%
%\begin{multline}\label{bisiminexample}
%u_{0} \sim_{\sigma} u_{1} \sim_{\sigma} u_{2} \sim_{\sigma} v_{0} \sim_{\sigma}
%v_{1},\\ 
%d_{0} \sim_{\sigma} d_{1} \sim_{\sigma} d_{2} \sim_{\sigma} c_{0} \sim_{\sigma} %c_{1}.
%\end{multline}
%
%Consider\nz{Ex~\ref{ex-fine1}?} the $\FOMS$-models $\mathfrak{M}_{1}$ and $\mathfrak{M}_{2}$ from Example~\ref{ex1} and $\sigma = \{A\}$. The pair $(W_1 \times W_1, D_1 \times D_1)$ is a $\sigma$-bisimulation between $\mathfrak M_1$ and $\mathfrak M_1$ that witnesses the relations 
%%
%%\begin{equation*}
%$(u_{i},d_{i})\sim_{\sigma}(u_{j},d_{j})$ and $ 
%(u_{k},d_{l})\sim_{\sigma}(u_{m},d_{n})$, 
%%\end{equation*}
%%
%for all $i,j,k,l,m,n \in \{1,2,3\}$, $k \ne l$, $m \ne n$. The pair $(W_1 \times W_2, D_1 \times D_2)$ is a $\sigma$-bisimulation between $\mathfrak M_1$ and $\mathfrak M_2$ witnessing
%%
%%\begin{equation*}
%$(u_{i},d_{i})\sim_{\sigma}(v_{j},e_{j})$ and  
%$(u_{k},d_{l})\sim_{\sigma}(v_{m},e_{n})$, 
%%\end{equation*}
%%
%for $i,k,l \in \{1,2,3\}$, $j,m,n \in \{1,2\}$, $k \ne l$, $m \ne n$. We also have 
%%
%\begin{multline}\label{bisiminexample}
%u_{1} \sim_{\sigma} u_{2} \sim_{\sigma} u_{3} \sim_{\sigma} v_{1} \sim_{\sigma}
%v_{2},\\ 
%d_{1} \sim_{\sigma} d_{2} \sim_{\sigma} d_{3} \sim_{\sigma} e_{1} \sim_{\sigma} e_{2}.
%\end{multline}
\end{exa}

Suppose that $\mathfrak{M}_{i}= (W_i,D_i,I_i)$, for $i = 1,2$, are $\sigma$-$\mathsf{S5}$-bisimilar
%$\omega$-saturated 
models with pairwise disjoint $W_i$ and $D_i$.
%By (the remark following) Lemma~\ref{bisim-lemma} and Theorem~\ref{S5bisim}, for any $i,j\in\{1,2\}$ and any $w_i\in W_{i}$, $w_j\in W_{j}$, $w_i\sim_\sigma w_j$ iff
%the same $\sigma$-formulas of the form $\exists\xi$ are true at $w_i$ in $\mathfrak M_i$ and at $w_j$ in $\mathfrak M_j$. Also, for any $i,j\in\{1,2\}$ and any $d_i\in D_{i}$, $d_j\in D_{j}$, $d_i\sim_\sigma d_j$ iff the same $\sigma$-formulas of the form $\Diamond\xi$ are true at $d_i$ in $\mathfrak M_i$ and at $d_j$ in $\mathfrak M_j$. Therefore,
By the definitions,
\begin{equation}\label{simequiv}
\mbox{$\sim_\sigma$ is an equivalence relation on $W_1\cup W_2$, and also on $D_1\cup D_2$.}
\end{equation}
%
%By \textbf{($\mathsf{s5}_W$)} and \textbf{($\mathsf{s5}_D$)} for $\mathfrak M_1$ and $\mathfrak M_2$, we also have that, 
Also, for all $w_1\in W_1$, $w_2\in W_2$, $d_1\in D_1$, $d_2\in D_2$,
\begin{align}
\nonumber
& \mbox{if $w_1\sim_\sigma w_2$ then there is $e_2\in D_2$ with
$d_1\sim_\sigma e_2$ and $\ell^\sigma_{\mathfrak M_1}(w_1,d_1) =\ell^\sigma_{\mathfrak M_2}(w_2,e_2)$,}\\
\label{simW}
& \hspace*{9.5cm}\mbox{and the other way round;}\\
\nonumber
& \mbox{if $d_1\sim_\sigma d_2$ then there is $v_2\in W_2$ with
$w_1\sim_\sigma v_2$ and $\ell^\sigma_{\mathfrak M_1}(w_1,d_1) =\ell^\sigma_{\mathfrak M_2}(
v_2,d_2)$,}\\
\label{simD}
& \hspace*{9.5cm}\mbox{and the other way round.}
\end{align}
Now we are in a position to define the necessary `data structures'. 
%We now introduce more complex `data structures' that allow us to extend the construction above from satisfiability to $\sigma$-bisimulation  consistency.
%
%Let $\mathfrak{M}_{i}= (W_{i},D_{i},I_{i})$, for $i = 1,2$, be $\FOMS$-models with pairwise disjoint $W_i$ and $D_i$. 
For any $w\in W_{1}\cup W_{2}$ and $i \in \{1,2\}$, we set
\begin{equation}\label{world-m}
T_{i}(w) = \{ \wt_{\mathfrak{M}_{i}}(v) \mid v \in W_{i}, \ v \sim_{\sigma} w\}
\end{equation}
and call $\wm(w) = (T_{1}(w),T_{2}(w))$ the \emph{world mosaic} of $w$ in $\mathfrak{M}_{1},\mathfrak{M}_{2}$. The pair of the form $\iwp(w) = (\wt_{\mathfrak{M}_{i}}(w),\wm(w))$, for $w\in W_{i}$, is called the \emph{$i$-world point} of $w$ in $\mathfrak{M}_{1}$, $\mathfrak{M}_{2}$. 
A \emph{world mosaic}, $\wm$, and an \emph{$i$-world point}, $\iwp$, in $\mathfrak{M}_{1}$, $\mathfrak{M}_{2}$ are defined as the world mosaic and $i$-world point of some $w\in W_{1} \cup W_2$ in $\mathfrak{M}_{1}$, $\mathfrak{M}_{2}$ (in the latter case, $w \in W_i$). 

Similarly, for any $d\in D_{1}\cup D_{2}$ and $i \in \{1,2\}$, we set
\begin{equation}\label{domain-m}
S_{i}(d) = \{ \dt_{\mathfrak{M}_{i}}(e) \mid e \in D_{i}, \ e \sim_{\sigma} d\}
\end{equation}
and call $\dm(d) = (S_{1}(d),S_{2}(d))$ the \emph{domain mosaic} of $d$ in $\mathfrak{M}_{1}$, $\mathfrak{M}_{2}$. If $d\in D_{i}$, the pair $\idp(d) = (\dt_{\mathfrak{M}_{i}}(d), \dm(d))$ is called the \emph{$i$-domain point} of $d$ in $\mathfrak{M}_{1}$, $\mathfrak{M}_{2}$. A \emph{domain mosaic}, $\dm$, and an  \emph{$i$-domain point}, $\idp$, in $\mathfrak{M}_{1}$, $\mathfrak{M}_{2}$ are defined as the domain mosaic and $i$-domain point of some $d\in D_{1} \cup D_2$. As follows from \eqref{world-m}, \eqref{domain-m} and \eqref{simequiv},
%the definitions, Lemma~\ref{bisim-lemma} and  Theorem~\ref{S5bisim},  
%
\begin{enumerate}[style=multiline,leftmargin=1.1cm]
\item[\textbf{(wm)}] $u \sim_{\sigma} v$ implies $\wm(u) = \wm(v)$,
 
\item[\textbf{(dm)}] $d \sim_{\sigma} e$ implies $\dm(d) = \dm(e)$.
\end{enumerate}
Observe that the number of distinct $\iwp$ and $\idp$ is at most double-exponential in  $|\varphi|+|\psi|$. 
%
%The following examples illustrate the definitions.

\begin{exa}\label{ex3}
(a) Take $\mathfrak{M}_{1}$, $\mathfrak{M}_{2}$ from Example~\ref{ex1}, $\sigma = \{\diag\}$ and $\tau = \{\diag,\pvar_0,\pvar_1,\pvar_2,\bvar_0,\bvar_1\}$. Then $\wt_{\mathfrak{M}_1}(u_i)$ and $\dt_{\mathfrak{M}_2}(c_i)$ contain, respectively, the sets 
$$
\{ \exists (\pvar_i \land \diag) \} \cup \{\exists \neg \pvar \mid \pvar \in \tau\} \cup \{ \neg \exists \pvar_j \mid j \ne i\},\ \  
\{ \Diamond (\pvar_i \land \diag) \} \cup \{\Diamond \neg \pvar \mid \pvar \in \tau\} \cup \{ \neg \Diamond \pvar_j \mid j \ne i\} .
$$
The $\sigma$-\BS s from Example~\ref{ex2} give $\wm(u_{0})=\wm(u_1) = \wm(u_{2})=\wm(v_{0})=\wm(v_{1})$, so $\mathfrak{M}_{1},\mathfrak{M}_{2}$ have  $\wm = (\{\wt_{\mathfrak M_1}(u_i) \mid i =0,1,2\}, \{\wt_{\mathfrak M_2}(v_i) \mid i = 0,1\})$ as the only world mosaic. 
$\mathfrak M_1$ has three distinct $1$-world points $(\wt_{\mathfrak M_1}(u_i), \wm)$, for $i=0,1,2$; $\mathfrak M_2$ has two $2$-world points. Similarly, $\mathfrak{M}_{1}$, $\mathfrak{M}_{2}$ define one domain mosaic, $\mathfrak M_1$ has three distinct $1$-domain points and $\mathfrak M_2$ has two $2$-domain points.
 	
(b) It can happen that non-bisimilar domain elements give the same domain-point.
Consider the models $\mathfrak M_1$ and $\mathfrak M_2$ below and suppose that $\sub_\Dh(\varphi,\psi)$ has no formulas with $\Dv$ \\
\centerline{
\begin{tikzpicture}[>=latex,line width=0.2pt]
\node[]  at (-1.8,.5) {{\footnotesize $\mathfrak M_1$}};
\node[point,fill=black,scale = 0.7,label=left:{\footnotesize $d$},label=above right:{\footnotesize $\avar$}] (00) at (-1,0) {};
\node[point,fill=black,scale = 0.7,label=above right:{\footnotesize $\avar$}] (10) at (0,0) {};
\node[]  at (-1.3,1) {{\footnotesize $d'$}};
\node[point,fill=black,scale = 0.7,label=above right:{\footnotesize $\avar$}] (11) at (0,1) {};
\draw[-] (-1,0) -- (1,0);
\draw[-] (-1,1) -- (1,1);
\draw[-] (-1,0) -- (-1,1);
\draw[-] (0,0) -- (0,1);
\draw[-] (1,0) -- (1,1);
\node[]  at (6.5,.5) {{\footnotesize $\mathfrak M_2$}};
\draw[-] (4,0) -- (6,0);
\draw[-] (4,1) -- (6,1);
\draw[-] (4,0) -- (4,1);
\draw[-] (5,0) -- (5,1);
\draw[-] (6,0) -- (6,1);
\node[point,fill=black,scale = 0.7,label=left:{\footnotesize $e$},label=above right:{\footnotesize $\avar$}] (40) at (4,0) {};
\node[point,fill=black,scale = 0.7,label=above right:{\footnotesize $\avar$}] (50) at (5,0) {};
\node[point,fill=white,scale = 0.7,label=left:{\footnotesize $e'$},label=above right:{\footnotesize $\pvar$}] (41) at (4,1) {};
\node[point,fill=black,scale = 0.7,label=above right:{\footnotesize $\avar$}] (51) at (5,1) {};
\node[point,fill=white,scale = 0.7,label=above right:{\footnotesize $\pvar$}] (61) at (6,1) {};
\node[point,fill=white,scale = 0.7,label=right:{\footnotesize $\pvar$}] (60) at (6,0) {};
\end{tikzpicture}}
in the scope of $\Dh$, $\sigma = \{\avar\}$ and $\sig(\varphi, \psi) = \{\avar,\pvar\}$. Then $\dt_{\mathfrak M_1}(d) = \dt_{\mathfrak M_1}(d')$ but $d \not\sim_{\sigma} d'$ because $\Dh (\avar \land \Dv \neg \avar)$ is true at $d$ and false at $d'$; likewise, we have $\dt_{\mathfrak M_2}(e) = \dt_{\mathfrak M_2}(e')$ but $e\not\sim_{\sigma} e'$. Since $d \sim_{\sigma} e$ and $d' \sim_{\sigma} e'$, we then have $\dm(d) = (\{ \dt_{\mathfrak M_1}(d) \}, \{ \dt_{\mathfrak M_2}(e) \} ) = \dm(e)$, $\dm(d') = (\{ \dt_{\mathfrak M_1}(d') \}, \{ \dt_{\mathfrak M_2}(e') \} ) = \dm(e')$, $\odp(d) = \odp(d')$, and $\tdp(e) = \tdp(e')$. \hfill $\dashv$ 
\end{exa}

%Consider\nz{do we need this?} the $\mathfrak M_i$ below and suppose $\sub_\exists(\varphi,\psi)$ has\\[-5pt]
%
%\centerline{\includegraphics[scale=0.76]{PICS/mos5}}\\[-5pt]
%
%no formulas with $\Diamond$ in the scope of $\exists$, $\sigma = \{\avar\}$ and $\sig(\varphi, \psi) = \{\avar,\pvar\}$. Then $\wt_{\mathfrak M_1}(v) = \wt_{\mathfrak M_1}(v')$ but $v \not\sim_{\sigma} v'$ as $\exists (\avar \land \Diamond \neg \avar)$ is true at $v'$ and false at $v$; likewise, $\wt_{\mathfrak M_2}(u) = \wt_{\mathfrak M_2}(u')$ but $u \not\sim_{\sigma} u'$. Since $v \sim_{\sigma} u$ and $v' \sim_{\sigma} u'$, we have $\wm(v) = (\{ \wt_{\mathfrak M_1}(v) \}, \{ \wt_{\mathfrak M_2}(u) \} ) = \wm(u)$, $\wm(v') = (\{ \wt_{\mathfrak M_1}(v') \}, \{ \wt_{\mathfrak M_2}(u') \} ) = \wm(u')$, $\owp(v) = \owp(v')$, and $\twp(u) = \twp(u')$. Thus, non-bisimilar worlds can give the same world-point.~$\dashv$

Suppose $\mathfrak{M}_{1},w_{1},d_{1} \sim_{\sigma} \mathfrak{M}_{2},w_{2},d_{2}$, $\mathfrak{M}_{1},w_{1},d_1 \models \varphi$ and $\mathfrak{M}_{2},w_{2}, d_2 \models\neg \psi$. We construct $\mathfrak{M}'_i = (W_i', D_i',I_i')$, $i = 1,2$, witnessing $\sigma$-bisimulation consistency of $\varphi$ and $\neg\psi$ and having at most double-exponential size in $|\varphi|+|\psi|$.
Intuitively, $W'_i$ and $D'_i$ consist of copies of the $i$-world and, respectively, $i$-domain points in $\mathfrak M_1$, $\mathfrak M_2$ rather than copies of the world- and domain-types as in Example~\ref{s5sat}. Then we obtain the required $\sigma$-\BS{} $(\bs_{W}, \bs_{D}$) by including in $\bs_{W}$ and $\bs_{D}$ exactly those 1/2-world and, respectively, 1/2-domain points that share the same world and domain mosaic.

Let $n$ be the number of full types occurring either in $\mathfrak M_1$ or $\mathfrak M_2$ (over $\sub(\varphi, \psi)$) and $[n] = \{1,\dots,n\}$. For $i = 1,2$, set 
\begin{equation*}
D_{i}' = \{ \idp^k \mid \idp \text{ an $i$-domain point in $\mathfrak{M}_{1}$, $\mathfrak{M}_{2}$, $k \in [n]$} \},
\end{equation*}
treating $\idp^k$ as the $k$th copy of $\idp$ and assuming all of the copies to be distinct. 
Next, we define $W_{i}'$, $i=1,2$, using the sets $\Pi_{\iwp,\idp}$ of all \emph{surjective} functions of the form
$$
\pi_{\iwp,\idp} \colon [n] \to \{ \ft_{\mathfrak{M}_{i}}(w,d) \mid \text{$(w,d)\in W_{i} \times D_{i}$},\quad \text{$\iwp=\iwp(w)$, $\idp = \idp(d)$}\}.
$$
Observe that, for any $\iwp = (\wt,\wm)$, $\idp = (\dt,\dm)$, $k\in [n]$, and
$\pi_{\iwp,\idp}\in \Pi_{\iwp,\idp}$ we have $\wt = \pi_{\iwp,\idp}(k) \cap \sub_\exists(\varphi,\psi)$ and $\dt =  \pi_{\iwp,\idp}(k) \cap \sub_\Diamond (\varphi,\psi)$. On the other hand, it might happen that $\pi_{\iwp,\idp}(k) = \ft_{\mathfrak{M}_{i}}(w,d)$,
but $\wm\ne \wm(w)$ or $\dm\ne\dm(d)$.  
%
%\begin{equation*}
%\wt = \pi_{\iwp,\idp}(k) \cap \sub_\exists(\varphi,\psi), \quad
%%
%\dt =  \pi_{\iwp,\idp}(k) \cap \sub_\Diamond (\varphi,\psi).
%\end{equation*}

Denote by $\Pi_{i}$ the set of all functions $\pi$ mapping every pair $\iwp,\idp$ to an element of $\Pi_{\iwp,\idp}$ and set $\pi_{\iwp,\idp}=\pi(\iwp,\idp)$.  
Let $\Pi_{i}^{\dag}\subseteq \Pi_{i}$ be a smallest set such that, for any full type $\ft$ in $\mathfrak M_i$ and any $k\in [n]$, if 
$\ft=\ft_{\mathfrak M_i}(w,d)$, for some $(w,d)\in W_i\times D_i$, then there exists $\pi\in \Pi_{i}^{\dag}$ with 
$\pi_{\iwp(w),\idp(d)}(k) = \ft$. 
%for any $\ft=\ft_{\mathfrak M_i}(w,d)$, $\iwp=\iwp(w)$, $\idp= \idp(d)$ with $(w,d)$ in $\mathfrak M_i$ and any $k\in [n]$, there is $\pi\in \Pi_{i}^{\dag}$ with $\pi_{\iwp,\idp}(k) = \ft$. 
%
We claim that $|\Pi^{\dag}_i|\leq n^{2}$, $i=1,2$. Indeed, for any $k\in [n]$ and any full type $\ft$ in $\mathfrak{M}_i$,  
we will include just a single function $f^{k,\ft}\in \Pi_i$ in $\Pi^\dag_i$.
Assume $k$ and $\ft=\ft_{\mathfrak{M}_i}(u,e)$ are given. Then we can choose $f^{k,\ft}$ to be any function mapping pairs $\iwp,\idp$ into $\Pi_{\iwp,\idp}$ such that, for any $w,d$ with $\ft=\ft_{\mathfrak{M}_{i}}(w,d)$,  we have  $f^{k,\ft}(\iwp(w),\idp(d))(k)=\ft$, where $\iwp(w)=(\wt_{\mathfrak{M}_{i}}(w),\wm(w))$ and
$\idp(d)=(\dt_{\mathfrak{M}_{i}}(d),\dm(d))$.
 Observe that there might be $w',d'$ with $\ft=\ft_{\mathfrak{M}_{i}}(w',d')$ but $\iwp(w')\not=\iwp(w)$
or $\idp(d')\not=\idp(d)$. Nevertheless, such pairs $(w,d)$, $(w',d')$ do not give conflicting requirements on the choice of $f^{k,\ft}$.
The constructed $\Pi^{\dag}_i$ is as required.

%Clearly, $|\Pi_{i}^{\dagger}|\leq n^{2}$.
	
Then we set
\begin{equation*}
W_{i}' = \{ \iwp^\pi \mid \iwp \text{ an $i$-world point in $\mathfrak{M}_{1}$, $\mathfrak{M}_{2}$, $\pi \in \Pi_{i}^{\dag}$}\}, 
\end{equation*}
treating $\iwp^\pi$ as a fresh $\pi$-copy of $\iwp$. Clearly, both $|D'_i|$ and $|W'_i|$ are double-exponential in $|\varphi|+|\psi|$. 
Finally, we set 
\begin{equation}\label{models}
\mathfrak M'_i, \iwp^\pi, \idp^k \models \pvar \quad \text{ iff } \quad \pvar \in \pi_{\iwp,\idp}(k)
\end{equation}
and define $\bs_W \subseteq W'_1 \times W'_2$ and $\bs_D \subseteq D'_1 \times D'_2$ by taking 
$(\owp^{\pi^1}, \twp^{\pi^2})\in\bs_W$ iff $\wm_1 = \wm_2$, where $\iwp = (\wt_i, \wm_i)$, for $i=1,2$; and similarly   
$(\odp^{k_1}, \tdp^{k_2})\in\bs_D$ iff $\dm_1 = \dm_2$, where $\idp = (\dt_i, \dm_i)$, for $i=1,2$.

%The next two lemmas show that the $\mathfrak{M}_{i}'$ are as required.

\begin{lem}\label{types}
$(i)$ $\mathfrak M'_i, \iwp^\pi, \idp^k \models \rho$ iff $\rho \in \pi_{\iwp,\idp}(k)$, for every $\rho \in \sub(\varphi, \psi)$. 

$(ii)$ The pair $(\bs_{W},\bs_{D})$ is a $\sigma$-\BS{} between $\mathfrak M'_1$ and $\mathfrak M'_2$.
\end{lem}
\begin{proof}
	$(i)$ The proof is by induction on the construction of $\rho$, with the basis given by~\eqref{models}. For the induction step, suppose first that $\rho = \exists \xi$.
	If $\mathfrak M'_i,\iwp^\pi, \idp^k \models \rho$, then there is $\idp'^{k'}$ such that $\mathfrak M'_i, \iwp^\pi, \idp'^{k'} \models \xi$.
	By IH, $\xi\in \pi_{\iwp,\idp'}(k')$ and $\exists\xi \in \pi_{\iwp,\idp'}(k')$. 
	Then $\exists \xi \in \wt$ for $\iwp=(\wt,\wm)$, and so $\rho \in \pi_{\iwp,\idp}(k)$.
	Conversely, suppose $\rho \in \pi_{\iwp,\idp}(k)$, where $\pi_{\iwp,\idp}(k)= \ft_{\mathfrak{M}_{i}}(w,d)$ with $\iwp=\iwp(w)$ and $\idp=\idp(d)$. Then there is $d'$ with $\mathfrak{M}_{i}, w,d' \models \xi$. Let $\idp'=\idp(d')$. As $\pi_{\iwp,\idp}$ is surjective, there is $k'$ with $\xi\in \pi_{\iwp,\idp'}(k')$. By IH, $\mathfrak M'_i, \iwp^\pi, \idp'^{k'} \models \xi$. It follows that $\mathfrak M'_i, \iwp^\pi, \idp^k \models \rho$.
	
	Next, let $\rho = \Diamond \xi$. Suppose $\mathfrak{M}_{i}',\iwp^\pi, \idp^k \models \rho$. Then there is $\iwp'^{\pi'}$ with $\mathfrak{M}_{i}',\iwp'^{\pi'}, \idp^k \models \xi$. By IH, $\xi \in \pi'_{\iwp',\idp}(k)$ and $\Diamond \xi \in \pi'_{\iwp',\idp}(k)$. Then $\Diamond \xi  \in \dt$ for $\idp=(\dt,\dm)$. It follows that $\rho \in \pi_{\iwp,\idp}(k)$.
	Conversely, let $\rho \in \pi_{\iwp,\idp}(k) = \ft_{\mathfrak{M}_{i}}(w,d)$ with $\iwp=\iwp(w)$ and $\idp=\idp(d)$. Then there is $w'$ with $\mathfrak{M}_{i}, w', d \models \xi$. By the choice of $\Pi_{i}^{\dag}$, there exists $\pi' \in \Pi_{i}^{\dag}$ such that $\pi'_{\iwp',\idp}(k) = \ft_{\mathfrak{M}_{i}}(w',d)$, where $\iwp' = \iwp(w')$. Then $\xi \in \pi'_{\iwp',\idp}(k)$, and so $\mathfrak M_i', \iwp'^{\pi'}, \idp \models \xi$  by IH, whence $\mathfrak M_i', \iwp^{\pi},\idp \models \Diamond \xi$.
	
	The induction step for the Booleans is straightforward.

$(ii)$ To check (\textbf{s5}$_W$), suppose $(\owp^{\pi^1}, \twp^{\pi^2})\in\bs_W$ with $\iwp = (\wt_i, \wm_i)$, for $i = 1,2$, so $\wm_1 = \wm_2$. Take any $\odp^{k_1} \in D'_1$ with $\odp = (\dt_1, \dm_1)$. We need to find $\tdp^{k_2} \in D'_2$ such that $\tdp = (\dt_2, \dm_2)$, $\dm_1 = \dm_2$ and $\pvar \in \pi^1_{\owp,\odp}(k_1)$ iff $\pvar \in \pi^2_{\twp,\tdp}(k_2)$, for all $\pvar \in \sigma$.

	Suppose $\pi^1_{\owp,\odp}(k_1) = \ft_{\mathfrak{M}_{1}}(u,d)$, for some $(u,d)\in W_{1} \times D_{1}$. Then $\owp = \owp(u) = (\wt_{\mathfrak M_1}(u), \wm(u) )$, $\wm_1 = \wm(u) = (T_1(u), T_2(u))$, and $\odp = \odp(d) = (\dt_{\mathfrak M_1}(d), \dm(d) )$ and $\dm_1 = \dm(d) = (S_1(d), S_2(d))$. 
	As $\wm_1 = \wm_2$, we have $\wt_2 \in T_2(u)$ by \eqref{simequiv}. Thus, by \eqref{world-m} and  {\bf (wm)},
	%and so 
	there is $v \in W_2$ with $u \sim_\sigma v$ and $\twp = (\wt_{\mathfrak M_2}(v), \wm(v) )$.
	
	%There exists $w_{2} \in W_{2}$ such that $w_{2}\sim_{\sigma} w_{1}$ and $\twp = (\wt_{\mathfrak M_2}(w_2), \wm(w_2) )$.
	%
	%(The reason for this is: $\wt_{2}\in T_{2}(w_{1})$, by definition as $\wm_{1}=\wm_{2}$. Hence, again by definition, there exists $w_{2}\in W_{2}$ with 
	%$w_{2}\sim_{\sigma} w_{1}$ and $\wt_{\mathfrak{M}_{2}}(w_{2})=\wt_{2}$.
	%But then from $w_{2}\sim_{\sigma} w_{1}$ also $\wm(w_{2}) = \wm_{1}$, as required.)
	
	%(A general principal we might want to formulate behind this: 
	%\begin{itemize}
	%	\item if $w_{1} \sim_{\sigma} w_{2}$, then $\wm(w_{1}) = \wm(w_{2})$; 
	%	\item if $d_{1} \sim_{\sigma} d_{2}$, then $\dm(d_{1}) = \dm(d_{2})$.)
	%\end{itemize}
	
	Now, by 
	%(\textbf{s5}$_1$) for $\mathfrak M_i$, $i=1,2$, 
	\eqref{simW}, there exists $e \in D_2$ with $d \sim_\sigma e$ and $\ell^\sigma_{\mathfrak M_1}(u,d)= \ell^\sigma_{\mathfrak M_2}(v,e)$. By {\bf (dm)}, $\dm(d) = \dm(e)$. Since all functions $\pi_{\iwp,\idp}$ are surjective, there exists $k_{2}\in [n]$ with $\pi^2_{\twp,\tdp}(k_2) = \ft_{\mathfrak M_2}(v,e)$, implying that $\pvar \in \pi^1_{\owp,\odp}(k_1)$ iff $\pvar \in \pi^2_{\twp,\tdp}(k_2)$, for all $\pvar \in \sigma$.\\
	\centerline{
		\begin{tikzpicture}[>=latex,line width=0.2pt,yscale=1]
			\node[scale = 0.9] (00) at (0,0) {$\twp^{\pi_2}$};
			\node[scale = 0.9] (10) at (3,0) {$\tdp^{k_2}$};
			\node[scale = 0.9] (01) at (0,2) {$\owp^{\pi_1}$};
			\node[scale = 0.9] (11) at (3,2) {$\odp^{k_1}$};
			\draw[-] (00) -- (10) node[midway,above] {{\footnotesize $\ft(v,e)$}};
			\draw[-] (01) -- (11) node[midway,above] {{\footnotesize $\ft(u,d)$}};
			\draw[-,dashed,bend left=25] (0,.3) to (0,1);
			\draw[-,dashed,bend right=25] (0,1) to (0,1.7);
			\node[]  at (-.3,.5) {$\sigma$};
			\draw[-,dashed,bend left=25] (3,.3) to (3,1);
			\draw[-,dashed,bend right=25] (3,1) to (3,1.7);
			\node[]  at (2.7,.5) {$\sigma$};
			\node[scale = 0.9] (00a) at (5,0) {$v$};
			\node[scale = 0.9] (10a) at (7,0) {$e$};
			\node[scale = 0.9] (01a) at (5,2) {$u$};
			\node[scale = 0.9] (11a) at (7,2) {$d$};
			\draw[-] (00a) -- (10a);
			\draw[-] (01a) -- (11a);
			\draw[-,dashed,bend left=25] (5,.3) to (5,1);
			\draw[-,dashed,bend right=25] (5,1) to (5,1.7);
			\node[]  at (4.7,.5) {$\sigma$};
			\draw[-,dashed,bend left=25] (7,.3) to (7,1);
			\draw[-,dashed,bend right=25] (7,1) to (7,1.7);
			\node[]  at (6.7,.5) {$\sigma$};
	\end{tikzpicture}}
	\\
	%\centerline{\includegraphics[scale=0.4]{pictureforbisimproof.pdf}}
	%
The other implication in (\textbf{s5}$_W$) is similar. 
	
	To check (\textbf{s5}$_D$), let $(\odp^{k_{1}}, \tdp^{k_{2}})\in\bs_D$ with $\idp = (\dt_i, \dm_i)$,  $i = 1,2$, so $\dm_1 = \dm_2$. Take any $\owp^{\pi^1} \in W'_1$ with $\owp = (\wt_1, \wm_1)$. We need to find $\twp^{\pi^{2}} \in W'_2$ with $\twp = (\wt_2, \wm_2)$ such that $\wm_1 = \wm_2$ and $\pvar \in \pi^1_{\owp,\odp}(k_1)$ iff $\pvar \in \pi^2_{\twp,\tdp}(k_2)$, for every $\pvar  \in \sigma$.
	
	Let $\pi^1_{\owp,\odp}(k_1) = \ft_{\mathfrak{M}_{1}}(u,d)$, for some $(u,d)\in W_{1} \times D_{1}$. Then $\owp = \owp(u) = (\wt_{\mathfrak M_1}(u), \wm(u) )$, $\wm_1 = \wm(u) = (T_1(u), T_2(u))$, and $\odp = \odp(d) = (\dt_{\mathfrak M_1}(d), \dm(d) )$ and $\dm_1 = \dm(d) = (S_1(d), S_2(d))$. 
As $\dm_1 = \dm_2$, we have $\dt_2 \in S_2(u)$ by \eqref{simequiv}. Thus, by \eqref{domain-m} and  {\bf (dm)}, 
	there exists $e \in D_{2}$ such that $e\sim_{\sigma} d$ and $\tdp = (\dt_{\mathfrak M_2}(e), \dm(e) )$.
	By \eqref{simD}, 
	%(\textbf{s5}$_2$) for $\mathfrak M_i$, $i=1,2$, 
	there is $v \in W_2$ with $u \sim_\sigma v$ and $\ell^\sigma_{\mathfrak M_1}(u,d)= \ell^\sigma_{\mathfrak M_2}(v,e)$. By {\bf (wm)}, we have $\wm(u) = \wm(v)$. By the choice of $\Pi_{2}^{\dag}$, there is $\pi^{2}\in \Pi_{2}^{\dag}$ such that $\pi^2_{\twp,\tdp}(k_2) = \ft_{\mathfrak M_2}(v,e)$, which implies that $\pvar \in \pi^1_{\owp,\odp}(k_1)$ iff $\pvar \in \pi^2_{\twp,\tdp}(k_2)$, for all $\pvar \in \sigma$.
	The other implication in (\textbf{s5}$_D$) is similar.
\end{proof}

The construction and lemmas above yield:

\begin{thm}\label{th:sizebound}
Any formulas $\varphi$ and $\neg\psi$ are $\sig(\varphi) \cap \sig(\psi)$-bisimulation consistent in $\FOMS$ iff there are witnessing $\FOMS$-models of size double-exponential in $|\varphi|+|\psi|$.
\end{thm}
\begin{proof}
	Let $\mathfrak M_1$, $\mathfrak M_2$ be 
	%$\omega$-saturated 
	$\FOMS$-models with	
	$\mathfrak{M}_{1},w_{1},d_{1} \sim_{\sigma} \mathfrak{M}_{2},w_{2},d_{2}$, $\mathfrak{M}_{1},w_{1}, d_1 \models \varphi$, $\mathfrak{M}_{2},w_{2}, d_2 \models\neg \psi$, where $\sigma = \sig(\varphi) \cap \sig(\psi)$.  Consider the models $\mathfrak{M}'_{1}$, $\mathfrak{M}'_{2}$ with 
$\sigma$-\BS{} $(\bs_W,\bs_D)$. For $i=1,2$, let $\iwp =\iwp(w_i) = (\wt_{\mathfrak M_i}(w_i), \wm(w_i))$ and let $\idp =\idp(d_i) = (\dt_{\mathfrak M_i}(d_i), \dm(w_i))$. By the choice of $\Pi_{i}^{\dag}$, we have $\pi^i$ with $\pi^i_{\iwp,\idp}(1) = \ft_{\mathfrak M_i}(w_i,d_i)$. Then $\mathfrak M'_1, \owp^{\pi^1}, \odp^1 \models \varphi$ and $\mathfrak M'_2, \twp^{\pi^2}, \tdp^1 \models\neg \psi$ by Lemma~\ref{types} $(i)$. Since $w_1 \sim_\sigma w_2$ and $d_1 \sim_\sigma d_2$, {\bf (wm)} and {\bf (dm)} imply $\wm(w_1) = \wm(w_2)$ and $\dm(d_1) = \dm(d_2)$. By Lemma~\ref{types} $(ii)$, $(\owp^{\pi^1},\twp^{\pi^2}) \in \bs_W$ and $ (\odp^{\pi^1},\tdp^{\pi^2}) \in \bs_D$, and so $\mathfrak M'_1, \owp^{\pi^1}, \odp^1 \sim_\sigma \mathfrak M'_2, \twp^{\pi^2}, \tdp^1$ by Theorem~\ref{S5bisim}.
\end{proof}

Theorems~\ref{th:sizebound},~\ref{int-crit} and~\ref{p:cipvspbdp} give the upper bound result in Theorem~\ref{thm1intro} for $\FOMS$, stating that both IEP and EDEP for $\FOMS$ are decidable in {\sc coN2ExpTime}.

\subsection{Lower bound}\label{ssec:s5xs5low}
We next prove the lower bound in Theorem~\ref{thm1intro} for $\FOMS$, stating that the IEP and EDEP for $\FOMS$ are both {\sc 2ExpTime}-hard. 

We reduce the word problem for languages recognised by exponentially space bounded alternating Turing machines (ATMs).
It is well-known that there are $2^n$-space bounded ATMs for which the recognised language is {\sc 2ExpTime}-hard~\cite{chandraAlternation1981}.
	
	%\begin{definition}
	%\rm
	A \emph{$2^n$-space bounded ATM} is a tuple $M=(Q,q_0,\Gamma,\Delta)$, 
	whose set $Q$ of states is partitioned to $\forall$-states and $\exists$-states, with the initial state $q_0$ being a $\forall$-state;
	$\Gamma$ is the tape alphabet containing the blank symbol $\blank$; 
	and 
	$
	\Delta \colon Q\times\Gamma\to\mathcal{P}\bigl(Q\times\Gamma\times\{L,R\}\bigr)
	$ 
	is the
	transition function such that $|\Delta(q,a)|$ is always either $0$ or $2$, and 
	$\forall$-states and $\exists$-states alternate on every computation path.
	$\forall$- and $\exists$-\emph{configurations} are represented by $2^n$-long sequences of symbols from $\Gamma\cup(Q\times\Gamma)$,
	with a single symbol in the sequence being from $Q\times\Gamma$.
	
	Similarly to~\cite{DBLP:conf/lics/JungW21},
	we use the following (slightly non-standard) acceptance condition.
	An \emph{accepting computation-tree} is an infinite tree of configurations such that $\forall$-configurations always have $2$ children, and $\exists$-configurations always have $1$ child (marked by $0$ or $1$).
	We say that $M$ \emph{accepts an input word} $\inpw=(\inpww_0,\inpww_1,\dots,\inpww_{n-1})$ if there
	is an accepting computation-tree with the configuration 
	$
	\cinit=\bigl((q_0,\inpww_0),\inpww_1,\dots,\inpww_{n-1},\blank,\dots,\blank\bigr)
	$
	at its root.
	Note that, starting from the standard ATM acceptance condition defined via accepting states, this can be achieved by assuming that the $2^n$-space bounded ATM terminates on every input and then modifying it to enter an infinite loop from the accepting state. 
	%\end{definition}
	
	Given a $2^n$-space bounded ATM $M$ and an input word $\inpw$ of length $n$, we will 
	%define formulas  $\varphi$ and $\psi$ of size polynomial in $|M|$ and $n$ 
	construct in polytime formulas  $\varphi$ and $\psi$
	such that 
	\begin{enumerate}
		%\label{follows}
		\item
		$\models_{\FOMS}\varphi\to\psi$, and
		
		\item
		%\label{bis}
		$M$ accepts $\inpw$ iff $\varphi$, $\neg\psi$ are $\sigma$-bisimulation consistent in $\FOMS$,
		where $\sigma=\sig(\varphi)\cap\sig(\psi)$.
	\end{enumerate}
	%
	%By Theorems~\ref{int-crit}, \ref{p:pbdpsem} and Proposition~\ref{p:cipvspbdp},
	By Theorems~\ref{p:cipvspbdp} and \ref{int-crit},
	it follows that both IEP and EDEP are {\sc 2ExpTime}-hard for $\FOMS$.
	
	One aspect of our construction is similar to that of ~\cite{DBLP:conf/aaai/ArtaleJMOW21,DBLP:conf/lics/JungW21}: we also represent accepting computation-trees as binary trees whose nodes are coloured by  predicates in $\sigma$. However, unlike the formalisms in the cited work, \FOMS{} cannot express the uniqueness of properties, and so the remaining ideas are novel. 
	One part of $\varphi$ `grows' $2^n$-many copies of $\sigma$-coloured binary trees, using a technique from 2D propositional modal logic~\cite{DBLP:conf/time/HodkinsonKKWZ03,DBLP:journals/tocl/GollerJL15}. Another part of $\varphi$ colours the tree-nodes with non-$\sigma$-symbols to ensure that, in the $m$th tree,  for each $m<2^n$, the content of the $m$th tape-cell is properly changing during the computation. Then we use ideas from Example~\ref{ex1} to make sure that the generated $2^n$-many trees are all $\sigma$-bisimilar, and so represent the same accepting computation-tree. 
	%In our reduction, we use the ideas of the aaai21,lics\nb{refs} to represent accepting computation-trees as binary node-coloured trees, with the colours being members of the common signature $\sigma$. Parts of the formula $\varphi$ `grow' $2^n$-many copies of trees of this kind, using techniques coming from ...\nb{product refs?} The remaining conjuncts in $\varphi$ colour the tree-nodes with symbols outside of $\sigma$ to ensure that in the $m$th tree, for each $m<2^n$, the content of the $m$th tape-cell is properly changing during the computation.
	%Then we significantly differ from aaai21,lics\nb{refs} by using the formula $\psi$ from Example~\ref{ex1} to enforce that the generated $2^n$-many trees are all $\sigma$-bisimilar, and so they indeed represent the same accepting computation-tree.
	
	We begin with defining the conjuncts \eqref{ffirst}--\eqref{diag} of $\varphi$.
	We will use three counters, $\cR$, $\cU$ and $\cV$,
	each counting modulo $2^n$ and implemented using $2n$-many unary predicate symbols:
	$\hvar_0^\cA,\dots,\hvar_{n-1}^\cA$, $\vvar_0^\cA,\dots,\vvar_{n-1}^\cA$ for $\cA\in \{\cR,\cU,\cV\}$.
	We write $\grid^\cA$ for $\bigwedge_{i<n}(\hvar_i^\cA\leftrightarrow \vvar_i^\cA)$, and write
	$\cvf{\cA}{=}{m}$
	for $m<2^n$, if $\grid^\cA$ holds and the $\hvar^\cA$- and $\vvar^\cA$-sequences represent $m$ in binary. 
	We use $\cvf{\cA}{<}{m}$ if $\cvf{\cA}{=}{k}$ for some $k<m$, and we use
	%$(A\ne m)$ 
	$\cvf{\cA}{\ne}{m}$ if $\cvf{\cA}{=}{k}$ for some $k\not=m$. 
	We write $\scc^\cA$ for expressing that `$\hvar^\cA$-value $=$ $\vvar^\cA$-value$+1$ (mod $2^n$)':
	$$
	\bigvee_{i<n}\Bigl(\hvar_i^\cA\land \neg \vvar_i^\cA\land
	\!\bigwedge_{j<i}\!(\neg \hvar_j^\cA\land \vvar_j^\cA)\land
	\!\!\!\!\bigwedge_{i<j<n}\!\!\!(\hvar_j^\cA\leftrightarrow \vvar_j^\cA)\Bigr) 
	\lor \bigwedge_{i<n}(\neg\hvar_i^\cA\land\vvar_i^\cA).
	$$
	%
	%In particular, $(\cA=0)$ stands for $\bigwedge_{i<n}(\neg\hvar_i^\cA\land\neg\vvar_i^\cA)$,
	%$(\cA=2^n-1)$ stands for $\bigwedge_{i<n}(\hvar_i^\cA\land\vvar_i^\cA)$, and
	%$(\cA\ne 2^{n-1})$ stands for $\bigvee_{i<n}\neg\hvar_i^\cA$.
	We express, for $\cA\in\{\cR,\cU,\cV\}$, that the $\hvar^\cA$-predicates are `modally-stable' and the $\vvar^\cA$-predicates are
	`FO-stable':
	\begin{align}
		\label{ffirst}
		& \Bh\Bv\bigwedge_{i<n}\bigl((\hvar_i^\cA\to\Bh \hvar_i^\cA)\land(\neg \hvar_i^\cA\to\Bh\neg \hvar_i^\cA)\bigr),\\
		\label{vstable}
		& \Bh\Bv\bigwedge_{i<n}\bigl((\vvar_i^\cA\to\Bv \vvar_i^\cA)\land(\neg \vvar_i^\cA\to\Bv\neg \vvar_i^\cA)\bigr).
	\end{align}
	We use the $\cR$-counter to generate $2^n$-many `special' $\grid^\cR$-points `coloured' by a fresh predicate $\start$ for the root-node of the trees representing the computation. The $\scc^\cR$-points used in generating the $\start$-points will be marked by a fresh predicate $\nextR$ (for `next $\cR$'):
	\begin{align}
		\label{root}
		& \cvf{\cR}{=}{0}\land\start,\\
		& \Bh\Bv \bigl(\start\land\cvf{\cR}{\ne}{2^n-1}\to \Dv\nextR\bigr),\\
		\label{flast}
		& \Bh\Bv (\nextR\to\Dh\start),\\
%	\end{align}
%	
%	\begin{align}	
		& \Bh\Bv(\start\to\grid^\cR),\\
		\label{genlast}
		& \Bh\Bv(\nextR\to\scc^\cR).
	\end{align}
	Then, at each $\start$-point, we `grow' an infinite binary rooted tree that we will use to represent the accepting computation-tree of $M$ on $\inpw$ as follows. The binary tree is divided into $2^n$-long `linear' levels
	(where each node has one child only): each linear $2^n$-long subpath within such a level represents a configuration.
	In addition, the infinite binary tree is branching to two at the last node of the linear subpath representing each $\forall$-configuration (see more details in the proof of Lemma~\ref{l:lbsound} below).
	
	We grow this infinite binary tree with the help of the $\cU$-counter.
	Nodes of this infinite `$\cU$-tree' are marked by a fresh predicate $\tree$, and
	the $\scc^\cU$-points used in generating the $\tree$-points will be marked by a fresh predicate $\nextU$.
	First, we generate a computation-tree `skeleton' of alternating $\forall$- and $\exists$-levels, and with appropriate branching. We use fresh predicates $\qA$ and $\qE^i$, $i=0,1$, to mark the levels, and an additional predicate $\Zvar$
	to enforce two different children at $\forall$-levels.
	Given any formula $\chi$, we write $\nextt{\chi}$ for
	$\Bv\bigl(\nextU\to\Bh(\tree\to\chi)\bigr)$. We add the following conjuncts, for $i=0,1$:
	\begin{align}
		\label{Utreefirst}
		& \Bh\Bv\bigl(\start\to \cvf{\cU}{=}{0}\land\qA\land\tree\bigr),\\
		& \Bh\Bv (\tree\to \Dv\nextU),\\
		& \Bh\Bv (\nextU\to\Dh\tree),\\
		& \Bh\Bv(\tree\to\grid^\cU),\\
		& \Bh\Bv(\nextU\to\scc^\cU),\\
		& \Bh\Bv\bigl(\tree\land \cvf{\cU}{\ne}{2^n-1}\land\qA\to\nextt{\qA}\bigr),\\
		& \Bh\Bv\bigl(\tree\land \cvf{\cU}{\ne}{2^n-1}\land\qE^i\to\nextt{\qE^i}\bigr),\\
		& \Bh\Bv\bigl(\tree\land \cvf{\cU}{=}{2^n-1}\land\qA 
		\to\Dv(\nextU\land\Zvar)\land\Dv(\nextU\land\neg\Zvar)\bigr),%\\
	\end{align}
	\begin{align}
		& \Bh\Bv\bigl(\tree\land\cvf{\cU}{=}{2^n-1}\land\qA\to\nextt{\qE^0\lor\qE^1}\bigr),\\
		\label{Utreelast}
		& \Bh\Bv\bigl(\tree\land\cvf{\cU}{=}{2^n-1}\land\qE^i\to\nextt{\qA}\bigr).
	\end{align}
	Next, for each $\gamma\in\Gamma\cup(Q\times\Gamma)$, we introduce a fresh predicate $\svar_\gamma$. 
	We initialise the computation on input $\inpw=(\inpww_0,\inpww_1,\dots, \inpww_{n-1})$, where $\inpww_i\ne\blank$ for $i<n$:
	\begin{align}\label{cinit}
		& \Bh\Bv\Bigl(\start\to\svar_{(q_0,\inpww_0)} \land\nextt{\svar_{\inpww_1}\land\dots\nextt{\svar_{\inpww_{n-1}}\land\nextt{\svar_\blank}}\dots}\Bigr),\\
		& \Bh\Bv\bigl(\tree\land\svar_\blank\land\cvf{\cU}{\ne}{2^n-1}\to\nextt{\svar_\blank}\bigr).
	\end{align}
	Next, we ensure that the subsequent configurations are properly represented. 
	Using the $\cV$-counter,
	we ensure that, for each $m<2^n$, the $\cU$-tree 
	%$T_m^{x_m}$ 
	that is grown at the $m$th $\start$-point 
	%$x_m$ 
	properly describes the `evolution' of the $m$th tape-cell's content during the accepting computation. 
	We begin with ensuring that the $\cV$-counter increases along the $\cU$-counter, and with initialising it as $2^n-1-m$ of the value $m$ of the $\cR$-counter:
	\begin{align}
		& \Bh\Bv(\tree\to\grid^\cV),\\
		%& \Bh\Bv\bigl(\tree\to\Bv(\nextU\to\nextV)\bigr)\\
		& \Bh\Bv(\nextU\to\nextV),\\
		\label{Vinit}
		& \Bh\Bv\Bigl[\start\to\bigwedge_{i<n}\bigl((\hvar_i^\cR\leftrightarrow \neg \hvar_i^\cV)\land(\vvar_i^\cR\leftrightarrow \neg \vvar_i^\cV)\bigr)\Bigr].
	\end{align}
	Below we enforce the proper evolution of the `middle' section of the $2^n$-long tape (when $0<m<2^n-1$), the two missing cells at the beginning and the end of the tape can be handled similarly. 
	
	In order to do this, we represent the transition function $\Delta$ of $M$
	by two partial functions
	\[
	f_i \colon\bigl(\Gamma\cup(Q\times\Gamma)\bigr)^3\to\bigl(\Gamma\cup(Q\times\Gamma)\bigr),\quad\mbox{for $i=0,1$},
	\] 
	giving the next content of the middle-cell for each triple of cells. 
	We ensure that the domain of the $f_i$ is proper by taking, for all $(q,\inpww)$ with $|\Delta(q,\inpww)|=0$, the conjunct
	\begin{equation}
		\label{accept}
		\Bh\Bv\neg\svar_{(q,\inpww)}.
	\end{equation}
	For each $\overline{\gamma}=(\gamma_0,\gamma_1,\gamma_2)\in \bigl(\Gamma\cup(Q\times\Gamma)\bigr)^3$ in the domain of any of the $f_i$, we write $\cell_{\overline{\gamma}}$ for
	\[
	\svar_{\gamma_0}\land\Dv\Bigl[\nextU\land\Dh\Bigl(\tree\land\bigl(\svar_{\gamma_1}\land\Dv(\nextU\land\Dh\svar_{\gamma_2})\bigr)\Bigr)\Bigr].
	\]
	In addition to the $\svar_\gamma$ variables,
	%(that are in $\Sigma$), 
	for some
	$\gamma\in\Gamma\cup(Q\times\Gamma)$, we will use
	additional variables $\svar_\gamma^0$, $\svar_\gamma^1$, and $\svar_\gamma^+$, and have the conjuncts,
	for $i=0,1$ and $\overline{\gamma}$ in the domain of any of the $f_i$:
	\begin{align}
		& \Bh\Bv\bigl(\tree\land\cvf{\cV}{=}{2^n-1}\land\cvf{\cU}{<}{2^n-2}\land\cell_{\overline{\gamma}}\land\qA 
		\to\nextt{\svar^0_{f_0(\overline{\gamma})}\land \svar^1_{f_1(\overline{\gamma})}}\bigr),\\
		& \Bh\Bv\bigl(\tree\land\cvf{\cV}{=}{2^n-1}\land\cvf{\cU}{<}{2^n-2}\land\cell_{\overline{\gamma}}\land\qE^i 
		\to\nextt{\svar^i_{f_i(\overline{\gamma})}}\bigr),\\
		& \Bh\Bv\bigl(\tree\land\cvf{\cU}{\ne}{2^n-1}\land\svar_\gamma^i\to\nextt{\svar_\gamma^i}\bigr),\\
		& \Bh\Bv\Bigl(\tree\land\cvf{\cU}{=}{2^n-1}\land\qA\land\svar_\gamma^0\to 
		\Bv\bigl(\nextU\land\Zvar\to\Bh(\tree\to\svar_\gamma^+)\bigr)\Bigr),\\
		& \Bh\Bv\Bigl(\tree\land\cvf{\cU}{=}{2^n-1}\land\qA\land\svar_\gamma^1\to 
		\Bv\bigl(\nextU\land\neg\Zvar\to\Bh(\tree\to\svar_\gamma^+)\bigr)\Bigr),\\
		& \Bh\Bv\bigl(\tree\land\cvf{\cU}{=}{2^n-1}\land\qE^i\land\svar_\gamma^i\to
		\nextt{\svar_\gamma^+}\bigr),\\
		& \Bh\Bv\bigl(\tree\land\cvf{\cV}{\ne}{2^n-1}\land\svar_\gamma^+\to
		\nextt{\svar_\gamma^+}\bigr),\\
		\label{treelast}
		& \Bh\Bv\bigl(\tree\land\cvf{\cV}{=}{2^n-1}\land\svar_\gamma^+\to
		\nextt{\svar_\gamma}\bigr).
	\end{align}
	Finally, we introduce a fresh predicate $\diag$ that will `interact' with the formula $\psi$. We add conjuncts to $\varphi$ ensuring that 
	each of the generated $\cU$-trees stays within the 
	%`$\diag$-square' 
	`$\cR$-domain' of its root $\start$-point
	(meaning every node of these trees is an $\diag$-point having the same $\cR$-value): 
	\begin{align}
		\label{startL}
		& \Bh\Bv(\tree\lor\nextU\to\diag),\\
		\label{diag}
		& \Bh\Bv(\diag\to\grid^\cR).
		%\label{samesqh}
		%& \Bh\Bv\bigwedge_{i<n}\bigl(\diag\land \hvar_i^\cR\to\Bv(\diag\to \hvar_i^\cR)\bigr)\\
		%\label{samesqv}
		%& \Bh\Bv\bigwedge_{i<n}\bigl(\diag\land \vvar_i^\cR\to\Bh(\diag\to \vvar_i^\cR)\bigr).
	\end{align}
	By this, we have completed the definition of $\varphi$.
	
	Next, using the second formula of Example~\ref{ex1}, we define the formula $\psi$ such that $\sig(\varphi)\cap\sig(\psi)$ is the set
	\[
	\sigma=\bigl\{\diag,\start,\nextU,\Zvar,\tree,\qA,\qE^0,\qE^1\bigr\}\cup
	\bigl\{\svar_\gamma\mid\gamma\in\Gamma\cup(Q\times\Gamma)\bigr\}.
	\]
	We let
	$$
	\psi ~=~  \chi\land\Bh\Bv (\diag \leftrightarrow  \bvar_0 \lor \bvar_1) \to  \Dh\Dv  \bigl(\bvar_0 \land \Dh (\neg \diag \land \Dv  \bvar_0)\bigr) \lor{} 
	\Dh\Dv  \bigl(\bvar_1 \land \Dh (\neg \diag \land \Dv  \bvar_1)\bigr),
	$$
	where %$\chi$ is a dummy containing the predicates in $\sigma\mathop{\setminus}\{\diag\}$, 
	$\chi=\bigwedge_{\pvar\in\sigma\mathop{\setminus}\{\diag\}}(\pvar\to\pvar)$
	and $\bvar_0$, $\bvar_1$ are fresh predicates.
	
	\begin{lem}\label{l:impvalid}
		If $n>1$ then $\models_{\FOMS}\varphi\to\psi$.
	\end{lem}
	\begin{proof}
		Suppose $\mathfrak M,w_0,d_0\models\varphi\land\Bh\Bv(\diag\leftrightarrow \bvar_0\lor \bvar_1)$ for some model $\mathfrak M=(W,D,I)$. Then, by \eqref{root}--\eqref{genlast}, we have at least $2^n\mathop{>} 3$ different $\start$-points $(w_0,d_0), \dots,(w_{2^n-1},d_{2^n-1})$ in $W\times D$, with the respective
		$\cR$-values $0,\dots,2^n\mathop{-}1$.
		As $\start$-points are also $\diag$-points by \eqref{Utreefirst} and \eqref{startL},
		the pigeonhole principle implies that there are $i\ne j<2^{n-1}$, $k\in\{0,1\}$ such that $\mathfrak M,w_i,d_i\models \bvar_k$ and $\mathfrak M,w_j,d_j\models \bvar_k$.
		Then $\mathfrak M,w_j,d_i\models\neg\diag$ by 
		%\eqref{samesqv}, 
		\eqref{ffirst}, \eqref{vstable} and \eqref{diag}, 
		and so  $\mathfrak M,w_0,d_0\models\psi$.
	\end{proof}
	
	\begin{lem}\label{l:lbsound}
		If $M$ accepts $\inpw$ then $\varphi$, $\neg\psi$ are $\sigma$-bisimulation consistent in $\FOMS$.
	\end{lem}
	\begin{proof}
		Let $\treemod=(\treedom,\treerel_0,\treerel_1,\qA,\qE^0,\qE^1,\svar_\gamma)_{\gamma\in\Gamma\cup(Q\times\Gamma)}$ be the infinite binary tree-shaped FO-structure with root $r\in\treedom$ and binary predicates $\treerel_0,\treerel_1$, that represents the accepting computation-tree of $M$ on $\inpw$ as discussed after formula \eqref{genlast} above, that is, configurations are represented by subpaths of $2^n$-many nodes linked by $\treerel_0$.
		Every node of the $2^n$-long subpath representing a $\forall$-configuration is coloured by $\qA$.
		The last node representing a $\forall$-configuration has one $S_i$-child, for each of $i=0,1$, where the representations of the two subsequent $\exists$-configurations start.
		For $i=0,1$, if it is the $i$-child of an $\exists$-configuration $\config$ that is present in the computation-tree, then
		every node of the $2^n$-long subpath representing $\config$ is coloured by $\qE^i$
		(see Fig.~\ref{f:ctree1} for an example).
		The last node representing an $\exists$-configuration has one $S_0$-child, where the representation of the next configuration starts.
		Nodes representing a configuration are also coloured
		with $\svar_\gamma$ for the corresponding symbol $\gamma$ from 
		$\Gamma\cup(Q\times\Gamma)$.
		
\begin{figure}
			%\vspace*{1cm}
			%\includegraphics[scale=.3]{PICS/ctree1.png}
			\begin{center}
				\begin{tikzpicture}[decoration={brace,mirror,amplitude=7},line width=0.6pt,xscale =.45,yscale =.45]
					\node[point,scale=0.7,label = below:$\vdots$] (l1) at (0,0) {};
					\node[point,scale=0.7,label = below:$\vdots$] (l2) at (2,0) {};
					\node[point,scale=0.7,label = below:$\vdots$] (l3) at (4,0) {};
					\node[point,scale=0.7,label = below:$\vdots$] (l4) at (6,0) {};
					\node[point,scale=0.7] (a1) at (1,2) {};
					\node[]  at (1,3.8) {$\vdots$};
					\node[point,scale=0.7] (a2) at (1,5) {};
					\node[point,scale=0.7] (a3) at (1,7) {};
					%\node[point,scale=0.7] (a4) at (1,9) {};
					\node[]  at (1,11) {$\vdots$};
					\node[]  at (1,9.5) {$\config_3$};
					\node[]  at (1,8.5) {$\vdots$};
					\node[point,scale=0.7] (a5) at (1,12) {};
					\node[point,scale=0.7] (a6) at (1,14) {};
					\node[point,scale=0.7] (b1) at (5,2) {};
					\node[]  at (5,3.8) {$\vdots$};
					\node[point,scale=0.7] (b2) at (5,5) {};
					\node[point,scale=0.7] (b3) at (5,7) {};
					%\node[point,scale=0.7] (b4) at (5,9) {};
					\node[]  at (5,11) {$\vdots$};
					\node[]  at (5,9.5) {$\config_4$};
					\node[]  at (5,8.5) {$\vdots$};
					\node[point,scale=0.7] (b5) at (5,12) {};
					\node[point,scale=0.7] (b6) at (5,14) {};
					\node[point,scale=0.7] (c1) at (1,16) {};
					\node[]  at (1,17.8) {$\vdots$};
					\node[point,scale=0.7] (c2) at (1,19) {};
					\node[point,scale=0.7] (c3) at (1,21) {};
					%\node[point,scale=0.7] (c4) at (1,23) {};
					\node[]  at (1,25) {$\vdots$};
					\node[]  at (1,23.5) {$\config_1$};
					\node[]  at (1,22.5) {$\vdots$};
					\node[point,scale=0.7] (c5) at (1,26) {};
					\node[point,scale=0.7] (c6) at (1,28) {};
					\node[point,scale=0.7] (d1) at (5,16) {};
					\node[]  at (5,17.8) {$\vdots$};
					\node[point,scale=0.7] (d2) at (5,19) {};
					\node[point,scale=0.7] (d3) at (5,21) {};
					%\node[point,scale=0.7] (d4) at (5,23) {};
					\node[]  at (5,25) {$\vdots$};
					\node[]  at (5,23.5) {$\config_2$};
					\node[]  at (5,22.5) {$\vdots$};
					\node[point,scale=0.7] (d5) at (5,26) {};
					\node[point,scale=0.7] (d6) at (5,28) {};
					\node[point,scale=0.7,label = left:$\svar_\blank$] (e1) at (3,30) {};
					\node[]  at (3,31.8) {$\vdots$};
					\node[point,scale=0.7,label = left:$\svar_\blank$] (e2) at (3,33) {};
					\node[point,scale=0.7,label = left:$\svar_{\inpww_{n-1}}$] (e3) at (3,35) {};
					%\node[point,scale=0.7] (e4) at (3,37) {};
					\node[]  at (3,36.5) {$\vdots$};
					\node[]  at (2.8,37.5) {$\cinit$};
					\node[]  at (3,39) {$\vdots$};
					\node[point,scale=0.7,label = left:$\svar_{\inpww_1}$] (e5) at (3,40) {};
					\node[point,scale=0.7,label = left:$\svar_{(q_0,\inpww_0)}$] (e6) at (3,42) {};
					\draw[->] (e6) to (e5);
					%\draw[->] (e4) to (e3);
					\draw[->] (e3) to (e2);
					\draw[->,left] (e1) to node[] {\scriptsize $S_0$} (c6);
					\draw[->] (c6) to (c5);
					%\draw[->] (c4) to (c3);
					\draw[->] (c3) to (c2);
					\draw[->,left] (c1) to node[] {\scriptsize $S_0$} (a6);
					\draw[->,right] (e1) to node[] {\scriptsize $S_1$} (d6);
					\draw[->] (d6) to (d5);
					%\draw[->] (d4) to (d3);
					\draw[->] (d3) to (d2);
					\draw[->,right] (d1) to node[] {\scriptsize $S_0$} (b6);
					\draw[->] (a6) to (a5);
					%\draw[->] (a4) to (a3);
					\draw[->] (a3) to (a2);
					\draw[->,left] (a1) to node[] {\scriptsize $S_0$} (l1);
					\draw[->,right] (a1) to node[] {\scriptsize $S_1$} (l2);
					\draw[->] (b6) to (b5);
					%\draw[->] (b4) to (b3);
					\draw[->] (b3) to (b2);
					\draw[->,left] (b1) to node[] {\scriptsize $S_0$} (l3);
					\draw[->,right] (b1) to node[] {\scriptsize $S_1$} (l4);
					\draw[-,thick] (1.4,2) -- (1.7,2) -- (1.7,14) -- (1.4,14);
					\draw[-,thick] (4.6,2) -- (4.3,2) -- (4.3,14) -- (4.6,14);
					%\node[]  at (3,8) {$\nwarrow\;\nearrow$};
					\node[]  at (3,6.2) {$\leftarrow\ \ \to$};
					\node[]  at (3,7) {$\qA$};
					\draw[-,thick] (1.4,16) -- (1.7,16) -- (1.7,28) -- (1.4,28);
					\draw[-,thick] (4.6,16) -- (4.3,16) -- (4.3,28) -- (4.6,28);
					%\node[]  at (3,22) {$\nwarrow\;\nearrow$};
					\node[]  at (3,20) {$\leftarrow\ \ \to$};
					\node[]  at (3,20.8) {$\qE^1$};
					\draw[-,thick] (3.4,30) -- (3.7,30) -- (3.7,42) -- (3.4,42);
					\node[]  at (4.7,35) {$\qA$};
					\node[]  at (-2,20) {$\leadsto$};
					\node[scale=0.7,label = below:$\vdots$] (tl1) at (-9,17) {};
					\node[scale=0.7,label = below:$\vdots$] (tl2) at (-7,17) {};
					\node[scale=0.7,label = below:$\vdots$] (tl3) at (-6,17) {};
					\node[scale=0.7,label = below:$\vdots$] (tl4) at (-4,17) {};
					\node[] (t0) at (-7.5,23) {$\cinit$};
					\node[scale=1] (t1) at (-9,21) {$\config_1$};
					\node[] (t2) at (-6,21) {$\config_2$};
					\node[] (t3) at (-8,19) {$\config_3$};
					\node[] (t4) at (-5,19) {$\config_4$};
					\node[]  at (-11,17) {$\exists$};
					\node[]  at (-11,19) {$\forall$};
					\node[]  at (-11,21) {$\exists$};
					\node[]  at (-11,23) {$\forall$};
					\draw[->,left] (t0) to node[] {\scriptsize $0$} (t1);
					\draw[->,above left] (t3) to node[] {\scriptsize $0$} (tl1);
					\draw[->,above right] (t3) to node[] {\scriptsize $1$} (tl2);
					\draw[->,above left] (t4) to node[] {\scriptsize $0$} (tl3);
					\draw[->,above right] (t4) to node[] {\scriptsize $1$} (tl4);
					\draw[->,right] (t0) to node[] {\scriptsize $1$} (t2);
					\draw[->,right] (t1) to node[] {\scriptsize $1$} (t3);
					\draw[->,right] (t2) to node[] {\scriptsize $1$} (t4);
					\draw [decorate] ([xshift = 34mm, yshift = 1mm]e1.east) --node[right=2mm]{\footnotesize $2^n\; S_0$-steps} ([xshift = 34mm, yshift = 1mm]e6.east);
					\draw [decorate] ([xshift = 14mm, yshift = 1mm]d1.east) --node[right=2mm]{\footnotesize $2^n\; S_0$-steps} ([xshift = 14mm, yshift = 1mm]d6.east);
					\draw [decorate] ([xshift = 14mm, yshift = 1mm]b1.east) --node[right=2mm]{\footnotesize $2^n\; S_0$-steps} ([xshift = 14mm, yshift = 1mm]b6.east);
					\node[]  at (5,42.5) {$\treemod$};
				\end{tikzpicture}
			\end{center}
			\caption{Representing accepting computation-trees.}\label{f:ctree1}
		\end{figure}
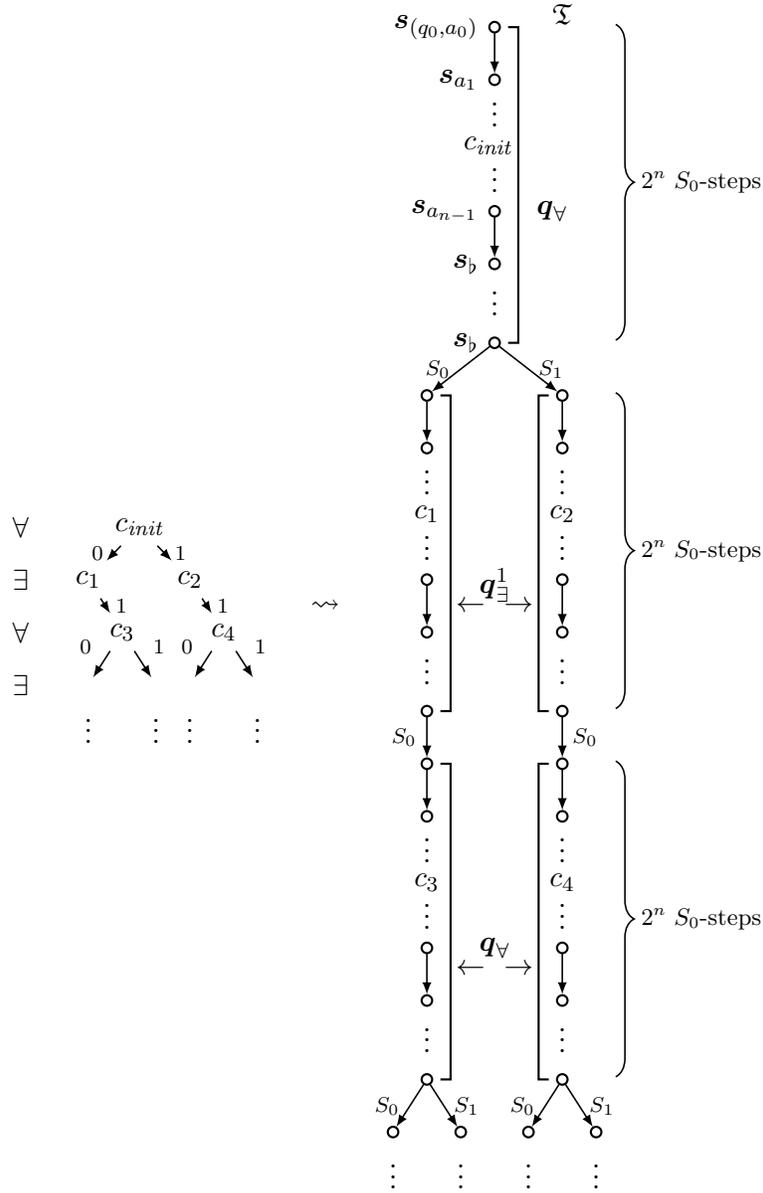

		We begin by defining a model $\mathfrak M=(W,D,I)$ making $\varphi$ true.
		Take $2\cdot 2^n$ many disjoint copies $W_m$ and $D_m$, $m<2^n$, of $\treedom$ and let
		$W=\bigcup_{m<2^n}W_m$ and $D=\bigcup_{m<2^n}D_m$. For each  $m<2^n$ and $\treet\in \treedom$, let
		$w_m^\treet$ and $d_m^\treet$ denote the copy of $\treet$ in $W_m$ and $D_m$, respectively.
		We define $I$ first for the symbols in $\sigma$. For all $m<2^n$, $\treet\in\treedom$ and
		$\pvar\in\{\qA,\qE^0,\qE^1\}\cup\{\svar_\gamma\mid\gamma\in\Gamma\cup(Q\times\Gamma)\}$, we let
		\begin{align}
			\label{Ifirst}
			\diag^{I(w_m^\treet)} & = \{d_m^{\treet'} \mid \treet' \in \treedom \},\\
			\start^{I(w_m^\treet)} & = \left\{
			\begin{array}{ll}
				\{d_m^{\treet}\}, & \mbox{if $\treet=r$,}\\[3pt]
				\emptyset, &\mbox{otherwise,}
			\end{array}
			\right.\\
			(\nextU)^{I(w_m^\treet)} & = \{d_m^{\treet'} \mid  \treerel_0(\treet,\treet')\mbox{ or } \treerel_1(\treet,\treet')\},%\\
\end{align}
\begin{align}
			\Zvar^{I(w_m^\treet)} & = \{d_m^{\treet'} \mid  \treerel_0(\treet,\treet') \},\\
			\tree^{I(w_m^\treet)} & = \{d_m^{\treet}\},\\
			\label{Ilast}
			\pvar^{I(w_m^\treet)} & = \left\{
			\begin{array}{ll}
				\{d_m^{\treet}\}, & \mbox{if $\pvar(\treet)$ holds in $\treemod$,}\\[3pt]
				\emptyset, &\mbox{otherwise.}
			\end{array}
			\right.
		\end{align}

		Next, we define $I$ for the symbols not in $\sigma$. The $\hvar^\cR_i$- and $\vvar^\cR_i$-predicates, for $i<n$, set up a binary counter counting from $0$ to $2^n\mathop{-}1$ on pairs $(w_0^r,d_0^r)$, $\dots$, $(w_{2^n-1}^r,d_{2^n-1}^r)$ in such a way
		that they are 
		\begin{itemize}
			\item[--]
			stable within each $W_m\times D_m$, $m<2^n$: 
			if $\mathfrak M,w_m^r,d_m^r\models\hvar^\cR_i$ then
			$\mathfrak M,w,d\models\hvar^\cR_i$ for all $w\in W_m$, $d\in D_m$;
			if $\mathfrak M,w_m^r,d_m^r\models\vvar^\cR_i$ then
			$\mathfrak M,w,d\models\vvar^\cR_i$ for all $w\in W_m$, $d\in D_m$;
			
			\item[--]
			the $\hvar^\cR_i$-predicates are modally-stable:  if $\mathfrak M,w,d\models\hvar^\cR_i$ for some
			$w\in W$ and $d\in D$ then
			$\mathfrak M,w',d\models\hvar^\cR_i$ for all $w'\in W$;
			
			\item[--]
			the $\vvar^\cR_i$-predicates are FO-stable:  if $\mathfrak M,w,d\models\vvar^\cR_i$ for some
			$w\in W$ and $d\in D$ then
			$\mathfrak M,w,d'\models\vvar^\cR_i$ for all $d'\in D$.
		\end{itemize}
		We let, for all $m<2^n$ and $\treet\in\treedom$,
		\[
		( \nextR)^{I(w_m^\treet)}  = \left\{
		\begin{array}{ll}
			\{d_{m+1}^{\treet}\}, & \mbox{if $m<2^n\mathop{-}1$ and $\treet=r$,}\\[3pt]
			\emptyset, &\mbox{otherwise.}
		\end{array}
		\right.
		\]
		For each $m<2^n$, the $\hvar^\cU_i$- and $\vvar^\cU_i$-predicates, for $i<n$, set up a binary counter counting from $0$ modulo $2^n$ infinitely along the levels of the tree $\treemod$, on pairs of the form $(w_m^t,d_m^t)$, for $t\in\treedom$. The $\hvar^\cU_i$-predicates are modally-stable, while the $\vvar^\cU_i$-predicates are FO-stable, in the above sense.

		Then, for each $m<2^n$, the modally-stable $\hvar^\cV_i$- and the FO-stable $\vvar^\cV_i$-predicates set up a binary counter counting from $2^n-1-m$ modulo $2^n$ infinitely along the levels of the tree $\treemod$, on pairs of the form $(w_m^t,d_m^t)$, for $t\in\treedom$. Also,
		we extend the FO-structure $\treemod$ to $\treemod_m^+$ by adding 
		%extensions for the 
		unary predicates
		%$\hvar^\cV_i,\vvar^\cV_i$, $i<n$, and 
		$\svar_\gamma^0$, $\svar_\gamma^1$, $\svar_\gamma^+$, for $\gamma\in\Gamma\cup(Q\times\Gamma)$, see Fig.~\ref{f:ctree2}. 
		For all $m<2^n$, 
		$\treet\in\treedom$, $\pvar\in\{\svar_\gamma^0,\svar_\gamma^1,\svar_\gamma^+\mid \gamma\in\Gamma\cup(Q\times\Gamma)\}$, we let 
		\[
		\pvar^{I(w_m^\treet)}  = \left\{
		\begin{array}{ll}
			\{d_m^{\treet}\},  & \mbox{if $\pvar(\treet)$ holds in $\treemod_m^+$,}\\[3pt]
			\emptyset, &\mbox{otherwise.}
		\end{array}
		\right.
		\]
		It is readily checked that $\mathfrak M,w_0^r,d_0^r\models\varphi$. 
		
		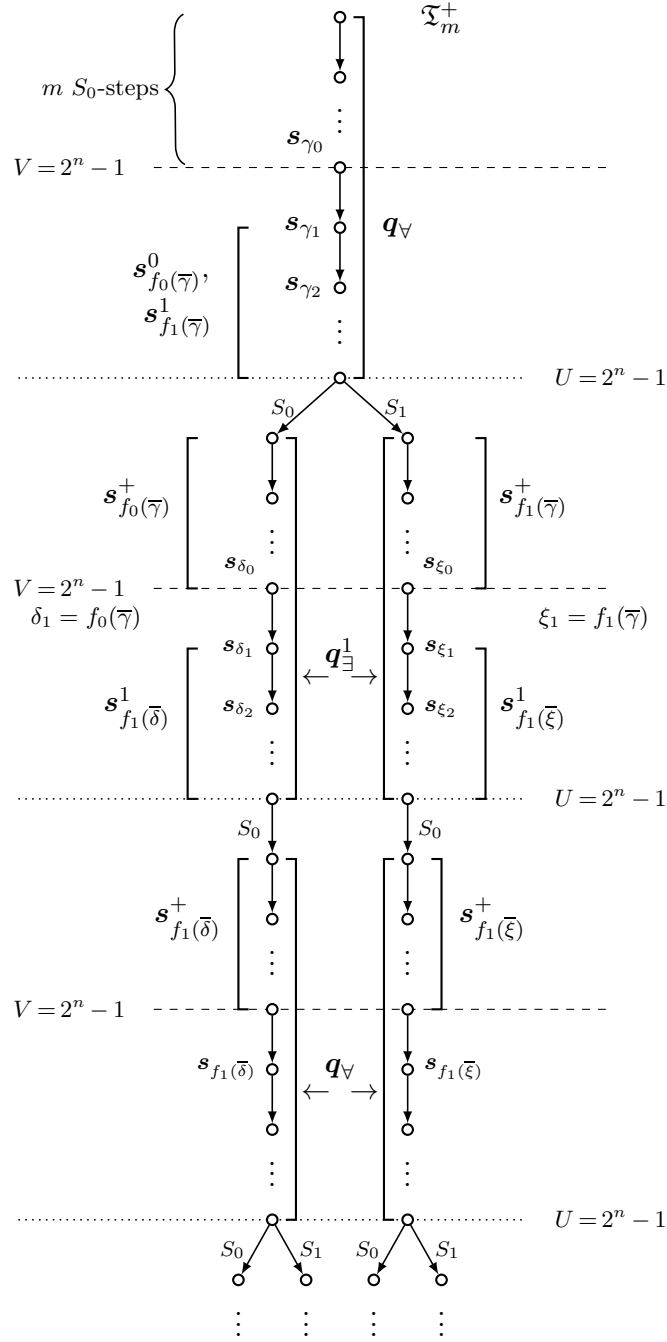
\begin{figure}
			%\hspace*{1cm}
			%\includegraphics[scale=.47]{PICS/ctree2.png}
			\begin{center}
				\begin{tikzpicture}[decoration={brace,mirror,amplitude=7},line width=0.6pt,xscale =.45,yscale =.45]
					\node[]  at (-5,37) {{\footnotesize $\cV\mathop{=}2^n\mathop{-}1$}};
					\draw[-,dashed,thin] (-2.5,37) -- (11,37);
					\node[]  at (-5,23) {{\footnotesize $\cV\mathop{=}2^n\mathop{-}1$}};
					\node[]  at (-4.5,22) {{\footnotesize $\delta_1=f_0(\overline{\gamma})$}};
					\node[]  at (10.5,22) {{\footnotesize $\xi_1=f_1(\overline{\gamma})$}};
					\draw[-,dashed,thin] (-2.5,23) -- (11,23);
					\node[]  at (-5,9) {{\footnotesize $\cV\mathop{=}2^n\mathop{-}1$}};
					\draw[-,dashed,thin] (-2.5,9) -- (11,9);
					\draw[-,dotted] (-6.5,30) -- (8.5,30);
					\node[]  at (11,30) {{\footnotesize $\cU\mathop{=}2^n\mathop{-}1$}};
					\draw[-,dotted] (-6.5,16) -- (8.5,16);
					\node[]  at (11,16) {{\footnotesize $\cU\mathop{=}2^n\mathop{-}1$}};
					\draw[-,dotted] (-6.5,2) -- (8.5,2);
					\node[]  at (11,2) {{\footnotesize $\cU\mathop{=}2^n\mathop{-}1$}};
					\node[point,scale=0.7,label = below:$\vdots$] (l1) at (0,0) {};
					\node[point,scale=0.7,label = below:$\vdots$] (l2) at (2,0) {};
					\node[point,scale=0.7,label = below:$\vdots$] (l3) at (4,0) {};
					\node[point,scale=0.7,label = below:$\vdots$] (l4) at (6,0) {};
					\node[point,fill=white,scale=0.7] (a1) at (1,2) {};
					\node[]  at (1,3.8) {$\vdots$};
					\node[point,scale=0.7] (a2) at (1,5) {};
					\node[point,scale=0.7,label = left:{\footnotesize $\svar_{f_1(\overline{\delta})}$}] (a3) at (1,7) {};
					\node[point,fill=white,scale=0.7] (a4) at (1,9) {};
					\node[]  at (1,10.8) {$\vdots$};
					\node[point,scale=0.7] (a5) at (1,12) {};
					\node[point,scale=0.7] (a6) at (1,14) {};
					\node[point,fill=white,scale=0.7] (b1) at (5,2) {};
					\node[]  at (5,3.8) {$\vdots$};
					\node[point,scale=0.7] (b2) at (5,5) {};
					\node[point,scale=0.7,label = right:{\footnotesize $\svar_{f_1(\overline{\xi})}$}] (b3) at (5,7) {};
					\node[point,fill=white,scale=0.7] (b4) at (5,9) {};
					\node[]  at (5,10.8) {$\vdots$};
					\node[point,scale=0.7] (b5) at (5,12) {};
					\node[point,scale=0.7] (b6) at (5,14) {};
					\node[point,fill=white,scale=0.7] (c1) at (1,16) {};
					\node[]  at (1,17.8) {$\vdots$};
					\node[point,scale=0.7,label = left:{\footnotesize $\svar_{\delta_2}$}] (c2) at (1,19) {};
					\node[point,scale=0.7,label = left:{\footnotesize $\svar_{\delta_1}$}] (c3) at (1,21) {};
					\node[point,fill=white,scale=0.7,label = above left:{\footnotesize $\svar_{\delta_0}$}] (c4) at (1,23) {};
					\node[]  at (1,24.8) {$\vdots$};
					\node[point,scale=0.7] (c5) at (1,26) {};
					\node[point,scale=0.7] (c6) at (1,28) {};
					\node[point,fill=white,scale=0.7] (d1) at (5,16) {};
					\node[]  at (5,17.8) {$\vdots$};
					\node[point,scale=0.7,label = right:{\footnotesize $\svar_{\xi_2}$}] (d2) at (5,19) {};
					\node[point,scale=0.7,label = right:{\footnotesize $\svar_{\xi_1}$}] (d3) at (5,21) {};
					\node[point,scale=0.7,fill=white,label = above right:{\footnotesize $\svar_{\xi_0}$}] (d4) at (5,23) {};
					\node[]  at (5,24.8) {$\vdots$};
					\node[point,scale=0.7] (d5) at (5,26) {};
					\node[point,scale=0.7] (d6) at (5,28) {};
					\node[point,fill=white,scale=0.7] (e1) at (3,30) {};
					\node[]  at (3,31.8) {$\vdots$};
					\node[point,scale=0.7,label = left:$\svar_{\gamma_2}$] (e2) at (3,33) {};
					\node[point,scale=0.7,label = left:$\svar_{\gamma_1}$] (e3) at (3,35) {};
					\node[point,fill=white,scale=0.7,label = above left:$\svar_{\gamma_0}$] (e4) at (3,37) {};
					\node[]  at (3,38.8) {$\vdots$};
					\node[point,scale=0.7] (e5) at (3,40) {};
					\node[point,scale=0.7] (e6) at (3,42) {};
					\draw[->] (e6) to (e5);
					\draw[->] (e4) to (e3);
					\draw[->] (e3) to (e2);
					\draw[->,left] (e1) to node[] {\scriptsize $S_0$} (c6);
					\draw[->] (c6) to (c5);
					\draw[->] (c4) to (c3);
					\draw[->] (c3) to (c2);
					\draw[->,left] (c1) to node[] {\scriptsize $S_0$} (a6);
					\draw[->,right] (e1) to node[] {\scriptsize $S_1$} (d6);
					\draw[->] (d6) to (d5);
					\draw[->] (d4) to (d3);
					\draw[->] (d3) to (d2);
					\draw[->,right] (d1) to node[] {\scriptsize $S_0$} (b6);
					\draw[->] (a6) to (a5);
					\draw[->] (a4) to (a3);
					\draw[->] (a3) to (a2);
					\draw[->,left] (a1) to node[] {\scriptsize $S_0$} (l1);
					\draw[->,right] (a1) to node[] {\scriptsize $S_1$} (l2);
					\draw[->] (b6) to (b5);
					\draw[->] (b4) to (b3);
					\draw[->] (b3) to (b2);
					\draw[->,left] (b1) to node[] {\scriptsize $S_0$} (l3);
					\draw[->,right] (b1) to node[] {\scriptsize $S_1$} (l4);
					\draw[-,thick] (1.4,2) -- (1.7,2) -- (1.7,14) -- (1.4,14);
					\draw[-,thick] (4.6,2) -- (4.3,2) -- (4.3,14) -- (4.6,14);
					%\node[]  at (3,8) {$\nwarrow\;\nearrow$};
					\node[]  at (3,6.2) {$\leftarrow\ \ \to$};
					\node[]  at (3,7) {$\qA$};
					\draw[-,thick] (1.4,16) -- (1.7,16) -- (1.7,28) -- (1.4,28);
					\draw[-,thick] (4.6,16) -- (4.3,16) -- (4.3,28) -- (4.6,28);
					%\node[]  at (3,22) {$\nwarrow\;\nearrow$};
					\node[]  at (3,20) {$\leftarrow\ \ \to$};
					\node[]  at (3,20.8) {$\qE^1$};
					\draw[-,thick] (3.4,30) -- (3.7,30) -- (3.7,42) -- (3.4,42);
					\node[]  at (4.7,35) {$\qA$};
					\draw[-,thick] (.3,9)-- (0,9) -- (0,14) -- (.3,14);
					\node[]  at (-1.5,12) {$\svar_{f_1(\overline{\delta})}^+$};
					\draw[-,thick] (5.7,9)-- (6,9) -- (6,14) -- (5.7,14);
					\node[]  at (7.5,12) {$\svar_{f_1(\overline{\xi})}^+$};
					\draw[-,thick] (-1.2,16) -- (-1.5,16) -- (-1.5,21) -- (-1.2,21);
					\node[]  at (-3,19) {$\svar_{f_1(\overline{\delta})}^1$};
					\draw[-,thick] (-1.2,23) -- (-1.5,23) -- (-1.5,28) -- (-1.2,28);
					\node[]  at (-3,26) {$\svar_{f_0(\overline{\gamma})}^+$};
					\draw[-,thick] (7,16) -- (7.3,16) -- (7.3,21) -- (7,21);
					\node[]  at (8.7,19) {$\svar_{f_1(\overline{\xi})}^1$};
					\draw[-,thick] (7,23) -- (7.3,23) -- (7.3,28) -- (7,28);
					\node[]  at (8.7,26) {$\svar_{f_1(\overline{\gamma})}^+$};
					\draw[-,thick] (.3,30) -- (0,30) -- (0,35) -- (.3,35);
					\node[]  at (-2,33.5) {$\svar_{f_0(\overline{\gamma})}^0$,};
					\node[]  at (-1.8,32) {$\svar_{f_1(\overline{\gamma})}^1$};
					\draw [decorate] ([xshift = -48mm, yshift = 1mm]e6.east) --node[left=2mm]{\footnotesize $m\ S_0$-steps} ([xshift = -44mm, yshift = 1mm]e4.west);
					\node[]  at (6,42) {$\treemod_m^+$};
				\end{tikzpicture}
			\end{center}
			\caption{Passing information from one configuration to the next.}\label{f:ctree2}
		\end{figure}

		Next, we define a model $\hat{\mathfrak M}=(\hat{W},\hat{D},\hat{I})$ making $\neg\psi$ true.
		We take {\color{red} four} disjoint copies $\hat{W}_0,\hat{W}_1$ and $\hat{D}_0,\hat{D}_1$ of $\treedom$ and let
		$\hat{W}=\hat{W}_0\cup \hat{W}_1$ and $\hat{D}=\hat{D}_0\cup \hat{D}_1$. 
		For each  $k<2$ and $\treet\in \treedom$, let
		$\hat{w}_m^{\treet}$ and $\hat{d}_m^\treet$ denote the copy of $\treet$ in $\hat{W}_k$ and $\hat{D}_k$, respectively.
		Now, for symbols in $\sigma$ we define $\hat{I}$ similarly to $I$ in \eqref{Ifirst}--\eqref{Ilast} above.
		For symbols not in $\sigma$ the only ones with non-empty $\hat{I}$-extensions are the $\bvar_i$, for $i<2$:
		For all $i,k<2$, $\treet\in\treemod$, we let
		\[
		\bvar_i^{\hat{I}(\hat{w}_k^\treet)} = \left\{
		\begin{array}{ll}
			\{\hat{d}_i^{\treet'}\mid \treet'\in\treedom\}, & \mbox{if $k=i$,}\\[3pt]
			\emptyset, &\mbox{otherwise.}
		\end{array}
		\right.
		\]
		It is readily checked that 
		$\hat{\mathfrak M},\hat{w}_0^r,\hat{d}_0^r\models\neg\psi$. \\[6pt]
		
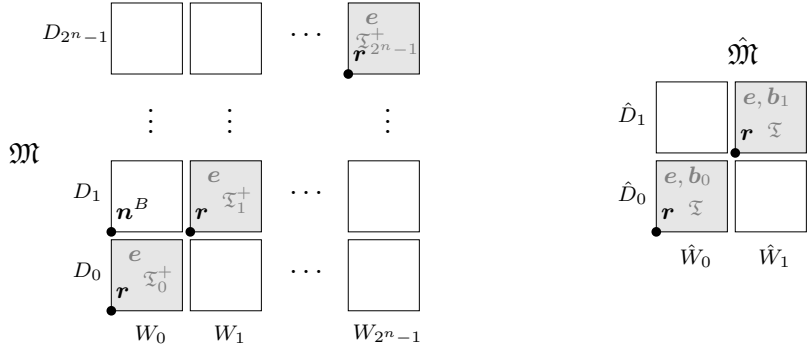
\centerline{
			\begin{tikzpicture}[>=latex,line width=0.2pt,scale = 1.05]
				\node[]  at (-1.1,2) {$\mathfrak M$};
				\node[]  at (-.3,.5) {{\scriptsize $D_0$}};
				\node[]  at (-.3,1.5) {{\scriptsize $D_1$}};
				\node[]  at (-.45,3.5) {{\scriptsize $D_{2^n-1}$}};
				\node[]  at (.5,-.3) {{\scriptsize $W_0$}};
				\node[]  at (1.5,-.3) {{\scriptsize $W_1$}};
				\node[]  at (3.5,-.3) {{\scriptsize $W_{2^n-1}$}};
				\node[]  at (2.5,.5) {$\dots$};
				\node[]  at (2.5,1.5) {$\dots$};
				\node[]  at (2.5,3.5) {$\dots$};
				\node[]  at (.5,2.5) {$\vdots$};
				\node[]  at (1.5,2.5) {$\vdots$};
				\node[]  at (3.5,2.5) {$\vdots$};
				\draw[fill=gray!20] (0,0) rectangle (.9,.9);
				\draw (1,0) rectangle (1.9,.9);
				\draw (3,0) rectangle (3.9,.9);
				\draw (0,1) rectangle (.9,1.9);
				\draw[fill=gray!20] (1,1) rectangle (1.9,1.9);
				\draw (3,1) rectangle (3.9,1.9);
				\draw (0,3) rectangle (.9,3.9);
				\draw (1,3) rectangle (1.9,3.9);
				\draw[fill=gray!20] (3,3) rectangle (3.9,3.9);
				\node[point,fill=black,scale = 0.5,label=above right:{\footnotesize $\!\!\start$}] (0) at (0,0) {};
				\node[point,fill=black,scale = 0.5,label=above right:{\footnotesize $\!\!\start$}] (1) at (1,1) {};
				\node[point,fill=black,scale = 0.5,label=above right:{\footnotesize $\!\!\start$}] (2) at (3,3) {};
				\node[point,fill=black,scale = 0.5,label=above right:{\footnotesize $\!\!\nextR$}] (n) at (0,1) {};
				\node[gray]  at (.3,.7) {{\footnotesize $\diag$}};
				\node[gray]  at (.6,.4) {{\scriptsize $\treemod_0^+$}};
				\node[gray]  at (1.3,1.7) {{\footnotesize $\diag$}};
				\node[gray]  at (1.6,1.4) {{\scriptsize $\treemod_1^+$}};
				\node[gray]  at (3.3,3.7) {{\footnotesize $\diag$}};
				\node[gray]  at (3.5,3.4) {{\scriptsize $\treemod_{2^n-1}^+$}};
				\node[]  at (8,3.3) {$\hat{\mathfrak M}$};
				\node[]  at (6.6,1.5) {{\scriptsize $\hat{D}_0$}};
				\node[]  at (6.6,2.5) {{\scriptsize $\hat{D}_1$}};
				\node[]  at (7.4,.7) {{\scriptsize $\hat{W}_0$}};
				\node[]  at (8.4,.7) {{\scriptsize $\hat{W}_1$}};
				\draw[fill=gray!20] (6.9,1) rectangle (7.8,1.9);  
				\draw (7.9,1) rectangle (8.8,1.9);
				\draw (6.9,2) rectangle (7.8,2.9);
				\draw[fill=gray!20] (7.9,2) rectangle (8.8,2.9);
				\node[point,fill=black,scale = 0.5,label=above right:{\footnotesize $\!\!\start$}] (00) at (6.9,1) {};
				\node[point,fill=black,scale = 0.5,label=above right:{\footnotesize $\!\!\start$}] (11) at (7.9,2) {};
				\node[gray]  at (7.3,1.7) {{\footnotesize $\diag,\bvar_0$}};
				\node[gray]  at (7.4,1.3) {{\scriptsize $\treemod$}};
				\node[gray]  at (8.3,2.7) {{\footnotesize $\diag,\bvar_1$}};
				\node[gray]  at (8.4,2.3) {{\scriptsize $\treemod$}};
				%
				%\draw[-,dashed,thin,bend right=15] (.6,.7) to (6.6,1.7); 
				%\draw[-,dashed,thin,bend right=15] (1.6,1.7) to (6.6,1.7); 
				%\draw[-,dashed,thin,bend right=15] (3.6,3.7) to (6.6,1.7); 
				%\draw[-,dashed,thin,bend right=15] (.6,.7) to (7.6,2.7); 
				%\draw[-,dashed,thin,bend right=15] (1.6,1.7) to (7.6,2.7); 
				%\draw[-,dashed,thin,bend right=15] (3.6,3.7) to (7.6,2.7); 
				%
				%\draw[-,dashed,thin,bend right=25] (1.2,.2) to (7.2,1.2); 
				%\draw[-,dashed,thin,bend right=35] (3.2,.2) to (7.2,1.2); 
				%\draw[-,dashed,thin,bend right=35] (3.2,1.2) to (7.2,1.2); 
			\end{tikzpicture}
		}
		
		Finally, we define a relation $\bs\subseteq (W\times D)\times(\hat{W}\times\hat{D})$ by taking, 
		for any $w,d,\hat{w},\hat{d}$, $\bigl((w,d),(\hat{w},\hat{d})\bigr)\in\bs$ iff 
		there exist $\treet,\treet'\in T$, $m,m'<2^n$, $k,k'<2$ such that $w=w_m^\treet$, $d=d_{m'}^{\treet'}$,
		$\hat{w}=\hat{w}_k^\treet$, $\hat{d}=\hat{d}_{k'}^{\treet'}$, and $m=m'$ iff $k=k'$.
		It is not hard to show that $\bs$ is a $\sigma$-bisimulation between $\mathfrak M$ and $\hat{\mathfrak M}$
		with $\bigl((w_0^r,d_0^r),(\hat{w}_0^r,\hat{d}_0^r)\bigr)\in\bs$. 
	\end{proof}
	
	\begin{lem}\label{l:lbcompl}
		If $n>1$ and $\varphi$, $\neg\psi$ are $\sigma$-bisimulation consistent, then $M$ accepts $\inpw$.
	\end{lem}
	\begin{proof}
		Let $\mathfrak M,w_0,d_0$ and $\mathfrak M',w_0',d_0'$ be models with
		$\mathfrak M,w_0,d_0\models\varphi$, $\mathfrak M',w_0',d_0'\models\neg\psi$, and 
		$\mathfrak M,w_0,d_0\sim_\sigma\mathfrak M',w_0',d_0'$.
		Let $(w_0,d_0),\dots,(w_{2^n-1},d_{2^n-1})$ be subsequent $\start$-points in $\mathfrak M$ generated by 
		\eqref{root}--\eqref{genlast}, with the respective
		$\cR$-values $0,\dots,2^n\mathop{-}1$. 
		By \eqref{Utreefirst} and \eqref{startL}, we have $\mathfrak M,w_i,d_i\models\diag$, for $i<2^n$.
		%
		%For each $i<2^n$, let $T_i^{x_i}$ denote the $\Sigma$-reduct of the tree we generate by
		We claim that for all $i,j<2^n$ there are $w_{ij},d_{ij}$ such that
		\begin{align}
			%\nonumber
			%& \mathfrak M,w_{ij},d_{ij}\models\grid^\cR\mbox{ and the $\cR$-value of $(w_{ij},d_{ij})$ is $i$}\\
			\label{cldsq}
			& \mathfrak M,w_{ij},d_{ij}\models\cvf{\cR}{=}{i},\\
			%& \hspace*{1cm} \mbox{the $\cR$-value of $(w_{ij},d_{ij})$ is $i$},\\
			\label{clsim}
			& \mathfrak M,w_{ij},d_{ij}\sim_\sigma \mathfrak M,w_j,d_j.
		\end{align}
		Indeed, to begin with, there are $w_i',d_i'$ such that $\mathfrak M,w_i,d_i\sim_\sigma\mathfrak M',w_i',d_i'$, and so
		$\mathfrak M',w_i',d_i'\models\diag$. Let $k<2^n$ and $k\ne i,j$.
		As the $\cR$-values of $(w_i,d_i)$ and $(w_k,d_k)$ are different, 
		$\mathfrak M,w_k,d_i\models\neg\diag$ follows by \eqref{diag}.
		Thus, there exist $w_k',d_k'$ such that 
		$\mathfrak M',w_k',d_i'\models\neg\diag$, 
		$\mathfrak M',w_k',d_k'\models\diag$, and 
		$\mathfrak M,w_k,d_k\sim_\sigma\mathfrak M',w_k',d_k'$.
		Similarly, there exist $w_j',d_j'$ such that 
		$\mathfrak M',w_j',d_k'\models\neg\diag$, 
		$\mathfrak M',w_j',d_j'\models\diag$, and 
		$\mathfrak M,w_j,d_j\sim_\sigma\mathfrak M',w_j',d_j'$, see Fig.~\ref{f:bis}.

		\begin{figure}[ht]
			\begin{tikzpicture}[>=latex,line width=0.2pt,scale = 1]
				\node[point,fill,scale = 0.7,label=below:{\footnotesize $w_{ij}$},label=left:{\footnotesize $d_i$},label=above right:{\footnotesize $\diag$}] (u0d0) at (0,0) {};
				\node[scale = 0.7,label=below:{\footnotesize $w_j$}] (u1d0) at (1,0) {};
				\node[point,fill=white,scale = 0.7,label=below:{\footnotesize $w_k$},label=above left:{\footnotesize $\neg\diag$}] (u2d0) at (2,0) {};
				\node[point,fill=black,scale = 0.7,label=below:{\footnotesize $w_i$},label=above left:{\footnotesize $\diag$}] (u3d0) at (3,0) {};
				\node[label=below:{\footnotesize $W$}] (w) at (3.5,0) {};
				\node[scale = 0.7,label=left:{\footnotesize $d_k$}] (u0d1) at (0,1) {};
				\node[point,fill=white,scale = 0.7,label=above left:{\footnotesize $\neg\diag$}] (u1d1) at (1,1) {};
				\node[point,fill=black,scale = 0.7,label=above left:{\footnotesize $\diag$}] (u2d1) at (2,1) {};
				\node[] (u3d1) at (3,1) {};
				\node[scale = 0.7,label=left:{\footnotesize $d_j$}] (u0d2) at (0,2) {};
				\node[point,fill=black,scale = 0.7,label=below left:{\footnotesize $\diag$}] (u1d2) at (1,2) {};
				%\node[label=above:{\footnotesize $\diag$},label=right:{\footnotesize $\pvar_2$}] (u2d2) at (2,2) {};
				\node[] (u3d2) at (3,2) {};
				\node[point,fill=black,scale = 0.7,label=left:{\footnotesize $d_{ij}$},label=above right:{\footnotesize $\diag$}] (u0d3) at (0,3) {};
				\node[label=left:{\footnotesize $D$}] (d) at (0,3.5) {};
				\draw[-] (u0d0) to (u2d0);
				\draw[->] (u2d0) to (w);
				\draw[-] (0,1) -- (u1d1);
				\draw[-] (u1d1) -- (3,1);
				\draw[-] (0,2) -- (3,2);
				\draw[-] (u0d3) -- (3,3);
				\draw[->] (u0d0) to (d);
				\draw[-] (1,0) -- (u1d1);
				\draw[-] (u1d1) -- (1,3);
				\draw[-] (u2d0) -- (2,3);
				\draw[-] (u3d0) -- (3,3);
				\node[point,fill=black,scale = 0.7,label=below:{\footnotesize $w_{i}'\ $},label=left:{\footnotesize $d_i'$},label=above right:{\footnotesize $\bvar_0$},label=below right:{\footnotesize $\diag$}] (Pu0d0) at (7,.5) {};
				\node[point,fill=white,scale = 0.7,label=below:{\footnotesize $w_{k}'$},label=above right:{\footnotesize $\neg\diag$}] (Pu1d0) at (8,.5) {};
				\node[point,fill=black,scale = 0.7,label=below:{\footnotesize $w_{j}'$},label=above right:{\footnotesize $\diag$}] (Pu2d0) at (9,.5) {};
				\node[label=below:{\footnotesize $W'$}] (Pw) at (9.7,.5) {};
				\node[scale = 0.7,label=left:{\footnotesize $d_k'$}] (Pu0d1) at (7,1.5) {};
				\node[point,fill=black,scale = 0.7,label=below right:{\footnotesize $\bvar_1$},label=above right:{\footnotesize $\diag$}] (Pu1d1) at (8,1.5) {};
				\node[point,fill=white,scale = 0.7,label=right:{\footnotesize $\neg\diag$}] (Pu2d1) at (9,1.5) {};
				\node[scale = 0.7,label=left:{\footnotesize $d_j'$}] (Pu0d2) at (7,2.5) {};
				\node[point,fill=black,scale = 0.7,label=below right:{\footnotesize $\bvar_0$},label=above right:{\footnotesize $\diag$}] (Pu2d2) at (9,2.5) {};
				\node[label=left:{\footnotesize $D'$}] (Pd) at (7,3) {};
				\draw[-] (Pu0d0) to (Pu1d0);
				\draw[->] (Pu1d0) to (Pw);
				\draw[-] (7,1.5) -- (Pu2d1);
				\draw[-] (7,2.5) -- (Pu2d2);
				\draw[->] (Pu0d0) -- (7,3);
				\draw[-] (Pu1d0) -- (8,2.5);
				\draw[-] (Pu2d0) to (Pu2d1);
				\draw[-] (Pu2d1) to (Pu2d2);
				\draw[-,dashed,thin] (u0d3) to (u1d2); 
				\draw[-,dashed,thin,bend left=20] (u0d3) to (Pu2d2); 
				\draw[-,dashed,thin,bend left=25] (u1d2) to (6,2); 
				\draw[-,dashed,thin,bend right=25] (6,2) to (Pu2d2); 
				\draw[-,dashed,thin,bend right=25] (u2d1) to (Pu1d1); 
				\draw[-,dashed,thin,bend right=32] (u3d0) to (Pu0d0); 
				\draw[-,dashed,thin,bend right=32] (u0d0) to (Pu2d0); 
			\end{tikzpicture}
			\caption{Enforcing $2^n$ $\sigma$-bisimilar trees.}\label{f:bis}
		\end{figure}
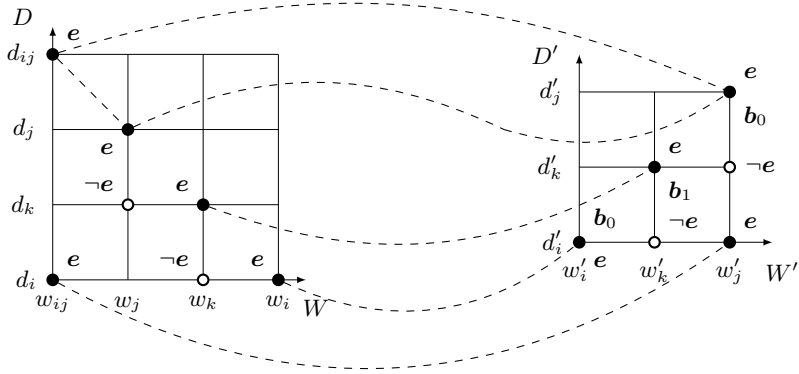

		As $\mathfrak M',w_0',d_0'\models\neg\psi$, we have 
		$\mathfrak M',w_i',d_i'\models\bvar_s$ for $s=0$ or $s=1$. 
		Suppose $s=0$ (the other case is similar). It follows from $\neg\psi$ that 
		$\mathfrak M',w_k',d_k'\models\bvar_1$ and $\mathfrak M',w_j',d_j'\models\bvar_0$.
		Then  $\mathfrak M',w_j',d_i'\models\diag$ also follows from $\neg\psi$.
		As $\mathfrak M,w_i,d_i\sim_\sigma\mathfrak M',w_i',d_i'$, there are $w_{ij},d_{ij}$ with
		$\mathfrak M,w_{ij},d_i\models\diag$, 
		$\mathfrak M,w_{ij},d_{ij}\models\diag$ and 
		$\mathfrak M,w_{ij},d_{ij}\sim_\sigma\mathfrak M',w_j',d_j'$. 
		Thus, \mbox{$\mathfrak M,w_{ij},d_{ij}\sim_\sigma\mathfrak M,w_j,d_j$}, and so \eqref{clsim} holds.
		We also have $\mathfrak M,w_{ij},d_{ij}\models\grid^\cR$ by \eqref{diag}, and so
		\eqref{cldsq} follows from 
		%\eqref{samesqv} and \eqref{vstable}.
		\eqref{vstable}.

		In particular, \eqref{cldsq} and \eqref{clsim} imply that, for each $m<2^n$, there exist $w_m^+,d_m^+$ such that 
		$\mathfrak M,w_m^+,d_m^+\models\cvf{\cR}{=}{m}$
		%$\mathfrak M,w_m^+,d_m^+\models\grid^\cR$ and the $\cR$-value of $(w_m^+,d_m^+)$ is $m$,
		and $\mathfrak M,w_m^+,d_m^+\sim_\sigma \mathfrak M,w_0,d_0$.
		Take the $\cU$-tree $\treemod_0$ grown from $(w_0,d_0)$ by \eqref{Utreefirst}--\eqref{Utreelast}. Then it has a $\sigma$-bisimilar copy $\treemod_m'$ grown from
		each $(w_m^+,d_m^+)$.
		Choose a computation-tree `skeleton' from $\treemod_0$ determined by its $\qE^i$ labels.
		As $\diag,\start,\Zvar,\nextU,\tree,\qA,\qE^i\in\sigma$,  the formulas \eqref{cinit}--\eqref{treelast}
		imply that the $\svar_\gamma$ labels in $\treemod_m'$
		properly describe the `evolution' of the $m$th tape-cell's content via the chosen `skeleton'\!. As all $\svar_\gamma$ are in $\sigma$,
		using its $\svar_\gamma$ labels and  \eqref{accept}, we can extract an accepting computation-tree from 
		$\treemod_0$ (with all tape-cell contents evolving properly).
	\end{proof}
	
	This completes the lower bound proof of Theorem~\ref{thm1intro} for $\FOMS{}$. 
The lower bound result of Theorem~\ref{thm2intro} for equality-free \FOT{} is proved in Appendix~\ref{s:FO}.

Note that the 2\textsc{ExpTime} lower bound results of Theorem~\ref{thm1intro} hold even if we want to decide, for any \FOM-formulas $\varphi$ and $\psi$, whether an interpolant or an explicit definition exists not only in \FOMS{} but in any finite-variable fragment of first-order $\SF$.
More precisely, we claim that,
for any $n,\ell<\omega$ with $2^n>\ell+1$, given a $2^n$-space bounded ATM $M$ and an input word $\inpw$ of length $n$, there exist
%define formulas  $\varphi$ and $\psi$ of size polynomial in $|M|$ and $n$ 
polytime \FOM-formulas  $\varphi$ and $\psi_\ell$ such that 
\begin{enumerate}
\item
$\models_{\FOMS}\varphi\to\psi_\ell$, and 

\item
$M$ accepts $\inpw$ iff $\varphi$, $\neg\psi_\ell$ are $\sigma$-bisimulation consistent in the $\ell$-variable fragment of quantified $\SF$, where $\sigma=\sig(\varphi)\cap\sig(\psi_\ell)$. 
\end{enumerate}
Indeed, $\varphi$ is the same as before and $\psi_\ell$ is similar to $\psi=\psi_1$ above: we just divide $\diag$ not into two
but $\ell+1$ parts, using fresh variables $\bvar_0,\dots,\bvar_{\ell}$. The proof that these modifications work is similar to the proof above.

%As in~\cite{DBLP:conf/lics/JungW21}, we represent accepting computation-trees as binary trees whose nodes are coloured by  predicates in $\sigma$. One part of $\varphi$ `grows' $2^n$-many copies of trees of this kind, in a way similar to~\cite{DBLP:conf/time/HodkinsonKKWZ03,DBLP:journals/tocl/GollerJL15}. The remainder of $\varphi$ colours the tree-nodes with non-$\sigma$-symbols to ensure that, in each $m$th tree,  $m<2^n$, the content of the $m$th tape-cell is properly changing during the computation. 
%A totally novel idea is the use of $\psi$ from Example~\ref{ex1} to make sure that the generated $2^n$-many trees are all $\sigma$-bisimilar, and so represent the same accepting computation-tree.
%

\section{The (S)CEP and UIEP in \QSF{} are Undecidable}\label{sec: CEP-S5}

We now turn to the (strong) conservative extension and uniform interpolant existence problems, which, in contrast to interpolant existence, turn out to be undecidable. We show the following refinement of Theorem~\ref{thm3intro}.

\begin{thm}\label{undec:cons}
$(i)$ The \textup{(}S\textup{)}CEP in \FOMS{} is undecidable.

$(ii)$ The UIEP in \QSF{} is undecidable.
\end{thm}	

The undecidability proof for the CEP is by adapting an undecidability proof for the CEP of \FOT{} in~\cite{DBLP:conf/icalp/JungLM0W17}. The main new idea here is the generation of arbitrary large binary trees  within \FOMS{}-models that can then be forced to be grids in case one does not have a (strong) conservative extension. The proof is by reduction of the following undecidable tiling problem.	By a \emph{tiling system} we mean a tuple $\tiling=(T,H,V,\ovar,\bbvarup,\bbvarright)$, where
$T$ is a finite set of \emph{tiles} with 
%the designated elements
$\ovar,\bbvarup,\bbvarright\in T$, and $H,V\subseteq T\times T$ are 	
horizontal and vertical \emph{matching relations\/}. 
%A \emph{solution} to $\tiling$ is a triple
We say that $\tiling$ \emph{has a solution} if there exists a triple
$(n,m,\tau)$, where $0<n,m < \omega$ and $\tau \colon \{0,\dots,n-1\}
\times \{0,\ldots,m-1\} \rightarrow T$, such that the following hold, for all 
$i< n$ and $j<m$:		
\begin{enumerate}[style=multiline,leftmargin=0.9cm]
	\item[\textbf{(t1)}] if $i<n-1$ then $(\tau(i,j),\tau(i+1,j)) \in H$;
	%for all $i+1<n$ and $j < m$;
	
	\item[\textbf{(t2)}] if $j<m-1$ then $(\tau(i,j),\tau(i,j+1)) \in V$;
	%for all $i< n$ and $j+1<m$;
	
	\item[\textbf{(t3)}] $\tau(i,j)=\ovar$ iff $i=j=0$;
	
	\item[\textbf{(t4)}] $\tau(i,j) = \bbvarright$ iff $i=n-1$, and $\tau(i,j) = \bbvarup$ iff $j=m-1$.
\end{enumerate}
%	
%It is undecidable whether a given tiling system has a solution~\cite{xx}.\nb{ref!}
The reader can easily show by reduction of the halting problem for Turing machines that it is undecidable whether a given tiling system has a solution; cf.~\cite{EmdeBoas97}.

For any tiling system $\tiling=(T,H,V,\ovar,\bbvarup,\bbvarright)$,
we show how to construct in polytime formulas $\varphi$ and $\psi$
such that $\tiling$ has a solution iff $\varphi \wedge \psi$ is not a (strong) conservative extension of $\varphi$. 
%In particular, models of formulas witnessing non-conservativity (satisfiable %w.r.t.~$\varphi$ but not w.r.t. $\varphi \wedge \psi$).
%
%Let $\mathfrak{T}=(T,H,V,\ovar,\bbvarup,\bbvarright)$ be a tiling system. 
For any model $\mathfrak{M}=(W,D,I)$, we mark the points on the finite grid to be tiled by a predicate $\gvar$, that is,
we let
$
\gvar^{\mathfrak M}=\{(w,d)\in W\times D\mid d\in\gvar^{I(w)}\}.
$
Then we define 
%Let $\sigma$ contain the predicate $\gvar$ denoting positions on the grid, $\xvar,\yvar$ for generating the grid, and the tiles $\tvar\in T$ (regarded as propositional variables).
%
%For any model $\mathfrak{M}=(W,D,I)$ 
the intended `horizontal' and `vertical' neighbour relations $R_{h}^{\mathfrak{M}}$ and $R_{v}^{\mathfrak{M}}$ on the grid by setting
\begin{align}\label{hrel}
	R_{h}^{\mathfrak{M}} & = \bigl\{\bigl((w,d),(w',d')\bigr) \in \gvar^{\mathfrak{M}}\times \gvar^{\mathfrak{M}} \mid  (w',d) \models \xvar,\ (w,d)\models \neg \bbvarright \bigr\}, \\
	\label{vrel}	
	R_{v}^{\mathfrak{M}} & = \bigl\{\bigl((w,d),(w',d')\bigr) \in \gvar^{\mathfrak{M}}\times \gvar^{\mathfrak{M}} \mid  (w,d') \models \yvar,\ (w,d)\models \neg \bbvarup\bigr\}.
\end{align}
We set, for any formula $\chi$: $\Bht{\chi} = \neg \Dht{\neg \chi}$, $\Bvt{\chi}=\neg \Dvt{\neg \chi}$, 
\[
\Dht{\chi} = \gvar \land \neg \bbvarright \land \Dh\bigl(\xvar \wedge \Dv (\gvar \wedge \chi)\bigr),\qquad
\Dvt{\chi}  = \gvar \land \neg \bbvarup\land \Dv\bigl(\yvar \wedge \Dh (\gvar \wedge \chi)\bigr).
\]
%
%\begin{align*}
%	\Dht{\chi} & = \gvar \land \neg \bbvarright \land \Dh\bigl(\xvar \wedge \Dv (\gvar \wedge \chi)\bigr),\\
%	\Dvt{\chi}  & = \gvar \land \neg \bbvarup\land \Dv\bigl(\yvar \wedge \Dh (\gvar \wedge \chi)\bigr).
%\end{align*}
%
Now $\varphi$ uses the following conjuncts to generate the grid: 
% (see the corrected whiteboard).
%
\begin{align}
	\nonumber
	& \ovar \land \gvar
	 \land \Bh\Bv \bigl(\gvar \wedge \neg (\bbvarup \wedge \bbvarright) \rightarrow \Dh \xvar\bigr)
	 \land \Bh\Bv (\xvar \rightarrow \Dv \gvar),\\
 		\label{hedge1}	
	& \Bh\Bv (\gvar \wedge \neg \bbvarup \rightarrow \Dv \yvar),\\
	\label{hedge2}		
	& \Bh\Bv (\yvar \rightarrow \Dh \gvar).
\end{align}
Next, we regard each tile $\tvar\in T$ as a fresh predicate, and we add the following conjuncts to $\varphi$, expressing the constraints for the tiles: 
\begin{align}
	\label{vedge}
	& \Bh\Bv (\gvar \wedge \neg \Dht{\top} \rightarrow \bbvarright),\\
	\label{tile1}	
	& \Bh\Bv (\gvar \leftrightarrow \bigvee_{\tvar\in T}\tvar),\\
%\end{align}
%
%\begin{align}		
\label{tile2}		
	&  \Bh\Bv \bigwedge_{\tvar\not=\tvar'}(\tvar\rightarrow \neg \tvar'),\\
       \label{tile3}	
	& \Bh\Bv \bigl(\tvar \rightarrow \Bht{\bigvee_{(\tvar,\tvar')\in H} \tvar'}\bigr),\\
	\label{tile4}	
	& \Bh\Bv \bigl(\tvar \rightarrow \Bvt{\bigvee_{(\tvar,\tvar')\in V} \tvar'}\bigr),\\	
	\label{ounique}
	& \Bh\Bv(\gvar\to\neg\Dht\ovar\land\neg\Dvt\ovar),\\
	\label{tilelast1}
	& \Bh\Bv \bigl((\bbvarright \rightarrow \Bvt{\bbvarright})
	\wedge (\Dvt{\bbvarright} \rightarrow \bbvarright)\bigr),\\
	\label{tilelast2}	
	& \Bh\Bv \bigl((\bbvarup \rightarrow \Bht{\bbvarup})
	\wedge (\Dht{\bbvarup}\rightarrow \bbvarup)\bigr).
\end{align}
%
%Consider any model $M=(W\times V,I)$ and let $((w,v),(w',v')\in R_{h}$ if $(w,v)\in m^{I}$, $(w',v)\in y^{I}$ and $(w',v') \in m^{I}$ and
%$((w,v),(w',v')\in R_{v}$ if $(w,v)\in m^{I}$, $(w,v')\in z^{I}$ and $(w',v') \in m^{I}$.
%Then intuitively $R_{h}$ is the `horizontal neighbor' relation and $R_{v}$ is the `vertical neighbor' relation on the grid points $m^{I}$.  
%
%\bigskip
%Let $\varphi$ be the conjunction of the formulas defined above. 
%Observe that $\sig(\varphi)=\sigma$. 
Let $\sigma=\sig(\varphi)=\{\gvar,\xvar,\yvar\}\cup\{\tvar\mid\tvar\in T\}$.

Note that we have not yet forced $R_{h}^\mathfrak M,R_{v}^\mathfrak M$ to form a grid-like structure on $\gvar^\mathfrak M$-points. We say that a $\gvar^\mathfrak M$-point $(w,d)$ is \emph{confluent} if,  
for every $R_h^\mathfrak M$-successor $(w_h,d_h)$ and every $R_v^\mathfrak M$-successor $(w_v,d_v)$ 
of $(w,d)$, there is $(w',d')$ that is both an $R_v^\mathfrak M$-successor of $(w_h,d_h)$ and an
$R_h^\mathfrak M$-successor of $(w_v,d_v)$. 
Forcing the grid to be finite and confluence of all grid-points are
achieved using the formula $\psi$, which contains two additional predicates, $\qvar$ and $\ssvar$, behaving like second-order variables over grid-points. We set
$$
\psi = \qvar\land\Bh\Bv\bigl(\qvar\to 
\Dht{\qvar}\lor\Dvt\qvar\lor(\Dvt{\Bht{\ssvar}}\land\Dht{\Bvt{\neg\ssvar}})\bigr).
$$
%
%$$
%\psi = \ovar \wedge Q \wedge \Bh\Bv (Q \rightarrow \psi_{1}\vee \psi_{2} \vee \psi_{3})
%$$ 
%where $\psi_{1}= \exists R_{h}.Q$, $\psi_{2} = \exists R_{v}.Q$, and
%$$
%\psi_{3}= \exists R_{v}\forall R_{h}.P \wedge \exists R_{h}\forall R_{v}.\neg P
%$$
%
It is readily seen that if $\mathfrak{M},w_0,d_0\models \varphi$, for some model $\mathfrak M$, then the following are equivalent: 
\begin{enumerate}[style=multiline,leftmargin=0.9cm]
	\item[\textbf{(c1)}] $\mathfrak{M}',w_0,d_0\models \psi$, for some model $\mathfrak M' = (W,D,I')$ with $I'$ the same as $I$ on all predicates save possibly $\qvar$ and $\ssvar$ (we call such an $\mathfrak M'$ a \emph{variant} of $\mathfrak M$);
	
	\item[\textbf{(c2)}] $\mathfrak M$ contains an infinite $R_{h}^\mathfrak M\cup R_{v}^\mathfrak M$-path starting at $(w_0,d_0)$ or a non-confluent $\gvar^\mathfrak M$-point accessible from $(w_0,d_0)$ via an $R_{h}^\mathfrak M\cup R_{v}^\mathfrak M$-path.
\end{enumerate}

%then
%we have\nz{single point where $\qvar$ is false: infinite path, $\neg\psi$ true}
%%
%\begin{align}
%\nonumber
%&  \mbox{$\mathfrak{M},w_0,d_0\models \neg\psi$\ \  iff}\\
%\nonumber
%& \mbox{ there is no infinite $R_{h}^\mathfrak M\cup R_{v}^\mathfrak M$-path starting at $(w_0,d_0)$,}\\
%\nonumber
%& \mbox{ and every $(w,d)$ that is accessible from $(w_0,d_0)$ via an}\\
% \label{negpsi}
%& \mbox{ $R_{h}^\mathfrak M\cup R_{v}^\mathfrak M$-path is confluent.}
%\end{align}

%from $(w,d)$ to some $(w',d')$ such that $(w',d')$ has an $R_{h}^\mathfrak M$- and an $R_{v}^\mathfrak M$-successor which do not have a common $R_{v}^\mathfrak M$- and $R_{h}^\mathfrak M$-successor, respectively.
%Consider a model $\mathfrak{M}= (W,D,I)$ with $\mathfrak{M},w,d\models \varphi$ not yet interpreting $P$ and $Q$. Then one can satisfy $\psi$ in $\mathfrak{M},w,d$ if either there is an infinite $R_{h}\cup R_{v}$-path in $\mathfrak{M}$ starting at $(w,d)$ or there is an $R_{h}\cup R_{v}$-path in  $\mathfrak{M}$ starting at $(w,d)$ to some $(w',d')$ such that $(w',d')$ has an $R_{h}$ and an $R_{v}$-successor which do not have a common $R_{v}$ and $R_{h}$-successor, respectively.

\begin{lem}\label{lem:ifsolthennotcons}
	If $\tiling$ has a solution, then $\varphi \wedge \psi$ is not a conservative extension of $\varphi$.
\end{lem}
\begin{proof}
	Let $(n,m,\tau)$ be a solution to $\tiling$. We enumerate the points of the $n\times m$-grid starting from the first horizontal row $(0,0),\dots, (n-1,0)$, then continuing with the second row $(0,1),\dots,(n-1,1)$, etc. 
	We define a model $\mathfrak{N}=(W,D,J)$ with $W=D=\{0,\dots,nm-1\}$ that represents this enumeration as follows (remember that $\ovar,\bbvarup,\bbvarright\in T$, $\bbvarup$ marks the tiles of last row, and  $\bbvarright$ marks the tiles of the last column). 
	For all $k<nm$ and $\tvar\in T$, 
	\begin{align}
		\label{Nfirst}
		\gvar^{J(k)} & =\{k\},\\
	\xvar^{J(k)} & = \left\{
		\begin{array}{ll}
			\{k-1\},  & \mbox{if $k>0$,}\\[3pt]
			\emptyset, &\mbox{otherwise,}
		\end{array}
		\right.\\
	\yvar^{J(k)} & = \left\{
		\begin{array}{ll}
			\{k+n\},  & \mbox{if $k<nm-n$,}\\[3pt]
			\emptyset, &\mbox{otherwise,}
		\end{array}
		\right.\\	
		\label{Nlast}
		\tvar^{J(k)} & = \left\{
		\begin{array}{ll}
			\{k\},  & \mbox{if $k=jn+i$ and $\tau(i,j)=\tvar$,}\\[3pt]
			\emptyset, &\mbox{otherwise.}
		\end{array}
		\right.
		%\\
	\end{align}
	For $n=m=3$, the model $\mathfrak N$ (without tiles other than $\ovar,\bbvarright,\bbvarup$) and the relations $R_h^\mathfrak N$, $R_v^\mathfrak N$ are illustrated in Fig.~\ref{f:modelN}.
	%the picture below.\\
	%
	It is easy to check that $\mathfrak{N},0,0\models \varphi$ and $\mathfrak{N}',0,0\models \neg\psi$, for any variant $\mathfrak{N}'$ of $\mathfrak{N}$. 
	
	%\centerline{
		\begin{figure}[ht]
			\begin{tikzpicture}[>=latex,line width=0.2pt,xscale = .9,yscale = .9]
				\node[point,fill=white,scale = 0.8,label=below right:{\footnotesize $\gvar$},label=below:{\footnotesize $\ovar$},label=left:{\footnotesize $0$}] (x0) at (0,0) {};
				\node[point,fill,scale = 0.8,label=below right:{\footnotesize $\xvar$}] (x1) at (1,0) {};
				\node[point,fill,scale = 0.5] (x2) at (2,0) {};
				\node[point,fill,scale = 0.5] (x3) at (3,0) {};
				\node[point,fill,scale = 0.5] (x4) at (4,0) {};
				\node[point,fill,scale = 0.5] (x5) at (5,0) {};
				\node[point,fill,scale = 0.5] (x6) at (6,0) {};
				\node[point,fill,scale = 0.5] (x7) at (7,0) {};
				\node[point,fill,scale = 0.5,label=right:{\footnotesize $W$}] (x8) at (8,0) {};
				\node[]  at (0,-.6) {\footnotesize $0$};
				\node[]  at (1,-.6) {\footnotesize $1$};
				\node[]  at (2,-.6) {\footnotesize $2$};
				\node[]  at (3,-.6) {\footnotesize $3$};
				\node[]  at (4,-.6) {\footnotesize $4$};
				\node[]  at (5,-.6) {\footnotesize $5$};
				\node[]  at (6,-.6) {\footnotesize $6$};
				\node[]  at (7,-.6) {\footnotesize $7$};
				\node[]  at (8,-.6) {\footnotesize $8$};
				\node[point,fill,scale = 0.5,label=left:{\footnotesize $1$}] (y1) at (0,1) {};
				\node[point,fill,scale = 0.5,label=left:{\footnotesize $2$}] (y2) at (0,2) {};
				\node[point,fill,scale = 0.8,label=left:{\footnotesize $3$},label=above right:{\footnotesize $\yvar$}] (y3) at (0,3) {};
				\node[point,fill,scale = 0.5,label=left:{\footnotesize $4$}] (y4) at (0,4) {};
				\node[point,fill,scale = 0.5,label=left:{\footnotesize $5$}] (y5) at (0,5) {};
				\node[point,fill,scale = 0.5,label=left:{\footnotesize $6$}] (y6) at (0,6) {};
				\node[point,fill,scale = 0.5,label=left:{\footnotesize $7$}] (y7) at (0,7) {};
				\node[point,fill,scale = 0.5,label=left:{\footnotesize $8$},label=above:{\footnotesize $D$}] (y8) at (0,8) {};
				\node[point,fill=white,scale = 0.8,label=below right:{\footnotesize $\gvar$}] (11) at (1,1) {};
				\node[point,fill=white,scale = 0.8,label=below right:{\footnotesize $\gvar$},label=above right:{\footnotesize $\bbvarright$}] (22) at (2,2) {};
				\node[point,fill=white,scale = 0.8,label=below right:{\footnotesize $\gvar$}] (33) at (3,3) {};
				\node[point,fill=white,scale = 0.8,label=below right:{\footnotesize $\gvar$}] (44) at (4,4) {};
				\node[point,fill=white,scale = 0.8,label=below right:{\footnotesize $\gvar$},label=above right:{\footnotesize $\bbvarright$}] (55) at (5,5) {};
				\node[point,fill=white,scale = 0.8,label=below:{\footnotesize $\gvar$},label=right:{\footnotesize $\bbvarup$}] (66) at (6,6) {};
				\node[point,fill=white,scale = 0.8,label=below:{\footnotesize $\gvar$},label=right:{\footnotesize $\bbvarup$}] (77) at (7,7) {};
				\node[point,fill=white,scale = 0.8,label=below:{\footnotesize $\gvar$},label=right:{\footnotesize $\bbvarup$},label=above:{\footnotesize $\bbvarright$}] (88) at (8,8) {};
				\node[point,fill,scale = 0.8,label=below right:{\footnotesize $\xvar$}] (21) at (2,1) {};
				\node[point,fill,scale = 0.8,label=below right:{\footnotesize $\xvar$}] (32) at (3,2) {};
				\node[point,fill,scale = 0.8,label=below right:{\footnotesize $\xvar$}] (43) at (4,3) {};
				\node[point,fill,scale = 0.8,label=below right:{\footnotesize $\xvar$}] (54) at (5,4) {};
				\node[point,fill,scale = 0.8,label=below right:{\footnotesize $\xvar$}] (65) at (6,5) {};
				\node[point,fill,scale = 0.8,label=below right:{\footnotesize $\xvar$}] (76) at (7,6) {};
				\node[point,fill,scale = 0.8,label=below right:{\footnotesize $\xvar$}] (87) at (8,7) {};
				\node[point,fill,scale = 0.8,label=above:{\footnotesize $\yvar$}] (14) at (1,4) {};
				\node[point,fill,scale = 0.8,label=above:{\footnotesize $\yvar$}] (25) at (2,5) {};
				\node[point,fill,scale = 0.8,label=above:{\footnotesize $\yvar$}] (36) at (3,6) {};
				\node[point,fill,scale = 0.8,label=above:{\footnotesize $\yvar$}] (47) at (4,7) {};
				\node[point,fill,scale = 0.8,label=above:{\footnotesize $\yvar$}] (58) at (5,8) {};
				\draw[-,thick] (x0) -- (x8);
				\draw[-,thick] (x0) -- (y8);
				\draw[->,dashed] (x0) to node[right] {\scriptsize $R_h^{\mathfrak N}$}  (11);
				\draw[->,dashed] (11) to node[right] {\scriptsize $R_h^{\mathfrak N}$}  (22);
				\draw[->,dashed] (33) to node[right] {\scriptsize $R_h^{\mathfrak N}$}  (44);
				\draw[->,dashed] (44) to node[right] {\scriptsize $R_h^{\mathfrak N}$}  (55);
				\draw[->,dashed] (66) to node[left] {\scriptsize $R_h^{\mathfrak N}$}  (77);
				\draw[->,dashed] (77) to node[left] {\scriptsize $R_h^{\mathfrak N}$}  (88);
				\draw[->,thick,left,dotted,bend left=35] (x0) to node[] {\scriptsize $R_v^{\mathfrak N}$} (33);
				\draw[->,thick,left,dotted,bend left=35] (11) to node[] {\scriptsize $R_v^{\mathfrak N}$} (44);
				\draw[->,thick,left,dotted,bend left=35] (22) to node[] {\scriptsize $R_v^{\mathfrak N}$} (55);
				\draw[->,thick,left,dotted,bend left=35] (33) to node[] {\scriptsize $R_v^{\mathfrak N}$} (66);
				\draw[->,thick,left,dotted,bend left=35] (44) to node[] {\scriptsize $R_v^{\mathfrak N}$} (77);
				\draw[->,thick,left,dotted,bend left=35] (55) to node[] {\scriptsize $R_v^{\mathfrak N}$} (88);
			\end{tikzpicture}
			%}
		%
		\caption{The model $\mathfrak N$.}\label{f:modelN}
	\end{figure}
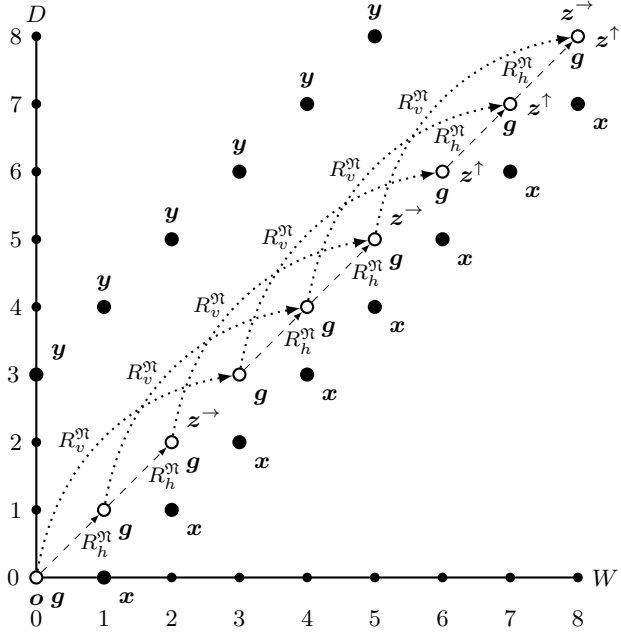

	We construct a formula $\chi$ such that $\sig(\chi) =\sigma$, $\models_\FOMS \psi\land \varphi \to \neg\chi$ but 
	$\not\models_\FOMS\varphi\to \neg\chi$, which means 
	that $\varphi \wedge \psi$ is not a (strong) conservative extension of $\varphi$. Intuitively, the formula $\chi$ characterises the model $\mathfrak{N}$ at $(0,0)$. 
	First, for every $(i,j)\in W\times D$, we construct a formula $\varphi_{i,j}$ such that, for all $(i',j')\in W\times D$,
	\begin{equation}
		\label{Mdesc}
		\mbox{$\mathfrak N,i',j'\models \varphi_{i,j}$ \quad iff} \quad  
		\mbox{$(i',j')=(i,j)$}.
	\end{equation}
	For instance, we can set inductively
	\begin{align*}
		\varphi_{0,0} & =\ovar, \\
		\varphi_{i+1,i+1}  & = \gvar \wedge \Dv (\xvar \wedge \Dh \varphi_{i,i}),\\
		\varphi_{i,j} & = \mbox{$\Dv \varphi_{i,i} \wedge \Dh \varphi_{j,j},\mbox{ for $i\not=j$.} $}
	\end{align*}
	%
	% and, inductively
	%	$$
	%	\varphi_{i+1,i+1} = \gvar \wedge \exists (\xvar \wedge \Diamond \varphi_{i,i}).
	%	$$
	%	Then $\varphi_{i,j}$ with $\varphi_{i,i}$ defined above and $\varphi_{i,j}= \Diamond \varphi_{i,i} \wedge \wedge \exists \varphi_{j,j}$ for $i\not=j$ 
	%	are as required. 
	Now let
	\begin{equation}\label{diagf}
		\chi_{i,j} = \varphi_{i,j} \wedge \bigwedge_{\pvar\in\sigma,\ \mathfrak N,i,j\models \pvar}\pvar\  \land
		\bigwedge_{\pvar\in\sigma,\ \mathfrak N,i,j\models \neg\pvar}\neg\pvar,
	\end{equation}
	and let $\chi$ be the conjunction of
	\begin{align}
		\label{origf}
		&\chi_{0,0}, \\
		\label{infiRh}
		&\Bh\Bv(\chi_{i,i} \rightarrow \Bht{\chi_{i+1,i+1}}), \text{ for $i<nm-1$}, \\
		\label{infiRv}
		&\Bh\Bv(\chi_{i,i} \rightarrow \Bvt{\chi_{i+n,i+n}}), \text{ for $i<nm-n$},\\
		\label{right}
		%&\Bh\Bv(\chi_{i,j} \rightarrow \Dh \chi_{i,l}), \text{ for $i,j,l <nm$}\\
		&\Bh\Bv(\chi_{i,i} \rightarrow \Dh \chi_{j,i}), \text{ for $i,j<nm$},\\
		\label{up}
		%&\Box \forall(\chi_{i,j} \rightarrow \Dv \chi_{l,j}), \text{ for $i,j,l<nm$}\\
		&\Box \forall(\chi_{i,i} \rightarrow \Dv \chi_{i,j}), \text{ for $i,j<nm$},\\
		\label{sol}
		&\Bh\Bv(\Dh \chi_{l,i} \wedge \Dv \chi_{i,j} \rightarrow \chi_{i,i}), \text{ for $i,j,l <nm$}.
	\end{align}
	Using \eqref{Mdesc}, it is easy to see that
	%Clearly 
	$\mathfrak{N},0,0\models \chi$, and so $\varphi \wedge \chi$ is satisfiable, i.e., $\not\models_\FOMS\varphi\to \neg\chi$. 
	Now suppose that $\mathfrak{M}$ is any model such that $\mathfrak{M},w_0,d_0 \models \varphi\wedge\chi$ for some $w_0,d_0$. Using the 
	%criterion 
	equivalence $\textbf{(c1)} \Leftrightarrow \textbf{(c2)}$, we show that $\mathfrak{M},w_0,d_0 \models\neg\psi$, which implies $\models_\FOMS \psi\land \varphi \to \neg\chi$. 
	%	Assume for a proof by contradiction 
	% By definition,
	%	$R_{h}$ and $R_{v}$ are defined on $\gvar$-nodes. Hence, by \eqref{eq:domain}, they are defined on $\bigvee\varphi_{i,i}$-nodes only. Hence, 
	
	To begin with, by \eqref{hrel}, \eqref{vrel}, the definition of $\mathfrak N$, and \eqref{diagf}--\eqref{infiRv},
	%By \eqref{infiRh} and \eqref{infiRv}, 
	there cannot exist an infinite $R_{h}^\mathfrak M\cup R_{v}^\mathfrak M$-chain.
	Now suppose there is an $R_{h}^\mathfrak M\cup R_{v}^\mathfrak M$-chain from $(w_0,d_0)$
	%a node satisfying $\ovar$ 
	to some node $(w,d)$ with an $R_{h}^\mathfrak M$-successor $(w_{1},d_{1})$ and $R_{v}^\mathfrak M$-successor $(w_{2},d_{2})$. Then $\mathfrak M,w,d\models \chi_{i,i}$ for some $i$, 
	$\mathfrak M,w_{1},d_{1}\models \chi_{i+1,i+1}$ and $\mathfrak M,w_{2},d_{2}\models \chi_{i+n,i+n}$, by \eqref{origf}--\eqref{infiRv}. 
	By \eqref{right} and \eqref{up},
	there exist $d_{1}'$ with $\mathfrak M,w_{1},d_{1}'\models \chi_{i+1,i+n+1}$, and $w_{2}'$ with $\mathfrak M,w_{2}',d_{2}\models \chi_{i+n+1,i+n}$. 
	By \eqref{sol}, $\mathfrak M,w_{2}',d_{1}'\models \chi_{i+n+1,i+n+1}$.	
	Moreover, as by \eqref{diagf}, $\bbvarup$ is not a conjunct of $\chi_{i+1,i+1}$, and $\bbvarright$  is not a conjunct of $\chi_{i+n,i+n}$, we have that $\yvar$ is a conjunct of $\chi_{i+1,i+n+1}$, and 
	$\xvar$ is a conjunct of $\chi_{i+n+1,i+n}$. Thus, by \eqref{hrel} and \eqref{vrel},
	$((w_{1},d_{1}),(w_{2}',d_{1}'))\in R_{v}^\mathfrak M$ and $((w_{2},d_{2})(w_{2}',d_{1}'))\in R_{h}^\mathfrak M$,
	and so $(w,d)$ is confluent.
	%Assume first that there is an $R_{v},R_{h}$-path in $m^{I}$ from  $(w_{0},v_{0})$ to some $(u,u')$ such that $(u,u')$ has an $R_{h}$-successor $(a,b)$ and an $R_{v}$-successor $(c,d)$ which do not have a common $R_{v}$ and $R_{h}$-successor, respectively. We then have $(u,u') \in \varphi_{v,v}^{I}$ for some $v$.
	%	Then $(a,b) \in \varphi_{v+1,v+1}^{I}$ and $(c,d)\in \varphi_{v+n,v+n}^{I}$. 
	%	Then there are $(a,b')\in (\varphi_{v+1+n,v+1} \wedge z)^{I}$ and $(c',d)\in  (\varphi_{v+n,v+n+1} \wedge z)^{I}$ and then $(c',b') \in \varphi_{v+1+n,v+n+1}^{I}$ since
	%	$\chi$ contains 
	%	$$
	%	\Box_{1}\Box_{2}(\Diamond_{2}\varphi_{v+1+n,v+1} \wedge \Diamond_{1} \varphi_{v+n,v+n+1} \rightarrow \varphi_{v+n+1,v+n+1})
	%	$$
	%	Moreover $(a,b),(c',b')\in R_{v}$ and $(c,d),(c',b')\in R_{v}$
	%	and we have derived a contradiction.
	%	
	%	That there is no infinite $R_{h},R_{v}$ path follows from the observation above.
\end{proof}

We say that a formula $\alpha$ is a \emph{model conservative extension} of a formula $\beta$
if $\models_{L} \alpha \rightarrow \beta$ and, for any model $\mathfrak{M},w,d$
with $\mathfrak{M},w,d\models \beta$, there exists a model $\mathfrak{M}'$
with $\mathfrak{M}',w,d\models \alpha$, which coincides with $\mathfrak{M}$ except for the interpretation of the predicates in $\sig(\alpha)\setminus\sig(\beta)$.
Clearly, if  $\alpha$ is a model conservative extension of $\beta$, then $\alpha$ is also a 
strong conservative extension of $\beta$.  Thus, if $\varphi \wedge \psi$ in our proof is not a conservative extension of $\varphi$, then $\varphi \wedge \psi$ is not a model conservative extension of $\varphi$. 

\begin{lem}\label{lem:ifnosolthencons}
	%If $\mathfrak{T}$ has no solution, then $\varphi \wedge \psi$ is a model conservative extension of $\varphi$ (and so also a signature independent conservative extension of $\varphi$).
	If $\varphi \wedge \psi$ is not a conservative extension of $\varphi$, then $\tiling$ has a solution.
\end{lem}
\begin{proof}
	Consider a model $\mathfrak{M}=(W,D,I)$ such that $\mathfrak{M},w,d\models \varphi$ but 
	$\mathfrak{M}',w,d\models \neg\psi$
	in any variant $\mathfrak M'$ of $\mathfrak{M}$.
	%We show that if $\psi$ is not satisfiable in an extension of $\mathfrak{M}$ obtained by interpreting the predicates $Q,P$, then $\mathfrak{T}$ has a solution. Recall that we can satisfy $\psi$ in an extension of $\mathfrak{M}$ if either there is an infinite $R_{h}\cup R_{v}$-path starting at $w,d$ or there is an $R_{h}\cup R_{v}$-path from  $(w,d)$ to some $(w',d')$ which
	%	has an $R_{h}$-successor and a $R_{v}$-successor which do not have a common $R_{v}$ and $R_{h}$-successor, respectively. 	
	%	So under the assumption that $\psi$ is not satisfiable in $\mathfrak{M}$ 
	Using the equivalence $\textbf{(c1)} \Leftrightarrow \textbf{(c2)}$, one can easily find within $\mathfrak{M}$ a finite grid-shaped (with respect to $R_{h}^\mathfrak M$ and 
	$R_{v}^\mathfrak M$) submodel, which gives a solution to $\tiling$. 
	
	For instance, we can start by taking an $R_{h}^\mathfrak M$-path of $\gvar^\mathfrak M$-points
	$$
	(w,d) = (w_0^0,d_0^0) R_{h}^\mathfrak M (w_1^0,d_1^0) R_{h}^\mathfrak M \dots R_{h}^\mathfrak M (w_{n-1}^0,d_{n-1}^0),
	$$
	for some $n>0$, that ends with the first point $(w_{n-1}^0,d_{n-1}^0)$ such that 
	$\mathfrak M,w_{n-1}^0,d_{n-1}^0\models \bbvarright$. Such a path must exist by $\neg$\textbf{(c2)} and \eqref{vedge}.  The chosen points $(w_i^0,d_i^0)$ form the first row of the required grid. 
	
	Next, observe that $\Bh\Bv (\gvar \wedge \neg \Dvt{\top} \rightarrow \bbvarup)$ follows from 
	\eqref{hedge1} and \eqref{hedge2}.
	So, similarly to the above, by $\neg$\textbf{(c2)} 
	%and \eqref{hedge}, 
	we can take an $R_{v}^\mathfrak M$-path
	$$
	(w_0^0,d_0^0) R_{v}^\mathfrak M (w_0^1,d_0^1) R_{v}^\mathfrak M \dots R_{h}^\mathfrak M (w_0^{m-1},d_0^{m-1}),
	$$
	for some $m>0$, that ends with the first point $(w^{m-1}_0,d^{m-1}_0)$ such that 
	$\mathfrak M,w^{m-1}_0,d^{m-1}_0 \models \bbvarup$. It forms the first column of the grid. 
	By $\neg$\textbf{(c2)} again, the point $(w_0^0,d_0^0)$ is confluent, and so we find $(w_1^1,d_1^1)$ with
	$$
	(w_1^0,d_1^0) R_{v}^\mathfrak M (w_1^1,d_1^1) \ \ \text{ and } \ \ (w^1_0,d^1_0) R_{h}^\mathfrak M (w_1^1,d_1^1).
	$$
	Similarly, we find the remaining $\gvar^\mathfrak M$-points $(w_i^j,d_i^j)$ for the whole $n\times m$-grid.
	By \eqref{tile1} and \eqref{tile2},
	each $(w_i^j,d_i^j)$ makes exactly one tile $\tvar\in T$ true.
	By \eqref{tile3} and \eqref{tile4}, the matching conditions of 
	\textbf{(t1)} and \textbf{(t2)} are satisfied. By \eqref{ounique}, we have \textbf{(t3)}.
	Finally, \textbf{(t4)} is satisfied by \eqref{tilelast1} and \eqref{tilelast2}.
\end{proof}
This completes the proof of Theorem~\ref{undec:cons} $(i)$.

The undecidability proof for the UIEP merges the counterexample to the UIP from Example~\ref{ex6} below with the formulas constructed to prove the undecidability of CEP above. 

\begin{exa}\label{ex6}
Suppose $\sigma=\{\avar,\pvar_{1},\pvar_{2}\}$ and 
$$
\varphi_{0} = \Bh\Bv \bigl(\avar \rightarrow \Dh (\pvar_{1} \wedge \bvar)\bigr) \land 
		\Bh\Bv \bigl(\pvar_{1} \wedge \bvar \rightarrow \Dv(\pvar_{2} \wedge \bvar)\bigr) \land{} 
\Bh\Bv \bigl(\pvar_{2} \wedge \bvar \rightarrow \Dh(\pvar_1 \wedge \bvar)\bigr).
$$
To show that $\avar \wedge \varphi_{0}$ has no $\sigma$-uniform interpolant in $\FOMS$, for every positive $r < \omega$, we define a formula $\chi_{r}$ inductively by taking $\chi_{0}= \top$ and $\chi_{r+1}= \pvar_{1} \wedge \exists (\pvar_{2} \wedge \Diamond \chi_{r})$. Then $\models_\FOMS\avar \land\varphi_0 \to \Diamond \chi_r$ for all $r>0$. Thus, if $\varrho$ were a $\sigma$-uniform interpolant of $\avar\land\varphi_0$, then
	$\models_\FOMS\varrho\to\Diamond\chi_r$ would follow for all $r>0$. Consider a model $\mathfrak{M}_{r}=(W_{r}, D_{r}, I_{r})$ with $W_{r}=D_{r}=\{0,\dots,r-1\}$, in which $\avar$ is true at $(0,0)$, $\pvar_1$ at $(k,k-1)$, and $\pvar_2$ at $(k,k)$, for $0<k<r$, as illustrated in the picture below: \\
%
%\begin{align*}
%		%&\avar^{\mathfrak{M}_{n}}= \{(0,0)\}\\
%		\avar^{I_r(k)} & = \left\{
%		\begin{array}{ll}
%			\{0\},  & \mbox{if $k=0$,}\\[3pt]
%			\emptyset, &\mbox{otherwise};
%		\end{array}
%		\right.
%		\\
%		%&\pvar_{1}^{\mathfrak{M}_{n}} =\{(k+1,k) \mid k<n\}\\
%		\pvar_{1}^{I_{r}(k)}  & = \left\{
%		\begin{array}{ll}
%			\{k-1\},  & \mbox{if $k>0$,}\\[3pt]
%			\emptyset, &\mbox{otherwise};
%		\end{array}
%		\right.
%		\\.   
%		%&\pvar_{2}^{\mathfrak{M}_{n}}=\{(k,k) \mid k<n\}
%		\pvar_{2}^{I_{r}(k)}  & = \left\{
%		\begin{array}{ll}
%			\{k\},  & \mbox{if $k>0$,}\\[3pt]
%			\emptyset, &\mbox{otherwise}.
%		\end{array}
%		\right.
%	\end{align*}
%	%
%	(See the picture below.)\\
%
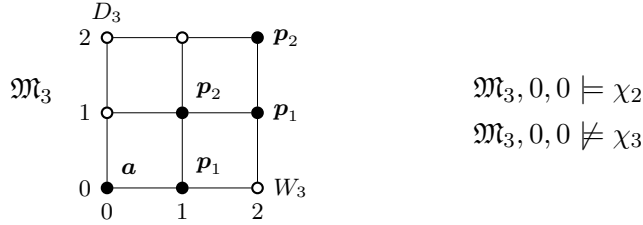
\centerline{
\begin{tikzpicture}[>=latex,line width=0.2pt,scale = 1]
\node[point,fill,scale = 0.7,label=below:{\footnotesize $0$},label=left:{\footnotesize $0$},label=above right:{\footnotesize $\avar$}] (u0d0) at (0,0) {};
\node[point,fill=black,scale = 0.7,label=below:{\footnotesize $1$},label=above right:{\footnotesize $\pvar_1$}] (u1d0) at (1,0) {};
\node[point,fill=white,scale = 0.7,label=below:{\footnotesize $2$},label=right:{\footnotesize $W_3$}] (u2d0) at (2,0) {};
\node[point,fill=white,scale = 0.7,label=left:{\footnotesize $1$}] (u0d1) at (0,1) {};
\node[point,fill=black,scale = 0.7,label=above right:{\footnotesize $\pvar_2$}] (u1d1) at (1,1) {};
\node[point,fill=black,scale = 0.7,label=right:{\footnotesize $\pvar_1$}] (u2d1) at (2,1) {};
\node[point,fill=white,scale = 0.7,label=above:{\footnotesize $D_3$},label=left:{\footnotesize $2$}] (u0d2) at (0,2) {};
\node[point,fill=white,scale = 0.7] (u1d2) at (1,2) {};
\node[point,fill=black,scale = 0.7,label=right:{\footnotesize $\pvar_2$}] (u2d2) at (2,2) {};
%\node[]  at (-.5,0) {{\footnotesize $d_0$}};
%\node[]  at (-.5,1) {{\footnotesize $d_1$}};
%\node[]  at (-.5,2) {{\footnotesize $d_2$}};
\node[]  at (-1,1.3) {$\mathfrak M_3$};
\node[]  at (6,1.3) {$\mathfrak M_3,0,0\models\chi_2$};
\node[]  at (6,.7) {$\mathfrak M_3,0,0\not\models\chi_3$};
\draw[-] (u0d0) to (u1d0);
\draw[-] (u1d0) to (u2d0);
\draw[-] (u0d1) to (u1d1);
\draw[-] (u1d1) to (u2d1);
\draw[-] (u0d2) to (u1d2);
\draw[-] (u1d2) to (u2d2);
\draw[-] (u0d0) to (u0d1);
\draw[-] (u0d1) to (u0d2);
\draw[-] (u1d0) to (u1d1);
\draw[-] (u1d1) to (u1d2);
\draw[-] (u2d0) to (u2d1);
\draw[-] (u2d1) to (u2d2);
\end{tikzpicture}}\\
%
%\centerline{\includegraphics[scale=0.7]{PICS/ex-u}}\\
%
Then $\mathfrak{M}_{r},0,0\not\models \Diamond \chi_r$, for any $r>0$, and so \mbox{$\mathfrak{M}_{r},0,0\not\models \varrho$}. On the other hand, $\mathfrak{M}_{r},0,0\models \Diamond \chi_{r'}$ for all $r'<r$. 
	%	Then $\mathfrak{M}_{n},0,0\not\models \varrho$ for all $n>0$. So see this consider the formula $\chi_{n}$ defined inductively by setting $\chi_{0}= \pvar_{2}$ and
	%	$\chi_{n+1}=\avar \wedge \Dh(\pvar_{1} \wedge \Dv \chi_{n})$. Then $\mathfrak{M}_{n},0,0\not\models \avar \wedge \chi_{n}$ but clearly
	%	$\models \varphi_{0}\rightarrow \chi_{n}$ for all $n>0$. 
	%Hence $\models \varrho \rightarrow \chi_{n}$ for all $n>0$. 
	%So $\mathfrak{M}_{n},0,0\not\models \varrho$ for all $n>0$. 
	Now consider the ultraproduct $\prod_{U}\mathfrak{M}_{r}$ with a non-principal ultrafilter $U$ on $\omega\setminus\{0\}$ (we refer the reader to~\cite{modeltheory} for the definition and relevant properties of ultrafilters and ultraproducts). As each $\Diamond \chi_{r'}$ is true at $(0,0)$ in almost all $\mathfrak{M}_{r}$, it follows from the properties of ultraproducts that, for a suitable $\overline{0}$ and all $r>0$, we have
$
\prod_{U}\mathfrak{M}_{r},\overline{0},\overline{0}\models \avar \land \neg \varrho \land \Diamond \chi_{r}.
$ 
One can interpret $\bvar$ in $\prod_{U}\mathfrak{M}_{r}$ so that  
	$\mathfrak{M},\overline{0},\overline{0}\models \varphi_{0}$ for the resulting model $\mathfrak{M}$. Then $\mathfrak{M}\models \avar \wedge \varphi_{0} \wedge \neg \varrho$, contrary to $\models_\FOMS\avar\land\varphi_0\to\varrho$, for any uniform interpolant $\varrho$ of $\avar\land\varphi_0$. \hfill $\dashv$
\end{exa}

We now come to the proof of Theorem~\ref{undec:cons} $(ii)$.
	Take $\tiling$, $\varphi$, and $\psi$ from the proof of Theorem~\ref{undec:cons} $(i)$. Using fresh predicates $\avar, \bvar, \pvar_{1},\pvar_{2}$, and $\varphi_{0}$ from Example~\ref{ex6},  set
		$$
		\varphi' =  \varphi\land (\pvar_{1}\rightarrow \pvar_{1})\land (\pvar_{2}\rightarrow \pvar_{2}),\qquad
		% \psi' & =  \bigl(\psi \vee (\avar \wedge \ovar)\bigr) \wedge \varphi_{0},
		\psi' =  (\psi \vee \avar) \wedge \varphi_{0}.
		$$
		%
		%$\varphi_{0}'$ defined as $\varphi_{0}$ from Example~\ref{ex6} with $\avar$ replaced by $A$.
		%
		%\begin{align}
		%&\Bh\Bv (A \rightarrow \Diamond (\pvar_{1} \wedge B))\\
		%&\Bh\Bv (\pvar_{1} \wedge B \rightarrow \exists(\pvar_{2} \wedge B))\\
		%&\Bh\Bv (\pvar_{2} \wedge B \rightarrow \Diamond(\pvar_1 \wedge B))
		%\end{align}
		Let $\sigma= \sig(\varphi')$. 
		We show that it is undecidable whether there exists a $\sigma$-uniform interpolant of $\varphi' \wedge \psi'$ in $\FOMS$.
		%In what follows we add conjuncts $\pvar_{1}\rightarrow \pvar_{1}$ and $\pvar_{2}\rightarrow \pvar_{2}$ to $\varphi$ so that $\sig(\varphi)=\sigma$. 
		
		\begin{lem}
			%	If $\mathfrak{T}$ has no solution, then $\varphi \wedge \psi'$ is a model conservative extension of $\varphi$ (and so also a 	signature independent conservative extension of $\varphi$ and a S5$_{1}(\sigma$)-uniform interpolant of $\varphi \wedge \psi'$).
			If there is no $\sigma$-uniform interpolant of $\varphi'\land\psi'$ in $\FOMS$, then $\tiling$ has a solution.
		\end{lem}
		\begin{proof}
			As the assumption implies that $\varphi'\land\psi'$ is not a model conservative extension of $\varphi'$,
			the proof is a straightforward variant of the proof of Lemma~\ref{lem:ifnosolthencons}.
			Consider a model $\mathfrak{M}=(W,D,I)$ with $\mathfrak{M},w,d\models \varphi'$ but 
			$\mathfrak{M}',w,d\models \neg\psi'$
			in any variant $\mathfrak M'$ of $\mathfrak{M}$ obtained by interpreting the predicates $\qvar,\ssvar,\avar,\bvar$. In particular, if $\avar$ and $\bvar$ are both interpreted as $\emptyset$ in all worlds, then 
			$\mathfrak{M}',w,d\models \varphi_0$, and so  
			$\mathfrak{M}',w,d\models \neg\psi$ follows.	 
			%		Consider a model $\mathfrak{M}=(W,D,I)$ with $\mathfrak{M},w,d\models \varphi$. We show that if $\psi'$
			%	is not satisfiable in an extension of $\mathfrak{M}$ obtained by interpreting the predicates $Q,P,A,B$, then $\mathfrak{T}$ has a solution. Thus assume that $\psi'$ is not satisfiable in any extension of $\mathfrak{M}$ obtained by interpreting $Q,P,A,B$. 
			%By setting $A^{\mathfrak{M}}=B^{\mathfrak{M}}=\emptyset$, we can satisfy $\varphi_{0}'$ in $\mathfrak{M}$. Hence it follows that we cannot satisfy $\psi$ in $\mathfrak{M}$ by interpreting $P,Q$ appropriately. 
			But this case is considered in the proof of Lemma~\ref{lem:ifnosolthencons}. 
		\end{proof}
		
		\begin{lem}
			If $\tiling$ has a solution, then there is no $\sigma$-uniform interpolant of $\varphi'\wedge \psi'$ in $\FOMS$.
		\end{lem}
		\begin{proof}
			Assume that $(n,m,\tau)$ is a solution to $\tiling$.
			Take the formulas $\chi$ and $\chi_{s}$ constructed in the proof of Lemma~\ref{lem:ifsolthennotcons} and in Example~\ref{ex6}, respectively. As it was shown, we have $\models_\FOMS\varphi\land\psi\to\neg\chi$ and 
			$\models_\FOMS\avar\land\varphi_0\to\chi_s$ for all $s>0$. Thus, 
			$\models_\FOMS \varphi' \wedge \psi' \rightarrow (\chi \rightarrow \chi_{s})$ for all $s>0$. 
			Therefore, if $\varrho$ were a uniform interpolant of $\varphi'\land\psi'$, then we would have 
			\begin{equation}
				\label{follows}
				\models_\FOMS\varrho\to(\chi \rightarrow \Diamond\chi_{s})\mbox{ for all $s>0$}.
			\end{equation}
			On the other hand, we combine $\mathfrak{N}=(W,D,J)$ from the proof of Lemma~\ref{lem:ifsolthennotcons} with the models $\mathfrak{M}_{s}=(W_{s},D_{s},I_{s})$ constructed in Example~\ref{ex6}.
			For every $s\geq nm$, we define a model $\mathfrak{N}_{s}=(W_{s},D_{s},J_{s})$ as follows. For every $k<s$, we let	
			\[
			\pvar^{J_s(k)} = \left\{
			\begin{array}{ll}
				\pvar^{J(k)},  & \mbox{if $k<nm$ and $\pvar\in\sig(\varphi)$,}\\[3pt]
				\pvar^{I_s(k)},  & \mbox{if $\pvar\in\{\avar,\pvar_1,\pvar_2\}$,}\\[3pt]
				\emptyset, &\mbox{otherwise.}
			\end{array}
			\right.
			\]
			%	Let $K= m\times n-1$ and define $\mathfrak{N}_{r}=(W_{r},D_{r},I_{r})$ with $W_{r}=D_{r}=\{0,\ldots,r\}$ for $r\geq k$, and set 
			%	\begin{align}
				%		& \gvar^{\mathfrak{N}_{r}} = \{(k,k) \mid k\leq K\}\\
				%		& \xvar^{\mathfrak{N}_{r}}=  \{(k+1,k) \mid k<K\}\\
				%		& \yvar^{\mathfrak{N}_{r}}= \{(k,k+n) \mid k<K-n\}\\
				%		& \tvar^{\mathfrak{N}_{r}} = \{ (jn+i,jn+i) \mid \tau(i,j) = \tvar\}\\
				%		& \avar^{\mathfrak{N}_{r}}= \{(0,0)\}\\
				%		& \pvar_{1}^{\mathfrak{N}_{r}}=\{(k+1,k) \mid k<r\}\\
				%		& \pvar_{2}^{\mathfrak{N}_{r}}=\{(k,k) \mid k<r\}
				%	\end{align}
			%	
			As shown in the proof of Lemma~\ref{lem:ifsolthennotcons}, $\mathfrak{N},0,0\models\varphi\land\chi$, so we have $\mathfrak{N}_s,0,0\models\varphi'\land\chi$.
			As it was shown in Example~\ref{ex6} that $\mathfrak{M}_{s},0,0\not\models \chi_s$ for all $s>0$, we have 
			$\mathfrak{N}_{s},0,0\not\models \chi_s$ for all $s\geq nm$.
			So it follows from \eqref{follows} that $\mathfrak{N}_{s},0,0\models\neg \varrho$.
			Note that also $\mathfrak{N}_{s},0,0\models \chi_{s'}$ for all $s>s'\geq nm$.
			Now consider the ultraproduct $\prod_{U}\mathfrak{N}_{s}$ with a non-principal ultrafilter $U$ on $\omega\setminus\{0,\dots,nm-1\}$. 
			As each $\chi_{s'}$ is true in almost all $\mathfrak{N}_{s},0,0$, it follows from the properties of ultraproducts~\cite{modeltheory} that 
			$\prod_{U}\mathfrak{N}_{s},\overline{0},\overline{0}\models \avar \land \neg \varrho \land\chi_{s'}$ for all $s'>0$, for a suitable $\overline{0}$. But then one can interpret $\bvar$ in $\prod_{U}\mathfrak{N}_{s}$ so that  
			$\mathfrak{N},\overline{0},\overline{0}\models \varphi_{0}$ for the resulting model $\mathfrak{N}$. Then $\mathfrak{N},\overline{0},\overline{0}\models \avar \land \varphi_{0} \land \varphi' \land \neg \varrho$ and so
			$\mathfrak{N},\overline{0},\overline{0}\models\varphi'\land\psi'\land\neg\varrho$.
			As $\models_\FOMS\varphi'\land\psi'\to\varrho$ should hold for
			a uniform interpolant $\varrho$ of $\varphi'\land\psi'$, we have derived a contradiction.	
		\end{proof}
		
		This completes the proof of Theorem~\ref{undec:cons} $(ii)$.

\begin{rem}\label{r:FOTUIP}
We can translate Example~\ref{ex6} into \FOT{} by taking $\sigma=\{A,R,P\}$ and 
$$
\varphi_{0} = \forall x (A(x) \rightarrow \exists y (R(x,y) \wedge B(y) \wedge P(y)) \land{} 
 \forall x (B(x) \wedge P(x) \rightarrow \exists y (R(x,y) \wedge B(y) \wedge P(y))).
$$
Then $A(x) \wedge \varphi_{0}$ has no $\sigma$-uniform interpolant in \FOT{}, which is shown similarly to the proof in Example~\ref{ex6}. It follows that \FOT{} does not have the UIP.
This example can be merged with the proof of the undecidability of the CEP in \FOT{} from~\cite{DBLP:conf/icalp/JungLM0W17}---in the same way as we combined Example~\ref{ex6} with the undecidability proof for the UIEP in \FOMS{}---to show that the UIEP is undecidable in \FOT{} \textup{(}with and without $=$\textup{)}. The latter problem has so far remained open.
\end{rem}

%******************

\section{The Modal Description Logic $\ALCIOS$}\label{sec:ALCIO}

Next, we extend the results of Sections~\ref{sec:s5xs5} and~\ref{sec: CEP-S5} to the modal description logic $\ALCIOS$, where $\ALCIO^u$ is the basic description logic $\ALC$ with the universal role~\cite{DBLP:books/daglib/0041477}. In other words, $\ALCIO^u$ is a notational variant of multimodal logic $\mathsf{K}$ with the universal modality and can be regarded as a fragment of $\FOT$. (An example showing that it does not enjoy the CIP will be given at the end of the section.) 

The \emph{concepts} of $\ALCIOS$ are constructed from \emph{concept names} $A \in \mathcal{C}$, \emph{role names} $R \in \mathcal{R}$, 
%\emph{inverse roles} $R^{-}$ with $R\in \mathcal{R}$, and \emph{individual names} $a\in \mathcal{N}$, 
for some countably-infinite and disjoint sets $\mathcal{C}$ and $\mathcal{R}$, 
%and $\mathcal{N}$, 
and a distinguished \emph{universal role} $U$ by means of the following grammar: 
%{\color{red} see comment by reviewer. Anything to do?} 
%
\[
C  ~:=~  A \ \mid  \ \top \ \mid \ C \sqcap C'  \ \mid \ \neg C \ \mid \ \exists R.C \ \mid \ \exists U.C \ \mid \ \Diamond C.
\]
%A \emph{role} $S$ is either a role name or the inverse thereof.
%
A \emph{signature} $\sigma$ is any finite set of concept and role names. As usual, the universal role is regarded as a logical symbol and not part of any signature. The signature $\sig(C)$ of a concept $C$ comprises the concept and role names in $C$, again excluding the universal role. If $\sig(C)\subseteq \sigma$, then $C$ is called a \emph{$\sigma$-concept}.
%The signature $\sig(S)$ of a role $S$ is the role name it uses.
%
We interpret $\ALCIOS$ in models $\mathfrak{M} = (W,\Delta,I)$, where $I(w)$ is an interpretation of the concept and role names at each world $w\in W$ over domain $\Delta \ne\emptyset$: $A^{I(w)}\subseteq \Delta$, $R^{I(w)} \subseteq \Delta \times \Delta$, and $U^{I(w)} = \Delta \times \Delta$. 
%We set $(R^{-})^{I(w)}= (R^{I(w)})^{-1}$.
%
The \emph{truth-relation} $\mathfrak{M},w,d\models C$ is defined by taking
\begin{itemize}
	\item[--] $\mathfrak M, w, d \models \top$, 
	
	\item[--] $\mathfrak M, w, d \models A$ iff $d \in A^{I(w)}$,  
	
%	\item[--] $\mathfrak M, w, d \models \{a\}$ iff $a^{I(w)}= d$,
	
	\item[--] $\mathfrak M, w, d \models \exists S.C$ iff there is $(d,d') \in S^{I(w)}$ such that $\mathfrak M, w, d' \models C$, 
	
	\item[--] $\mathfrak M, w, d \models \Diamond C$ iff there is $w' \in W$ with $\mathfrak M, w', d \models C$,
\end{itemize}
and the standard clauses for Boolean $\sqcap$ and $\neg$. An expression of the form $C \sqsubseteq D$ is called a \emph{concept inclusion}.
We sometimes use more conventional  
$C^{I(w)} = \{ d\in \Delta \mid \mathfrak{M},w,d\models C\}$,  
writing $\mathfrak{M},w\models C \sqsubseteq D$ if $C^{I(w)} \subseteq D^{I(w)}$, and \mbox{$\models C \sqsubseteq D$} if $\mathfrak{M},w\models C \sqsubseteq D$ for all $\mathfrak{M}$ and $w$. The problem of deciding whether $\models C \sqsubseteq D$, for given $C$ and $D$, is \textsc{coNExpTime}-complete~\cite{GabEtAl03}. 

An \emph{interpolant for $C \sqsubseteq D$ in $\ALCIOS$} is a concept $E$ such that $\sig(E) \subseteq \sig(C) \cap \sig(D)$, $\models C \sqsubseteq E$, and \mbox{$\models E \sqsubseteq D$}.
The IEP for $\ALCIOS$ is to decide whether a given concept inclusion $C \sqsubseteq D$ has an interpolant in $\ALCIOS$.

\begin{rem}\label{rem:dl}
Typical applications of description logics use reasoning modulo \emph{ontologies}, which are finite sets, $\mathcal{O}$, of concept inclusions. We then set $\mathcal{O}\models C\sqsubseteq D$ iff whenever $\mathfrak{M},w\models \alpha$ for all $\alpha \in \mathcal{O}$, then $\mathfrak{M},w\models C \sqsubseteq D$. Reasoning modulo ontologies is reducible to the ontology-free case by the following equivalence:
$$
\mathcal{O}\models C \sqsubseteq D \quad \text{iff} \quad
 \models \top \sqsubseteq  \bigsqcup_{C' \sqsubseteq D' \in \mathcal{O}}  \exists U. (C' \sqcap \neg D') \sqcup \forall U.(\neg C \sqcup D). 
$$
where $\sqcup$ is dual to $\sqcap$. 
The following problems can easily be reduced to the IEP in polynomial time (see Appendix~\ref{s:sec6proofs}): 
\begin{description}
\item[\emph{IEP modulo ontologies}] Given an ontology $\mathcal{O}$, a signature $\sigma$, and a concept inclusion $C \sqsubseteq D$, does there exist a $\sigma$-concept $E$ such that $\mathcal{O}\models C \sqsubseteq E$ and $\mathcal{O}\models E \sqsubseteq D$? 
	
\item[\emph{ontology interpolant existence \textup{(}OIEP\textup{)}}] Given an ontology $\mathcal{O}$, a signature $\sigma$, and a concept inclusion $C \sqsubseteq D$, is there an ontology $\mathcal{O}'$ with $\sig(\mathcal{O}') \subseteq \sigma$, $\mathcal{O}\models \mathcal{O}'$, and $\mathcal{O}' \models C \sqsubseteq D$? 
	
\item[\emph{EDEP modulo ontologies}] Given an ontology $\mathcal{O}$, a signature $\sigma$, and a concept name $A$, does there exist a concept $C$ such that $\sig(C)\subseteq \sigma$ and $\mathcal{O}\models A \equiv C$? 
\end{description}
\end{rem}

Explicit definitions have been proposed for query rewriting in ontology-based data access~\cite{FraEtAl13,TomWed21}, developing and maintaining ontology alignments~\cite{DBLP:conf/ekaw/GeletaPT16}, and ontology engineering~\cite{TenEtAl06}. The IEP is fundamental for robust  modularisations and decompositions of ontologies~\cite{DBLP:series/lncs/KonevLWW09,DBLP:conf/rweb/BotoevaKLRWZ16}. 

Our main result in this section is the following:

\begin{thm}\label{thm:ALC-comp}
The	IEP, EDEP \textup{(}modulo ontologies\textup{)}, and the OIEP are all decidable in {\sc coN2ExpTime}, being $2$\textsc{ExpTime}-hard.
\end{thm}

We begin by formulating a model-theoretic characterisation of interpolant existence in $\ALCIOS$ in terms of the following generalisation of $\sigma$-\BS s for $\FOMS$ from Section~\ref{ssec:s5xs5up}. Similar bisimulations have been used~\cite{DBLP:conf/ijcai/WildS17} 
%\cite{DBLP:journals/corr/WildS17} 
to characterise the fragment $\ALCIOSnou$ of $\ALCIOS$ without the universal role as a  bisimulation-invariant fragment of full first-order modal logic $\SF$ over finite and unrestricted models. 

A $\sigma$-\emph{bisimulation} between models $\mathfrak{M}_{i}=(W_{i},\Delta_{i},I_{i})$, $i=1,2$,  is any triple 
$(\bs_{W},\bs_{\Delta},\bs)$ with $\bs_{W}\subseteq W_{1} \times W_{2}$, $\bs_{\Delta}\subseteq \Delta_{1} \times \Delta_{2}$, and $\bs \subseteq (W_{1}\times \Delta_{1}) \times (W_{2} \times \Delta_{2})$ if 
\begin{enumerate}[style=multiline,leftmargin=0.7cm]
	\item[\textbf{(w)}] for any $(w_{1},w_{2})\in \bs_{W}$ and $d_{1}\in \Delta_{1}$, there is $d_{2}\in \Delta_{2}$ with $((w_{1},d_{1}),(w_{2},d_{2}))\in \bs$ and similarly for $d_{2}\in \Delta_{2}$,
	
	\item[\textbf{(d)}] for any $(d_{1},d_{2})\in \bs_{\Delta}$ and $w_{1}\in W_{1}$, there is $w_{2}\in W_{2}$ with $((w_{1},d_{1}),(w_{2},d_{2}))\in \bs$ and similarly for \mbox{$w_{2}\in W_{2}$},
	
	\item[\textbf{(c)}] $((w_{1},d_{1}),(w_{2},d_{2})) \in \bs$ implies both	$(w_{1},w_{2}) \in \bs_{W}$ and $(d_{1},d_{2})\in \bs_{\Delta}$,
\end{enumerate}	
and the following hold for all $((w_{1},d_{1}),(w_{2},d_{2})) \in \bs$:
\begin{enumerate}[style=multiline,leftmargin=0.7cm]
	\item[\textbf{(a)}] $\mathfrak M_{1},w_{1},d_{1} \models A $ iff 
	$\mathfrak M_{2},w_{2},d_{2} \models A$, for all $A\in \sigma$;
	
	\item[\textbf{(r)}] if $(d_{1},e_{1})\in S^{I(w_1)}$ and $\sig(S)\in \sigma$, then there is $e_{2}\in \Delta_{2}$ with $(d_{2}, e_{2})\in S^{I(w_2)}$ and
	$((w_{1},e_{1}),(w_{2},e_{2}))\in \bs$, and the other way round.
   
%    \item[\textbf{(i)}] $\mathfrak M_{1},w_{1},d_{1} \models \{a\}$ iff 
%    $\mathfrak M_{2},w_{2},d_{2} \models \{a\}$, for all $a\in \sigma$.\nz{?} 
\end{enumerate}
We write $\mathfrak{M}_{1},w_{1},d_{1}\sim_{\sigma} \mathfrak{M}_{2},w_{2},d_{2}$ to say that there is a $\sigma$-bisimulation $(\bs_{W},\bs_{\Delta},\bs)$ between $\mathfrak{M}_1$ and $\mathfrak{M}_{2}$ for which $((w_{1},d_{1}),(w_{2},d_{2}))\in \bs$; 
$\mathfrak{M}_{1},w_{1}\sim_{\sigma} \mathfrak{M}_{2},w_{2}$ says that there is a $\sigma$-bisimulation $(\bs_{W},\bs_{\Delta},\bs)$ with $(w_{1},w_{2})\in \bs_{W}$; and $\mathfrak{M}_{1},d_{1}\sim_{\sigma} \mathfrak{M}_{2},d_{2}$ that there is $(\bs_{W},\bs_{\Delta},\bs)$ with $(d_{1},d_{2})\in \bs_{\Delta}$. 
The usage of $\mathfrak{M}_{1},w_{1},d_{1} \equiv_{\sigma} \mathfrak{M}_{2},w_{2},d_{2}$
is as in Section~\ref{Sec:prelims} but for any $\sigma$-concepts,
$\mathfrak{M}_{1},w_{1}\equiv_{\sigma} \mathfrak{M}_{2},w_{2}$ means that the same $\sigma$-concepts of the form  $\exists U.C$ are true at $w_1$ in $\mathfrak M_1$ and at $w_2$ in $\mathfrak M_2$, 
and $\mathfrak{M}_{1},d_{1}\equiv_{\sigma} \mathfrak{M}_{2},d_{2}$ means that the same $\sigma$-concepts of the form  $\Diamond C$ are true at $d_1$ in $\mathfrak M_1$ and at $d_2$ in $\mathfrak M_2$.
%
%\ $\mathfrak{M}_{1},w_{1}\equiv_{\sigma} \mathfrak{M}_{2},w_{2}$, and $\mathfrak{M}_{1},w_{1}\equiv_{\sigma} \mathfrak{M}_{2},w_{2}$ is as in Section~\ref{Sec:prelims} but  for any $\sigma$-concepts and those of the form $\exists U.C$ and $\Diamond C$, respectively. 
%
The following is an $\ALCIOS$-analogue of Lemma~\ref{bisim-lemma}: 

\begin{lem}\label{bisim-lemma-ALC}
	For any $\omega$-saturated $\ALCIOS$-models $\mathfrak{M}_{1}$ with $w_1,d_1$ and $\mathfrak{M}_{2}$ with $w_2,d_2$, 
	\begin{itemize}
		\item[--] \mbox{$\mathfrak{M}_{1},w_{1},d_{1}  \equiv_{\sigma}  \mathfrak{M}_{2},w_{2},d_{2}$ iff \ $\mathfrak{M}_{1},w_{1},d_{1}  \sim_{\sigma} \mathfrak{M}_{2},w_{2},d_{2}$},
		
		\item[--] $\mathfrak{M}_{1},w_{1} \equiv_{\sigma} \mathfrak{M}_{2},w_{2}$ \ iff \ $
		\mathfrak{M}_{1},w_{1} \sim_{\sigma} \mathfrak{M}_{2},w_{2}$,
		
		\item[--] $\mathfrak{M}_{1},d_{1}  \equiv_{\sigma} \mathfrak{M}_{2},d_{2}$ \ iff \ $
		\mathfrak{M}_{1},d_{1}  \sim_{\sigma} \mathfrak{M}_{2},d_{2}$.
	\end{itemize}
	The implication from right to left holds for arbitrary models.
\end{lem}
\begin{proof}
	The proof of the direction from right to left is by a straightforward induction on the construction of concepts $C$, using {\bf (a)} for concept names, ${\bf (r)}$ for $\exists R.C$, {\bf (c)} and {\bf (w)} for $\exists U.C$, and {\bf (c)} and {\bf (d)} for $\Diamond C$. 
	For the converse direction, we define $\bs_{W}$, $\bs_{\Delta}$, and $\bs$ via $\equiv_{\sigma}$ in the obvious way. Then we observe that 
	$\mathfrak{M}_{1},w_{1},d_{1}  \equiv_{\sigma} \mathfrak{M}_{2},w_{2},d_{2}$ implies $\mathfrak{M}_{1},w_{1} \sim_{\sigma} \mathfrak{M}_{2},w_{2}$ and
	$\mathfrak{M}_{1},d_{1}  \equiv_{\sigma} \mathfrak{M}_{2},d_{2}$. Hence we obtain {\bf (c)}. We obtain {\bf (w)} and {\bf (d)} using saturatedness, {\bf (a)} is trivial, and {\bf (r)} follows again from saturatedness.
\end{proof}

The criterion below---in which $\sigma$-bisimulation consistency is defined as in Section~\ref{main-notions} with concepts $C$, $D$ in place of formulas $\varphi$, $\psi$---is an $\ALCIOS$-analogue of Theorem~\ref{int-crit}:

\begin{thm}\label{int-critdl}
	The following conditions are equivalent for any concepts $C$ and $D$\textup{:}
	\begin{itemize}
		\item[--] there does not exist an interpolant for $C \sqsubseteq D$ in $\ALCIOS$\textup{;}
		
		\item[--] $C$ and $\neg D$ are $\sig(C) \cap \sig(D)$-bisimulation consistent.
	\end{itemize}
\end{thm}

We now extend the construction of Section~\ref{ssec:s5xs5up} from $\FOMS$ to $\ALCIOS$. In contrast to $\FOMS$, we now have to deal with more involved $\sigma$-bisimulations between the respective first-order models $I(w_{1})$ and $I(w_{2})$ (satisfying conditions $(\mathbf{a})$ and $(\mathbf{r})$). To this end we introduce \emph{full mosaics} (sets of full types realised in $\sigma$-bisimilar pairs $(w,d)$) and \emph{full points} (full mosaics with a distinguished full type). The range of the surjections $\pi$ used to construct $W'$ and $D'$ then consists of full points rather than full types. This provides us with the data structure to define $\sigma$-bisimilar first-order models $I(w)$ when required. This construction establishes an upper bound on the size of models witnessing bisimulation consistency:

\begin{thm}\label{th:sizeboundALC}
	For any concepts $C$ and $D$, there does not exist an interpolant for $C \sqsubseteq D$ in $\ALCIOS$ iff there are  models witnessing that $C$ and $\neg D$ are $\sig(C)\cap \sig(D)$-bisimulation consistent of size double-exponential in $|C|+|D|$. 
\end{thm}
\begin{proof}
Given concepts $C$ and $D$, we define the sets $\sub(C,D)$, $\sub_{\Diamond}(C,D)$, and $\sub_{\exists}(C,D)$ as in Section~\ref{ssec:s5xs5up} regarding $\exists U$ as the $\ALCS$-counterpart of $\exists$ in $\FOMS$. 
The \emph{world-type} $\wt_{\mathfrak{M}}(w)$ of $w \in W$ in $\mathfrak{M} = (W,\Delta,I)$, the  \emph{domain-type} $\dt_{\mathfrak{M}}(d)$ of $d \in \Delta$, and the \emph{full type} $\pet_{\mathfrak{M}}(w,d)$ of $(w,d)$ in $\mathfrak{M}$ are also defined as in Section~\ref{ssec:s5xs5up}. 
%We let $\wt$, $\dt$ and $\ft$ range over the world, domain and full types, respectively. 
%
Observe that we have $\wt_{\mathfrak{M}}(w) = \pet_{\mathfrak{M}}(w,d)^{\wt}$ and $\dt_{\mathfrak{M}}(d) = \pet_{\mathfrak{M}}(w,d)^{\dt}$, where
$$
\pet_{\mathfrak{M}}(w,d)^{\wt}= \sub_{\exists}(C,D) \cap \pet_{\mathfrak{M}}(w,d),\quad 
\pet_{\mathfrak{M}}(w,d)^{\dt} = \sub_{\Diamond}(C,D) \cap \pet_{\mathfrak{M}}(w,d). 
$$

Now, suppose that $\sigma = \sig(C) \cap \sig(D)$, $\mathfrak{M}_{i}= (W_{i},\Delta_{i},I_{i})$, for $i = 1,2$, have pairwise disjoint $W_i$ and $\Delta_i$, $\mathfrak{M}_{1},w_{1},d_{1} \sim_{\sigma} \mathfrak{M}_{2},w_{2},d_{2}$  with $\mathfrak{M}_{1},w_{1},d_1 \models C$, and $\mathfrak{M}_{2},w_{2}, d_2 \models\neg D$. 
For $w\in W_{1}\cup W_{2}$, we  define the \emph{world mosaic} $\wm(w) = (T_{1}(w),T_{2}(w))$ by~\eqref{world-m} and the \emph{$i$-world point} $\iwp(w) = (\wt_{\mathfrak{M}_{i}}(w), \wm(w))$ of $w$ in $\mathfrak{M}_{1}$, $\mathfrak{M}_{2}$. Using~\eqref{domain-m} with $\Delta_i$ in place of $D_i$, we define the \emph{domain mosaic} $\dm(d)=(S_{1}(d),S_{2}(d))$ and  \emph{$i$-domain point} $\idp(d) = (\dt_{\mathfrak{M}_{i}}(d),\dm(d))$ of $d \in \Delta_1 \cup \Delta_2$ in $\mathfrak{M}_{1}$, $\mathfrak{M}_{2}$.
Then, for $(w,d) \in (W_{1} \times \Delta_{1}) \cup (W_{2} \times \Delta_{2})$, we set
$$
F_{i}(w,d) =\{ \pet_{\mathfrak{M}_{i}}(v,e) \mid (v,e)\in W_{i}\times \Delta_{i}, (v,e) \sim_{\sigma} (w,d)\}
$$
calling $\pam(w,d)=(F_{1}(w,d),F_{2}(w,d))$ the \emph{full mosaic} and $\ipp(w,d) = (\pet_{\mathfrak{M}_{i}}(w,d),\pam(w,d))$ the \emph{$i$-full point} of $(w,d)$
in $\mathfrak{M}_{1}$, $\mathfrak{M}_{2}$. 
Given $\pam=(F_{1},F_{2})$, we set
$$
\pam^{\wt} = (\{ \pet^{\wt} \mid \pet\in F_{1}\},\{ \pet^{\wt} \mid \pet\in F_{2}\}),\ \quad 
\pam^{\dt} = (\{ \pet^{\dt} \mid \pet\in F_{1}\},\{ \pet^{\dt} \mid \pet\in F_{2}\}).
$$
%
%Conditions\nz{HERE} {\bf (w)}, {\bf (d)}, {\bf (c)} are captured by conditions relating world and domain mosaics to full mosaics.

\begin{lem}\label{lem:surj}
Suppose $\pam = \pam(w,d)$, $\wm=\wm(w)$, and $\dm= \dm(d)$. Then $\pam^{\wt} =\wm$ and $\pam^{\dt}=\dm$. 
\end{lem}
\begin{proof}
The inclusions $\wm \subseteq \pam^{\wt}$ and $\dm \subseteq \pam^{\dt}$ follow from 
{\bf (w)} and {\bf (d)}. Indeed, suppose $\wm = (T_{1}(w),T_{2}(w))$, $\wt\in T_{i}(w)$, and $\pam=(F_{1}(w,d),F_{2}(w,d))$. By definition, there exists $v\in W_{i}$ with $v\sim_{\sigma} w$ such that
$\wt=\wt_{\mathfrak{M}_{i}}(v)$. By {\bf (w)}, there exists $e\in \Delta_{i}$ with $(w,d) \sim_{\sigma} (v,e)$. But then $\pet_{\mathfrak{M}_{i}}(v,e)\in F_{i}(w,d)$ and $\pet_{\mathfrak{M}_{i}}(v,e)^{\wt}=\wt$, as required.
The second claim is proved in the same way using {\bf (d)}.   

The converse inclusions $\pam^{\wt} \subseteq \wm$ and $\pam^{\dt} \subseteq \dm$ follow from 
{\bf (c)}. To see this, let $\pam=(F_{1}(w,d),F_{2}(w,d))$, $\wm=(T_{1}(w),T_{2}(w))$, and $\dm=(S_{1}(d),S_{2}(d))$.
Let $\pet \in F_{i}(w,d)$. Then there are $(v,e)\in W_{i}\times \Delta_{i}$ with $(v,e)\sim_{\sigma} (w,d)$. Therefore, by {\bf (c)}, $v\sim_{\sigma} w$ and $e\sim_{\sigma} d$, and so $\wt \in T_{i}(w)$ and $\dt\in S_{i}(d)$, as required.
\end{proof}

As in Section~\ref{ssec:s5xs5up}, we construct  models $\mathfrak{M}_{1}'= (W'_1,\Delta'_1,I'_1)$ and $\mathfrak{M}'_{2} = (W'_2, \Delta'_2, I'_2)$ 
from copies of $i$-world points $\iwp$ and $i$-domain points $\idp$ in $\mathfrak{M}_{1}$, $\mathfrak{M}_{2}$.
Let $n=m_{1}\times m_{2}$, where $m_{1}$ and $m_{2}$ are the number of full types and, respectively, full mosaics over $\sub(C,D)$ in $\mathfrak{M}_{1}$, $\mathfrak{M}_{2}$. For $i=1,2$, we set
$$
\Delta_{i}'= \{ \idp^{k} \mid \text{$\idp$ an $i$-domain point in $\mathfrak{M}_{1},\mathfrak{M}_{2}$,  $k\in [n]$}\}.
$$
For an $i$-world point $\iwp$ and an $i$-domain point $\idp$, let
$$
L^{i}_{\iwp,\idp} = \{ \ipp(w,d) \mid
\iwp=\iwp(w), \ \idp=\idp(d),\ (w,d)\in W_{i}\times \Delta_{i}\}.
$$
We define $W_{i}'$ using the set $\Pi_{\iwp,\idp}$ of surjective functions of the form 
$$
\pi_{\iwp,\idp} \colon [n] \rightarrow L^{i}_{\iwp,\idp}.
$$
Let $\Pi_{i}$ denote the set of all functions $\pi$ mapping any pair $\iwp,\idp$ to an element of $\Pi_{\iwp,\idp}$. We set $\pi_{\iwp,\idp}=\pi(\iwp,\idp)$ and then let $\Pi_{i}^{\dag}\subseteq \Pi_{i}$ be a smallest set such that, for any $\ipp= \ipp(w,d)=(\pet_{\mathfrak{M}_{i}}(w,d),\pam(w,d))$, $\iwp=\iwp(w)$, and $\idp=\idp(d)$ with $(w,d)\in \mathfrak{M}_{i}$ and any $k\in [n]$, there is $\pi\in \Pi_{i}^{\dag}$ with $\pi_{\iwp,\idp}(k) = \ipp$. 
It can be seen in the same way as in the proof for $\QSF$ in Section~\ref{ssec:s5xs5up} that $|\Pi^{\dag}_i| \leq n^{2}$, for $i=1,2$.

Now let 
$$
W_{i}' = \{ \iwp^{\pi} \mid \text{ $\iwp$ an $i$-world point in $\mathfrak{M}_{1},\mathfrak{M}_{2}$ and $\pi\in \Pi_{i}^{\dag}$}\}.
$$
Note that $|\Delta'_i|$ and $|W'_i|$ are double-exponential in $|C|+|D|$. 
It remains to define the extensions of concept and role names in $\mathfrak{M}_{1}'$ and $\mathfrak{M}_{2}'$. 
When defining them and investigating their properties, we use the following notation: 
$$
C^{\mathfrak M'_i} = \{ (w,d) \mid d \in C^{I'_i(w)}\},\quad 
R^{\mathfrak M'_i} = \{ ((w,d),(w,d')) \mid (d,d')\models R^{I'_i(w)}\}
$$
and similarly for $C^{\mathfrak{M}_{i}}$ and $R^{\mathfrak{M}_{i}}$. We set 
$$
\text{$(\iwp^{\pi},\idp^{k}) \in A^{\mathfrak{M}_{i}'}$ \ iff \ $A\in \pet$ \ for $\pi_{\iwp,\idp}(k)=(\pet,\pam)$.}
$$
The definition of $R^{\mathfrak{M}_{i}'}$ is more involved. Call a pair $\pet_{1},\pet_{2}$  \emph{$R$-coherent} if $\exists R.C\in \pet_{1}$ whenever $C\in \pet_{2}$, for all $\exists R.E\in \sub(C,D)$, and call $\pet_{1},\pet_{2}$ \emph{$R$-witnessing} if they are $R$-coherent and $\pet_{1}^{\wt}=\pet_{2}^{\wt}$. 
For full mosaics $\pam=(F_{1},F_{2})$ and $\pam'=(F_{1}',F_{2}')$ and a role $R \in \sigma$, we set $\pam \preceq_{R} \pam'$ if there exist functions $f_{i} \colon F_{i} \rightarrow F_{i}'$, $i=1,2$, such that, for all $\pet\in F_{i}$, the pair $\pet$, $f_{i}(\pet)$ is $R$-witnessing.
Now suppose that $\pi_{\iwp,\idp}(k)=(\pet, \pam)$ and $\pi_{\iwp,\idp'}(k')=(\pet',\pam')$. For $R\in \sigma$, we set 
\begin{equation}\label{defR}
((\iwp^{\pi},\idp^{k}),(\iwp^{\pi},\idp^{k'})) \in R^{\mathfrak{M}_{i}'} \quad \text{iff}\quad
\text{$\pam \preceq_{R} \pam'$ and $\pet$, $\pet'$ is $R$-witnessing.}
\end{equation}
For $R \notin \sigma$, we omit the condition $\pam \preceq_{R} \pam'$ from~\eqref{defR}.

\begin{lem}\label{final-lemma}
Suppose $E \in \sub(C,D)$. Then we have $(\iwp^{\pi},\idp^{k}) \in E^{\mathfrak{M}_{i}'}$ iff $E\in \pet$, for $\pi_{\iwp,\idp}(k) = (\pet,\pam)$.
\end{lem}
\begin{proof}
The proof is by induction on the construction of $E$. The basis of induction follows from the  definition, and the inductive step for the Booleans is trivial.

Let $E = \exists U.E'$. Suppose first $(\iwp^{\pi},\idp^{k}) \in E^{\mathfrak{M}_{i}'}$ and $\pi_{\iwp,\idp}(k)= (\pet, \pam)$.
By definition, there exists ${\idp'}^{k'}$ with  
$(\iwp^{\pi},{\idp'}^{k'}) \in E'^{\mathfrak{M}_{i}'}$. By IH,
$E'\in \pet'$ for $\pi_{\iwp,\idp'}(k')=
(\pet',\pam')$. Then $\exists U.E'\in \pet'$, and so $\exists U.E'\in {\pet'}^{\wt}$. As $\pet^{\wt}={\pet'}^{\wt}$, we obtain $E\in \pet$, as required.
Conversely, let $E = \exists U.E' \in \pet$, for $\pi_{\iwp,\idp}(k)=
(\pet, \pam)$. Take $(w,d)$ with $\pet=\pet(w,d)$ and $\pam=\pam(w,d)$. By definition, there is $d'$ with $E'\in \pet'$ for
$\pet'=\pet(w,d')$. Let $\idp'=\idp(d')$ and $\pam'=\pam(w,d')$. As $\pi_{\iwp,\idp'}$ is surjective, there is
$k'$ with $\pi_{\iwp,\idp'}(k')=(\pet',\pam')$.
Then, by IH, we obtain $(\iwp^\pi, \idp'^{k'}) \in E'^{\mathfrak{M}_{i}'}$, and so $(\iwp^\pi, \idp^k) \in E^{\mathfrak{M}_{i}'}$.

Let $E = \Diamond E'$. Suppose first that $(\iwp^\pi, \idp^k) \in E^{\mathfrak{M}_{i}'}$. Let $\pi_{\iwp,\idp}(k)= (\pet, \pam)$ and $\idp = (\dt,\dm)$. By definition, there is $\iwp'^{\pi'}$ with
$(\iwp'^{\pi'},\idp^k) \in E'^{\mathfrak{M}_{i}'}$. By IH,
$E'\in \pet'$ for $\pi'_{\iwp',\idp}(k)=
(\pet',\pam')$. By definition, $\Diamond E'\in \pet'$, and so
$\Diamond E'\in {\pet'}^{\dt}=\dt$. But then, again by definition, $E\in \pet$,
as required.

Conversely, let $E = \Diamond E' \in \pet$ for $\pi_{\iwp,\idp}(k)=
(\pet,\pam)$. Take $(w,d)$ with $\pet=\pet(w,d)$ and $\pam=\pam(w,d)$. By definition, there is $w'$ with $E'\in \pet'$ for
$\pet'=\pet(w',d)$. Let $\iwp'=\iwp(w')$ and $\pam'=\pam(w',d)$. By definition of $\Pi_{i}^{\dag}$, there is $\pi'\in \Pi_{i}^{\dag}$ with $\pi'_{\iwp',\idp}(k) =(\pet',\pam')$. By IH, $(\iwp'^{\pi'},\idp^k) \in E'^{\mathfrak{M}_{i}'}$, and so $(\iwp^\pi,\idp^k) \in E^{\mathfrak{M}_{i}'}$.

Let $E = \exists R.E'$. Suppose that $(\iwp^\pi,\idp^k) \in E^{\mathfrak{M}_{i}'}$ and $R\in \sigma$. Let $\pi_{\iwp,\idp}(k)= (\pet, \pam)$. By definition, there exists $\idp'^{k'}$ with
$((\iwp^\pi,\idp^k)(\iwp^\pi,\idp'^{k'})) \in R^{\mathfrak{M}_{i}'}$
and $(\iwp^\pi,\idp'^{k'}) \in E'^{\mathfrak{M}_{i}'}$. By IH,
$E'\in \pet'$ for $\pi_{\iwp,\idp'}(k')=
(\pet',\pam')$. By the definition of $R^{\mathfrak{M}_{i}'}$, we obtain that $\pet,\pet'$ are $R$-coherent. But then $E\in \pet$, as required.
The case $(\iwp^\pi,\idp^k) \in E^{\mathfrak{M}_{i}'}$ and $R\not\in \sigma$ is similar.

Conversely, let $E=\exists R.E' \in \pet$ for $\pi_{\iwp,\idp}(k)=
(\pet,\pam)$. Take $(w,d)$ with $\pet=\pet(w,d)$ and $\pam=\pam(w,d)$. By definition, there exists $e$ with $E'\in \pet'$ for
$\pet'=\pet(w,e)$ and $((w,d),(w,e))\in R^{\mathfrak{M}_{i}}$. Let $\idp'=\idp(e)$ and $\pam'=\pam(w,e)$. 
Assume first that $R\in \sigma$.
We define functions $f_{j}$, $j=1,2$ witnessing $\pam \preceq_{R} \pam'$. 

Suppose $\et \in F_{j}$, where $\pam=(F_{1},F_{2})$.  
Then we find $(w',d')\sim_{\sigma} (w,d)$ with  
$\et = \pet(w',d')$. 
By {\bf (r)}, $(w,d),(w,e) \in R^{\mathfrak{M}_{i}}$ and $(w',d')\sim_{\sigma} (w,d)$ give us $e'$ with $(w',d'),(w',e') \in R^{\mathfrak{M}_{i}}$ and $(w,e)\sim_{\sigma} (w',e')$. Let $\et'= \pet(w',e')$. We define the required $f_{j}$ by taking $f_{j}(\et)=(\et')$.
%
%e')$ generate $(t'',\ell''',s''')$. Then set $f_{i}(t'',\ell'', s'') = %(t'',\ell''',s''')$.
%The pair $\pet,\pet'$ is $R$-witnessing.

Since $\pi_{\iwp,\idp'}$ is surjective, there exists 
$k'$ such that $\pi_{\iwp,\idp'}(k')=(\pet',\pam')$.
By IH, $(\iwp^\pi,\idp'^{k'}) \in E'^{\mathfrak{M}_{i}'}$. By the definition of $R^{\mathfrak{M}_{i}'}$,
$
((\iwp^\pi,\idp^k),(\iwp^\pi,\idp'^{k'})) \in R^{\mathfrak{M}_{i}'},
$
so $(\iwp^\pi,\idp^k)\in E^{\mathfrak{M}_{i}'}$.
The case $R\not\in\sigma$ is similar. 
\end{proof}

We define $\bs_W \subseteq W'_1 \times W'_2$, $\bs_\Delta \subseteq \Delta'_1 \times \Delta'_2$, and $\bs \subseteq (W'_{1}\times \Delta'_{1}) \times (W'_{2} \times \Delta'_{2})$ by taking  
\begin{itemize}
\item[--] $((\wt,\wm)^{\pi},(\wt',\wm')^{\pi'})\in \bs_{W}$ iff $\wm=\wm'$;

\item[--] $((\dt,\dm)^{k},(\dt',\dm')^{k'}) \in \bs_{\Delta}$ iff $\dm=\dm'$;

\item[--] $((\iwp^{\pi}, \idp^{k}),({\iwp'}^{\pi'},{\idp'}^{k'}))\in \bs$ iff $\iwp= (\wt,\wm)$, $\iwp'=(\wt',\wm')$, $\idp=(\dt,\dm)$, and $\idp'=(\dt',\dm')$ with $\wm=\wm'$, $\dm=\dm'$, $\pam=\pam'$, and $\pi_{\iwp,\idp}(k) = (\pet,\pam)$ and $\pi'_{\iwp',\idp'}(k') = (\pet',\pam')$. 
\end{itemize}
%	
%	\item $((t,p,\pi),(s,q,k)) \sim_{\tau} ((t',p',\pi'),(s',q',k'))$ if
%	$p=p'$, $q=q'$ and for $\pi_{t,p,s,q}(k) = ((F_{1},F_{2}),(t,\ell,s))$ and $\pi'_{t',p',s',q'}(k) = ((F_{1}',F_{2}'),(t',\ell',s'))$ we have $(F_{1},F_{2})=(F_{1}',F_{2}')$.
%	\item $((t,p,\pi) \sim_{\tau} (t',p',\pi')$ if $p=p'$;
%	\item $(s,q,k)) \sim_{\tau} (s',q',k'))$ if $q=q'$.
%\end{itemize}

\begin{lem}\label{ALC-bis-l}
The triple $(\bs_{W},\bs_{\Delta},\bs)$ is a $\sigma$-bisimulation between $\mathfrak{M}_{1}'$ and 
$\mathfrak{M}_{2}'$.
\end{lem}
\begin{proof}
We show that $(\bs_{W},\bs_{\Delta},\bs)$ satisfies conditions {\bf (w)}, {\bf (d)}, {\bf (c)}, {\bf (a)},  and {\bf (r)}.
 
{\bf (w)} Suppose $((\wt,\wm)^{\pi},(\wt',\wm')^{\pi'})\in \bs_{W}$ and $(\dt,\dm)^{k}\in \Delta_{i}'$. We need to find $(\dt',\dm')^{k'}$ with 
$
(((\wt,\wm)^{\pi},(\dt,\dm)^{k}),((\wt',\wm')^{\pi'},(\dt',\dm')^{k'}))\in \bs.
$ 
We have $\wm=\wm'$ and set $(\pet,\pam)= \pi_{\wt,\wm,\dt,\dm}(k)$. Assume $\pam=(F_{1},F_{2})$. 
By Lemma~\ref{lem:surj}, there exists $\pet'\in F_{i}$ with ${\pet'}^{\wt}=\wt'$. 
As the component $\pi'_{\wt',\wm,{\pet'}^{\dt},\dm}$ of $\pi'$ is surjective,
there exists $k'\in [n]$ such that
$
\pi'_{\wt',\wm,{\pet'}^{\dt},\dm}(k') = (\pet',\pam). 
$
Then $({\pet'}^{\dt},\dm)^{k'}$ is as required; see the picture below.

\centerline{
\begin{tikzpicture}[>=latex,line width=0.2pt,scale=.9]
\node[scale = 0.9] (00) at (0,0) {$(\wt',\wm)^{\pi'}$};
\node[scale = 0.9] (10) at (4,0) {$({\pet'}^{\dt},\dm)^{k'}$};
\node[scale = 0.9] (01) at (0,2) {$(\wt,\wm)^{\pi}$};
\node[scale = 0.9] (11) at (4,2) {$(\dt,\dm)^{k}$};
\draw[-] (00) -- (10) node[midway,below] {{\footnotesize $(\pet',\pam)$}};
\draw[-] (01) -- (11) node[midway,above] {{\footnotesize $(\pet,\pam)$}};
\draw[-,dashed] (00) -- (01) node[midway,left] {{\footnotesize $\bs_W$}};
\draw[-,dashed] (2,.2) -- (2,1.8) node[midway,right] {{\footnotesize $\bs$}};
\end{tikzpicture}}

%\includegraphics[scale=0.7]{PICS/wM.pdf}
%\centerline{\includegraphics[scale=0.5]{PICS/wF.pdf}}

{\bf (d)} Suppose that $((\dt,\dm)^{k},(\dt',\dm')^{k'})\in \bs_{\Delta}$ and $(\wt,\wm)^{\pi}\in W_{i}'$. We need to construct 
$(\wt',\wm')^{\pi'}$ with
$
(((\wt,\wm)^{\pi},(\dt,\dm)^{k}),((\wt',\wm')^{\pi'},(\dt',\dm')^{k'}))\in \bs.
$ 
We have $\dm=\dm'$ and set $(\pet,\pam)= \pi_{\wt,\wm,\dt,\dm}(k)$. Assume $\pam=(F_{1},F_{2})$. 
By Lemma~\ref{lem:surj}, there exists $\pet'\in F_{i}$ with ${\pet'}^{\dt}=\dt'$. 
By the construction of $\Pi_{i}^{\dag}$, there exists $\pi'\in \Pi_{i}^{\dag}$ such that, for the component $\pi'_{{\wt'}^{\wt},\wm,\pet',\dm}$ of $\pi'$, we have
$
\pi'_{{\wt'}^{\wt},\wm,\pet',\dm}(k') = (\pet',\pam).
$
Then $({\ft'}^{\wt},\wm)^{\pi'}$ is as required; see the picture below.\\
\centerline{
\begin{tikzpicture}[>=latex,line width=0.2pt,scale=.9]
\node[scale = 0.9] (00) at (0,0) {$(\dt',\dm)^{k'}$};
\node[scale = 0.9] (10) at (4,0) {$({\pet'}^{\wt},\wm)^{\pi'}$};
\node[scale = 0.9] (01) at (0,2) {$(\dt,\dm)^{k}$};
\node[scale = 0.9] (11) at (4,2) {$(\wt,\wm)^{\pi}$};
\draw[-] (00) -- (10) node[midway,below] {{\footnotesize $(\pet',\pam)$}};
\draw[-] (01) -- (11) node[midway,above] {{\footnotesize $(\pet,\pam)$}};
\draw[-,dashed] (00) -- (01) node[midway,left] {{\footnotesize $\bs_\Delta$}};
\draw[-,dashed] (2,.2) -- (2,1.8) node[midway,right] {{\footnotesize $\bs$}};
\end{tikzpicture}}
\\
Condition {\bf (c)} follows from the definition of $\bs_{W}$, $\bs_{\Delta}$, $\bs$; and condition {\bf (a)} follows from the definition of $\bs$. 

{\bf (r)} Suppose $R\in \sigma$, $\bs$ contains 
$
(((\wt_{0},\wm)^{\pi^{0}}\!\!,(\dt_{0},\dm)^{k_{0}}),((\wt_{1},\wm)^{\pi^{1}}\!\!,(\dt_{1},\dm)^{k_{1}}))
$
and $R^{\mathfrak{M}_{i}}$ contains
$
(((\wt_{0},\wm)^{\pi^{0}},(\dt_{0},\dm)^{k_{0}}),((\wt_{0},\wm)^{\pi^{0}},(\dt_{2},\dm_{2})^{k_{2}})).
$
Assume also that $(\pet,\pam) = \pi^{0}_{\wt_{0},\wm,\dt_{0},\dm}(k_{0})$. Then there exists
$\pet_{1}$ such that $(\pet_{1},\pam)= \pi^{1}_{\wt_{1},\wm,\dt_{1},\dm}(k_{1})$.
Moreover, for $(\pet_{2},\pam_{2}) = \pi^{0}_{\wt_{0},\wm,\dt_{2},\dm_{2}}(k_{2})$, we have  
$\pam \preceq_{R} \pam_{2}$ and we may assume that $f_{i}(\pet)=\pet_{2}$, for the function $f_{i}$ witnessing $\pam \preceq_{R} \pam_{2}$. 
Then $(f_{i}(\pet_{1})^{\dt},\dm_{2})^{k_{3}}$ with $k_{3}\in [n]$ such that 
$
\pi^{1}_{\wt_{1},\wm,f_{i}(\pet_{1})^{\dt},\dm_{2}}(k_{3})= (f_{i}(\pet_{1}),\pam_{2})
$
is as required; see the picture below.\\
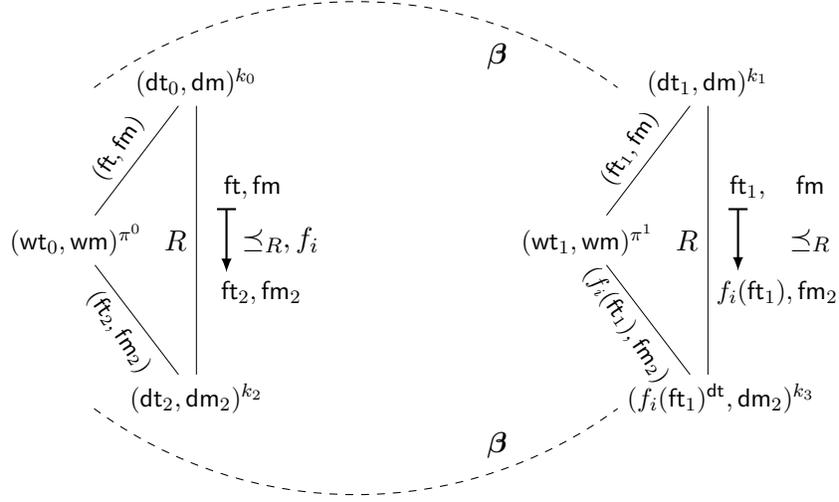
\centerline{
\begin{tikzpicture}[>=latex,line width=0.2pt,xscale=.8,yscale=.7]
\node[scale = 0.9] (a) at (0,2) {$(\wt_{0},\wm)^{\pi^{0}}$};
\node[scale = 0.9] (b) at (2,5) {$(\dt_{0},\dm)^{k_{0}}$};
\node[scale = 0.9] (c) at (2,-1) {$(\dt_{2},\dm_{2})^{k_{2}}$};
\node[scale = 0.9] (x) at (2.5,3) {$\qquad\pet,\pam$};
\node[]  at (3.4,2) {$\preceq_R,f_i$};
\node[scale = 0.9] (y) at (2.5,1) {$\ \ \qquad\pet_2,\pam_2$};
\draw[-] (a) -- (b) node[midway,above,sloped] {{\footnotesize $(\pet,\pam)$}};
\draw[-] (a) -- (c) node[midway,below,sloped] {{\footnotesize $(\pet_2,\pam_2)$}};
\draw[-] (b) -- (c) node[midway,left] {$R$};
\draw[|->,thick] (x) to (y);
\node[scale = 0.9] (aa) at (8.5,2) {$(\wt_{1},\wm)^{\pi^{1}}$};
\node[scale = 0.9] (bb) at (10.5,5) {$(\dt_{1},\dm)^{k_{1}}$};
\node[scale = 0.9] (cc) at (10.5,-1) {$\quad(f_i(\pet_1)^{\dt},\dm_{2})^{k_{3}}$};
\node[scale = 0.9] (xx) at (11,3) {$\quad\qquad\pet_1,\quad\pam$};
\node[]  at (12.2,2) {$\preceq_R$};
\node[scale = 0.9] (yy) at (11,1) {$\quad\qquad f_i(\pet_1),\pam_2$};
\draw[-] (aa) -- (bb) node[midway,above,sloped] {{\footnotesize $(\pet_1,\pam)$}};
\draw[-] (aa) -- (cc) node[midway,below,sloped] {{\footnotesize $(f_i(\pet_1),\pam_2)\ \ $}};
\draw[-] (bb) -- (cc) node[midway,left] {$R$};
\draw[|->,thick] (xx) to (yy);
\draw[-,dashed,thin,bend left=40] (.3,4.9) to (9,4.9); 
\node[]  at (7,5.5) {$\bs$};
\draw[-,dashed,thin,bend right=40] (.3,-1.2) to (9,-1.2); 
\node[]  at (7,-1.9) {$\bs$};
\end{tikzpicture}}
This completes the proof of the lemma.
\end{proof}

Theorem~\ref{th:sizeboundALC} follows.
\end{proof}

This result gives the upper bound of Theorem~\ref{thm:ALC-comp}. The lower one follows from the same lower bound for $\FOMS{}$ by treating $\FOM$-formulas $\varphi$, $\psi$ as role-free $\ALCIOS$-concepts and, using Theorems~\ref{int-crit} and~\ref{int-critdl}, one can readily show that $\varphi$ and $\psi$ have an interpolant in $\FOMS$ iff they have an interpolant in $\ALCIOS$. 

The (strong) conservative extension problem, (S)CEP,  and the uniform interpolant
existence problem, UIEP, in $\ALCIOS$ are defined in the obvious way. Using the same argument as for interpolation, the undecidability of the (S)CEP and UIEP in $\ALCIOS$ follows directly from the undecidability of both problems for \FOMS.
Note that, for the component logics---propositional $\mathsf{S5}$ and description logic $\mathcal{ALC}^{u}$---the CEP is \textsc{coNExpTime} and \textsc{2ExpTime}-complete, respectively~\cite{DBLP:conf/aiml/GhilardiLWZ06,DBLP:conf/icalp/JungLM0W17}.

%*************

\section{First-Order Modal Logic $\FOMK$}\label{sec:K}

We consider the one-variable first-order modal logic $\FOMK$ and show Theorem~\ref{thm4intro}. By the \emph{modal depth} $\md(\varphi)$ of a $\FOM$-formula $\varphi$ we mean the maximal  number of nestings of $\Diamond$ in $\varphi$; if $\varphi$ has no modal operators, then $\md(\varphi) = 0$. Formulas of modal depth $k$ can be characterised using a finitary version of bisimulations, called $k$-bisimulations, that are defined below.

For a signature $\sigma$ and two models $\mathfrak{M}= (W,R,D,I)$ and $\mathfrak{M}'= (W',R',D',I')$, a sequence $\bs_{0},\dots,\bs_{k}$ of relations  
$
\bs_{i} \subseteq (W\times D) \times (W'\times D')
$
is a $\sigma$-\emph{$k$-bisimulation} between $\mathfrak M$ and $\mathfrak M'$ if the following conditions hold for all $\pvar \in \sigma$ and  $((w,d),(w',d'))\in \bs_{i}$: {\bf (a)} and {\bf (d)} from Section~\ref{Sec:prelims} as well as
\begin{enumerate}[style=multiline,leftmargin=0.9cm]
%\item[(a)] $\mathfrak M,w,d \models \pvar$ iff $\mathfrak M',w',d' \models \pvar$;\nz{same}
	
\item[\textbf{(w$'$)}] if $i>0$ and $(w,v)\in R$, then there is $v'$ with  \mbox{$(w',v')\in R'$} and $((v,d),(v',d'))\in \bs_{i-1}$, and the other way round.
	
%\item[(d)] for every $e\in D$, there is $e'\in D'$ such that $((w,e),(w',e'))\in \bs_{i}$, and the other way round.\nz{same}
\end{enumerate}
We say that $\mathfrak{M},w,d$ and $\mathfrak{M}',w',d'$ are $\sigma$-\emph{$k$-bisimilar} and write 
$\mathfrak{M},w,d  \sim_{\sigma}^{k} \mathfrak{M}',w',d'$
if there is a $\sigma$-$k$-bisimulation $\bs_{0},\dots,\bs_{k}$ between $\mathfrak{M}$ and $\mathfrak{M}'$  with $((w,d),(w',d')) \in \bs_{k}$. We write
$\mathfrak{M},w,d  \equiv_{\sigma}^{k} \mathfrak{M}',w',d'$ when 
$\mathfrak{M},w, d \models \varphi$ iff $\mathfrak{M}',w', d'\models \varphi$, for every $\sigma$-formula $\varphi$ with $\md(\varphi)\leq k$.
%

%\begin{lemma}\label{bisim-k-lemma}
%	For any signature $\sigma$ and any {\color{red} $\omega$-saturated} models $\mathfrak{M}$ with $w,d$ and $\mathfrak{M}'$ with $w',d'$, we have\marginpar{can we omit saturated?}	%
%	$$
%	\mathfrak{M},w,d  \equiv_{\sigma}^{k} \mathfrak{M}',w',d'
%	\quad \text{ iff } \quad
%	\mathfrak{M},w,d  \sim_{\sigma}^{k} \mathfrak{M}',w',d'.
%	$$
%\end{lemma}

We now define formulas $\tau^k_{\mathfrak M,\sigma}$ generalising the characteristic formulas of~\cite{goranko20075}, that describe every model $\mathfrak{M}$ up to $\sigma$-$k$-bisimulations.
For $\mathfrak{M}=(W,R,D,I)$ and signature $\sigma$, let
\begin{align*}
	& t_{\mathfrak{M},\sigma}^{0}(w,d)= \bigwedge \{\pvar \in\sigma \mid \mathfrak{M},w,d\models \pvar\} \land{} 
	\bigwedge \{\neg \pvar \mid \pvar \in\sigma,\ \mathfrak{M},w,d\not\models \pvar\},\\
	& \tau_{\mathfrak{M},\sigma}^{0}(w,d)=t_{\mathfrak{M},\sigma}^{0}(w,d) \wedge{}
	\bigwedge_{e\in D} \exists t_{\mathfrak{M},\sigma}^{0}(w,e) \wedge \forall \bigvee_{e\in D} t_{\mathfrak{M},\sigma}^{0}(w,e) ,
\end{align*}
and let $\tau_{\mathfrak{M},\sigma}^{k+1}(w,d)$ be the conjunction of the formulas below: 
\begin{align*}
	& t_{\mathfrak{M},\sigma}^{0}(w,d) \wedge \hspace*{-2mm}  \bigwedge_{(w,v) \in R} \hspace*{-2mm} \Diamond \tau_{\mathfrak{M},\sigma}^{k}(v,d) \wedge \Box \hspace*{-2mm} \bigvee_{(w,v) \in R} \hspace*{-2mm}  \tau_{\mathfrak{M},\sigma}^{k}(v,d),\\
	& \bigwedge_{e\in D} \exists \big(t_{\mathfrak{M}}^{0}(w,e) \wedge \hspace*{-2mm} \bigwedge_{(w,v) \in R} \hspace*{-2mm} \Diamond \tau_{\mathfrak{M}}^{k}(v,e) \wedge 
	\Box \hspace*{-2mm} \bigvee_{(w,v) \in R} \hspace*{-2mm} \tau_{\mathfrak{M}}^{k}(v,e)\big),\\
	& \forall \bigvee_{e\in D} \big(t_{\mathfrak{M}}^{0}(w,e) \wedge \hspace*{-2mm} \bigwedge_{(w,v) \in R} \hspace*{-2mm} \Diamond \tau_{\mathfrak{M}}^{k}(v,e) \wedge 
	\Box \hspace*{-2mm} \bigvee_{(w,v) \in R} \hspace*{-2mm} \tau_{\mathfrak{M}}^{k}(v,e)\big).
\end{align*}
The following lemma says that $\tau^{k}_{\mathfrak{M},\sigma}(w,d)$
is the strongest formula of modal depth $k$ that is true at $w,d$ in $\mathfrak{M}$:

\begin{lem}\label{lem:new3}
	For any models $\mathfrak{M}$ with $w,d$ and $\mathfrak{N}$ with $v,e$, and any $k < \omega$,  the following conditions are equivalent\textup{:} 
	\begin{enumerate}[style=multiline,leftmargin=0.8cm]
		\item[$(i)$] $\mathfrak{N},v,e \equiv_{\sigma}^{k} \mathfrak{M},w,d$\textup{;}
		\item[$(ii)$] $\mathfrak{N},v,e \models \tau_{\mathfrak{M},\sigma}^{k}(w,d)$\textup{;}
		\item[$(iii)$] $\mathfrak{N},v,e \sim_{\sigma}^{k} \mathfrak{M},w,d$.
	\end{enumerate}
\end{lem} 
\begin{proof}
	The proof is by induction over $k$. For $k=0$ the equivalences hold by definition. Assume the equivalences have been shown for $k$. For $(i) \Rightarrow (ii)$ assume that $\mathfrak{N},v,e \equiv_{\sigma}^{k+1} \mathfrak{M},w,d$. Obviously $\mathfrak{M},w,d \models \tau_{\mathfrak{M},\sigma}^{k+1}(w,d)$ and $\tau_{\mathfrak{M},\sigma}^{k+1}(w,d)$ has modal depth $k+1$. Hence
	$\mathfrak{N},v,e \models \tau_{\mathfrak{M},\sigma}^{k+1}(w,d)$, as required.
	$(ii) \Rightarrow (iii)$
	Assume $\mathfrak{N},v,e\models \tau_{\mathfrak{M},\sigma}^{k+1}(w,d)$. Then define $\bs_i$ for $i\leq k+1$ by taking
	$$
	\bs_{i} = \bigl\{\bigl((v',e'),(w',d')\bigr) \mid \mathfrak{N},v',e' \models \tau_{\mathfrak{M},\sigma}^{i}(w',d')\bigr\}.
	$$
	It is readily seen that $\bs_{0},\dots,\bs_{k+1}$ is a $\sigma$-$(k+1)$-bisimulation. Hence $\mathfrak{N},v,e \sim_{\sigma}^{k+1} \mathfrak{M},w,d$.
	$(iii) \Rightarrow (i)$ holds by definition
	of $\sigma$-$k$-bisimulations. 
\end{proof}	

For any $k\geq 0$ and formula $\varphi$ with $\md(\varphi) \leq k$, we now set 
$$
\exists^{\sim\sigma,k}\varphi ~= \bigvee_{\mathfrak{M},w,d \models \varphi} \tau_{\mathfrak{M},\sigma}^{k}(w,d).
$$
Thus, for any $\mathfrak{N},v,e$, we have $\mathfrak{N},v,e \models \exists^{\sim\sigma,k}\varphi$ iff there is $\mathfrak{M},w,d$ with $\mathfrak{M},w,d \models \varphi$ and $\mathfrak{N},v,e \sim_{\sigma}^{k} \mathfrak{M},w,d$, i.e., $\exists^{\sim\sigma,k}$ is an existential depth restricted bisimulation quantifier~\cite{DBLP:journals/japll/DAgostinoL06,French06}. Clearly, $\models_\FOMK \varphi \rightarrow \exists^{\sim\sigma,k}\varphi$. 

\begin{thm}\label{thm:intK} 
The following conditions are equivalent, for any formulas $\varphi$ and $\psi$ and $k,k'\geq 0$ with $\md(\varphi)=k$, $\md(\psi) = k'$, and $n = \max{\{k,k'\}}$\textup{:}
\begin{enumerate}[style=multiline,leftmargin=0.7cm]
\item[\rm\textup{(a)}] there is $\chi$ such that $\sig(\chi) \subseteq \sigma$, $\models_\FOMK \varphi \rightarrow \chi$, and \mbox{$\models_\FOMK \chi \rightarrow \psi$\textup{;}}

\item[\rm\textup{(b)}] $\models_\FOMK \exists^{\sim\sigma,n}\varphi \rightarrow \psi$.
\end{enumerate}
\end{thm}
\begin{proof}
${\rm(a)} \Rightarrow {\rm (b)}$ If $\not\models_\FOMK \exists^{\sim\sigma,n}\varphi \rightarrow \psi$, then there is $\mathfrak{M},w, d$ with $\mathfrak{M}, w, d \models \exists^{\sim\sigma,n}\varphi$ and $\mathfrak{M},w, d \models \neg\psi$. By the definition of $\exists^{\sim\sigma,n}\varphi$, we then have $\mathfrak{M},w,d\models \tau_{\mathfrak{M}',\sigma}^{n}(w',d')$ and $\mathfrak{M}',w',d'\models \varphi$, for some model $\mathfrak{M}'$, $w',d'$. By Lemma~\ref{lem:new3}, $\mathfrak{M}',w',d' \sim_{\sigma}^{n} \mathfrak{M},w,d$. Using a standard unfolding argument, we may assume that $(W,R)$ in $\mathfrak M$ and $(W',R')$ in $\mathfrak M'$ are tree-shaped with respective roots $w,w'$. As $\varphi$ and $\psi$ have modal depth $\leq n$, we may also assume that the depth of $(W,R)$ and $(W',R')$ is $\leq n$. But then $\mathfrak{M}',w',d'\sim_{\sigma}\mathfrak{M},w,d$, contrary to (a). The implication ${\rm(b)} \Rightarrow {\rm (a)}$ is trivial.
\end{proof}

We do not know whether $\exists^{\sim\sigma,k}\varphi$ is equivalent to a formula whose size can be bounded by an elementary function in $|\sigma|$, $|\varphi|$, $k$. For pure $\mathcal{ALC}$, it is indeed equivalent to an exponential-size concept~\cite{TenEtAl06}.

Condition (b) in Theorem~\ref{thm:intK} gives an obvious non-elementary algorithm for checking whether given formulas have an interpolant in $\FOMK$. Thus, by Theorem~\ref{p:cipvspbdp}, we obtain:

\begin{thm}
The IEP and EDEP for $\FOMK$ are decidable in non-elementary time.
\end{thm}

The proof above seems to give a hint that the UIEP for \FOMK{} might also be decidable as (an analogue of) $\exists^{\sim\sigma,k}\varphi$ of modal depth $\md(\varphi)$ is a uniform interpolant of any propositional modal formula $\varphi$ in $\mathsf{K}$~\cite{Visser96}. The next example illustrates why this is not the case for `two-dimensional' $\FOMK$.

\begin{exa}\label{exK}
Suppose $\sigma = \{\avar,\bvar\} =\text{sig}(\psi)$,  
\begin{align*}
& \varphi = \forall (\avar \leftrightarrow \bvar   \leftrightarrow \hvar) \land \forall  (\hvar \leftrightarrow \Box \hvar \leftrightarrow \Diamond \hvar) \land{}
 \Diamond \forall (\bvar \leftrightarrow \hvar),\\ %\land \Box \forall \neg \avar,\\
& \psi = \forall (\avar \leftrightarrow \Box\Box \avar  \leftrightarrow \Diamond\Diamond \avar) \land \Box\Diamond\top  \to \Diamond \forall (\bvar \leftrightarrow \Diamond \avar).
\end{align*}
Intuitively, $\varphi$ at a world $w$ says (using the `help' predicate $\hvar$) that there is an $R$-successor $v$ 
such that, for every $e$, we have $v,e \models \bvar$ iff $w,e \models \bvar$. 
The premise of $\psi$ at $w$ says two things: first, at distance two, $R$-successors $u$ of $w$, 
for every $e$, we have $u,e \models\avar$ iff $w,e \models\avar$; and second, every $R$-successor of $w$, in particular $v$, also has an $R$-successor $u$. These conditions imply that, for every $e$, we have $v,e \models \bvar$ iff $u,e\models\avar$, and so $\models_{\FOMK} \varphi \rightarrow \psi$. 
On the other hand, $\not\models_{\FOMK} \exists^{\sim\sigma,1}\varphi \rightarrow \psi$ because, for the models $\mathfrak{M}$ and $\mathfrak{M}'$ below, we have $\mathfrak{M},w,d\models \varphi$\\
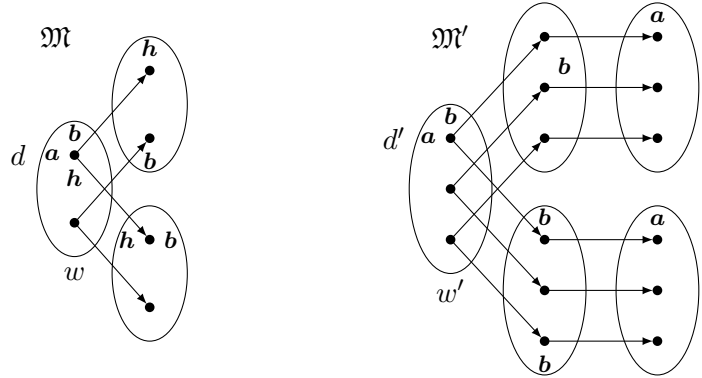
\centerline{
\begin{tikzpicture}[>=latex,line width=0.4pt,xscale = .5,yscale = .45]
\node[]  at (-2.5,8.5) {$\mathfrak M$};
\node[]  at (-3.5,5) {$d$};
\node[]  at (-2,1.5) {$w$};
\node[point,fill=black,scale = 0.5] (w1) at (-2,3) {};
\node[point,fill=black,scale = 0.5,label=left:{\footnotesize $\avar$},label=above:{\footnotesize $\bvar$},label=below:{\footnotesize $\hvar$}] (w2) at (-2,5) {};
\draw[] (-2,4) ellipse (1 and 2);
\node[point,fill=black,scale = 0.5,label=above:{\footnotesize $\hvar$}] (u2) at (0,7.5) {};
\node[point,fill=black,scale = 0.5,label=below:{\footnotesize $\bvar$}] (u1) at (0,5.5) {};
\draw[] (0,6.5) ellipse (1 and 2);
\node[point,fill=black,scale = 0.5,label=left:{\footnotesize $\hvar$},label=right:{\footnotesize $\bvar$}] (v2) at (0,2.5) {};
\node[point,fill=black,scale = 0.5] (v1) at (0,.5) {};
\draw[] (0,1.5) ellipse (1 and 2);
\node[]  at (8,8.5) {$\mathfrak M'$};
\node[]  at (6.5,5.5) {$d'$};
\node[]  at (8,1) {$w'$};
\node[point,fill=black,scale = 0.5,label=left:{\footnotesize $\avar$},label=above:{\footnotesize $\bvar$}] (ww3) at (8,5.5) {};
\node[point,fill=black,scale = 0.5] (ww2) at (8,4) {};
\node[point,fill=black,scale = 0.5] (ww1) at (8,2.5) {};
\draw[] (8,4) ellipse (1.1 and 2.5);
\node[point,fill=black,scale = 0.5] (uu3) at (10.5,8.5) {};
\node[point,fill=black,scale = 0.5,label=above right:{\footnotesize $\bvar$}] (uu2) at (10.5,7) {};
\node[point,fill=black,scale = 0.5] (uu1) at (10.5,5.5) {};
\draw[] (10.5,7) ellipse (1.1 and 2.5);
\node[point,fill=black,scale = 0.5,label=above:{\footnotesize $\bvar$}] (vv3) at (10.5,2.5) {};
\node[point,fill=black,scale = 0.5] (vv2) at (10.5,1) {};
\node[point,fill=black,scale = 0.5,label=below:{\footnotesize $\bvar$}] (vv1) at (10.5,-.5) {};
\draw[] (10.5,1) ellipse (1.1 and 2.5);
\node[point,fill=black,scale = 0.5] (xx1) at (13.5,5.5) {};
\node[point,fill=black,scale = 0.5] (xx2) at (13.5,7) {};
\node[point,fill=black,scale = 0.5,label=above:{\footnotesize $\avar$}] (xx3) at (13.5,8.5) {};
\draw[] (13.5,7) ellipse (1.1 and 2.5);
\node[point,fill=black,scale = 0.5] (yy1) at (13.5,-.5) {};
\node[point,fill=black,scale = 0.5] (yy2) at (13.5,1) {};
\node[point,fill=black,scale = 0.5,label=above:{\footnotesize $\avar$}] (yy3) at (13.5,2.5) {};
\draw[] (13.5,1) ellipse (1.1 and 2.5);
\draw[->] (w1) to (u1);
\draw[->] (w2) to (u2);
\draw[->] (w1) to (v1);
\draw[->] (w2) to (v2);
\draw[->] (ww1) to (uu1);
\draw[->] (ww2) to (uu2);
\draw[->] (ww3) to (uu3);
\draw[->] (ww1) to (vv1);
\draw[->] (ww2) to (vv2);
\draw[->] (ww3) to (vv3);
\draw[->] (uu1) to (xx1);
\draw[->] (uu2) to (xx2);
\draw[->] (uu3) to (xx3);
\draw[->] (vv1) to (yy1);
\draw[->] (vv2) to (yy2);
\draw[->] (vv3) to (yy3);
\end{tikzpicture}
}
\\
and $\mathfrak{M}',w',d'\not\models \psi$ but
$\mathfrak{M},w,d\sim_{\sigma}^{1}\mathfrak{M}',w',d'$ 
(a $\sigma$-$1$-bisimulation connects all points in the roots $w$ and $w'$ that agree on $\sigma$, and all points in the depth 1 worlds that agree on $\sigma$). 
\hfill $\dashv$
\end{exa} 

In fact, by adapting the undecidability proof for \FOMS{}, we prove the following.

\begin{thm}\label{undec:cons1}
$(i)$ The \textup{(}S\textup{)}CEP for \FOMK{} is undecidable.

$(ii)$ The	UIEP for \FOMK{} is undecidable.
\end{thm}	

For the proof of $(i)$, for any tiling system $\tiling$, 
we show how to construct in polytime formulas $\varphi$ and $\psi$
such that $\tiling$ has a solution iff $\varphi \wedge \psi$ is not a (strong) conservative extension of $\varphi$.
	%In particular, models of formulas witnessing non-conservativity (satisfiable %w.r.t.~$\varphi$ but not w.r.t. $\varphi \wedge \psi$).
	
	Let $\tiling=(T,H,V,\ovar,\bbvarup,\bbvarright)$ be a tiling system. 
	To prove undecidability of strong conservative extensions we work with models $\mathfrak{M}=(W,R,D,I)$ of modal depth 1 
	having a root $r\in W$ and $R$-successors $W'=W\setminus\{r\}$ of $r$. 
	We encode the finite grid to be tiled on 
	%the $R$-successors 
	$W'\times D$ in essentially 
	the same way as previously on the whole 
	%domain 
	$W\times D$. In particular,
	$\gvar^{\mathfrak{M}}\subseteq W'\times D$ and $R_{h}^{\mathfrak{M}},R_{v}^{\mathfrak{M}}\subseteq \gvar^\mathfrak M\times\gvar^\mathfrak M$
	%(W'\times D)\times (W'\times D)$ 
	are defined as in \eqref{hrel} and \eqref{vrel} before. We cannot, however, define the modalities $\Diamond_{h}\chi$ and $\Diamond_{v}\chi$ using \FOM-formulas,  as we cannot directly refer from $(w,d)$ to $(w',d)$ in our model $\mathfrak M$. Instead, we have to `speak about' $R_{h}^\mathfrak M$ and $R_{v}^\mathfrak M$ from the viewpoint of points of the form $(r,d)$, for the root $r$ of $(W,R)$.
	For instance, $\Bh\Bv (\chi_{1} \rightarrow \Dht{\chi_{2}})$
	is expressed using 
	$$
	\Bv \bigl[\Dh (\gvar \wedge \chi_{1}\wedge \neg \Zvar^{\rightarrow}) \rightarrow \Dh \bigl(\xvar \wedge \Dv (\gvar \wedge \chi_{2})\bigr)\bigr]
	$$ 
	and $\Bh\Bv (\chi_{1} \rightarrow \Dvt{\chi_{2}})$
	using
	$$
	\Bv \bigl[\Dh \bigl(\yvar \wedge \Dv (\gvar \wedge \chi_{1} \wedge \neg \Zvar^{\uparrow})\bigr) \rightarrow \Dh (\gvar \wedge \chi_{2})\bigr].
	$$
	We now define the new formula $\varphi$ in detail. The following conjuncts generate the grid:
	\begin{align*}
		& \Dh(\ovar \wedge \gvar),\\
		& \Bv \bigl[\Dh \bigl(\gvar \wedge \neg (\bbvarup \wedge \bbvarright)\bigr) \rightarrow \Dh \xvar\bigr],\\
		& \Bh\Bv (\xvar \rightarrow \Dv \gvar),\\
		& \Bh\Bv (\gvar \wedge \neg \bbvarup \rightarrow \Dv \yvar),\\
		& \Bv (\Dh \yvar \rightarrow \Dh \gvar).
	\end{align*}
	The constraints on the tiles are expressed by the following conjuncts: 
	\begin{align*}
		&\Bv \bigl(\Dh \gvar \wedge \neg \Dh \xvar \rightarrow \Bh (\gvar \rightarrow \bbvarright)\bigr),\\
		& \Bh\Bv (\gvar \leftrightarrow \bigvee_{\tvar\in T}\tvar) \land
		    \Bh\Bv \bigwedge_{\tvar\not=\tvar'}(\tvar \rightarrow \neg \tvar'),\\
		&\Bv \bigl[\Dh (\tvar \wedge \neg \bbvarright) \rightarrow \Bh \bigl(\xvar\rightarrow \Bv (\gvar \rightarrow \bigvee_{(\tvar,\tvar')\in H} \tvar')\bigr)\bigr],\\
		&
		\Bv \bigl[\Dh \bigl(\yvar \wedge \Dv (\tvar\wedge \neg \bbvarup)\bigr) \rightarrow \Bh (\gvar \rightarrow \bigvee_{(\tvar,\tvar')\in V} \tvar')\bigr],
		\\
		&\Bv \bigl(\Dh (\yvar \wedge \Dv \Zvar^{\rightarrow}) \rightarrow \Bh (\gvar \rightarrow \bbvarright)\bigr),\\
		&\Bv \bigl[\Dh\bbvarright \rightarrow \Bh \bigl(\yvar \rightarrow \Bv (\gvar \rightarrow \bbvarright)\bigr)\bigr],\\
		& \Bv\bigl[\Dh\bbvarup\to\Bh\bigl(\xvar\to\Bv(\gvar\to\bbvarup)\bigr)\bigr],\\
		& \Bv\bigl(\Dh(\xvar\land\Dv\bbvarup)\to\Bh(\gvar\to\bbvarup)\bigr).
	\end{align*}
	Finally, we take a fresh predicate $\pvar_0$ and add the conjunct $\pvar_0\to\pvar_0$ to $\varphi$.
	%This finished the definition of $\varphi$.

	We now aim to construct a formula $\psi$ for which, as previously, 
	we have $\textbf{(c1)} \Leftrightarrow \textbf{(c2)}$.
	This is slightly more involved, as we cannot directly express case distinctions using disjunction and nested `modalities'.
	%formulas. 
	We require auxiliary predicates to achieve this:
	we use $\avar_{h}$ to encode that $\qvar$ is true in an $R_{h}$-successor of a $\qvar$-node,
	$\avar_{v}$ to encode that $\qvar$ is true in an $R_{v}$-successor of a $\qvar$-node,
	and $\bvar',\bvar''$ are used to encode that a $\qvar$-node is not confluent. In detail, $\psi$ starts with the conjunct
	$$
	\Diamond (\gvar \wedge \qvar).
	$$
	Next we add a conjunct making a case distinction between $\avar_{h}$, $\avar_{v}$, and $\bvar'$:
	$$
	\Bh\Bv\bigl(\qvar\to
	(\neg\bbvarright\land\avar_h)\lor
	(\neg\bbvarup\land\Dv(\yvar\land\avar_v)) 
	\lor(\neg\bbvarright\land\neg\bbvarup\land\Dv(\yvar\land\bvar'))
	\bigr).
	$$
	%
	%\begin{eqnarray*}
	%\forall \Box (\qvar & \rightarrow & (\neg \Zvar^{\rightarrow} \wedge \avar_{1}) \vee \\ 
	% &  & (\neg \Zvar^{\uparrow}\wedge \exists (\avar_{2} \wedge \yvar)) \vee\\
	% &  & (\neg \Zvar^{\rightarrow} \wedge \neg \Zvar^{\uparrow} \wedge \bvar \wedge \exists (\bvar \wedge \yvar)))
	%\end{eqnarray*}
	%
	%(It seems important for the desired behavior that $\avar_{1}$ is placed on the $\q$-node and $\avar_{2}$ is placed on the $\yvar$-node. Otherwise their might be conflicting instructions on cycles.)
	%
	The next two conjuncts state the consequences of $\avar_{h}$ and $\avar_{v}$, respectively:
	\begin{align*}
		& \Bv\bigl[\Dh(\qvar \wedge \avar_{h}) \rightarrow \Dh\bigl(\xvar \wedge \Dv (\gvar \wedge \qvar)\bigr)\bigr], \\
		&
		\Bv\bigl[ \Dh \bigl(\yvar \wedge\Dv(\qvar\land\avar_{v})\bigr) \rightarrow \Dh (\gvar \wedge \qvar)\bigr].
	\end{align*}
	The next conjunct forces $\ssvar$ to be true in the horizontal successors of a vertical successor:
	$$
	\Bv \bigl[\Dh (\yvar \wedge \Dv\bvar' )\rightarrow \Bh \bigl(\xvar \rightarrow \Bv (\gvar \rightarrow \ssvar)\bigr)\bigr].
	$$ 
	Finally, the following formulas force $\neg\ssvar$ in the horizontal successors of a vertical successor:
	\begin{align*}
		&\Bv \bigl[\Dh (\qvar \wedge \bvar')\rightarrow \Dh \bigl(\xvar \wedge \Bv (\yvar \rightarrow \bvar'')\bigr)\bigr],\\
		&
		\Bv\bigl(\Dh (\yvar \wedge \bvar'') \rightarrow \Bh (\gvar \rightarrow \neg \ssvar)\bigr).
	\end{align*} 
	It is not difficult to show that the equivalence $\textbf{(c1)} \Leftrightarrow \textbf{(c2)}$ holds
	for $\varphi$ and $\psi$.
	
	%\tagi{Maybe I overlook sg, but I don't see any problems with cycles. Say, if we have an infinite $R_h\cup R_v$-chain
		%$u_0,u_1,\dots,u_n,\dots$ then we can always assume wlog that for all $i,j$, if $u_i=u_j$ then
		%$u_{i+1}=u_{j+1}$. Then there are no conflicting instructions on $\avar_h$ and $\avar_v$.
		%}
	
	\begin{lem}\label{lem:ifsolthennotcons2}
		If $\tiling$ has a solution, then $\varphi \wedge \psi$ is not a 
		%strong 
		conservative extension of $\varphi$.
	\end{lem}
	\begin{proof}
		The proof is obtained by modifying the proof of 
		Lemma~\ref{lem:ifsolthennotcons}. 
		To begin with, we modify the model $\mathfrak{N}=(W,D,J)$ defined in that proof by 
		adding a root world to $W$ from where every other world is accessible in one $R$-step.
		More precisely, we define a new model $\mathfrak{N}=(W,R,D,J)$ as follows.
		We let $D=W'=\{0,\dots,nm-1\}$, $W=W'\cup \{r\}$, $R=\{(r,w)\mid w\in W'\}$, and $J$ is defined
		by \eqref{Nfirst}--\eqref{Nlast}, plus having $\pvar^{J(r)}=\emptyset$ for 
		$\pvar\in\{\gvar,\xvar,\yvar\}\cup T$, and $\pvar_0^{J(w)}=\emptyset$ for all $w\in W$.
		
		%	Assume that $(n,m,\tau)$ is a solution to $\tiling$. We enumerate the points of the $n\times m$-grid
		%	by starting with the first horizontal row $(0,0),\dots (n-1,0)$, then continuing with $(0,1),\dots,(n-1,1)$, and so on.
		%	Instead taking exactly the grid-points as $W\times D$ we now ad a root world $r$ to $W$. In detail, let $D=W'=\{0,\dots,nm-1\}$, $W=W'\cup \{r\}$, and
		%	define a model $\mathfrak{N}=(W,R,D,J)$ representing this enumeration as follows. Let $(r,w)\in R$ for all $w\in W'$.
		%	For all $k<nm$, let
		%We take the model $\mathfrak{M}=(W,D,I)$ depicted in the corrected whiteboard figure. Let $K= m\times n-1$, $W=D=\{0,\ldots,K\}$, and set 
		%	
		%	\begin{align*}
			%		%& \gvar^{\mathfrak{M}} = \{(k,k) \mid k\leq K\}\\
			%		\gvar^{J(k)} & =\{k\}\\
			%		%& \xvar^{\mathfrak{M}}=  \{(k+1,k) \mid k<K\}\\
			%		\xvar^{J(k)} & = \left\{
			%		\begin{array}{ll}
				%			\{k-1\},  & \mbox{if $k>0$,}\\[3pt]
				%			\emptyset, &\mbox{otherwise}
				%		\end{array}
			%		\right.
			%		\\
			%		%& \yvar^{\mathfrak{M}}= \{(k,k+n) \mid k<K-n\}\\
			%		\yvar^{J(k)} & = \left\{
			%		\begin{array}{ll}
				%			\{k+n\},  & \mbox{if $k<nm-n$,}\\[3pt]
				%			\emptyset, &\mbox{otherwise}
				%		\end{array}
			%		\right.
			%		\\
			%		%& \tvar^{\mathfrak{M}} = \{ (jn+i,jn+i) \mid \tau(i,j) = \tvar\}
			%		\tvar^{J(k)} & = \left\{
			%		\begin{array}{ll}
				%			\{k\},  & \mbox{if $k=jn+i$ and $\tau(i,j)=\tvar$,}\\[3pt]
				%			\emptyset, &\mbox{otherwise}
				%		\end{array}
			%		\right.
			%	\end{align*}
		%	%
		%	for $\tvar\in T$.		
		%	By construction, 
		It is straightforward to check that $\mathfrak{N},r,0\models \varphi$ and $\mathfrak{N},r,0\models \neg\psi$.
		%	and $\mathfrak{M},0,0,\not\models \psi$. 
		First, we show that $\varphi \wedge \psi$ is not a 
		\emph{strong} conservative extension of $\varphi$.
		We construct a formula $\chi$ with $\sig(\chi) \cap \sig(\psi)\subseteq \sig(\varphi)$ such that $\varphi \wedge \chi$ is satisfiable but 
		$\models_\FOMS\varphi\land\chi\to\neg\psi$.
		%$\varphi \wedge \psi \wedge \chi$ is not satisfiable. 
		It then follows 
		%by definition 
		that $\varphi \wedge \psi$ is not a strong conservative extension of $\varphi$. The formula $\chi$ provides a description of the model $\mathfrak{N}$ at $(r,0)$. We take, for every $(i,j)\in W\times D$, a fresh predicate $\pvar_{i,j}$ and extend $\mathfrak{N}$ to $\mathfrak{N}'$ by setting,
		for all $(i',j')\in W\times D$,	
		\begin{equation}
			\label{Mdesc2}
			\mbox{$\mathfrak N',i',j'\models \pvar_{i,j}$\quad iff\quad $(i',j')=(i,j)$.}
		\end{equation}
		Now let $\sigma'=\sig(\varphi) \cup \{\pvar_{i,j} \mid (i,j)\in W\times D\}$, and
		set
		\begin{equation}\label{Kdiagf}
			\chi_{i,j} = \bigwedge_{\pvar\in\sigma',\ \mathfrak N,i,j\models \pvar}\pvar\  \land
			\bigwedge_{\pvar\in\sigma',\ \mathfrak N,i,j\models \neg\pvar}\neg\pvar.
		\end{equation}
		Let $\chi$ be the conjunction of
		of the following formulas:
		\begin{align}
			\label{ref00}
			&\chi_{r,0}\land\Bh(\ovar\land\gvar\to\chi_{0,0}),\\
			\label{1ref}
			&\Bv \bigl[\Dh \chi_{i,i} \rightarrow \Bh \bigl(\xvar \rightarrow (\chi_{i+1,i} \wedge \Bv (\gvar \rightarrow \chi_{i+1,i+1})\bigr)\bigr],
			\text{ for $i<nm-1$}, \\
			\label{2ref}
			&\Bv \bigl[\Dh(\yvar\land\Dv \chi_{i,i}) \rightarrow \Bh(\yvar \rightarrow \chi_{i,i+n})\land\Bh(\gvar \rightarrow \chi_{i+n,i+n})\bigr], 
			\text{ for $i<nm-n$},\\
			\label{3refa}
			&\Bh\Bv (\chi_{i,i} \rightarrow \Dv \chi_{i,j}),\text{ for $i,j<nm$,}\\
			\label{3refb}
			&\Bv(\Dh\chi_{i,i} \rightarrow \Dh\chi_{j,i}),\text{ for $i,j<nm$,}\\		
			\label{3refc}
			&\Bv(\Dh\chi_{i,j} \rightarrow \chi_{r,j}),\text{ for $i,j<nm$},\\        
			\label{3ref}
			&\Bv \bigl(\chi_{r,j} \rightarrow \Bh( \Dv \chi_{l,k} \rightarrow \chi_{l,j})\bigr),\text{ for $j,k,l<nm$}.	
		\end{align}
		It is easy to see that
		%Clearly 
		$\mathfrak{N}',r,0\models \chi$, and so $\varphi \wedge \chi$ is satisfiable. 
		Now suppose that $\mathfrak{M}$ is any model such that $\mathfrak{M},w_0,d_0 \models \varphi\wedge\chi$ for some $w_0,d_0$. We show that $\mathfrak{M},w_0,d_0 \models\neg\psi$.
		%	Assume for a proof by contradiction 
		% By definition,
		%	$R_{h}$ and $R_{v}$ are defined on $\gvar$-nodes. Hence, by \eqref{eq:domain}, they are defined on $\bigvee\varphi_{i,i}$-nodes only. Hence, 
		Observe that if $\bigl((w,d),(w',d')\bigr)\in R_{h}^{\mathfrak{M}}$
		and $\mathfrak{M},w,d\models \chi_{i,i}$, then $\mathfrak{M},w',d'\models \chi_{i+1,i+1}$, by \eqref{1ref}, and if $\bigl((w,d),(w',d')\bigr)\in R_{v}^{\mathfrak{M}}$
		and $\mathfrak{M},w,d\models \chi_{i,i}$, then $\mathfrak{M},w',d'\models \chi_{i+n,i+n}$, by \eqref{2ref}. Hence, there cannot be an infinite $R_{h}^{\mathfrak{M}} \cup R_{v}^{\mathfrak{M}}$-chain. 
		
		Now suppose there is an $R_{h}^\mathfrak M\cup R_{v}^\mathfrak M$-chain from $(w_0,d_0)$
		%a node satisfying $\ovar$ 
		to some node $(w,d)$ which has an $R_{h}^\mathfrak M$-successor $(w_{1},d_{1})$ and $R_{v}^\mathfrak M$-successor $(w_{2},d_{2})$. Then $\mathfrak M,w,d\models \chi_{i,i}$ for some $i$, 
		$\mathfrak M,w_{1},d_{1}\models \chi_{i+1,i+1}$ and $\mathfrak M,w_{2},d_{2}\models \chi_{i+n,i+n}$, by \eqref{ref00}--\eqref{2ref}.

		There exist $d_{1}'$ with $\mathfrak M,w_{1},d_{1}'\models \chi_{i+1,i+n+1}$, and $w_{2}'$ with $\mathfrak M,w_{2}',d_{2}\models \chi_{i+n+1,i+n}$, by \eqref{3refa} and \eqref{3refb}.
		Then $\mathfrak M,w_0,d'\models\chi_{r,i+n+1}$ by \eqref{3refc}.
		By \eqref{3ref} for $l=j=i+n+1$ and $k=i+n$,
		%	applied to
		%$$
		%\Bv (\chi_{r,i+n+1} \rightarrow \Bh( \Dv \chi_{i+n+1,i+n} \rightarrow \chi_{i+n+1,i+n+1}))
		%$$
		we obtain $\mathfrak M,w_{2}',d_{1}'\models \chi_{i+n+1,i+n+1}$.	
		Moreover, as $\bbvarup$ is not a conjunct of $\chi_{i+1,i+1}$, and $\bbvarright$ is not a conjunct of $\chi_{i+n,i+n}$, we have that $\yvar$ is a conjunct of $\chi_{i+1,i+n+1}$, and 
		$\xvar$ is a conjunct of $\chi_{i+n+1,i+n}$. Thus, 
		$\bigl((w_{1},d_{1}),(w_{2}',d_{1}')\bigr)\in R_{v}^\mathfrak M$ and $\bigl((w_{2},d_{2})(w_{2}',d_{1}')\bigr)\in R_{h}^\mathfrak M$,
		and so $(w,d)$ is confluent.
		%Assume first that there is an $R_{v},R_{h}$-path in $m^{I}$ from  $(w_{0},v_{0})$ to some $(u,u')$ such that $(u,u')$ has an $R_{h}$-successor $(a,b)$ and an $R_{v}$-successor $(c,d)$ which do not have a common $R_{v}$ and $R_{h}$-successor, respectively. We then have $(u,u') \in \varphi_{v,v}^{I}$ for some $v$.
		%	Then $(a,b) \in \varphi_{v+1,v+1}^{I}$ and $(c,d)\in \varphi_{v+n,v+n}^{I}$. 
		%	Then there are $(a,b')\in (\varphi_{v+1+n,v+1} \wedge z)^{I}$ and $(c',d)\in  (\varphi_{v+n,v+n+1} \wedge z)^{I}$ and then $(c',b') \in \varphi_{v+1+n,v+n+1}^{I}$ since
		%	$\chi$ contains 
		%	$$
		%	\Box_{1}\Box_{2}(\Diamond_{2}\varphi_{v+1+n,v+1} \wedge \Diamond_{1} \varphi_{v+n,v+n+1} \rightarrow \varphi_{v+n+1,v+n+1})
		%	$$
		%	Moreover $(a,b),(c',b')\in R_{v}$ and $(c,d),(c',b')\in R_{v}$
		%	and we have derived a contradiction.
		%	
		%	That there is no infinite $R_{h},R_{v}$ path follows from the observation above.
		%
		
		We next aim to prove that $\varphi\land\psi$ is not a (necessarily strong) conservative extension of $\varphi$. In this case, we are not allowed to use the fresh predicates 
		$\pvar_{i,j}$ in the formula $\chi$ to achieve \eqref{Mdesc2}.
		We instead  use the predicate $\pvar_0\in\sig(\varphi)$ to uniquely characterise the points of $\mathfrak N$.
		Take a bijection $f$ from $\{0,\ldots,n-1\}\times \{0,\ldots,m-1\}$ to $\{0,\ldots,nm-1\}$ and 
		set, for all $(i,j)\in W\times D$:
		\[ 
		\varphi_{i,j}=\Dh^{f(i,j)+1}\pvar_0,\quad \varphi_{r,j}=\neg \Dv \gvar \wedge \Dh \varphi_{0,j},\ \ 
		\mbox{for $i,j<nm$.}
		\]
		Modify the model $\mathfrak N=(W,R,D,J)$ constructed above to $\mathfrak N^+=(W^+,R^+,D,J^+)$ by adding an $nm$-long $R$-chain to each
		leaf $i<nm$ of $(W,R)$, and define $\pvar_0^{J^+(w)}$ for each $w\in W^+\setminus W$ such that we still have, for all `old' points $(i',j')\in W\times D$, the analogue of \eqref{Mdesc}:
		%
		%	\begin{equation}		\label{MdescP}
			\[
			\mbox{$\mathfrak N^+,i',j'\models \varphi_{i,j}$\quad iff\quad $(i',j')=(i,j)$.}
			%	\end{equation}
		\]
		Now let $\sigma^-=\sig(\varphi)\setminus\{\pvar_0\}$ and set, for all $(i,j)\in W\times D$,
		%
		%	\begin{equation}\label{KdiagfP}
			\[
			\chi_{i,j}' = \varphi_{i,j}\land\bigwedge_{\pvar\in\sigma^-,\ \mathfrak N,i,j\models \pvar}\pvar\  \land
			\bigwedge_{\pvar\in\sigma^-,\ \mathfrak N,i,j\models \neg\pvar}\neg\pvar.
			%	\end{equation}
		\]
		Finaly, define $\chi'$ with $\sig(\chi')\subseteq\sig(\varphi)$ by replacing $\chi_{i,j}$ in \eqref{ref00}--\eqref{3ref} by $\chi_{i,j}'$.	
	\end{proof}
	
	\begin{lem}\label{lem:ifnosolthencons2}
		%If $\mathfrak{T}$ has no solution, then $\varphi \wedge \psi$ is a model conservative extension of $\varphi$ (and so also a signature independent conservative extension of $\varphi$).
		If $\varphi \wedge \psi$ is not a model conservative extension of $\varphi$, then $\tiling$ has a solution.
	\end{lem}
	\begin{proof}
		Consider a model $\mathfrak{M}=(W,R,D,I)$ such that $\mathfrak{M},w,d\models \varphi$ but 
		$\mathfrak{M}',w,d\models \neg\psi$
		in any extension $\mathfrak M'$ of $\mathfrak{M}$ obtained by interpreting the predicates $\qvar,\ssvar$.
		Similarly, to the proof of Lemma~\ref{lem:ifnosolthencons}, 
		by using the equivalence $\textbf{(c1)} \Leftrightarrow \textbf{(c2)}$, one can easily find within $\mathfrak{M}$ a finite grid-shaped (with respect to $R_{h}^\mathfrak M$ and 
		$R_{v}^\mathfrak M$) submodel, which gives a solution to $\tiling$.
	\end{proof}
	
This completes the proof of Theorem~\ref{undec:cons1} $(i)$.
%We next aim to prove undecidable for not necessarily strong conservative extensions. In this case, we are not allowed to use the fresh predicates 
%$\pvar_{i,j}$. We instead proceed as follows. We take a single fresh predicate $\avar$
%and add $\avar \rightarrow \avar$ as a conjunct for $\varphi$ to obtain $\varphi'$. We claim that $\varphi' \wedge \psi$ is a conservative extension of $\varphi'$ iff $\varphi \wedge \psi$ is a strong conservative extension of $\varphi$. To show this, it suffices to replace predicates $\pvar_{i,j}$ by suitable formulas only using $\avar$. Take a bijection $f$ between $\{0,\ldots,n-1\}\times \{0,\ldots,m-1\}$ and $\{0,\ldots,nm-1\}$ and 
%replace $\pvar_{i,j}$, $i\not=r$, in the construction of $\chi$ by
%$\varphi_{i,j}=\Diamond^{f^{-1}(i,j)}\avar$ and replace $\pvar_{r,j}$ by $\varphi_{r,j}=\neg \exists \gvar \wedge \Diamond \varphi_{0,j}$.  
We next prove Theorem~\ref{undec:cons1} $(ii)$ based on a modification of Example~\ref{ex6} for the case of $\FOMK{}$.
\begin{exa}\label{ex62}
	Let $\varphi_{0}$ %be the conjunction of
	$$
	\varphi_{0} = \Bv \bigl(\Dh\avar \rightarrow \Dh (\pvar_{1} \wedge \bvar)\bigr) \land 
	\Bh \Bv \bigl(\pvar_{1} \wedge \bvar \rightarrow \Dv (\pvar_{2} \wedge \bvar)\bigr)\land
	\Bv \bigl(\Dh (\pvar_{2} \wedge \bvar) \rightarrow \Dh (\pvar_1 \wedge \bvar)\bigr)
	$$
	and let $\sigma=\{\avar,\pvar_{1},\pvar_{2}\}$.  
	We show that there is no $\sigma$-uniform interpolant of
	$\Diamond \avar \wedge \varphi_{0}$ in $\FOMK{}$.  
	
	For every $s>0$, we define a formula $\chi_{s}$ as follows. Take fresh predicates $\avar_{s}^h$ and $\avar_{s}^v$ and construct
	$\chi_{s}$ stating that $s$-many steps of the $\pvar_{1},\pvar_{2}$-ladder have been constructed by forcing $\avar_{1}^h,\ldots,\avar_{s}^h$ to be true
	horizontally and $\avar_{1}^v,\ldots,\avar_{s}^v$ to be true vertically in the corresponding steps.  
	In detail, we let
	\begin{align*}
		\chi_{1}' & = \Bv(\Dh\avar \rightarrow \Box \avar_{1}^h) \wedge \Bh\Bv (\avar_{1}^h \wedge \pvar_{1} \rightarrow \Bv \avar_{1}^v),\\
		\chi_{1} & = \chi_{1}' \rightarrow \exists \Diamond (\pvar_{2} \wedge \avar_{1}^v),
	\end{align*}	
	and for $s>0$,
	\begin{align*}
		\chi_{s+1}' & = \chi'_s\land\Bv\bigl(\Dh(\avar_s^v\land\pvar_2) \rightarrow \Box \avar_{s+1}^h\bigr)  
		\wedge \Bh\Bv (\avar_{s+1}^h \wedge \pvar_{1} \rightarrow \Bv \avar_{s+1}^v),\\
		\chi_{s+1} & = \chi_{s+1}' \rightarrow \exists \Diamond (\pvar_{2} \wedge \avar_{s+1}^v).
	\end{align*}	
	%
	%	Define
	%	$$
	%	\chi_{2} =  \chi_{2}' \rightarrow \exists \Diamond (\pvar_{2} \wedge \avar_{2}')
	%	$$
	%	where 
	%	$$
	%	\chi_{2}'= \chi_{1}' \wedge \forall \Diamond (\pvar_{2} \wedge \avar_{1}' \rightarrow \Box \avar_{2}) \wedge \forall \Box (\avar_{2} \wedge \pvar_{1} \rightarrow \forall \avar_{2}')
	%	$$
	%	And define in general:
	%	$$
	%	\chi_{r+1} =  \chi_{r+1}' \rightarrow \exists \Diamond (\pvar_{2} \wedge \avar_{r+1}')
	%	$$
	%	where 
	%	$$
	%	\chi_{r+1}'= \chi_{r}' \wedge \forall \Diamond (\pvar_{2} \wedge \avar_{r}' \rightarrow \Box \avar_{r+1}) \wedge \forall \Box (\avar_{r+1} \wedge \pvar_{1} \rightarrow \forall \avar_{r+1}')%
	%	$$
	%	
	Then $\models_\FOMK \Diamond \avar\land \varphi_0\to \chi_s$ for all $s>0$.
	Thus, if $\varrho$ were a $\sigma$-uniform interpolant of $\Diamond \avar\land\varphi_0$, then
	$\models_\FOMK\varrho\to \chi_s$ would follow, for all $s>0$.

	On the other hand, for $s>0$, 
	we modify the model $\mathfrak{M}_s=(W_s,D_s,I_s)$ defined in Example~\ref{ex6}
	by adding a root world to $W_s$ from where every other world is accessible in one $R_s$-step.
	More precisely, we define a new model $\mathfrak{M}_s=(W_s,R_s,D_s,I_s)$ as follows.
	We let $W_s'=D_s=\{0,\dots,s-1\}$, $W_s=W_s'\cup \{r\}$, $R_s=\{(r,w)\mid w\in W_s'\}$, and $I_s$ is defined
	by 
	\begin{align}
		\label{uiep1}
		%&\avar^{\mathfrak{M}_{n}}= \{(0,0)\}\\
		\avar^{I_s(k)} & = \left\{
		\begin{array}{ll}
			\{0\},  & \mbox{if $k=0$,}\\[3pt]
			\emptyset, &\mbox{otherwise};
		\end{array}
		\right.
		\\
		%&\pvar_{1}^{\mathfrak{M}_{n}} =\{(k+1,k) \mid k<n\}\\
		\pvar_{1}^{I_{s}(k)}  & = \left\{
		\begin{array}{ll}
			\{k-1\},  & \mbox{if $k>0$,}\\[3pt]
			\emptyset, &\mbox{otherwise};
		\end{array}
		\right.
		\\
		\label{uiep3}   
		%&\pvar_{2}^{\mathfrak{M}_{n}}=\{(k,k) \mid k<n\}
		\pvar_{2}^{I_{s}(k)}  & = \left\{
		\begin{array}{ll}
			\{k\},  & \mbox{if $k>0$,}\\[3pt]
			\emptyset, &\mbox{otherwise};
		\end{array}
		\right.
	\end{align}
	plus having $\pvar^{I_s(r)}=\emptyset$ for 
	$\pvar\in\{\avar,\pvar_1,\pvar_2,\avar_1^h,\dots,\avar_s^h,\avar_1^v,\dots,\avar_s^v\}$ and, for all 
	$0<i\le s$ and all $k<s$,
	%	
	%	take $W_{r}'=D_{r}=\{0,\ldots,r-1\}$, $W_{r}=W_{r}'\cup \{w_{0}\}$ and define a model
	%		$\mathfrak{M}_{r}=(W_{r},R,D_{r},I_{r})$ as follows. Let $(w_{0},k)\in R$ for every $k<r$ and let for every $k<r$ 
	%	Assume there is a uniform interpolant $\rho$. 
	%Then consider the following models $\mathfrak{M}_{n}=(W_{n},D_{n},I_{n})$ with $W_{n}=D_{n}=\{0,\ldots,n-1\}$ and
	%
%	\begin{align*}
%
\[
				(\avar_{i}^h)^{I_{s}(k)}   = \{i-1\},
		\qquad
		(\avar_{i}^v)^{I_{s}(k)}   = \left\{
		\begin{array}{ll}
			\{0,\ldots,s-1\},  & \mbox{if $0<k=i<s$,}\\[3pt]
			\emptyset, &\mbox{otherwise.}
		\end{array}
		\right.
\]
%		
%	\end{align*}
	%
	Then, for every $s>0$, $\mathfrak{M}_{s},r,0\not\models \chi_s$ and so $\mathfrak{M}_{s},r,0\models\neg \varrho$. Also, $\mathfrak{M}_{s},r,0\models \chi_{s'}$ for all $s'<s$. 
	%	Then $\mathfrak{M}_{n},0,0\not\models \varrho$ for all $n>0$. So see this consider the formula $\chi_{n}$ defined inductively by setting $\chi_{0}= \pvar_{2}$ and
	%	$\chi_{n+1}=\avar \wedge \Dh(\pvar_{1} \wedge \Dv \chi_{n})$. Then $\mathfrak{M}_{n},0,0\not\models \avar \wedge \chi_{n}$ but clearly
	%	$\models \varphi_{0}\rightarrow \chi_{n}$ for all $n>0$. 
	%Hence $\models \varrho \rightarrow \chi_{n}$ for all $n>0$. 
	%So $\mathfrak{M}_{n},0,0\not\models \varrho$ for all $n>0$. 
	Now consider the ultraproduct $\prod_{U}\mathfrak{M}_{s}$ with $U$ a non-principal ultrafilter on $\omega\setminus\{0\}$. As each $ \chi_{s'}$ is true in almost all $\mathfrak{M}_{s},r,0$, it follows from the properties of ultraproducts \cite{modeltheory} that $\prod_{U}\mathfrak{M}_{s},\overline{r},\overline{0}\models \Dh \avar \land \neg \varrho \land \chi_{s'}$ for all $s'>0$, for some suitable $\overline{r},\overline{0}$. But then one can interpret $\bvar$ in $\prod_{U}\mathfrak{M}_{s}$ such that  
	$\mathfrak{M},\overline{r},\overline{0}\models \varphi_{0}$ for the resulting model $\mathfrak{M}$. Then $\mathfrak{M}\models \Diamond \avar \wedge \varphi_{0} \wedge \neg \varrho$ and as $\models_\FOMK\Diamond\avar\land\varphi_0\to\varrho$ should hold for a uniform interpolant $\varrho$ of $\Diamond \avar\land\varphi_0$, we have derived a contradiction. \hfill $\dashv$
\end{exa}

The proof of Theorem~\ref{undec:cons1} $(ii)$ is now by combining the construction of Theorem~\ref{undec:cons1} and Example~\ref{ex62} in exactly the same way as the construction of Theorem~\ref{undec:cons} and Example~\ref{ex6} were combined in the proof of Theorem~\ref{undec:cons} $(ii)$.

%*************

\section{Outlook}\label{outlook}

Craig interpolation and Beth definability have been studied extensively for most logical systems, let alone those with applications in computing such as KR, verification, and  databases. In fact, one of the first questions typically asked about a logic $L$ of interest is whether $L$ has interpolants for \emph{all} valid implications $\varphi \to \psi$. Some $L$ enjoy this property, while others miss it. 
This paper and preceding~\cite{DBLP:conf/lics/JungW21,DBLP:journals/tocl/ArtaleJMOW23} open a new, \emph{non-uniform} perspective on interpolation/definability for the latter type of $L$ by regarding formulas $\varphi$ and $\psi$ as input and deciding whether they have an interpolant in $L$. We refer the reader to ~\cite{chapter:separation} for an overview of current research and open questions from this perspective. 

%{\color{blue}Interpolation, explicit definition, conservative extension are among the most important concepts for  applications of logic in KR, verification, databases, and other areas. However, a typical problem  considered so far has been whether or not a logic of interest \emph{uniformly}  enjoys, say, Craig interpolation:  a classical example is Maksimova's classification\nz{maybe it's not complete:-)} of modal logics containing $\mathsf{S4}$ with respect to interpolation~\cite{MGabbay2005-MGAIAD}. 
%
%The results of this paper and preceding~\cite{DBLP:conf/lics/JungW21,DBLP:conf/aaai/ArtaleJMOW21} open a new perspective on investigating and  using interpolants, definitions, etc.\ for logics where their existence is not guaranteed.}

In the context of first-order modal logics, 
challenging open questions that arise from this work are: 
\begin{itemize}
\item[--] What is the tight complexity of the IEP for \FOMS{} and $\ALCIOS$? 
\item[--] Is the non-elementary upper bound for the IEP in \FOMK{} optimal? (Note that just like \FOMS-validity, \FOMK-validity is known to be \textsc{coNExpTime}-complete~\cite{DBLP:journals/logcom/Marx99}.)
\item[--] Is the IEP decidable for $\mathsf{K}_{\mathcal{ALC}^{u}}$?
\item[--] More generally, what happens if we replace $\mathsf{S5}$ and $\mathsf{K}$ by other standard modal logics, e.g., $\mathsf{S4}$, multimodal $\mathsf{S5}$, or the linear temporal logic \LTL, and/or use in place of $\mathcal{ALC}^{u}$ more expressive DLs containing, for instance, nominals or role inclusions, and/or consider other monodic fragments of first-order modal logics such as the monodic guarded or two-variable fragment?
\item[--] It would also be of interest to investigate variants of the IEP in more detail. For instance, for the logics considered here, does a Lyndon interpolant exist iff an arbitrary interpolant exists? If not, is the existence of Lyndon interpolants also decidable and of the same complexity? Is global deductive interpolant existence decidable for \FOMK{}? 
\end{itemize}
A different line of research is computing interpolants. For logics with the CIP, this is typically done using resolution, tableau, or sequent calculi as , e.g., in~\cite{DBLP:books/daglib/0082098,DBLP:conf/lpar/KovacsV17,DBLP:journals/apal/Kuznets18,DBLP:journals/jar/Wernhard21}. It remains to be seen if interpolants in the logics without the CIP considered in this article can be extracted from tableau or resolution proofs in these calculi or those designed specifically for monodic fragments~\cite{DBLP:conf/cade/LutzSWZ01,DBLP:journals/sLogica/LutzSWZ02,DBLP:journals/tocl/DegtyarevFK06}. A more recent approach is based on type-elimination known from complexity proofs for modal and guarded logics~\cite{DBLP:journals/tocl/BenediktCB16,baldernote}. The question whether these proofs can be turned into an algorithm computing interpolants in, say, \FOMS{} is non-trivial and open. More generally, one can try to develop calculi for the consequence relation `$\varphi \models \psi$ iff there are no $\sig(\varphi)\cap\sig(\psi)$-bisimilar models satisfying $\varphi$ and $\neg \psi$' and use them to compute interpolants; see~\cite{DBLP:journals/jsyml/BarwiseB99} for a model-theoretic account of such consequence relations for infinitary logics without the CIP. 

\section*{Acknowledgments}

This research was supported by the EPSRC UK grants EP/S032207 and EP/S032282 for the project `quant$^\text{MD}$: Ontology-Based Management for Many-Dimensional Quantitative Data'\!.

%***************

%\bigskip

\begin{appendix}

%***************

%\centerline{\Large\bf Appendix}

%**************

\section{Connections with \FOT}\label{s:FO}

We begin by discussing the connections between \FOMS{} and \FOT. Then we prove the lower bound result of Theorem~\ref{thm2intro}, stating that interpolant and definition existence in \FOT{} without equality
are 2\textsc{ExpTime}-hard. 

The \emph{atoms} of \FOT{} are of the form $x\mathop{=}y$, $\pvar(x,y)$, $\pvar(y,x)$,
$\pvar(x,x)$, and $\pvar(y,y)$, with $\pvar$ ranging over binary predicate symbol in $\varset$.\footnote{It is shown in~\cite{DBLP:journals/bsl/GradelKV97} that, in \FOT{}, one can replace relations of arbitrary arity by binary relations as far the complexity of satisfiability is concerned. This has been extended to bisimulation consistency in~\cite{DBLP:conf/lics/JungW21}.
	We therefore consider \FOT{} with binary relations only.}
A \emph{signature} is any finite set $\sigma\subseteq\varset$.
\emph{\FOT-formulas} 
are built up from atoms using $\neg,\land,\exists x,\exists y$. 
We consider two proper fragments of \FOT: in the \emph{equality-free} fragment, we do not have atoms of the form
$x\mathop{=}y$,
and in the \emph{equality- and substitution-free} fragment, the only available atoms are of the form
$\pvar(y,x)$.
We interpret \FOT-formulas in usual \emph{\FO-models} of the form $\mathfrak A=(A^\mathfrak A,\pvar^\mathfrak A)_{\pvar\in\varset}$, where $A^\mathfrak A$ is a nonempty set and $\pvar^\mathfrak A\subseteq A^\mathfrak A\times A^\mathfrak A$, for each
$\pvar\in\varset$.

Now, fix some signature $\sigma$.
We connect two \FO-models $\mathfrak A,\mathfrak B$ with three different kinds of $\sigma$-bisimulations,
depending on the chosen fragment $\mathcal L$ of \FOT{}, as follows.
Given $\mathcal{L}$, let $\textit{Lit}_{\mathcal{L}(\sigma)}$ denote the set of available \emph{literals} for $\pvar\in\sigma$ (atoms and negated atoms) in $\mathcal{L}$.
Given $\mathfrak A$ and $a,a'\in A^\mathfrak A$, we define 
\[
\ell_{\mathfrak A}^{\mathcal{L}(\sigma)}(a,a')=\bigl\{\ell\in \textit{Lit}_{\mathcal{L}(\sigma)} \mid \mathfrak A\models\ell[a/y,a'/x]\bigr\}.
\]
A relation $\bs\subseteq (A^\mathfrak A\times A^\mathfrak A)\times (B^\mathfrak B\times B^\mathfrak B) $
is a $\sigma$-\emph{bisimulation between $\mathfrak A$ and $\mathfrak B$ in $\mathcal{L}$} if the following hold,
for all $\bigl((a,a'),(b,b')\bigr)\in\bs$:
\begin{enumerate}
\item
$\ell_{\mathfrak A}^{\mathcal{L}(\sigma)}(a,a')=\ell_{\mathfrak B}^{\mathcal{L}(\sigma)}(b,b')$;
\item
for every $a''\in A^\mathfrak A$ there is $b''\in B^\mathfrak B$ such that $\bigl((a,a''),(b,b'')\bigr)\in\bs$, and the other way round;
\item
for every $a''\in A^\mathfrak A$ there is $b''\in B^\mathfrak B$ such that $\bigl((a'',a'),(b'',b')\bigr)\in\bs$, and the other way round.
\end{enumerate}
If $\bigl((a,a'),(b,b')\bigr)\in\bs$ for some $\bs$, then we say that $\mathfrak A,a,a'$ and $\mathfrak B,b,b'$ are \emph{$\sigma$-bisimilar in $\mathcal{L}$\/}.

%Let $\mathcal L$ be any of the three fragments above and $\sigma$ a signature. 
Given \FO-models $\mathfrak A,\mathfrak B$ and
$a,a'\in A^\mathfrak A$, $b,b'\in B^\mathfrak B$, we write
$\mathfrak A,a,a'\equiv_{\mathcal{L}(\sigma)}\mathfrak B,b,b'$ whenever
$\mathfrak A\models\varphi[a/y,a'/x]$ iff $\mathfrak B\models\varphi[b/y,b'/x]$
hold for all $\mathcal L(\sigma)$-formulas $\varphi$. 
Then we have the following well-known equivalence, for any pair $\mathfrak A,\mathfrak B$ of saturated models:
$$
\mathfrak A,a,a'\equiv_{\mathcal{L}(\sigma)}\mathfrak B,b,b'\quad \text{iff}\quad \mathfrak A,a,a' \text{ and } \mathfrak B,b,b' \text{ are $\sigma$-bisimilar in $\mathcal{L}$}.
$$
Clearly, if $\mathfrak A,a,a'$ and $\mathfrak B,b,b'$ are $\sigma$-bisimilar in \FOT{}, then they are $\sigma$-bisimilar in
the equality-free fragment, and if they are $\sigma$-bisimilar in the equality-free fragment, then they are 
$\sigma$-bisimilar in the equality- and substitution-free fragment. However, as the models below show, 
the converse directions do not always hold.\\
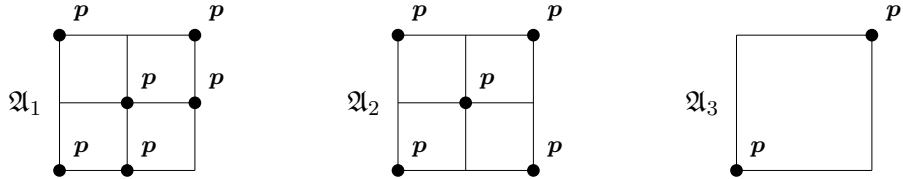
\centerline{
\begin{tikzpicture}[>=latex,line width=0.2pt,scale=.9]
\node[]  at (-.5,1) {$\mathfrak A_1$};
\node[point,fill=black,scale = 0.7,label=above right:{\footnotesize $\pvar$}] (aa1) at (0,0) {};
\node[point,fill=black,scale = 0.7,label=above right:{\footnotesize $\pvar$}] (ba1) at (1,0) {};
\node[point,fill=black,scale = 0.7,label=above right:{\footnotesize $\pvar$}] (bb1) at (1,1) {};
\node[point,fill=black,scale = 0.7,label=above right:{\footnotesize $\pvar$}] (cb1) at (2,1) {};
\node[point,fill=black,scale = 0.7,label=above right:{\footnotesize $\pvar$}] (ac1) at (0,2) {};
\node[point,fill=black,scale = 0.7,label=above right:{\footnotesize $\pvar$}] (cc1) at (2,2) {};
\draw[-] (0,0) -- (2,0);
\draw[-] (0,1) -- (2,1);
\draw[-] (0,2) -- (2,2);
\draw[-] (0,0) -- (0,2);
\draw[-] (1,0) -- (1,2);
\draw[-] (2,0) -- (2,2);
\node[]  at (4.5,1) {$\mathfrak A_2$};
\node[point,fill=black,scale = 0.7,label=above right:{\footnotesize $\pvar$}] (aa2) at (5,0) {};
\node[point,fill=black,scale = 0.7,label=above right:{\footnotesize $\pvar$}] (ca2) at (7,0) {};
\node[point,fill=black,scale = 0.7,label=above right:{\footnotesize $\pvar$}] (bb2) at (6,1) {};
\node[point,fill=black,scale = 0.7,label=above right:{\footnotesize $\pvar$}] (ac2) at (5,2) {};
\node[point,fill=black,scale = 0.7,label=above right:{\footnotesize $\pvar$}] (cc2) at (7,2) {};
\draw[-] (5,0) -- (7,0);
\draw[-] (5,1) -- (7,1);
\draw[-] (5,2) -- (7,2);
\draw[-] (5,0) -- (5,2);
\draw[-] (6,0) -- (6,2);
\draw[-] (7,0) -- (7,2);
\node[]  at (9.5,1) {$\mathfrak A_3$};
\node[point,fill=black,scale = 0.7,label=above right:{\footnotesize $\pvar$}] (aa3) at (10,0) {};
\node[point,fill=black,scale = 0.7,label=above right:{\footnotesize $\pvar$}] (bb3) at (12,2) {};
\draw[-] (10,0) -- (12,0);
\draw[-] (10,2) -- (12,2);
\draw[-] (10,0) -- (10,2);
\draw[-] (12,0) -- (12,2);
\end{tikzpicture}
}\\
Here, $\mathfrak A_1$ and $\mathfrak A_3$ are $\{\pvar\}$-bisimilar in the equality- and substitution-free fragment of \FOT, but not in the
equality-free fragment: 
$\forall x\forall y\bigl(\pvar(x,y)\leftrightarrow\pvar(y,x)\bigr) \land \forall x \pvar(x,x)$ is true
in $\mathfrak A_3$, while it is not true in $\mathfrak A_1$.
$\mathfrak A_2$ and $\mathfrak A_3$ are $\{\pvar\}$-bisimilar in the equality-free fragment, but not in \FOT:
$ \forall x\forall y\bigl(\pvar(x,y)\to x\mathop{=}y\bigr)$ is true
in $\mathfrak A_3$, while it is not true in $\mathfrak A_2$.

Next, we discuss connections between $\FOMS$ and the three fragments of \FOT{} introduced above.
With a slight abuse of notation, we consider the predicate symbols in $\varset$ (and thus in any signature $\sigma$)
as unary symbols when dealing with \FOM{} and binary ones when dealing with (fragments of) \FOT.
We translate each \FOM-formula $\varphi$ to an \FOT-formula $\varphi^\dag$ 
by taking 
$$
\pvar(x)^\dag = \pvar(y,x), \quad
(\neg\varphi)^\dag  = \neg\varphi^\dag, \quad
(\varphi\land\psi)^\dag  = \varphi^\dag\land\psi^\dag, \quad
(\Dh\varphi)^\dag  = \exists y\varphi^\dag, \quad
(\Dv\varphi)^\dag  = \exists x\varphi^\dag.
$$
Observe that the image of this translation is in the equality- and substitution-free fragment
of \FOT{}.
%(where we do not have atomic formulas of the form $\pvar(x,y)$ or $\pvar(x,x)$).
%Given a $\FOM$-signature $\sigma$, we define the $\FOT$-signature 
%$\sigma^\dag=\{\pvar^\dag\mid\pvar\in\sigma\}$.
It is easy to show the following:
\begin{lem}\label{l:fot}
%Suppose $\varphi$, $\psi$ are 
For all  \FOM-formulas $\varphi$, $\psi$ and $\FOM$-signature $\sigma$,
\begin{enumerate}[style=multiline,leftmargin=0.7cm]
\item[$(i)$]
$\models_\FOMS\varphi$ iff $\models_\FOT\varphi^\dag$\textup{;}

\item[$(ii)$]
$\varphi$ and $\psi$ are $\sigma$-bisimulation consistent in \FOMS{} iff
$\varphi^\dag$ and $\psi^\dag$ are $\sigma$-bisimulation consistent in the equality- and substitution-free fragment of \FOT{}.
\end{enumerate}
\end{lem}
\begin{proof}
Follows straightforwardly from the observations $(a)$ and $(b)$ below:

%\begin{itemize}
%\item
$(a)$ \FO-models for the
equality- and substitution-free fragment of \FOT{} and \emph{square} \FOMS-models $(W,D,I)$ with $|W|=|D|$ are in 1--1 correspondence in the following sense:
\begin{itemize}
\item[--]
 For every \FO-model $\mathfrak A$, take the square
 \FOMS-model $\mathfrak M_\mathfrak A=(A^\mathfrak A,A^\mathfrak A,I)$ where, for all $a,b\in A^\mathfrak A$ and $\pvar\in\varset$, $b\in\pvar^{I(a)}$ iff $(a,b)\in\pvar^\mathfrak A$. 
 Then we have
 \[
\mbox{$\mathfrak M_\mathfrak A,a,b\models\varphi$\quad iff\quad $\mathfrak A\models\varphi^\dag[a/y,b/x]$.}
\]
 %
 %Further, for any relation $\bs$, $\bs$ is a $\sigma$-bisimulation between $\mathfrak A$ and $\mathfrak B$ in the equality- and substitution-free fragment of \FOT{} iff
% $\bs$ is a $\sigma$-bisimulation between $\mathfrak M_\mathfrak A$ and $\mathfrak M_\mathfrak B$ in \FOMS.
 
 \item[--]
 For every square \FOMS-model $\mathfrak M=(W,D,I)$ and every bijection $f\colon D\to W$, take
 the \FO-model $\mathfrak A_{\mathfrak M,f}=(D,\pvar^{\mathfrak A_{\mathfrak M,f}})$ where, for all $a,b\in D$ and $\pvar\in\varset$, $(a,b)\in\pvar^{\mathfrak A_{\mathfrak M,f}}$ iff $b\in\pvar^{I(f(a))}$.
 Then we have
 \begin{equation}
 \label{sqm}
 \mbox{$\mathfrak M,f(a),b\models\varphi$\quad iff\quad $\mathfrak A_{\mathfrak M,f}\models\varphi^\dag[a/y,b/x]$.}
\end{equation}
\end{itemize}

%\item
$(b)$ For all \FOMS-models $\mathfrak M$, $w,d$, there exists a square
\FOMS-model $\mathfrak M'$, $w',d'$  such that 
%$|W|=|D|$ and 
$\mathfrak M,w,d$ and $\mathfrak M',w',d'$ are bisimilar in \FOMS{}
(see, e.g.,~\cite[Prop.3.12]{GabEtAl03}).
%\end{itemize}
%
\end{proof}

\noindent
\begin{proofof}{Theorem~\ref{thm2intro}, lower bound}
For the equality- and substitution-free fragment of \FOT{}, 2\textsc{ExpTime}-hardness is now a straightforward consequence of Lemma~\ref{l:fot}. Indeed, we can simply use
the $\dag$-translations of the formulas used in Section~\ref{ssec:s5xs5low}.
%the proof of Theorem~\ref{thm:upperS5}~$(ii)$.
In order to prove 2\textsc{ExpTime}-hardness 
%Corollary~\ref{co:lowerbound} 
for the equality-free fragment, we need an additional step. Namely, 
we need to show that there are suitable bijections between the FO- and modal domains of each of the two (square) \FOMS-models constructed in the proof of Lemma~\ref{l:lbsound} such
that the resulting \FO-models are not only $\sigma$-bisimilar in the equality- and substitution-free fragment of \FOT{},
but are also $\sigma$-bisimilar in the equality-free fragment. In fact, we claim that they are $\sigma$-bisimilar in full
\FOT, and so 
%Corollary~\ref{co:lowerbound} 
the lower bound result of Theorem~\ref{thm2intro} generalises that of~\cite{DBLP:conf/lics/JungW21}.
To this end, take the $\sigma$-bisimulation $\bs$ in \FOMS{} between the  \FOMS-models $\mathfrak M=(W,D,I)$ and $\hat{\mathfrak M}=(\hat{W},\hat{D},\hat{I})$ defined in the proof of Lemma~\ref{l:lbsound}.
Now define a bijection $f\colon D\to W$ by taking $f(d_m^\treet)=w_m^\treet$ for all $m<2^n$, $\treet\in T$,
and a bijection $\hat{f}\colon \hat{D}\to \hat{W}$ by taking $\hat{f}(\hat{d}_k^\treet)=\hat{w}_k^\treet$ for all $k<2$, $\treet\in T$. 
Using that $(i)$ the respective restrictions of the FO-models $\mathfrak A_{\mathfrak M,f}$ to $D_m$ and $\mathfrak A_{\hat{\mathfrak M},\hat{f}}$ to $\hat{D}_k$ are $\sigma$-isomorphic for any $m<2^n$, $k<2$, and 
$(ii)$ for all $\pvar\in\sigma$, we have $\mathfrak A_{\mathfrak M,f}\not\models\pvar[y/d_m^\treet,x/d_{m'}^{\treet'}]$ if $m\ne m'$, and $\mathfrak A_{\hat{\mathfrak M},\hat{f}}\not\models\pvar[y/\hat{d}_k^\treet,x/\hat{d}_{k'}^{\treet'}]$
if $k\ne k'$,
it is straightforward to see that the relation
\[
\bs^{f,\hat{f}}= \bigl\{\bigl((a,b),(\hat{a},\hat{b})\bigr)\in (D\times D)\times(\hat{D}\times\hat{D}) \mid 
\bigl((f(a),b),\hat{f}(\hat{a}),\hat{b})\bigr)\in\bs\bigr\}
\]
is a $\sigma$-bisimulation between $\mathfrak A_{\mathfrak M,f}$ and $\mathfrak A_{\hat{\mathfrak M},\hat{f}}$
in \FOT.
\end{proofof}

\section{Proofs for Section~\ref{sec:ALCIO}}\label{s:sec6proofs}

Here, we give polynomial-time reductions of the interpolant existence problems modulo ontologies (introduced in Remark~\ref{rem:dl}) to the IEP. Call a concept $E$ with $\sig(E)\subseteq \sigma$ a \emph{$\sigma$-interpolant for a concept inclusion $C \sqsubseteq D$} if $\models C \sqsubseteq E$ and $\models E \sqsubseteq D$. The problem to decide whether a $\sigma$-interpolant exists for $C \sqsubseteq D$ can be reduced in polytime to interpolant existence, as the following two conditions are easily seen to be equivalent, for any $\sigma$ and concept inclusion $C \sqsubseteq D$:
\begin{itemize}
\item[--] there exists a $\sigma$-interpolant for $C \sqsubseteq D$;
\item[--] there exists an interpolant for $C'\sqsubseteq D'$, where $C',D'$ are obtained from $C,D$ by replacing in $C$ all symbols not in $\sigma$ by fresh symbols not in $C$ and $D$ and adding, for all $A\in \sigma$ and $R\in \sigma$, the conjuncts $A \sqcup \neg A$ and $\exists R.\top \sqcup \neg \exists R.\top$ to $C$ and $D$. 
\end{itemize}  
For an ontology $\mathcal{O}$, take the concept 
$$
\mathcal{O}^{c}=\bigwedge_{C\sqsubseteq D\in \mathcal{O}} \forall U.(\neg C \sqcup D).
$$
Recall that the IEP modulo ontologies is to decide, given an ontology $\mathcal{O}$, a signature $\sigma$, and a concept inclusion $C \sqsubseteq D$, whether there exists a $\sigma$-concept $E$ such that $\mathcal{O}\models C \sqsubseteq E$ and $\mathcal{O}\models E \sqsubseteq D$.  
For the reduction, assume $\mathcal{O}$, a signature $\sigma$, and a concept inclusion $C \sqsubseteq D$ are given.
Then the following conditions are equivalent:
\begin{itemize}
	\item[--] there exists a $\sigma$-interpolant for $\mathcal{O}^{c} \sqcap C \sqsubseteq \neg \mathcal{O}^{c} \sqcup D$;
	\item[--] there exists a $\sigma$-concept $E$ such that $\mathcal{O}\models C \sqsubseteq E$ and $\mathcal{O}\models E \sqsubseteq D$.
\end{itemize}		
Next, recall that ontology interpolant existence is to decide, given an ontology $\mathcal{O}$, a signature $\sigma$, and a concept inclusion $C \sqsubseteq D$, whether there is an ontology $\mathcal{O}'$ with $\sig(\mathcal{O}') \subseteq \sigma$, $\mathcal{O}\models \mathcal{O}'$, and $\mathcal{O}' \models C \sqsubseteq D$.  
For the reduction, assume $\mathcal{O}$, a signature $\sigma$, and a concept inclusion $C \sqsubseteq D$ are given.
Then the following conditions are equivalent:
\begin{itemize}
		\item[--] there exists a $\sigma$-interpolant for $\mathcal{O}^{c} \sqsubseteq \neg C \sqcup D$;
		\item[--] there exists an ontology $\mathcal{O}'$ with $\sig(\mathcal{O}') \subseteq \sigma$, $\mathcal{O}\models \mathcal{O}'$, and $\mathcal{O}' \models C \sqsubseteq D$.
\end{itemize}

Finally, we recall that the DEP modulo ontologies is to decide, given an ontology $\mathcal{O}$, a signature $\sigma$, and a concept name $A$, whether there exists a $\sigma$-concept $C$ such that $\mathcal{O}\models A \equiv C$. We reduce this problem to the IEP modulo ontologies.
Suppose an ontology $\mathcal{O}$, a signature $\sigma$, and a concept name $A$
are given. We may assume that $A\not\in\sigma$ since otherwise the problem is trivial.
Now let $\mathcal{O}'$ be the result of replacing in  $\mathcal{O}$ all symbols
$X$ that are not in $\sigma$ by fresh $X'$. Then the following conditions are equivalent:
\begin{itemize}
	\item[--] there exists a $\sigma$-concept $C$ such that $\mathcal{O}\cup \mathcal{O}'\models A \sqsubseteq C$ and $\mathcal{O}\cup \mathcal{O}'\models C \sqsubseteq A'$; 
	
	\item[--] there exists a $\sigma$-concept $C$ such that $\mathcal{O}\models A \equiv C$.
\end{itemize}

\end{appendix}

%***************

\bibliographystyle{alphaurl}
\bibliography{bibliobeth,local}

\newcommand{\etalchar}[1]{$^{#1}$}
\providecommand{\noopsort}[1]{}
\begin{thebibliography}{{\noopsort{Cate}{ten Cate}}JK{\etalchar{+}}26}

\bibitem[ABM03]{DBLP:journals/apal/ArecesBM03}
Carlos Areces, Patrick Blackburn, and Maarten Marx.
\newblock Repairing the interpolation theorem in quantified modal logic.
\newblock {\em Ann. Pure Appl. Log.}, 124(1-3):287--299, 2003.

\bibitem[AJM{\etalchar{+}}21]{DBLP:conf/aaai/ArtaleJMOW21}
Alessandro Artale, Jean~Christoph Jung, Andrea Mazzullo, Ana Ozaki, and Frank Wolter.
\newblock Living without {B}eth and {C}raig: Definitions and interpolants in description logics with nominals and role inclusions.
\newblock In {\em Proceedings of the 35th {AAAI} Conference on Artificial Intelligence, {AAAI} 2021}, pages 6193--6201. {AAAI} Press, 2021.

\bibitem[AJM{\etalchar{+}}23]{DBLP:journals/tocl/ArtaleJMOW23}
Alessandro Artale, Jean~Christoph Jung, Andrea Mazzullo, Ana Ozaki, and Frank Wolter.
\newblock Living without {B}eth and {C}raig: Definitions and interpolants in description and modal logics with nominals and role inclusions.
\newblock {\em {ACM} Trans. Comput. Log.}, 24(4):34:1--34:51, 2023.

\bibitem[AKK{\etalchar{+}}17]{DBLP:conf/time/ArtaleKKRWZ17}
Alessandro Artale, Roman Kontchakov, Alisa Kovtunova, Vladislav Ryzhikov, Frank Wolter, and Michael Zakharyaschev.
\newblock Ontology-mediated query answering over temporal data: {A} survey (invited talk).
\newblock In Sven Schewe, Thomas Schneider, and Jef Wijsen, editors, {\em 24th International Symposium on Temporal Representation and Reasoning, {TIME} 2017}, volume~90 of {\em LIPIcs}, pages 1:1--1:37. Schloss Dagstuhl - Leibniz-Zentrum f{\"{u}}r Informatik, 2017.

\bibitem[AMO24]{DBLP:journals/tocl/ArtaleMO24}
Alessandro Artale, Andrea Mazzullo, and Ana Ozaki.
\newblock First-order temporal logic on finite traces: Semantic properties, decidable fragments, and applications.
\newblock {\em {ACM} Trans. Comput. Log.}, 25(2):13:1--13:43, 2024.

\bibitem[{\'{A}}RS22]{DBLP:conf/semweb/AlvarezRS22}
Luc{\'{\i}}a~G{\'{o}}mez {\'{A}}lvarez, Sebastian Rudolph, and Hannes Strass.
\newblock How to agree to disagree - managing ontological perspectives using standpoint logic.
\newblock In Ulrike Sattler, Aidan Hogan, C.~Maria Keet, Valentina Presutti, Jo{\~{a}}o Paulo~A. Almeida, Hideaki Takeda, Pierre Monnin, Giuseppe Pirr{\`{o}}, and Claudia d'Amato, editors, {\em The Semantic Web - {ISWC} 2022 - 21st International Semantic Web Conference, Proceedings}, volume 13489 of {\em Lecture Notes in Computer Science}, pages 125--141. Springer, 2022.

\bibitem[BG07]{DBLP:books/el/07/BraunerG07}
Torben Bra{\"{u}}ner and Silvio Ghilardi.
\newblock First-order modal logic.
\newblock In Patrick Blackburn, Johan van Benthem, and Frank Wolter, editors, {\em Handbook of Modal Logic}, volume~3 of {\em Studies in logic and practical reasoning}, pages 549--620. Elsevier, 2007.

\bibitem[BHLS17]{DBLP:books/daglib/0041477}
Franz Baader, Ian Horrocks, Carsten Lutz, and Ulrike Sattler.
\newblock {\em An Introduction to Description Logic}.
\newblock Cambridge University Press, 2017.

\bibitem[BKL{\etalchar{+}}16]{DBLP:conf/rweb/BotoevaKLRWZ16}
Elena Botoeva, Boris Konev, Carsten Lutz, Vladislav Ryzhikov, Frank Wolter, and Michael Zakharyaschev.
\newblock Inseparability and conservative extensions of description logic ontologies: {A} survey.
\newblock In Jeff~Z. Pan, Diego Calvanese, Thomas Eiter, Ian Horrocks, Michael Kifer, Fangzhen Lin, and Yuting Zhao, editors, {\em Reasoning Web: Logical Foundation of Knowledge Graph Construction and Query Answering - 12th International Summer School 2016, Tutorial Lectures}, volume 9885 of {\em Lecture Notes in Computer Science}, pages 27--89. Springer, 2016.

\bibitem[BKM10]{DBLP:conf/cav/BasinKM10}
David~A. Basin, Felix Klaedtke, and Samuel M{\"{u}}ller.
\newblock Policy monitoring in first-order temporal logic.
\newblock In Tayssir Touili, Byron Cook, and Paul~B. Jackson, editors, {\em Computer Aided Verification, 22nd International Conference, {CAV} 2010, Proceedings}, volume 6174 of {\em Lecture Notes in Computer Science}, pages 1--18. Springer, 2010.

\bibitem[BL09]{DBLP:journals/ai/BelardinelliL09}
Francesco Belardinelli and Alessio Lomuscio.
\newblock Quantified epistemic logics for reasoning about knowledge in multi-agent systems.
\newblock {\em Artif. Intell.}, 173(9-10):982--1013, 2009.

\bibitem[BLR{\etalchar{+}}19]{DBLP:journals/ai/BotoevaLRWZ19}
Elena Botoeva, Carsten Lutz, Vladislav Ryzhikov, Frank Wolter, and Michael Zakharyaschev.
\newblock Query inseparability for {$\mathcal{ALC}$} ontologies.
\newblock {\em Artif. Intell.}, 272:1--51, 2019.

\bibitem[BLtCT16]{DBLP:series/synthesis/2016Benedikt}
Michael Benedikt, Julien Leblay, Balder ten Cate, and Efthymia Tsamoura.
\newblock {\em Generating Plans from Proofs: The Interpolation-based Approach to Query Reformulation}.
\newblock Synthesis Lectures on Data Management. Morgan {\&} Claypool Publishers, 2016.

\bibitem[BtCV16]{DBLP:journals/tocl/BenediktCB16}
Michael Benedikt, Balder ten Cate, and Michael {Vanden Boom}.
\newblock Effective interpolation and preservation in guarded logics.
\newblock {\em {ACM} Trans. Comput. Log.}, 17(2):8, 2016.

\bibitem[BvB99]{DBLP:journals/jsyml/BarwiseB99}
Jon Barwise and Johan van Benthem.
\newblock Interpolation, preservation, and pebble games.
\newblock {\em J. Symb. Log.}, 64(2):881--903, 1999.

\bibitem[{\noopsort{Cate}{ten Cate}}JK{\etalchar{+}}26]{taci}
Balder {\noopsort{Cate}{ten Cate}}, Jean~Christoph Jung, Patrick Koopmann, Christoph Wernhard, and Frank Wolter, editors.
\newblock {\em Theory and Applications of Craig Interpolation}.
\newblock Ubiquity Press, 2026.
\newblock To appear, preprints accessible from \url{https://cibd.bitbucket.io/taci/}.

\bibitem[CK98]{modeltheory}
{Chen Chung} Chang and H.~Jerome Keisler.
\newblock {\em Model Theory}.
\newblock Elsevier, 1998.

\bibitem[CKS81]{chandraAlternation1981}
Ashok~K. Chandra, Dexter~C. Kozen, and Larry~J. Stockmeyer.
\newblock Alternation.
\newblock {\em J. ACM}, 28:114--133, 1981.

\bibitem[CT18]{DBLP:reference/db/ChomickiT18a}
Jan Chomicki and David Toman.
\newblock Temporal logic in database query languages.
\newblock In {\em Encyclopedia of Database Systems, Second Edition}. Springer, 2018.

\bibitem[DFK06]{DBLP:journals/tocl/DegtyarevFK06}
Anatoli Degtyarev, Michael Fisher, and Boris Konev.
\newblock Monodic temporal resolution.
\newblock {\em {ACM} Trans. Comput. Log.}, 7(1):108--150, 2006.

\bibitem[DL06]{DBLP:journals/japll/DAgostinoL06}
Giovanna D'Agostino and Giacomo Lenzi.
\newblock On modal mu-calculus with explicit interpolants.
\newblock {\em J. Appl. Log.}, 4(3):256--278, 2006.

\bibitem[DLN{\etalchar{+}}98]{DBLP:journals/ai/DoniniLNNS98}
Francesco~M. Donini, Maurizio Lenzerini, Daniele Nardi, Werner Nutt, and Andrea Schaerf.
\newblock An epistemic operator for description logics.
\newblock {\em Artif. Intell.}, 100(1-2):225--274, 1998.

\bibitem[EK19]{DBLP:journals/ki/EiterK19}
Thomas Eiter and Gabriele Kern{-}Isberner.
\newblock A brief survey on forgetting from a knowledge representation and reasoning perspective.
\newblock {\em K{\"{u}}nstliche Intell.}, 33(1):9--33, 2019.

\bibitem[Fin79]{DBLP:journals/jsyml/Fine79}
Kit Fine.
\newblock Failures of the interpolation lemma in quantified modal logic.
\newblock {\em J. Symb. Log.}, 44(2):201--206, 1979.

\bibitem[Fit96]{DBLP:books/daglib/0082098}
Melvin Fitting.
\newblock {\em First-Order Logic and Automated Theorem Proving, Second Edition}.
\newblock Graduate Texts in Computer Science. Springer, 1996.

\bibitem[Fit02]{DBLP:journals/jsyml/Fitting02}
Melvin Fitting.
\newblock Interpolation for first order {S5}.
\newblock {\em J. Symb. Log.}, 67(2):621--634, 2002.

\bibitem[FKN13]{FraEtAl13}
Enrico Franconi, Volha Kerhet, and Nhung Ngo.
\newblock Exact query reformulation over databases with first-order and description logics ontologies.
\newblock {\em J. Artif. Intell. Res.}, 48:885--922, 2013.

\bibitem[FKW22]{DBLP:journals/corr/abs-2202-07186}
Marie Fortin, Boris Konev, and Frank Wolter.
\newblock Interpolants and explicit definitions in extensions of the description logic $\mathcal{EL}$.
\newblock In {\em Proceedings of the 19th International Conference on Principles of Knowledge Representation and Reasoning, {KR} 2022}, 2022.

\bibitem[FM12]{FitMen12}
Melvin Fitting and Richard~L. Mendelsohn.
\newblock {\em First-order {M}odal {L}ogic}.
\newblock Springer Science \& Business Media, 2012.

\bibitem[Fre06]{French06}
Tim French.
\newblock {\em Bisimulation quantifiers for modal logics}.
\newblock {PhD} thesis, The University of Western Australia, 2006.

\bibitem[Fus26]{chapter:nonclassical}
Wesley Fussner.
\newblock Interpolation in non-classical logics.
\newblock In {\noopsort{Cate}{ten Cate}} et~al. \cite{taci}.
\newblock To appear, preprints accessible from \url{https://cibd.bitbucket.io/taci/}.

\bibitem[GHKS08]{GrauHKS08}
Bernardo~Cuenca Grau, Ian Horrocks, Yevgeny Kazakov, and Ulrike Sattler.
\newblock Modular reuse of ontologies: Theory and practice.
\newblock {\em J.\ of Artifical Intelligence Research}, 31:273--318, 2008.

\bibitem[GJL15]{DBLP:journals/tocl/GollerJL15}
Stefan G{\"{o}}ller, Jean~Christoph Jung, and Markus Lohrey.
\newblock The complexity of decomposing modal and first-order theories.
\newblock {\em {ACM} Trans. Comput. Log.}, 16(1):9:1--9:43, 2015.

\bibitem[GKL23]{DBLP:journals/tplp/GoncalvesKL23}
Ricardo Gon{\c{c}}alves, Matthias Knorr, and Jo{\~{a}}o Leite.
\newblock Forgetting in answer set programming - {A} survey.
\newblock {\em Theory Pract. Log. Program.}, 23(1):111--156, 2023.

\bibitem[GKV97]{DBLP:journals/bsl/GradelKV97}
Erich Gr{\"{a}}del, Phokion~G. Kolaitis, and Moshe~Y. Vardi.
\newblock On the decision problem for two-variable first-order logic.
\newblock {\em Bulletin of Symbolic Logic}, 3(1):53--69, 1997.

\bibitem[GKWZ03]{GabEtAl03}
Dov~M. Gabbay, Agi Kurucz, Frank Wolter, and Michael Zakharyaschev.
\newblock {\em Many-dimensional Modal Logics: Theory and Applications}, volume 148 of {\em Studies in Logic and The Foundations of Mathematics}.
\newblock Elsevier, 2003.

\bibitem[GLW06]{DBLP:conf/kr/GhilardiLW06}
Silvio Ghilardi, Carsten Lutz, and Frank Wolter.
\newblock Did {I} damage my ontology? {A} case for conservative extensions in description logics.
\newblock In Patrick Doherty, John Mylopoulos, and Christopher~A. Welty, editors, {\em Proceedings of the 10th International Conference on Principles of Knowledge Representation and Reasoning, KR 2006}, pages 187--197. {AAAI} Press, 2006.

\bibitem[GLWZ06]{DBLP:conf/aiml/GhilardiLWZ06}
Silvio Ghilardi, Carsten Lutz, Frank Wolter, and Michael Zakharyaschev.
\newblock Conservative extensions in modal logic.
\newblock In Guido Governatori, Ian~M. Hodkinson, and Yde Venema, editors, {\em Advances in Modal Logic, Volume 6}, pages 187--207. College Publications, 2006.

\bibitem[GM05]{MGabbay2005-MGAIAD}
Dov~M. Gabbay and Larisa Maksimova.
\newblock {\em Interpolation and Definability: Modal and Intuitionistic Logics}.
\newblock Oxford University Press, 2005.

\bibitem[GO07]{goranko20075}
Valentin Goranko and Martin Otto.
\newblock Model theory of modal logic.
\newblock In Patrick Blackburn, Johan van Benthem, and Frank Wolter, editors, {\em Handbook of Modal Logic}, volume~3 of {\em Studies in logic and practical reasoning}, pages 249--329. Elsevier, 2007.

\bibitem[GPT16]{DBLP:conf/ekaw/GeletaPT16}
David Geleta, Terry~R. Payne, and Valentina A.~M. Tamma.
\newblock An investigation of definability in ontology alignment.
\newblock In Eva Blomqvist, Paolo Ciancarini, Francesco Poggi, and Fabio Vitali, editors, {\em Knowledge Engineering and Knowledge Management - 20th International Conference, {EKAW} 2016, Proceedings}, volume 10024 of {\em Lecture Notes in Computer Science}, pages 255--271, 2016.

\bibitem[GZ95]{DBLP:journals/sLogica/GhilardiZ95}
Silvio Ghilardi and Marek~W. Zawadowski.
\newblock Undefinability of propositional quantifiers in the modal system {S4}.
\newblock {\em Stud Logica}, 55(2):259--271, 1995.

\bibitem[Hen88]{henkell1}
Karsten Henckell.
\newblock Pointlike sets: the finest aperiodic cover of a finite semigroup.
\newblock {\em J. Pure Appl. Algebra}, 55(1-2):85--126, 1988.

\bibitem[HK15]{DBLP:journals/tocl/HampsonK15}
Christopher Hampson and Agi Kurucz.
\newblock Undecidable propositional bimodal logics and one-variable first-order linear temporal logics with counting.
\newblock {\em {ACM} Trans. Comput. Log.}, 16(3):27:1--27:36, 2015.

\bibitem[HKK{\etalchar{+}}03]{DBLP:conf/time/HodkinsonKKWZ03}
Ian~M. Hodkinson, Roman Kontchakov, Agi Kurucz, Frank Wolter, and Michael Zakharyaschev.
\newblock On the computational complexity of decidable fragments of first-order linear temporal logics.
\newblock In {\em 10th International Symposium on Temporal Representation and Reasoning / 4th International Conference on Temporal Logic {(TIME-ICTL} 2003)}, pages 91--98. {IEEE} Computer Society, 2003.

\bibitem[HP18]{DBLP:conf/rv/HavelundP18}
Klaus Havelund and Doron Peled.
\newblock Runtime verification: From propositional to first-order temporal logic.
\newblock In {\em Runtime Verification - 18th International Conference, {RV} 2018, Proceedings}, volume 1123 of {\em Lecture Notes in Computer Science}, pages 90--112. Springer, 2018.

\bibitem[HRS10]{henkell2}
Karsten Henckell, John Rhodes, and Benjamin Steinberg.
\newblock Aperiodic pointlikes and beyond.
\newblock {\em Internat. J. Algebra Comput.}, 20(2):287--305, 2010.

\bibitem[HWZ00]{DBLP:journals/apal/HodkinsonWZ00}
Ian~M. Hodkinson, Frank Wolter, and Michael Zakharyaschev.
\newblock Decidable fragment of first-order temporal logics.
\newblock {\em Ann. Pure Appl. Log.}, 106(1-3):85--134, 2000.

\bibitem[JLM{\etalchar{+}}17]{DBLP:conf/icalp/JungLM0W17}
Jean~Christoph Jung, Carsten Lutz, Mauricio Martel, Thomas Schneider, and Frank Wolter.
\newblock Conservative extensions in guarded and two-variable fragments.
\newblock In Ioannis Chatzigiannakis, Piotr Indyk, Fabian Kuhn, and Anca Muscholl, editors, {\em 44th International Colloquium on Automata, Languages, and Programming, {ICALP} 2017}, volume~80 of {\em LIPIcs}, pages 108:1--108:14. Schloss Dagstuhl - Leibniz-Zentrum f{\"{u}}r Informatik, 2017.

\bibitem[JLM22]{DBLP:conf/kr/JungLM22}
Jean~Christoph Jung, Carsten Lutz, and Jerzy Marcinkowski.
\newblock Conservative extensions for existential rules.
\newblock In Gabriele Kern{-}Isberner, Gerhard Lakemeyer, and Thomas Meyer, editors, {\em Proceedings of the 19th International Conference on Principles of Knowledge Representation and Reasoning, {KR} 2022}, pages 195--204, 2022.

\bibitem[JLPW22]{DBLP:journals/ai/JungLPW22}
Jean~Christoph Jung, Carsten Lutz, Hadrien Pulcini, and Frank Wolter.
\newblock Logical separability of labeled data examples under ontologies.
\newblock {\em Artif. Intell.}, 313:103785, 2022.
\newblock URL: \url{https://doi.org/10.1016/j.artint.2022.103785}, \href {https://doi.org/10.1016/J.ARTINT.2022.103785} {\path{doi:10.1016/J.ARTINT.2022.103785}}.

\bibitem[JW21]{DBLP:conf/lics/JungW21}
Jean~Christoph Jung and Frank Wolter.
\newblock Living without {B}eth and {C}raig: Definitions and interpolants in the guarded and two-variable fragments.
\newblock In {\em Proceedings of the 36th Annual {ACM/IEEE} Symposium on Logic in Computer Science, {LICS} 2021}, pages 1--14. {IEEE}, 2021.

\bibitem[KKWZ07]{DBLP:reference/spatial/KontchakovKWZ07}
Roman Kontchakov, Agi Kurucz, Frank Wolter, and Michael Zakharyaschev.
\newblock Spatial logic + temporal logic = ?
\newblock In Marco Aiello, Ian Pratt{-}Hartmann, and Johan van Benthem, editors, {\em Handbook of Spatial Logics}, pages 497--564. Springer, 2007.

\bibitem[KLWW09]{DBLP:series/lncs/KonevLWW09}
Boris Konev, Carsten Lutz, Dirk Walther, and Frank Wolter.
\newblock Formal properties of modularisation.
\newblock In Heiner Stuckenschmidt, Christine Parent, and Stefano Spaccapietra, editors, {\em Modular Ontologies: Concepts, Theories and Techniques for Knowledge Modularization}, volume 5445 of {\em Lecture Notes in Computer Science}, pages 25--66. Springer, 2009.

\bibitem[Koo20]{DBLP:journals/ki/Koopmann20}
Patrick Koopmann.
\newblock {LETHE:} forgetting and uniform interpolation for expressive description logics.
\newblock {\em K{\"{u}}nstliche Intell.}, 34(3):381--387, 2020.

\bibitem[Kr{\"{o}}87]{DBLP:series/eatcs/Kroger87}
Fred Kr{\"{o}}ger.
\newblock {\em Temporal Logic of Programs}, volume~8 of {\em {EATCS} Monographs on Theoretical Computer Science}.
\newblock Springer, 1987.

\bibitem[KS15]{DBLP:conf/aaai/KoopmannS15}
Patrick Koopmann and Renate~A. Schmidt.
\newblock Uniform interpolation and forgetting for $\mathcal{ALC}$ ontologies with aboxes.
\newblock In {\em Proceedings of the 29th {AAAI} Conference on Artificial Intelligence, {AAAI} 2015}, pages 175--181. {AAAI} Press, 2015.

\bibitem[Kur07]{DBLP:books/el/07/Kurucz07}
Agi Kurucz.
\newblock Combining modal logics.
\newblock In Patrick Blackburn, Johan van Benthem, and Frank Wolter, editors, {\em Handbook of Modal Logic}, volume~3 of {\em Studies in logic and practical reasoning}, pages 869--924. Elsevier, 2007.

\bibitem[Kuz18]{DBLP:journals/apal/Kuznets18}
Roman Kuznets.
\newblock Multicomponent proof-theoretic method for proving interpolation properties.
\newblock {\em Ann. Pure Appl. Log.}, 169(12):1369--1418, 2018.

\bibitem[KV17]{DBLP:conf/lpar/KovacsV17}
Laura Kov{\'{a}}cs and Andrei Voronkov.
\newblock First-order interpolation and interpolating proof systems.
\newblock In Thomas Eiter and David Sands, editors, {\em LPAR-21, 21st International Conference on Logic for Programming, Artificial Intelligence and Reasoning}, volume~46 of {\em EPiC Series in Computing}, pages 49--64. EasyChair, 2017.

\bibitem[KWW09]{DBLP:conf/ijcai/KonevWW09}
Boris Konev, Dirk Walther, and Frank Wolter.
\newblock Forgetting and uniform interpolation in large-scale description logic terminologies.
\newblock In {\em Proceedings of the 21st International Joint Conference on Artificial Intelligence, {IJCAI} 2009}, pages 830--835, 2009.

\bibitem[KWZ10]{AIJ10}
Roman Kontchakov, Frank Wolter, and Michael Zakharyaschev.
\newblock Logic-based ontology comparison and module extraction with an application to {D}{L}-{L}ite.
\newblock To appear in Journal of Artificial Intelligence, 2010.

\bibitem[KWZ23a]{DBLP:conf/kr/KuruczWZ23}
Agi Kurucz, Frank Wolter, and Michael Zakharyaschev.
\newblock Definitions and (uniform) interpolants in first-order modal logic.
\newblock In Pierre Marquis, Tran~Cao Son, and Gabriele Kern{-}Isberner, editors, {\em Proceedings of the 20th International Conference on Principles of Knowledge Representation and Reasoning, {KR} 2023}, pages 417--428, 2023.

\bibitem[KWZ23b]{DBLP:journals/corr/abs-2312-05929}
Agi Kurucz, Frank Wolter, and Michael Zakharyaschev.
\newblock A non-uniform view of {C}raig interpolation in modal logics with linear frames.
\newblock {\em CoRR}, abs/2312.05929, 2023.

\bibitem[KWZ24]{DBLP:journals/corr/abs-2403-11255}
Agi Kurucz, Frank Wolter, and Michael Zakharyaschev.
\newblock The interpolant existence problem for weak {K4} and difference logic.
\newblock {\em CoRR}, abs/2403.11255, 2024.

\bibitem[KWZ26]{chapter:separation}
Agi Kurucz, Frank Wolter, and Michael Zakharyaschev.
\newblock From interpolating formulas to separating languages and back again.
\newblock In {\noopsort{Cate}{ten Cate}} et~al. \cite{taci}.
\newblock To appear, preprints accessible from \url{https://cibd.bitbucket.io/taci/}.

\bibitem[LPRW22]{liu_et_al:LIPIcs.MFCS.2022.70}
Mo~Liu, Anantha Padmanabha, R.~Ramanujam, and Yanjing Wang.
\newblock {Generalized Bundled Fragments for First-Order Modal Logic}.
\newblock In Stefan Szeider, Robert Ganian, and Alexandra Silva, editors, {\em 47th International Symposium on Mathematical Foundations of Computer Science (MFCS 2022)}, volume 241 of {\em Leibniz International Proceedings in Informatics (LIPIcs)}, pages 70:1--70:14. Schloss Dagstuhl -- Leibniz-Zentrum f{\"u}r Informatik, 2022.

\bibitem[LPRW23]{DBLP:journals/iandc/LiuPRW23}
Mo~Liu, Anantha Padmanabha, R.~Ramanujam, and Yanjing Wang.
\newblock Are bundles good deals for first-order modal logic?
\newblock {\em Inf. Comput.}, 293:105062, 2023.

\bibitem[LR94]{Lin94a}
Fangzhen Lin and Ray Reiter.
\newblock Forget it!
\newblock In {\em Proceedings of AAAI Fall Symposium on Relevance}, 1994.

\bibitem[LSW12]{DBLP:conf/kr/LutzSW12}
Carsten Lutz, Inan{\c{c}} Seylan, and Frank Wolter.
\newblock An automata-theoretic approach to uniform interpolation and approximation in the description logic {$\mathcal{EL}$}.
\newblock In {\em Proceedings of the 13th International Conference on Principles of Knowledge Representation and Reasoning, {KR} 2012}. {AAAI} Press, 2012.

\bibitem[LSWZ01]{DBLP:conf/cade/LutzSWZ01}
Carsten Lutz, Holger Sturm, Frank Wolter, and Michael Zakharyaschev.
\newblock Tableaux for temporal description logic with constant domains.
\newblock In Rajeev Gor{\'{e}}, Alexander Leitsch, and Tobias Nipkow, editors, {\em Automated Reasoning, First International Joint Conference, {IJCAR} 2001, Proceedings}, volume 2083 of {\em Lecture Notes in Computer Science}, pages 121--136. Springer, 2001.

\bibitem[LSWZ02]{DBLP:journals/sLogica/LutzSWZ02}
Carsten Lutz, Holger Sturm, Frank Wolter, and Michael Zakharyaschev.
\newblock A tableau decision algorithm for modalized {$\mathcal{ALC}$} with constant domains.
\newblock {\em Stud Logica}, 72(2):199--232, 2002.

\bibitem[LW11]{DBLP:conf/ijcai/LutzW11}
Carsten Lutz and Frank Wolter.
\newblock Foundations for uniform interpolation and forgetting in expressive description logics.
\newblock In {\em Proceedings of the 22nd International Joint Conference on Artificial Intelligence, {IJCAI} 2011}, pages 989--995. {IJCAI/AAAI}, 2011.

\bibitem[LWZ08]{DBLP:conf/time/LutzWZ08}
Carsten Lutz, Frank Wolter, and Michael Zakharyaschev.
\newblock Temporal description logics: {A} survey.
\newblock In St{\'{e}}phane Demri and Christian~S. Jensen, editors, {\em 15th International Symposium on Temporal Representation and Reasoning, {TIME} 2008}, pages 3--14. {IEEE} Computer Society, 2008.

\bibitem[MA98]{DBLP:journals/ndjfl/MarxA98}
Maarten Marx and Carlos Areces.
\newblock Failure of interpolation in combined modal logics.
\newblock {\em Notre Dame J. Formal Log.}, 39(2):253--273, 1998.

\bibitem[Mar99]{DBLP:journals/logcom/Marx99}
Maarten Marx.
\newblock Complexity of products of modal logics.
\newblock {\em J. Log. Comput.}, 9(2):197--214, 1999.

\bibitem[McM18]{DBLP:reference/mc/McMillan18}
Kenneth~L. McMillan.
\newblock Interpolation and model checking.
\newblock In Edmund~M. Clarke, Thomas~A. Henzinger, Helmut Veith, and Roderick Bloem, editors, {\em Handbook of Model Checking}, pages 421--446. Springer, 2018.

\bibitem[NR14]{DBLP:journals/ai/NikitinaR14}
Nadeschda Nikitina and Sebastian Rudolph.
\newblock ({N}on-)succinctness of uniform interpolants of general terminologies in the description logic $\mathcal{EL}$.
\newblock {\em Artif. Intell.}, 215:120--140, 2014.

\bibitem[PZ16]{DBLP:journals/corr/PlaceZ14}
Thomas Place and Marc Zeitoun.
\newblock Separating regular languages with first-order logic.
\newblock {\em Log. Methods Comput. Sci.}, 12(1), 2016.

\bibitem[tC22]{baldernote}
Balder ten Cate.
\newblock Lyndon interpolation for modal logic via type elimination sequences.
\newblock Technical report, ILLC, Amsterdam, 01 2022.

\bibitem[tCCMV06]{TenEtAl06}
Balder ten Cate, Willem Conradie, Maarten Marx, and Yde Venema.
\newblock Definitorially complete description logics.
\newblock In {\em Proceedings of the 10th International Conference on Principles of Knowledge Representation and Reasoning, {KR} 2006}, pages 79--89. {AAAI} Press, 2006.

\bibitem[TW11]{DBLP:series/synthesis/2011Toman}
David Toman and Grant~E. Weddell.
\newblock {\em Fundamentals of Physical Design and Query Compilation}.
\newblock Synthesis Lectures on Data Management. Morgan {\&} Claypool Publishers, 2011.

\bibitem[TW21]{TomWed21}
David Toman and Grant~E. Weddell.
\newblock {{FO} Rewritability for {OMQ} using Beth Definability and Interpolation}.
\newblock In {\em Proceedings of the 34th International Workshop on Description Logics, {DL} 2021}. CEUR-WS.org, 2021.

\bibitem[vEB97]{EmdeBoas97}
Peter van Emde~Boas.
\newblock The convenience of tilings.
\newblock In A.~Sorbi, editor, {\em Complexity, Logic and Recursion Theory}, volume 187 of {\em Lecture Notes in Pure and Applied Mathematics}, pages 331--363. Marcel Dekker Inc., 1997.

\bibitem[Vis96]{Visser96}
Albert Visser.
\newblock Uniform interpolation and layered bisimulation.
\newblock In {\em G\"{o}del'96: Logical foundations of mathematics, computer science and physics--Kurt G\"{o}del's legacy}, pages 139--164. Association for Symbolic Logic, 1996.

\bibitem[Waj33]{Wajsberg33}
Mordchaj Wajsberg.
\newblock Ein erweiterter {K}lassenkalk{\"u}l.
\newblock {\em Monatsh Math. Phys.}, 40:113--126, 1933.

\bibitem[Wan17]{DBLP:journals/corr/Wang17d}
Yanjing Wang.
\newblock A new modal framework for epistemic logic.
\newblock In J{\'{e}}r{\^{o}}me Lang, editor, {\em Proceedings of the 16th Conference on Theoretical Aspects of Rationality and Knowledge, {TARK} 2017}, volume 251 of {\em {EPTCS}}, pages 515--534, 2017.

\bibitem[Wer21]{DBLP:journals/jar/Wernhard21}
Christoph Wernhard.
\newblock Craig interpolation with clausal first-order tableaux.
\newblock {\em J. Autom. Reason.}, 65(5):647--690, 2021.

\bibitem[WS17]{DBLP:conf/ijcai/WildS17}
Paul Wild and Lutz Schr{\"{o}}der.
\newblock A characterization theorem for a modal description logic.
\newblock In Carles Sierra, editor, {\em Proceedings of the 26th International Joint Conference on Artificial Intelligence, {IJCAI} 2017}, pages 1304--1310. ijcai.org, 2017.

\bibitem[WWS22]{DBLP:journals/apal/WangWS22}
Yanjing Wang, Yu~Wei, and Jeremy Seligman.
\newblock Quantifier-free epistemic term-modal logic with assignment operator.
\newblock {\em Ann. Pure Appl. Log.}, 173(3):103071, 2022.

\bibitem[WZ01]{DBLP:journals/jsyml/WolterZ01}
Frank Wolter and Michael Zakharyaschev.
\newblock Decidable fragments of first-order modal logics.
\newblock {\em J. Symb. Log.}, 66(3):1415--1438, 2001.

\bibitem[WZ24]{DL24:undecidability}
Frank Wolter and Michael Zakharyaschev.
\newblock Interpolant existence is undecidable for two-variable first-order logic with two equivalence relations.
\newblock In {\em Proceedings of the 37th International Workshop on Description Logics, DL 2024}, {CEUR} Workshop Proceedings. CEUR-WS.org, 2024.

\bibitem[ZFA{\etalchar{+}}18]{DBLP:conf/dlog/ZhaoFADS18}
Yizheng Zhao, Hao Feng, Ruba Alassaf, Warren Del{-}Pinto, and Renate~A. Schmidt.
\newblock The {FAME} family: {A} family of reasoning tools for forgetting in expressive description logics.
\newblock In Magdalena Ortiz and Thomas Schneider, editors, {\em Proceedings of the 31st International Workshop on Description Logics co-located with 16th International Conference on Principles of Knowledge Representation and Reasoning, {KR} 2018}, volume 2211 of {\em {CEUR} Workshop Proceedings}. CEUR-WS.org, 2018.

\bibitem[ZS16]{DBLP:conf/ijcai/ZhaoS16}
Yizheng Zhao and Renate~A. Schmidt.
\newblock Forgetting concept and role symbols in $\mathcal{ALCOIH}$-ontologies.
\newblock In {\em Proceedings of the 25th International Joint Conference on Artificial Intelligence, {IJCAI} 2016}, pages 1345--1353. {IJCAI/AAAI} Press, 2016.

\end{thebibliography}

\end{document}